\documentclass[11pt,a4paper]{article}
\usepackage[]{graphicx}
\usepackage{epsfig,amssymb,amsmath,psfrag,appendix}
\begin{document}


\def\BF{\bf \large}
\def\a{\alpha}
 \def\b{\beta}
\def\C{\Gamma}
 \def\d{\delta}\def\D{\Delta}
 \def\e{\epsilon}\def\vare{\varepsilon}
 \def\f{\phi}\def\F{\Phi}\def\vf{\varphi}
 \def\h{\eta}\def\z{\zeta}
 \def\k{\kappa}
 \def\l{\lambda}\def\L{\Lambda}
 \def\m{\mu}
 \def\n{\nu}
 \def\P{\Pi}
 \def\r{\rho}
 \def\s{\sigma}\def\S{\Sigma}
 \def\t{\tau}
 \def\th{\theta}\def\Th{\Theta}\def\vth{\vartheta}
 \def\X{\Xeta}
 \def\adt{\dot \alpha}
 \def\bdt{\dot \beta}
 \def\cdt{\dot\gamma}
 \def\ddt{\dot\delta}
 \def\edt{\dot\epsilon}
 \def\hdt{\dot\eta}
 \def\del{\partial}
 \def\delb{\bar\partial}
\def \matrix #1 {\left(\begin{array}{cc} #1 \end{array}\right)}
\def \Tr {\mathop{\rm Tr}\nolimits}
\def \tr {\mathop{\rm tr}\nolimits}
\def \Im {\mathop{\rm Im}\nolimits}
\def \Re {\mathop{\rm Re}\nolimits}
\def \res{\mathop{\rm res}\nolimits}
\def \e  {\mathop{\rm e}\nolimits}
\newcommand\lr[1]{{\left({#1}\right)}}
\newcommand \widebar [1] {\overline{#1}}
\newcommand\bin[2]{\left({#1}\atop{#2}\right)}
\newcommand \vev [1] {\langle{#1}\rangle}
\newcommand \VEV [1] {\left\langle{#1}\right\rangle}
\newcommand \ket [1] {|{#1}\rangle}
\newcommand \bra [1] {\langle {#1}|}

\newcommand{\ft}[2]{{\textstyle\frac{#1}{#2}}}


\newcommand{\al}{\alpha}
\newcommand{\Be}{\mbox{B}}
\newcommand{\gm}{\gamma}
\newcommand{\Gm}{\Gamma}
\newcommand{\dl}{\delta}
\newcommand{\Dl}{\Delta}
\newcommand{\eps}{\epsilon}
\newcommand{\ep}{\epsilon}
\newcommand{\kp}{\kappa}
\newcommand{\ka}{\kappa}
\newcommand{\lm}{\lambda}
\newcommand{\Lm}{\Lambda}
\newcommand{\om}{\omega}
\newcommand{\pr}{\partial}
\newcommand{\pa}{\partial}
\newcommand{\dd}{\mbox{d}}
\newcommand{\dr}{{\rm d}}
\newcommand{\la}{\langle}
\newcommand{\ra}{\rangle}
\newcommand{\MS}{\mbox{MS}}
\newcommand{\nn}{\nonumber}
 \def\bea{\begin{eqnarray}}\def\eea{\end{eqnarray}}
\def \be  {\begin{equation}}
\def \ee  {\end{equation}}
\def \ba  {\begin{eqnarray}}
\def \ea  {\end{eqnarray}}
\def \pa  {\partial}
\def \baa {\begin{eqnarray*}}
\def \eaa {\end{eqnarray*}}

\newcommand \ci [1] {\cite{#1}}
\newcommand \bi [1] {\bibitem{#1}}
\newcommand\re[1]{(\ref{#1})}

\newcommand \LN [1] {\ln\left( #1 \right)}
\newcommand {\LNP}[2] {{\ln^{#2}\left( #1 \right)}}
\newcommand {\PLN}[2] {{\rm Li}_{#1}\left( #2 \right)}


\newcommand\fr[1]{\fbox{ ${#1}$ }}
\def \vep {\epsilon}
\newcommand{\p}[1]{(\ref{#1})}
\newcommand{\cN}{{\cal N}}


\renewcommand{\title}[1]{\vbox{\center\LARGE{#1}}\vspace{5mm}}
\renewcommand{\author}[1]{\vbox{\center#1}\vspace{5mm}}
\newcommand{\address}[1]{\vbox{\center\em#1}}
\newcommand{\email}[1]{\vbox{\center\tt#1}\vspace{5mm}}

\begin{titlepage}

\begin{center}
\vspace{5mm}
\hfill {LAPTH-THESE-1276/08}\\
\center{{\LARGE Duality between Wilson loops and gluon amplitudes}\footnote{Based on the author's PhD thesis at the university Lyon 1 (France) and prepared at LAPTH, Annecy-le-Vieux (France).}\vspace{5mm}}

\author{Johannes Henn}
\email{henn@physik.hu-berlin.de} 
\address{Humboldt-Universit\"at zu Berlin, Institut f\"ur Physik, Newtonstr. 15, 12489 Berlin, Germany}
\end{center}

\begin{abstract}
  An intriguing new duality between
planar MHV gluon amplitudes and light-like Wilson loops in
${\mathcal N}=4$ super Yang-Mills is investigated.
We extend previous checks of the duality by performing
a two-loop calculation of the rectangular and pentagonal
Wilson loop.
Furthermore, we derive an all-order broken conformal
Ward identity for the Wilson loops and analyse its consequences.
Starting from six points, the Ward identity allows for an
arbitrary function of conformal invariants to appear in the expression for the Wilson loop. We compute
this function at six points and two loops and discuss its
implications for the corresponding gluon amplitude.
It is found that the duality disagrees with a conjecture
for the gluon amplitudes by Bern {\it et al}.
A recent calculation by Bern {\it et al} indeed shows
that the latter conjecture breaks down at six gluons and
at two loops. By doing a numerical comparison with their
results we find that the duality between gluon amplitudes
and Wilson loops is preserved.
This review is based on the author's PhD thesis and includes developments until May 2008.
\end{abstract}

\setcounter{tocdepth}{2}
\tableofcontents  
\thispagestyle{empty}

\end{titlepage}


\newpage

{N° d'ordre 139-2008 \hfill Ann\'ee 2008}
\vspace{2cm}
\begin{center}
\center{THESE}
\center{pr\'esent\'ee}
\center{devant l'UNIVERSITE CLAUDE BERNARD - LYON 1}
\center{pour l'obtention}
\center{du DIPLOME DE DOCTORAT}
\center{(arr\^{e}t\'e du 7 ao\^{u}t 2006)}
\vspace{1cm}
\center{pr\'esent\'ee et soutenue publiquement le 29.09.2008}
\vspace{1cm}
\center{par}
\vspace{1cm}
\center{\Large \bf Johannes HENN}
\vspace{1cm}
\center{\bf \Huge Dualit\'e entre boucles de Wilson}
\center{\bf \Huge  et amplitudes de gluons}
\vspace{2cm}
\center{Directeur de th\`{e}se~: Prof. Emery SOKATCHEV}
\\
\vspace{2cm}
\hspace{-1cm}
\begin{tabular}{llll}
  JURY~: & Prof. Costas BACHAS, &Ecole Normale Sup\'{e}rieure, & Rapporteur  \\
   & Prof. Fran\c{c}ois GIERES,& Universit\'e Lyon 1, & \\
   & Prof. Jan PLEFKA, &Humboldt Universit\"at Berlin,& Rapporteur\\
   & Prof. Emery SOKATCHEV, &Universit\'e de Savoie & \\
   & Prof. Kellogg STELLE, &Imperial College London &
\end{tabular}

\end{center}
\thispagestyle{empty}

\newpage

\setcounter{page}{1}

\section{Introduction}

In the standard model of elementary particles, the strong
interactions are described by quantum chromodynamics (QCD). It is
a non-Abelian Yang-Mills gauge theory, with the quarks being
in the fundamental representation of the gauge group SU(3).
In contrast to quantum electrodynamics, the non-Abelian nature of
the gauge group allows the gauge bosons to interact with each
other. It is this property which leads to asymptotic freedom and
the confinement of quarks. Despite the simplicity and elegance of
the QCD Lagrangian, many open problems remain, such as
understanding the transition from the high-energy region (short
distances), where perturbation theory is valid, to the low-energy
region (long distances) of confined quarks.\\

`t Hooft proposed to consider Yang-Mills theories with gauge group
$SU(N)$, for $N$ very large, while keeping the `t Hooft coupling
$g^2 N$ (with $g$ being the Yang-Mills coupling constant) fixed
\cite{tHooft:1973jz}. In this way, one may hope to see
simplifications for $N$ large, and eventually get insight into QCD
by computing terms in an $1/N$ expansion. The latter resembles
that of the genus expansion of string theory. For $N \rightarrow
\infty$, non-planar diagrams are suppressed, which is why the `t
Hooft limit is also referred to as the planar limit. Hints that
simplifications may occur in the `t Hooft limit appeared in
\cite{Lipatov:1993qn,Faddeev:1994zg}, where integrable structures
were found when studying high-energy scattering in gauge theory.
Nevertheless, QCD in the `t Hooft limit is still far from being
solved. It seems natural to study supersymmetric
Yang-Mills theories, which are much simpler.
In particular, the maximally supersymmetric Yang-Mills theory,
$\cN=4$ SYM, has many special
properties. It was shown long ago that its $\beta$ function
vanishes, and hence its coupling constant does not run. It is an
interacting superconformal field theory, where the coupling
constant is a free parameter. Moreover, through the AdS/CFT
correspondence it is expected to be dual to type IIB superstring
theory on $AdS_{5} \times S_{5}$ \cite{Maldacena:1997re}. This is
a realisation of `t Hooft's idea of a gauge/string duality in the
$N \rightarrow \infty$ limit. The nature of this duality, which
relates field theory at strong coupling with string theory at weak
coupling, implies that the perturbative series of `observables'
(such as e.g. correlation functions of gauge-invariant operators)
in $\cN=4$ SYM has to reproduce perturbative string theory
results. In order for this to happen, one may suspect that the
perturbative expansion of these quantities must have special
properties.
Indeed, remarkably simple structures have been observed in $\cN=4$ SYM, mainly in
two domains: firstly, anomalous dimensions of composite operators and
secondly, on-shell $n$-particle scattering amplitudes.\\

The AdS/CFT correspondence identifies string states with
composite, gauge-invariant operators on the gauge theory side. In
testing the correspondence, computing scaling dimensions of such
operators in $\cN=4$ SYM plays an important role. The first tests
were done for `protected' operators, whose scaling dimension can
be shown to be coupling independent. It therefore equals their
classical (tree-level) dimension, which has to agree with the
string theory prediction. For generic operators, the scaling
dimension depends on the coupling constant, and a comparison with
string theory is difficult because of the weak/strong nature of
the duality, i.e. it would require summing the complete
perturbation series.\\

Important progress in computing anomalous dimensions in $\cN=4$
SYM was made with the discovery of integrability in the planar
(large $N$) limit. In QCD, integrable structures at one loop
were observed some time ago \cite{Belitsky:1999qh,Belitsky:1999ru,Belitsky:1999bf,
Braun:1998id,Braun:1999te,Derkachov:1999ze}.
In certain cases, the dilatation operator, which measures the conformal dimension of
a given operator, turns out to be described by an integrable spin
chain. In $\cN=4$ SYM, the integrability of the complete dilatation
operator at one loop was found in \cite{Minahan:2002ve,
Beisert:2002ff,Beisert:2003tq,Beisert:2003yb}. Most importantly,
in some sectors it was shown to extend to higher loop levels
\cite{Beisert:2003ys,Belitsky:2003ys,Beisert:2003jj,Beisert:2003jb,
Dolan:2003uh,Arutyunov:2003rg,Dolan:2004ps,Belitsky:2004yg,
Belitsky:2004sc,Ryzhov:2004nz,Frolov:2005ty}. If this
integrability persists to arbitrary loop levels, then one can hope
to make contact with strong coupling results. Indeed, on the
string theory side of the AdS/CFT correspondence, the classical
Green-Schwarz superstring action for $AdS_{5} \times S_{5}$,
constructed in \cite{Metsaev:1998it}, is integrable
\cite{Bena:2003wd}. If the integrability survives quantisation
\cite{Vallilo:2003nx,Berkovits:2004xu}, it might be possible to
identify the integrable structures on both sides of the correspondence.\\

Integrability, or in certain cases the assumption of integrability,
allowed the computation of the anomalous dimensions of various operators in $\cN=4$ SYM.
Spectacular progress was made by the all-order
conjecture of Beisert, Eden and Staudacher \cite{Beisert:2006ez}
(see also \cite{Staudacher:2004tk,Arutyunov:2004vx,Beisert:2005fw,Janik:2006dc,Beisert:2006ib}) for the asymptotic Bethe ansatz
\footnote{Asymptotic means that it is valid only for `long'
operators, i.e. valid to $O(g^{2L-2})$, where $L$ is the number of
elementary fields constituting the composite operator.} describing
the anomalous dimensions of composite operators in the $SL(2)$
sector of $\cN=4$ SYM. Their proposal leads to an integral
equation (BES equation) for the cusp anomalous dimension \cite{Polyakov:1980ca,Korchemsky:1985xj,Korchemsky:1987wg}, valid to
all orders in the coupling constant. It correctly reproduces the
known perturbative values of the cusp anomalous dimension \cite{Korchemsky:1987wg,Belitsky:2003ys,Bern:2005iz,Bern:2006ew}
and also agrees with results at strong coupling \cite{Gubser:2002tv,Frolov:2002av,Roiban:2007dq} (for the strong-coupling
expansion of the BES equation see \cite{Benna:2006nd,Basso:2007wd,Kostov:2008ax}).
\\

The second domain where unexpected simplicity was found is that
of on-shell gluon scattering. Indeed, the scattering amplitudes
turn out to be much simpler than one would expect on general grounds.
Even at tree-level, the number of Feynman diagrams contributing
to a given amplitude increases factorially with the number of external
gluons.
Nevertheless, some classes of tree-level amplitudes are known for an
arbitrary number of legs. For example, the maximally helicity-violating (MHV)
amplitudes are given by very simple one-line expressions \cite{Parke:1986gb,Berends:1987me},
and all tree-level amplitudes satisfy recursion relations derived by Britto, Cachazo and Feng (and Witten) \cite{Britto:2004ap,Britto:2005fq}.
Moreover, tree-level amplitudes have simple properties in twistor space \cite{Witten:2003nn}.
\\

Furthermore, enormous progress has been achieved to compute loop level
amplitudes, mainly using unitarity-based techniques
\cite{Bern:1994zx,Bern:1994cg}, which are inspired by the Cutkosky rules
\cite{Cutkosky:1960sp}. These have allowed the computation of a
large class of one-loop amplitudes, and in the case of $\cN=4$ SYM
even some two-, three-, and four-loop amplitudes
\cite{Bern:1997nh,Anastasiou:2003kj,Bern:2005iz,Bern:2006ew}.
Anastasiou, Bern, Dixon and Kosower (ABDK) noticed that the four-gluon amplitude at two loops
has a remarkable iterative structure \cite{Anastasiou:2003kj}. For
the infrared divergent terms of the amplitude, such an iteration
is expected on general grounds, but a similar iteration was found to
hold also for the finite part. Following years of intensive
studies of gluon scattering amplitudes
\cite{Bern:1994zx,Bern:1997nh,Bern:2007dw} and based on the observation
of ABDK, the conjecture was put
forward by Bern, Dixon and Smirnov \cite{Bern:2005iz} that the maximally
helicity-violating (MHV) planar gluon amplitudes in
$\mathcal{N}=4$ SYM have a remarkably simple all-loop iterative
structure. In general, these amplitudes have the following form:
\begin{equation}\label{intro-BDS}
\ln\mathcal{M}_n^{\rm
  (MHV)} = \text{[IR divergences]} + F_n^{\rm
  (MHV)}(p_1,\ldots,p_n; a) + O(\epsilon)\,.
\end{equation}
{Here $\mathcal{M}_n^{\rm (MHV)}$ is the colour-ordered planar
gluon amplitude, divided by the tree amplitude.} The first term on
the right-hand side describes the infrared (IR) divergences and
the second term is the finite contribution dependent on the gluon
momenta $p_i$ and on the `t Hooft coupling $a= {g^2 N}/(8\pi^2)$.
The structure of IR divergences is well understood in any gauge
theory~\cite{Akhoury:1978vq,Mueller:1979ih,Collins:1980ih,
Collins:1989gx,Collins:1989bt,Sen:1981sd,Sen:1982bt,Sterman:1986aj,Catani:1989ne,
Catani:1990rp,
Magnea:1990zb,
Catani:1998bh,Sterman:2002qn,
Botts:1989kf,Korchemsky:1988hd,Korchemsky:1988pn}.
In particular, in theories with a vanishing $\beta$ function like
${\cal N}=4$ SYM, the leading IR singularity in dimensional
regularisation is a double pole, whose coefficient is the
universal cusp anomalous dimension appearing in many physical
processes
\cite{Korchemsky:1987wg,Korchemsky:1993uz,Korchemskaya:1996je,Belitsky:2003ys}.
Interestingly, the latter is also predicted by integrable models,
as mentioned above. The BDS conjecture provides an explicit
expression for the finite part, $F_n^{\rm (MHV)}=F_n^{\rm (BDS)}$,
for an arbitrary number $n$ of external gluons, to all orders in
the coupling $a$. Remarkably, the dependence of $F_n^{\rm (BDS)}$
on the kinematical invariants is described by a function which is
coupling independent and, therefore, can be determined at one
loop. At present, the BDS conjecture has been tested up to three
loops for $n=4$ \cite{Bern:2005iz} and up to two loops for $n=5$
\cite{Bern:2006vw}. The results of this thesis are relevant for
the case $n=6$ at two
loops.
\\

The aforementioned scattering amplitudes in $\cN=4$ have even more
interesting features. Drummond, Sokatchev, Smirnov, and the
present author found that all integrals appearing in the
four-gluon amplitude up to three loops are of a special type
\cite{Drummond:2006rz}. In dual momentum variables defined by
 \be \label{intro-pseudoconformal}
p_{i} = x_{i+1} - x_{i}\,, \ee all integrals can be seen to have
broken conformal properties. This is unexpected since this broken
dual conformal symmetry is not, at least not in an obvious way, a
consequence of the usual (super-)conformal symmetry of $\cN=4$
SYM. This surprising observation was confirmed at four loops \cite{Bern:2006ew} and
was used as an assumption for constructing the four-gluon
amplitude at five loops \cite{Bern:2007ct}.
\\

In an important recent development in the study of the AdS/CFT
correspondence, Alday and Maldacena proposed \cite{Alday:2007hr}
the strong coupling description of planar gluon scattering
amplitudes in $\mathcal{N}=4$ SYM and were able to make a direct
comparison with the BDS prediction based on weak coupling
results for the same amplitudes.
According to their proposal, certain planar gluon
amplitudes at strong coupling are related to the area of a minimal surface in AdS${}_5$
space attached to a specific closed contour $C_n$,
\begin{equation}\label{intro-AM-prescription}
\ln \mathcal{M}_n = -\frac{\sqrt{g^2 N}}{2\pi} A_{\rm min}(C_n)\,.
\end{equation}
The contour $C_{n}$ is a polygon with light-like edges $[x_i,x_{i+1}]$ defined by the gluon
momenta through relation (\ref{intro-pseudoconformal}), and with the cyclicly condition $x_{n+1} \equiv x_1$.
Notice that the $x_{i}$, which correspond to the $n$ cusps of the polygon $C_{n}$, are
the same dual momentum variables
that were introduced previously for discussing the broken conformal properties of
integrals appearing in the gluon amplitudes! The prescription of \cite{Alday:2007hr} is insensitive to
the helicity configuration of the gluon amplitude under consideration.
\\

 For $n=4$ the minimal surface $A_{\rm
min}(C_4)$ was found explicitly in \cite{Alday:2007hr}, by making
use of the conformal symmetry of the problem. With the appropriate
AdS equivalent of dimensional regularisation, the divergent part
of $\ln \mathcal{M}_4$ has the expected pole structure, with the
coefficient in front of the double pole given by the known strong
coupling value of the cusp anomalous dimension. Most importantly,
the finite part of $\ln \mathcal{M}_4$ is in perfect agreement
with $F_4^{\rm (BDS)}$ from the BDS ansatz. For $n\ge 5$ the
practical evaluation of the solution of the classical string
equations turns out to be difficult,  but it becomes possible for
$n$ large \cite{Alday:2007he}. In the limit $n\to\infty$ the
strong coupling prediction for $\ln\mathcal{M}_n$ disagrees with
the BDS ansatz. This indicates \cite{Alday:2007he} that the BDS
conjecture should fail for a sufficiently large number of gluons
and/or at sufficiently high loop level.\\

Alday and Maldacena also pointed out \cite{Alday:2007hr} that their
prescription (\ref{intro-AM-prescription}) is mathematically
equivalent to the strong coupling calculation of the expectation
value of a Wilson loop $W(C_n)$, defined on the light-like contour
$C_n$ \cite{Maldacena:1998im,Kruczenski:2002fb}, \be
\label{intro-WL}
 W(C_{n})  = \frac{1}{N}\langle 0 |\,  {\rm Tr}\, {\rm P} \exp \left(i g \oint_{C_{n}} dx^{\mu} A_{\mu}\right)   |0\rangle\,.
\ee
 This should not come as a total surprise, since the intimate
relationship between the infrared divergences of the scattering of
massless particles and the ultraviolet divergences of Wilson loops
with cusps is well known in QCD
\cite{Korchemsky:1987wg,Korchemsky:1993uz,Korchemskaya:1996je}.
Inspired by this, Drummond, Korchemsky
and Sokatchev conjectured that a similar duality relation between
planar MHV gluon amplitudes and light-like Wilson loops also exists at
weak coupling \cite{Drummond:2007aua}. They illustrated this duality by an explicit
one-loop calculation in the simplest case $n=4$. This was later
extended to the case of arbitrary $n$ at one loop by Brandhuber, Heslop and Travaglini
\cite{Brandhuber:2007yx}. 
The duality relation identifies the two
objects up to an additive constant and non-planar corrections,
\begin{equation}\label{intro-duality}
\ln \mathcal{M}_n^{\rm (MHV)} = \ln W(C_n) + \mbox{const} + O(\epsilon,1/N)\,.
\end{equation}
This means that upon a specific identification of the regularisation parameters and the kinematical
invariants, the infrared divergences of the logarithm of the scattering amplitude
$\ln\mathcal{M}_n$,  match the ultraviolet divergences of the light-like Wilson loop $\ln W(C_n)$,
and, most importantly, the finite parts of the two objects also coincide (up to an additive
constant and non-planar corrections),
\begin{equation}\label{intro-finiteduality}
F_n^{\rm (MHV)} = F_n^{\rm (WL)} +\text{const}+O(1/N)\ .
\end{equation}
While the former property follows from the known
structure of divergences of scattering amplitudes and of Wilson
loops in generic gauge theories
\cite{Korchemsky:1987wg,Korchemskaya:1992je,Korchemsky:1993uz},
the property
\re{intro-finiteduality} is extremely non-trivial.\\

In this report we present work in collaboration with Drummond,
Korchemsky and Sokatchev that provided further evidence in favour
of the duality relation (\ref{intro-finiteduality}). To begin with, we carried
out explicit two-loop calculations of $\ln W(C_n)$ for $n=4$
\cite{Drummond:2007cf} and $n=5$ \cite{Drummond:2007au}. Our
results are in perfect agreement with the two-loop MHV gluon
amplitude calculations \cite{Anastasiou:2003kj,Bern:2006vw}, and
hence with the BDS ansatz for $n=4,5$.
\\

Furthermore, in \cite{Drummond:2007cf} we argued that we can
profit from the (broken) conformal symmetry of the light-like
Wilson loops. Due to the presence of a cusp anomaly in the Wilson
loops, conformal invariance manifests itself in the form of
anomalous Ward identities. In \cite{Drummond:2007cf} we proposed and
in \cite{Drummond:2007au} we proved an anomalous
conformal Ward identity for the finite part of the Wilson loops,
valid to all orders in the coupling  constant \footnote{Later,
similar Ward identities were also obtained at strong coupling
using the AdS/CFT correspondence in
Refs.~\cite{AM-unpubished,Komargodski:2008wa}.}. It reads 
\begin{equation}\label{intro-cwi}
  K^{\mu} \, {F^{\rm (WL)}_n} =  \sum^n_{i=1} \left( 2\, x_i^\mu x_i^{\nu} \frac{\partial}{\pa x_{i}^{\nu}} - x_i^2 \frac{\partial}{\pa x_{i \mu}} \right) \, {F^{\rm (WL)}_n}  =  \frac{1}{2} \Gamma_{\rm cusp}(a) \sum_{i=1}^n  \ln \frac{x_{i,i+2}^2}{x_{i-1,i+1}^2}
    x^\mu_{i,i+1}\,,
\end{equation}
where $x_{i,j}=x_{i}-x_{j}$ and $\Gamma_{\rm cusp}(a)$ is the (coupling-dependent) cusp
anomalous dimension.
 This identity uniquely fixes the functional form of the finite part
of $\ln W(C_n)$ for $n=4$ and $n=5$, up to an additive constant,
to agree with the conjectured BDS form for the corresponding MHV gluon
amplitudes. For $n \ge 6$, (\ref{intro-cwi}) gives partial
restrictions on the functional dependence on the kinematical
variables. Quite remarkably, the BDS ansatz (\ref{intro-BDS}) for
the $n$-gluon MHV amplitudes satisfies the conformal Ward identity
for arbitrary n \cite{Drummond:2007cf}.
\\

However, for $n\ge 6$ the conformal Ward identity allows ${F^{\rm (WL)}_n}$ to differ from the BDS ansatz ${F^{\rm (BDS)}_n}$ 
by an arbitrary function of conformal invariants (for
$n=6$ there are three such invariants) \cite{Drummond:2007cf}.
This result provided a possible explanation of the BDS conjecture
for $n=4,5$ (assuming that the MHV amplitudes have the same
conformal properties as the Wilson loop), but left the door open
for potential deviations from it for $n\geq 6$. To verify whether
the BDS conjecture and/or the proposed duality relation
(\ref{intro-finiteduality}) still hold for $n=6$ to two loops, it
was necessary to perform explicit two-loop calculations of the
finite parts of the six-gluon MHV amplitude $F_6^{\rm (MHV)}$, and
of the hexagon Wilson loop $F_6^{\rm (WL)}$.
\\

By doing an explicit two-loop calculation we were able to derive a
(multiple) parameter integral representation for $F_6^{\rm (WL)}$,
which we evaluated numerically. We found that $F_6^{\rm
(WL)}$ differs from the BDS ansatz at two loops by a non-trivial
function of conformal cross-ratios. We studied this function
numerically and found that it was consistent with the collinear
limit (of gluon amplitudes) \cite{Drummond:2007bm}. Later, when the results of the
two-loop calculation of the six-gluon MHV amplitude became
available, we compared the two results numerically and found agreement
with the duality relation (\ref{intro-finiteduality}) \cite{Drummond:2008aq,Bern:2008ap}.
The BDS ansatz fails at two loops and at $n=6$, but it fails just in such a
way as to preserve the duality with Wilson loops! We take this as
strong evidence that the duality relation
(\ref{intro-finiteduality}) holds for arbitrary $n$ and to an
arbitrary number of loops.
\\

We would like to point out that a weaker form of the duality
(\ref{intro-finiteduality}) has already been observed in QCD in
the special, high-energy (Regge) limit $s \gg -t > 0$ for the
four-gluon amplitude up to two loops \cite{Korchemskaya:1996je}. The same relationship
holds in any gauge theory ranging from QCD to $\cN = 4$ SYM. The
essential difference between these theories is that in the former
case the duality is only valid in the Regge limit, whereas in the
latter case it is exact in general kinematics.
We would like to mention that a thorough analysis of the Regge limit of planar
multi-gluon amplitudes in $\cN = 4$ SYM was recently performed in
\cite{Brower:2008nm,Bartels:2008ce} and it provided further
evidence that the BDS ansatz needs to be corrected
\cite{Bartels:2008ce}.

\newpage

This report is organised as follows:
\\

Sections \ref{ch-conformal}--\ref{ch-duality} are mainly introductory. In section \ref{ch-conformal}, we recall some notions of conformal symmetry and its implications for correlation functions of gauge invariant operators in $\cN=4$ SYM. section \ref{ch-amplitudes} briefly introduces the reader to gluon scattering amplitudes, focusing on multi-loop results for maximally helicity-violating (MHV) amplitudes and the BDS conjecture.
Following that, section \ref{ch-AM} summarises the prescription of Alday and Maldacena to compute gluon scattering amplitudes at strong coupling. This will lead us to consider Wilson loops, and after introducing them
in section \ref{ch-WL} and summarising their renormalisation properties, we discuss in section \ref{ch-duality} the main point of study of this thesis: the duality between gluon amplitudes and Wilson loops.
The remaining sections present original work by the author and collaborators. In particular sections \ref{section-conformalfourpoint} and \ref{ch-pseudoconformal} are based on \cite{Drummond:2006rz},
and sections \ref{ch-four-point}, \ref{section-CWI}, \ref{ch-six-point} and appendix \ref{ch-WL-appendix} are based on \cite{Drummond:2007cf,Drummond:2007au,Drummond:2007bm,Drummond:2008aq,DHKSunpublished}.
\\

The only exceptions to the introductory nature of sections \ref{ch-conformal}--\ref{ch-duality}
are sections \ref{section-conformalfourpoint} and
\ref{ch-pseudoconformal}, where we present original work on
off-shell conformal four-point integrals, and their relevance for
on-shell gluon scattering amplitudes, respectively. We study the
known (to four loops) four-gluon scattering amplitude in $\cN=4$
SYM and find that it is given in terms of `pseudo-conformal'
integrals.
\\

In section \ref{ch-four-point}, we verify the Wilson loop/gluon amplitude duality in the non-trivial cases
of four and five points/gluons at two loops by an explicit Feynman graph calculation of the rectangular and pentagonal Wilson loops.\\

In section \ref{section-CWI}, we derive all-order Ward identities
for the light-like Wilson loops and discuss their consequences.\\

From this discussion of the Ward identities in section \ref{section-CWI} it will become clear that
a crucial test of the duality is at two loops and for six points/gluons.
We therefore compute the hexagonal Wilson loop at two loops in section \ref{ch-six-point} and study its
properties in the collinear limit (of gluon amplitudes). We present a numerical comparison with
recently available results for the six-gluon MHV amplitude at two loops.
\\

In section \ref{ch-conclusions}, we present our conclusions.
\\

There are two appendices. Appendix \ref{ch-MB} contains an alternative proof of
an identity between two conformal integrals derived in section
\ref{section-conformalfourpoint}, using the Mellin--Barnes
technique. In appendix \ref{ch-WL-appendix} we
present the previously unpublished details of the calculation of the four-point Wilson loop at
two loops.


%
%

\section{(Super-)conformal symmetry}
\label{ch-conformal} Conformal symmetry in quantum field theory
was extensively studied beginning from the late 1960's, see e.g.
\cite{Todorov:1978rf,Fradkin:1978pp} and references therein. The
conformal properties of correlation functions were studied.
Conformal symmetry has very strong consequences in two dimensions,
where the conformal group has an infinite number of generators. In
four-dimensional field theories, which are more relevant for
particle physics, the conformal group has only $15$ parameters,
and as a consequence, it is less restrictive than in two
dimensions. Moreover, in generic field theories, conformal
symmetry is broken by quantum corrections. An example is
(massless) QCD. Nevertheless, studying the deviation from
conformal invariance can be useful in practice, as discussed in
the review \cite{Braun:2003rp}. The situation is much better in a
field theory which is conformally invariant also at the quantum
level. As we will see in section \ref{sect-N=4}, the maximally
supersymmetric Yang-Mills theory in four dimensions, $\cN=4$ SYM,
which was discovered in \cite{Gliozzi:1976qd,Brink:1976bc}, has
this remarkable property.

\subsection{Definition of conformal transformations}
The presentation of conformal symmetry follows roughly chapter four of
the book \cite{DiFrancesco:1997nk}, where
the reader can find more details. Let us consider a $d$-dimensional space with flat metric $\eta_{\mu\nu}$ (the treatment for Euclidean and Minkowski space is identical). By definition, conformal transformations leave the metric invariant up to a local rescaling
\be \label{conf-def}
x^{\mu} \rightarrow x{'}^{\mu}\,,\qquad dx_{\mu} dx^{\mu} \rightarrow \Lambda(x) dx{'}_{\mu} dx{'}^{\mu}\,.
\ee
Geometrically, (\ref{conf-def}) means that conformal transformations preserve angles.
In order to find the most general solution to (\ref{conf-def}), consider
an arbitrary infinitesimal coordinate transformation
\be
x^{\mu} \rightarrow x{'}^{\mu} = x^{\mu} + \epsilon^{\mu}(x)\,.
\ee
It is then easy to show that in $d>2$ dimensions, the most general form
of $\epsilon_{\mu}$ compatible with (\ref{conf-def}) is given by
\be \label{conf-inftrans}
\epsilon_{\mu} = a_{\mu} + m_{\mu \nu} x^{\nu} + \lambda x_{\mu} + 2 (b \cdot x) x_{\mu} -b_{\mu} x^2\,,\qquad m_{\mu\nu}=-m_{\nu\mu}\,,
\ee
where we used the notations $b \cdot x = b^{\nu} x_{\nu}$ and $x^2 = x^{\nu} x_{\nu}$.
The two-dimensional case is special and we will not treat it here,
since we are interested in $d=4$. For more information on conformal symmetry
in two dimensions, see for example \cite{DiFrancesco:1997nk}.
The infinitesimal transformations in (\ref{conf-inftrans}) corresponding to the parameters $a^{\mu}$ and $m^{\mu\nu}$
are translations and rotations, respectively. Thus the Poincar\'{e} group is a subgroup of the conformal
group.
The transformations corresponding to $\lambda$ and $b^{\nu}$ are dilatations, and special conformal transformations, respectively.
Let us define generators for the infinitesimal transformations according to
\be \label{conf-generators}
x{'}^{\rho} = (1 + i a^{\mu}P_{\mu} + i m^{\mu\nu} M_{\mu\nu}+i\lambda D + i b^{\mu}K_{\mu}) x^{\rho}\,.
\ee
By comparing (\ref{conf-generators}) and (\ref{conf-inftrans}) it is
easy to see that the generators of the conformal group
are given by
\ba \label{conf-generators2}
P_{\mu} &=& -i\partial_{\mu}\,,\nonumber \\
M_{\mu\nu} &=&i (x_{\mu}\partial_{\nu}-x_{\nu}\partial_{\mu})\,, \nonumber \\
D &=& -i x^{\mu}\partial_{\mu}\,, \nonumber \\
K_{\mu} &=& -i (2 x_{\mu}x^{\nu}\partial_{\nu} - x^2 \partial_{\mu})\,.
\ea
Here we used the shorthand notation $\partial_{\mu} = \frac{\partial}{\partial x^{\mu}}$.
From (\ref{conf-generators2}) we can see that the conformal group in four dimensions has
15 generators, i.e. four translation generators $P_{\mu}$, six rotations $M_{\mu\nu}$, a dilatation $D$
and four special conformal boosts $K_{\mu}$.
In addition to the usual commutation relations of the Poincar\'{e} algebra,
\begin{eqnarray} \label{Poincare-algebra}
&& \lbrack P_{\mu}, P_{\nu} \rbrack = 0 \,, \qquad \lbrack M_{\mu
\nu},
P_{\rho} \rbrack = i(\eta_{\nu\rho} P_{\mu}-\eta_{\mu\rho} P_{\nu}) \,, \nonumber \\
&& \lbrack M_{\mu \nu}, M_{\rho \sigma} \rbrack = i(-\eta_{\mu \sigma}
M_{\nu \rho} + \eta_{\nu \sigma} M_{\mu \rho} +\eta_{\mu \rho} M_{\nu
\sigma} - \eta_{\nu \rho} M_{\mu \sigma})\,,
\end{eqnarray}
the generators (\ref{conf-generators2}) have
the following commutation relations:
\begin{eqnarray}\label{Conformal-algebra}
\lbrack D, P_{\mu} \rbrack = i P_{\mu}\,, &\qquad&  \lbrack K_{\rho},
M_{\mu \nu} \rbrack =i (\eta_{\rho
\mu}K_{\nu} - \eta_{\rho \nu}K_{\mu})\,,\nonumber \\
\lbrack D, K_{\mu} \rbrack = -i K_{\mu}\,, &\qquad& \lbrack
K_{\mu},P_{\nu} \rbrack =2i (\eta_{\mu \nu}D - M_{\mu\nu})\,.
\end{eqnarray}
Relations (\ref{Poincare-algebra}) and (\ref{Conformal-algebra}) define the conformal algebra.
\\

Let us now consider finite conformal transformations.
For translations $P^{\mu}$, rotations $M^{\mu\nu}$ and dilatations $D$,
the form of the finite transformations are easy to find.
Furthermore, it is possible to show that the finite transformation corresponding
to the infinitesimal conformal boosts is given by
\be \label{Kspecial-conf}
x{'}^{\mu} = \frac{x^{\mu} - b^{\mu} x^2}{1- 2 b\cdot x+b^2 x^2}\,.
\ee
For practical purposes it is convenient to introduce another finite transformation,
the inversion,
\be \label{inversion}
I : x^{\mu}\rightarrow \frac{x^{\mu}}{x^2}\,.
\ee
The special conformal transformation (\ref{Kspecial-conf}) can be obtained
by composing an inversion, a translation by $-b^{\mu}$, and another inversion.
Because of this, it will often be convenient to check the conformal invariance
(or covariance) of a given translation invariant expression by doing inversions (\ref{inversion})
rather than the more complicated special conformal transformations (\ref{Kspecial-conf}).
In particular, inversions will be useful
to us in order to determine the consequences of conformal symmetry on
correlation functions.
Note that the inversion (\ref{inversion}) is an element of the conformal group
not connected to the identity. In other words, there is no infinitesimal
generator corresponding to (\ref{inversion}).
\\

So far we discussed the action of the conformal group on coordinates.
In a quantum field theory, one has to define the action of the
conformal generators on fields as well.
The conformal generators acting on fundamental fields $\phi_I(x)$ with conformal
weight $\Delta$ and Lorentz indices $I$ are~\footnote{The generators $\mathbb{G}$
determine the infinitesimal transformations with parameters $\varepsilon$:
$\phi'(x) = \phi(x) + \varepsilon\cdot \mathbb{G} \phi(x)$.}
\begin{eqnarray} \label{conf-trans-fields}
  &\mathbb{M}^{\m\n}  \phi_I &= (x^\m \pa^\n - x^\n \pa^\m)  \phi_I + (m^{\m\n})_I{}^J \phi_J \,,  \nn\\
  &\mathbb{D} \,\phi_I &= x\cdot\pa\ \phi_I + \Delta\, \phi_I \,,   \nn  \\
  &\mathbb{P}^\m  \phi_I &= \pa^\m\ \phi_I \,, \nn\\
  &\mathbb{K}^\m \phi_I &= \left(2 x^\m x\cdot\pa -  x^2 \pa^\m \right) \phi_I + 2x^\m \, \Delta\, \phi_I + 2x_\nu (m^{\m\n})_I{}^J \phi_J \,.
\end{eqnarray}
Here $m^{\m\n}$ is the generator of spin rotations, e.g., $m^{\m\n}=0$ for a
scalar field and $(m^{\m\n})_\l{}^\rho = g^{\n\rho} \delta^\m_\l - g^{\m\rho}
\delta^\n_\l$ for a gauge field.
From (\ref{conf-trans-fields}) one can in principle determine the variation of
fields under finite transformations.
For example, a scalar conformal field has the transformation property
\be \label{conf-prim}
\phi(x) \rightarrow \phi{'}(x{'}) = \left| \frac{\partial x{'}}{\partial x}\right|^{-\Delta/d} \phi(x)\,,
\ee
where $\Delta$ is the conformal dimension of $\phi(x)$. Here
\be
 \left| \frac{\partial x{'}}{\partial x}\right| = \Lambda(x)^{-d/2}
\ee
is the Jacobian of the conformal transformation of coordinates, c.f. (\ref{conf-def}).
Fields obeying (\ref{conf-prim}) are called {\it primary}.
\\

As was already said, we will consider conformal symmetry in four dimensions only.
Examples of classically conformal field theories are theories which have only dimensionless
parameters in the action (i.e. dimensionless coupling constant, no masses). An explicit example
is the scalar $\phi^4$ model in four dimensions, whose action reads
\be \label{cs-1-1}
{S} =\int d^{4}x \left[\frac{1}{2}\pa^{\mu} \phi(x) \pa_{\mu} \phi(x) + \frac{g}{4!} \phi^4(x) \right]\,.
\ee
Under translations and rotations, the scalar field transforms as $\phi{'}(x{'}) = \phi(x)$,
so that (\ref{cs-1-1}) is invariant under Poincar\'{e} transformations.
Further, one can see that (\ref{cs-1-1}) is invariant under
dilatations and special conformal transformations, if $\phi(x)$ transforms
according to (\ref{conf-prim}) with weight $\Delta = 1$.
Note that the coupling constant $g$ in (\ref{cs-1-1}) is dimensionless.
Dimensional parameters in the Lagrangian, as for example
$m$ corresponding to a mass term $m^2 \phi^2(x)$,
would spoil conformal symmetry.
\\

Let us stress that the conformal invariance of (\ref{cs-1-1}) is
only valid classically, and it is broken by quantum corrections.
The reason for this is that due to ultraviolet divergences, the coupling
constant $g$ and the elementary field $\phi$ have to be
renormalised. This is done in the usual way by introducing
appropriate renormalisation factors $Z_{\phi}, Z_{g}$
\cite{Collins:1984xc}. The net result is that the renormalised
coupling constant $g$ depends on the renormalisation scale $\mu$,
i.e. the $\beta$ function is nonvanishing, \be \mu \frac{d
g}{d\mu}  = \beta(g) \neq 0 \,. \ee This is the generic situation
in quantum field theory: the conformal symmetry is lost in the
quantum theory, and may reappear only at certain fixed points
$g_{c}$ where $\beta(g_{c}) = 0$.

\subsection{Conformal correlation functions}
\label{section-conformalcorr}
In this section, we briefly review the implications of conformal symmetry
for correlation functions. For a comprehensive review of conformal symmetry
 in quantum field theory, see e.g. \cite{Todorov:1978rf,Fradkin:1978pp}.
\\

In a conformal field theory one usually considers primary operators because they have simple transformation
properties under conformal transformations, see (\ref{conf-prim}). Requiring conformal
covariance of these operators under conformal transformations according to (\ref{conf-prim})
restrains the functional form that their correlation functions can have.
As we will see, conformal symmetry is most restrictive for two- and three-point functions of scalar primary operators.
\subsubsection{Two-point functions}
Take for example two scalar operators $O_1$ and $O_2$ possessing conformal dimension $\Delta_1$ and $\Delta_2$, respectively. Under conformal transformations their two-point
function has to transform according to (\ref{conf-prim}), i.e.
\be \label{conf-twopoint}
\VEV{O_{1}(x_{1})\, O_{2}(x_{2})} = \left| \frac{\partial x{'}}{\partial x}\right|^{\Delta_{1}/d}_{x=x_{1}}
\left| \frac{\partial x{'}}{\partial x}\right|^{\Delta_{2}/d}_{x=x_{2}} \VEV{O_{1}(x'_{1})\, O_{2}(x'_{2})}\,,
\ee
where $d$ is the space-time dimension.
From invariance under the Poincar\'{e} group (i.e. under translations and rotations)
we immediately deduce that their two-point function
can only depend on the translation and rotation invariant variable $x_{12}^2 = (x_1^\mu - x_2^\mu)^2$, i.e.
\be \label{cs-1}
\VEV{O_{1}(x_{1})\, O_{2}(x_{2})} = f(x_{12}^2)\,.
\ee
Requiring covariance under dilatations with conformal weights $\Delta_{1}$ and $\Delta_{2}$ (c.f. equation (\ref{conf-prim})),
respectively, we find
\be
\VEV{O_{1}(x_{1})\, O_{2}(x_{2})} ={C} \left[x_{12}^2\right]^{-(\Delta_{1}+\Delta_{2})/2}  \,,
\ee
where $C$ is an arbitrary normalisation constant.
Finally, covariance under conformal boosts requires $\Delta_{1}$ to be equal to $\Delta_{2}$ (this is easiest seen
by doing an inversion), and therefore we have
\be \label{two-point}
\VEV{O_{1}(x_{1})\, O_{2}(x_{2})} =  C \left[x_{12}^2\right]^{-\Delta} \,,\quad \mathrm{for}\;\Delta_{1}=\Delta_{2}\equiv \Delta\,,\quad 0\; \mathrm{otherwise.}
\ee
Thus, the functional form of  $\VEV{O_{1}(x_{1})\, O_{2}(x_{2})}$ is completely fixed, and the only dynamically determined quantities are the overall normalisation $C$ and scaling dimensions $\Delta_1$ and $\Delta_2$.

\subsubsection{Three-point functions}

The same reasoning as in the previous section leads to the most general form of the three-point
function of scalar operators,
\be \label{conf-3pt}
\VEV{O_{1}(x_{1})\, O_{2}(x_{2})\, O_{3}(x_{3}) } =  C \left[x_{12}^2\right]^{-(\Delta_{1}+\Delta_{2}-\Delta_{3})/2} \left[x_{23}^2\right]^{-(\Delta_{2}+\Delta_{3}-\Delta_{1})/2}\left[x_{13}^2\right]^{-(\Delta_{1}+\Delta_{3}-\Delta_{2})/2}
\ee
We can immediately check that under an inversion (\ref{inversion}), (\ref{conf-3pt}) has the correct conformal
weight $\Delta_{i}$ at points $i=1,2,3$.
An example of relation (\ref{conf-3pt}) is the well-known star-triangle identity \cite{Kazakov:1984km}.
\\

Three-point functions of operators with spin are also constrained.
For example, consider a spin one operator $V^{\mu}$ of conformal dimension $\Delta_{1}=3$ (for example a conserved current).
Its three-point function with two scalar fields of conformal dimension $\Delta_{2}=\Delta_{3}=2$ is
\be \label{correlator-VSS}
\VEV{V^{\mu}(x_{1})\,O(x_{2})\,O(x_{3})} = \frac{1}{x_{13}^2 x_{12}^2 x_{23}^2} Y^{\mu}_{1;23}\,,
\ee
where the vector
\be
Y^{\mu}_{1;23}= \frac{x_{12}^{\mu}}{x_{12}^2} -  \frac{x_{13}^{\mu}}{x_{13}^2}
\ee
is conformally covariant at point $1$ and invariant
at points $2$ and $3$.
The generalisation of (\ref{correlator-VSS}) to tensor fields of arbitrary spin
and to arbitrary conformal dimensions is straightforward (see for example \cite{Makeenko:1980bh}). An application to twist-two operators in $\cN=4$ SYM can be found in
\cite{Henn:2005mw}.
Three-point functions of spin one operators \cite{Schreier:1971um} were studied in the context of anomalies in \cite{Erdmenger:1996yc}.

\subsubsection{Four-point functions}
An interesting new phenomenon occurs starting from four points.
It is then possible to write down conformal invariants in the form of
cross-ratios, which in general read
\be
\frac{x_{ij}^2 x_{kl}^2}{x_{ik}^2 x_{jl}^2}\,.
\ee
At four points, there are two independent cross-ratios,
\be\label{cro}
u = \frac{x_{12}^2 x_{34}^2}{x_{13}^2 x_{24}^2}\,,\qquad v = \frac{x_{14}^2 x_{23}^2}{x_{13}^2 x_{24}^2}\,.
\ee
Thus a four-point correlation function contains in general an arbitrary function
of $u$ and $v$,
\be \label{conf-4point}
\VEV{O_{1}(x_{1})\,O_{2}(x_{2})\,O_{3}(x_{3})\,O_{4}(x_{4})}=\frac{1}{x_{13}^2 x_{24}^2 x_{12}^2 x_{34}^2} f(u,v)\,,
\ee
and the prefactor of $f$ on the r.h.s. of (\ref{conf-4point}) carries the overall conformal weight
of the external points (we have chosen $\Delta_{1} = \Delta_{2} = \Delta_{3} =\Delta_{4} = 1$ for simplicity).
\\

Let us remark that correlation functions of operators with spin are also constrained,
see e.g. \cite{Sotkov:1976xe,Sotkov:1980qh}.
As an example, the most general form of the four-point function of two scalars and two spin-one
operators is found to be
\begin{eqnarray} \label{conf-4point-spin}
\VEV{O_{1}(x_{1})\,O_{2}(x_{2})\,V^{\mu}_{3}(x_{3})\,V^{\nu}_{4}(x_{4})}&=&\frac{1}{x_{12}^2}
\bigg[ I^{\mu\nu}_{34} f_{1}(u,v) +Y^{\mu}_{3;12} Y^{\nu}_{4;12} f_{2}(u,v) \nonumber \\
&&\hspace{-3.5cm}+Y^{\mu}_{3;14} Y^{\nu}_{4;12} f_{3}(u,v)
+Y^{\mu}_{3;12} Y^{\nu}_{4;13} f_{4}(u,v) +Y^{\mu}_{3;14} Y^{\nu}_{4;13} f_{5}(u,v) \bigg]\,.
\end{eqnarray}
Here, the conformal tensor
\be
I^{\mu\nu}_{12} = \frac{\eta^{\mu\nu}}{x_{12}^2} - 2 \frac{x_{12}^{\mu}x_{12}^{\nu}}{x_{12}^4}
\ee
carries spin $1$ and dimension $1$ at points one and two.

\subsection{Supersymmetric extension of the algebra}
One can extend the Poincar\'e algebra (\ref{Poincare-algebra}) by supplementing
it with fermionic generators satisfying the anti-commutation relations (for an introduction to supersymmetry, see e.g. \cite{Wess:1992cp,Bailin:1994qt})
\be \label{susy-gen}
\left\{ Q_{\alpha} , \bar{Q}_{\dot{\alpha}} \right\} = -2 \sigma_{\alpha \dot{\alpha}}^{\mu} P_{\mu}\,.
\ee
The supersymmetry generators $Q_{\alpha}$ and $\bar{Q}_{\dot{\alpha}}$ commute with $P_{\mu}$, and
they transform under the representation $(1/2,0)$ and $(0,1/2)$ of the Lorentz group, respectively.
It is possible to give a representation of the Super-Poincar\'e algebra on a superspace.
This space consists, in addition to the usual commuting variables $x^{\mu}$,
of two anticommuting spinors $\theta^{\alpha}$ and $\bar{\theta}^{\dot{\alpha}}$.
Then, we can define a supersymmetry transformation with
anticommuting parameters $\xi^{\alpha}$, $\bar{\xi}^{\dot{\alpha}}$ by
\be \label{superpoincare-repr}
\delta_{\xi} x_{\mu} = i \left( \xi \sigma_{\mu} \bar{\theta}-\theta\sigma_{\mu}\bar{\xi}\right)\,,\qquad \delta_{\xi} \theta = \xi\,,\qquad \delta_{\xi} \bar{\theta} = \bar{\xi}\,.
\ee
From (\ref{superpoincare-repr}) we can see that the infinitesimal supersymmetry generators are defined by
\be
Q_{\alpha} = \frac{\partial}{\partial \theta^{\alpha}} - i \left(\bar{\theta}\sigma^{\mu}\right)_{\alpha}\partial_{\mu}\,,\qquad \bar{Q}_{\dot{\alpha}} = -\frac{\partial}{\partial \bar{\theta}^{\dot{\alpha}}}+i\left(\bar{\theta}\sigma^{\mu}\right)_{\dot{\alpha}}\partial_{\mu}\,.
\ee
They satisfy the anticommutation relation (\ref{susy-gen}).
\\

If one combines supersymmetry generators and conformal generators
(note that from commuting them one gets additional generators),
one obtains the superconformal group.
In section \ref{sect-N=4}, we will study a field theory
which has an $\cN=4$ extended superconformal symmetry, with symmetry algebra $PSU(2,2|4)$, see e.g. \cite{Sohnius:1985qm} for more details on the algebra.
\\

We remark that superconformal symmetry constrains correlation functions of
(superconformal) primary operators in a similar way as discussed in section \ref{section-conformalcorr}
for conformal symmetry.
For more details, see e.g. \cite{Park:1999pd}.

\subsection{The superconformal field theory $\cN=4$ SYM}\label{sect-N=4}
Let us introduce the field theory studied in this thesis.
$\cN=4$ SYM is the maximally supersymmetric Yang-Mills theory in four dimensions.

\subsubsection{Action in components and supersymmetry transformations}
Its action was first found by dimensional reduction
of $\cN=1$ SYM in ten dimensions \cite{Brink:1976bc}.
The original action in ten dimensions is
\be \label{N=1D10SYM}
\mathcal{L} = {\rm Tr}\, \left( -\frac{1}{4} F_{MN} F^{MN} + i g\frac{1}{2} \bar{\Psi} \Gamma^{N} \mathcal{D}_{N} \Psi \right)\,.
\ee
The (trivial) dimensional reduction consists in requiring that the fields in (\ref{N=1D10SYM})
do not depend on six of the ten spacetime dimensions, i.e.
\be
\partial^{4+m} A^{N} = 0\,,\quad \partial^{4+m} \Psi = 0\,,\quad \partial^{4+m} \bar{\Psi} = 0\,,\quad m=1,\ldots 6\,.
\ee Then, one splits the ten-dimensional indices $M,N$ up into
four- and six-dimensional ones and makes the following
definitions. One has to make a choice for the representation of
the ten dimensional $\Gamma$ matrices in terms of four and six
dimensional $\gamma$ matrices, e.g. (see \cite{Sohnius:1985qm})
\begin{eqnarray}
\Gamma_{\mu} = \gamma_{\mu} \otimes {1} \,,\quad \textrm{for}\; \mu=1,\ldots,4\,, \\
\Gamma_{4+m} = \gamma_{5} \otimes \tilde{\Gamma}_{m}\,,\quad \textrm{for}\; m=1,\ldots 6\,,
\end{eqnarray}
where $\gamma_{\mu}$ and $\gamma_{5}$ are the standard Dirac matrices in four-dimensional
Minkowski space, and $\tilde{\Gamma}_{m}$ are Dirac matrices in a six-dimensional
Euclidean space, which can be written as
\be
\tilde{\Gamma}_{m} = \left[
                              \begin{array}{cc}
                                0 & \tilde{\sigma}_{m} \\
                                \tilde{\sigma}^{-1}_{m} & 0 \\
                              \end{array}
                            \right]\,.
\ee
The scalars are defined as the last six components of the ten dimensional gauge field $A_{N}$:
\be
\phi_{m} \equiv A_{4+m} \quad \textrm{for}\; m=1,\ldots 6\,.
\ee
Finally, with the help of the six dimensional $\tilde{\sigma}_{m}$ matrices we can
define
\be\label{four-six}
\phi_{ij} = -\frac{1}{2} \left(\tilde{\sigma}_{m}\right)_{ij} \phi_{m}\,.
\ee
This leads to the following form of the four-dimensional action
\begin{eqnarray}\label{N=4 action}
\mathcal{L} &=& {\rm Tr} \bigg( -\frac{1}{4} F_{\mu\nu} F^{\mu\nu} + i \lambda_{i} \sigma^{\mu} \mathcal{D}_{\mu} \bar{\lambda}^{i} + \frac{1}{2} \mathcal{D}_{\mu} \phi_{ij} \mathcal{D}^{\mu} \phi^{ij}\nonumber \\
&& + i g \lambda_{i} \left[ \lambda_{j} , \phi^{ij}\right] + i g \bar{\lambda}^{i} \left[ \bar{\lambda}^{j},\phi_{ij}\right] +g^2 \frac{1}{4} \left[ \phi_{ij},\phi_{kl}\right] \left[ \phi^{ij} , \phi^{kl}\right]  \bigg) \,.
\end{eqnarray}
The theory contains a gauge
field $A^{\mu}$, four complex fermions $\lambda^{\alpha
i}$ ($i=1,2,3,4$), and six real scalars $\phi^{ij}=-\phi^{ji}$.
All fields are in the adjoint representation
of the gauge group $SU(N)$.
\\

The action following from (\ref{N=4 action}) can be seen to be invariant under $\cN=4$ on-shell supersymmetry transformations (which follow from the $\cN=1$ supersymmetry of (\ref{N=1D10SYM}),
\begin{eqnarray}\label{N=4susy}
\delta A_{\mu} &=& i \xi_{i} \sigma_{\mu} \bar{\lambda}^{i} - i \lambda_{i} \sigma_{\mu} \bar{\xi}^{i}\nonumber \\
\delta \phi_{ij} &=& \xi_{i} \lambda_{j} - \xi_{j} \lambda_{i} + \epsilon_{ijkl} \bar{\xi}^{k} \bar{\lambda}^{k}\\
\delta \lambda_{i} &=& -\frac{1}{2}i \sigma^{\mu\nu} \xi_{i} F_{\mu\nu} + 2i \sigma^{\mu} \mathcal{D}_{\mu} \phi_{ij}
\bar{\xi}^{j} + 2i g \left[ \phi_{ij},\phi^{jk}\right] \xi_{k}\nonumber \,,
\end{eqnarray}
where the transformation parameters $\xi$ and $\bar{\xi}$ are chiral and antichiral spinors, respectively.
The algebra of the supersymmetry transformations (\ref{N=4susy}) closes up to a gauge transformation and
on-shell (i.e. using the equations of motion).

\subsubsection{Finiteness}
The form and relative factors in (\ref{N=4 action}) are completely fixed by $\cN=4$ supersymmetry,
 and for the same reason there is just one coupling constant $g$.
A special property of $\cN=4$ SYM is that its superconformal symmetry is not broken by quantum corrections.
The $\beta$ function of $\cN=4$ SYM was
shown to vanish up to three loops by direct calculations \cite{Grisaru:1980nk,Caswell:1980yi,Avdeev:1981ew}.
Furthermore, there exist several arguments for the vanishing of the $\beta$ function to all loops \cite{Howe:1982tm,Mandelstam:1982cb,Brink:1982wv,Howe:1983wj,Howe:1983sr}.
For more details and further references, see the review \cite{Sohnius:1985qm}.
\\

Let us remark on different formalisms in which $\cN=4$ SYM can be
studied. Unfortunately, an off-shell $\cN=4$ formalism is not
available to date, and there are reasons to believe that it does
not exist. For this reason one has to resort to less
supersymmetric formalisms. For example, one can very well use the
action (\ref{N=4 action}), after the usual gauge fixing procedure,
for practical calculations. However, it may be advantageous  to
utilise a manifestly supersymmetric setup, either for
calculational convenience, or because of conceptual advantages.
For example, one can take a formulation of $\cN=4$ SYM in terms of
$\cN=1$ superfields. We will give the necessary definitions in the
next section, and present a sample calculation in section
\ref{N=1four-point1loop}. It is also possible to use a manifestly
$\cN=2$ supersymmetric formalism, which employs harmonic
superspace \cite{Galperin:2001uw}. For examples of recent
calculations in this setup, see \cite{Eden:2000bk,Eden:2005bt}.
Further, there is even a $\cN=3$ harmonic superspace formalism
\cite{Galperin:2001uw}, and its quantisation was considered in
\cite{Delduc:1988cp}. However up to now, no supergraph
calculations were carried out in this formalism.
\\

Let us clarify a point that might otherwise lead to confusion. $\cN=4$ SYM is often referred to as a `finite' field theory. In a supersymmetric formalism, it is indeed true that propagators and the coupling constant acquire no or only
finite corrections in perturbative calculations. On the other hand, in a non-supersymmetric gauge (e.g. the Wess-Zumino gauge), the gauge dependent propagators do get divergent corrections and need to be renormalised by appropriate wavefunction renormalisations, and the same is true for the coupling constant. It is only the $\beta$ function (which is gauge independent) that vanishes.
\\

Furthermore, even in a finite superspace setup, divergences can and usually do arise when one studies composite operators.
The latter are traces of products of several operators at the same space-time point. Such operators generically have short distance singularities which have to be regularised. We will see an example in section \ref{ch-conf-pert}.

\subsubsection{Action of $\cN=4$ SYM in $\cN=1$ superspace}
\label{N=4inN=1}
The field content of $\cN=4$ SYM can be realised in $\cN=1$ superspace
by introducing three chiral superfields $\Phi$ and one real gauge superfield $V$.
The gauge fixed action in the Feynman gauge, using the conventions of \cite{Bianchi:2000hn} \footnote{Except
for the definition of the coupling constant, which is $2g$ here compared to $g$ in \cite{Bianchi:2000hn}.}, reads
\begin{eqnarray}\label{N=1action-exp}
S &=& \int d^{4}x d^{2}\theta d^{2}\bar{\theta} \bigg\{ V^{a} \square V_{a} - \Phi^{a}_{I} \square \Phi_{a}^{\dagger I}-i 2 g f_{abc} \Phi^{\dagger a}_{I} V^{b} \Phi^{I c} +{2 g^2} f_{abe} f_{ecd} \Phi^{\dagger a}_{I}V^{b} V^{c}\Phi^{I d} \nonumber \\
&& \hspace{2cm} -\frac{ig \sqrt{2}}{3!}f^{abc} \left[\epsilon_{IJK} \Phi^{I}_{a} \Phi^{J}_{b}\Phi^{K}_{c} \delta(\bar{\theta})-\epsilon^{IJK}\Phi^{\dagger}_{aI}\Phi^{\dagger}_{bJ}\Phi^{\dagger}_{cK} \right] +\ldots \bigg\}\,.
\end{eqnarray}
Here the dots stand for other vertices involving three and more gluons and also ghosts.
We do not display them since they will not appear
in the one-loop calculation we present later.
The $f_{abc}$ are the structure constants of the gauge group $SU(N)$. 
\\

For the study of conformal field theories, it is suitable to write correlators
in coordinate (super-)space, rather than in momentum space.
The coordinate space propagators following from (\ref{N=1action-exp}) are
\be\label{N=1Phiprop}
\VEV{\Phi^{\dagger}_{Ia}(x_{i},\theta_{i},\bar{\theta}_{i})\,\Phi_{b}^{J}(x_{j},\theta_{j},\bar{\theta}_{j})} = -\frac{ \delta_{I}^{J} \delta_{ab}}{4 \pi^2} e^{i (\xi_{ii}+\xi_{jj}-2\xi_{ji})\cdot \partial_{i}} \frac{1}{x_{ij}^2}\,,
\ee
and
\be\label{N=1Vprop}
\VEV{V_{a}(x_{i},\theta_{i},\bar{\theta}_{i})\,V_{b}(x_{j},\theta_{j},\bar{\theta}_{j})} = \frac{\delta_{ab}}{8\pi^2} \frac{\delta(\theta_{ij}) \delta(\bar{\theta}_{ij})}{x_{ij}^2}\,,
\ee
where
\be
x_{ij}= x_{i}-x_{j}\,,\qquad \theta_{ij} = \theta_{i}-\theta_{j}\,, \qquad \xi^{\mu}_{ij}=\theta^{\alpha}_{i}\sigma^{\mu}_{\alpha \dot{\alpha}}\bar{\theta}^{\dot{\alpha}}_{j}\,.
\ee
The Feynman rules can be read off from (\ref{N=1action-exp}).

\subsection{Consequences of conformal symmetry for perturbative calculations}
\label{ch-conf-pert}

In this section, we give a number of examples to show how the abstract
conformal correlation functions discussed in section \ref{section-conformalcorr} are implemented in $\cN=4$ SYM.
A subtlety arises because the gauge fixing procedure usually (e.g. when using a covariant gauge) breaks the conformal
invariance of the action following from (\ref{N=4 action}).
Therefore, the conformal predictions for correlation functions of section \ref{section-conformalcorr}
only apply to gauge invariant quantities, in which the gauge dependent variation of the
gauge fixing term drops out.
\\

In the next section we give examples of gauge-invariant
operators. Then, we introduce the concept of anomalous dimensions and
briefly discuss operator mixing in a conformal field theory.
Finally, in section \ref{N=1four-point1loop} we show how a
particular four-point function can be used to compute anomalous
dimensions using the operator product expansion.

\subsubsection{Gauge invariant operators}
From what was just said it is clear that in order for conformal properties to
be manifest, one should consider gauge independent quantities.
There are different possibilities.
For example, one can take a trace
over several elementary fields multiplied together at the same space-time point.
For the gauge group $SU(N)$, the minimal number of fields is two, since ${\rm Tr}(t_{a})=0$.
\\

Let us give two examples which often appear in the literature on $\cN=4$ SYM. The first,
\be\label{o20}
\mathcal{Q}^{ij}_{\bf 20'} = {\rm Tr}\, \left( \phi^{i} \phi^{j} - \frac{1}{6} \delta^{ij} \phi_{k} \phi^{k} \right)\,,
\ee
is in the representation $\mathbf{20'}$ of $SU(4)$.
It is the lowest component of the energy momentum supermultiplet.
By applying supersymmetry transformations, one can
generate the conserved currents of $\cN=4$ SYM. For example, in the supersymmetry
variation of (\ref{o20}) the $SU(4)$ current of $\cN=4$ SYM appears. It reads
\begin{equation}
J^{\mu B}_{A} = \mbox{Tr} \big\{ {\cal D}^{\mu} \phi_{AC} \phi^{CB}
-
  \phi_{AC} {\cal D}^{\mu} \phi^{CB} - \frac{i}{2} \big( \bar{\lambda}_{A} \sigma^{\mu} \lambda^{B} -\frac{1}{4} \delta_{A}^{B} \bar{\lambda}_{C} \sigma^{\mu} \lambda^{C} \big) \big\}\,,
\end{equation}
and satisfies the conservation equation $\partial_{\mu} J^{\mu B}_{A} = 0$.
From the point of view of representation theory of $PSU(2,2|4)$, $\mathcal{Q}^{ij}_{\bf 20'}$
belongs to a short representation of fixed conformal dimension $\Delta=2$ \cite{Andrianopoli:1999vr}.
Thus, its conformal dimension stays at its classical value to all orders in perturbation theory.
Furthermore, it is known (see e.g. the review \cite{Bianchi:2000vh}) that two- and three-point function of (\ref{o20}) do not receive quantum corrections.
\\

The second example,
\be\label{konishi}
\mathcal{K} = \frac{1}{3}  \mathrm{Tr}\;\left(\phi^{i} {\phi}_{i}\right)\,,
\ee
is a singlet of $SU(4)$. It is the lowest component of the Konishi supermultiplet
\cite{Clark:1978jx,Konishi:1983hf}.
\\

Finally, one can also define non-local gauge invariant operators. An example is the
Wilson loop
\be\label{wilsonloop}
W[C] = \frac{1}{N} {\rm Tr}\,{\rm P}\, \textrm{exp}\;\left(i g \oint_{C} dx_{\mu} A^{\mu} \right)\,,
\ee
which is a functional of the (closed) contour $C$. Wilson loops defined on particular
polygonal contours are in fact one of the main objects of study of this
thesis. We will discuss them in much more detail in section \ref{ch-WL}.

\subsubsection{Anomalous dimensions}\label{section-andim}

In a quantum field theory, the conformal dimension $\Delta$ of an
operator depends in general on the coupling constant, and one writes
\be \label{defandim}
\Delta(g) = \Delta_{0} + \gamma(g)\,,
\ee
where $\Delta_{0}$ is the classical, i.e. {\it tree-level} dimension and
$\gamma(g)$ is the {\it anomalous} dimension of the operator.
\\

Let us start with the operators $\mathcal{Q}^{ij}_{\bf 20'}$,
for which $\Delta(g) = \Delta_{0}$.
For practical calculations, it is convenient to employ the
$\cN=1$ superfield formalism presented in section \ref{N=4inN=1}.
Since the $SU(4)$ symmetry of $\cN=4$ SYM is not manifest in
this formalism, we have to decompose $SU(4) \rightarrow SU(3) \times U(1)$.
Under this decomposition, $\mathcal{Q}^{ij}_{\bf 20'}$ gives rise to the following operators (among others):
\be \label{o20defsu3}
\mathcal{C}^{IJ} = {\rm Tr} \left( \Phi^{I} \Phi^{J} \right) \,,\qquad \mathcal{C}^{\dagger}_{IJ} =
{\rm Tr} \left( \Phi^{\dagger}_{I} \Phi^{\dagger}_{J} \right) \,.
\ee
Here $I,J=1,2,3$. It is convenient to use them for quantum calculations since
they are gauge invariant without the need to include gauge links like $\exp(2g V)$.
Their lowest component is
\be \label{o20defsu3l}
{C}^{IJ} = {\rm Tr} \left( \phi^{I} \phi^{J} \right) \,,\qquad {C}^{\dagger}_{IJ} =
{\rm Tr} \left( \phi^{\dagger}_{I} \phi^{\dagger}_{J} \right) \,.
\ee
Let us compute the two-point function of $C^{11}$ and $C^{\dagger}_{11}$
in order to check the absence of perturbative corrections at one loop.
\begin{figure}[htbp]
\psfrag{a}[cc][cc]{$(a)$}
\psfrag{b}[cc][cc]{$(b)$}
\psfrag{c}[cc][cc]{$(c)$}
\centerline{\epsfxsize 5.0 truein \epsfbox{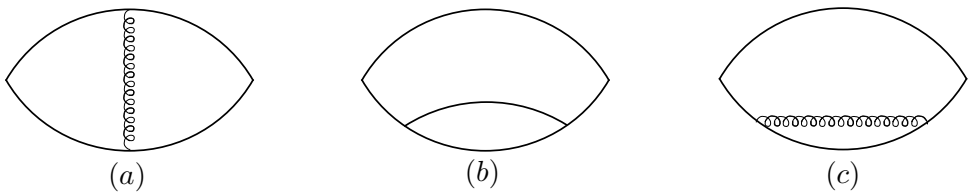}}
\caption{\small Feynman graphs contributing to $\vev{ C^{11}(x_{1}) \,{C}^{\dagger}_{11}(x_{2})}$ and $\VEV{ \mathcal{K}(x_{1}) \mathcal{K}(x_{2})}$ at
one loop. Solid lines denote the chiral superfield propagator (\ref{N=1Phiprop}), wiggly lines the gluon
superfield propagator (\ref{N=1Vprop}).}
\label{Fig:O20-two-point-label}
\end{figure}
We find that the sum of the graphs contributing to the one-loop propagator corrections, Fig. \ref{Fig:O20-two-point-label} (b,c), vanishes identically\footnote{This is true in the Feynman gauge.}.
The graph in Fig. \ref{Fig:O20-two-point-label} (a) leads to an integral that can be shown to vanish for $x_{12}\neq 0$.
Thus, up to contact terms, we have
\be \label{o20twopoint}
\vev{ C^{11}(x_{1}) \,{C}^{\dagger}_{11}(x_{2})} = \frac{c}{(x^2_{12})^2} +O(g^4)\,, \qquad c= \frac{N^2-1}{2 (2\pi)^4}\,.
\ee
As we already said, (\ref{o20twopoint}) is in fact valid to all orders in the coupling
constant. Comparing (\ref{o20twopoint}) to (\ref{two-point}),
one can see that $C^{11}$ has the conformal dimension
$\Delta=2$, which is just the sum of the classical dimensions of its constituents $\phi^{1}$.
\\

The second example is the Konishi operator (\ref{konishi}).
In the $\cN=1$ formalism it is given by
\be \label{Konishi-N=1}
\mathcal{K} = \frac{2}{3}{\rm tr}\left( e^{-2 g V} \Phi^{\dagger}_{I} e^{2 g V} \Phi^{I} \right)\,.
\ee
The exponentials in (\ref{Konishi-N=1}) are needed to make $\mathcal{K}$ gauge invariant.
It has the same classical
dimension $\Delta_{0}=2$ as $\mathcal{Q}^{ij}_{\bf 20'}$, but unlike $\mathcal{Q}^{ij}_{\bf 20'}$, it
does get renormalised.
\begin{figure}[htbp]
\psfrag{a}[cc][cc]{$(a)$}
\psfrag{b}[cc][cc]{$(b)$}
\psfrag{c}[cc][cc]{$(c)$}
\centerline{\epsfxsize 5.0 truein \epsfbox{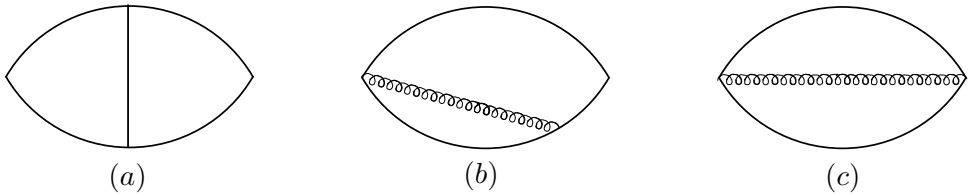}}
\caption{\small Additional Feynman graphs contributing to $\VEV{ \mathcal{K}(x_{1}) \mathcal{K}(x_{2})}$ at
one loop. Solid lines denote the chiral superfield propagator (\ref{N=1Phiprop}), wiggly lines the gluon
superfield propagator (\ref{N=1Vprop}).}
\label{Fig:Konishi}
\end{figure}
Starting from one loop \footnote{By a
calculation to $n$ loops, we mean up to order $g^{2n}$ in the coupling constant. For composite
operators, the pictures would suggest a different loop order.},
$\mathcal{K}$ has UV divergences typical for composite operators.
They come from diagrams (a) and (b) in Figure \ref{Fig:Konishi}.
Evaluating them
in dimensional regularisation, one finds (for simplicity, we will compute the lowest, i.e. $\theta = \bar{\theta}=0$, component of $\VEV{\mathcal{K}(x_{1})\,\mathcal{K}(x_{2})}$ only)
\be \label{Konishi-div}
{\rm Fig.}\ref{Fig:Konishi}(a) + {\rm Fig.}\ref{Fig:Konishi}(b) = \frac{1}{(x_{12}^2)^{2-2\epsilon}} g^2 (x_{12}^2 \mu^2)^{\epsilon} \left[ \frac{1}{\epsilon} \frac{(N^2-1) N}{(2 \pi)^6}+ O(\epsilon^{0}) \right]\,.
\ee
The first factor on the r.h.s. of (\ref{Konishi-div}) accounts for the engineering dimension of
$\VEV{\mathcal{K}(x_{1})\,\mathcal{K}(x_{2})}$ in dimensional regularisation, and in addition one gets a
 factor of $g^2 (x_{12}^2 \mu)^{\epsilon}$ for each loop, with $\mu$ being the dimensional regularisation scale.
Furthermore, diagrams of the same topology as those in Fig.~$\ref{Fig:O20-two-point-label}$ contribute, but the
only difference is the orientation of some matter lines, and they still vanish for $x_{12}\neq 0$.
The graph in Fig. \ref{Fig:Konishi}(c) leads to a finite contribution.
\\

In dimensional regularisation, one defines renormalised operators by multiplying
by a suitable renormalisation factor $Z$ \cite{Collins:1984xc}. This means that the renormalised operators $\left[ \mathcal{K} \right]$
are defined by
\begin{equation}
\left[ \mathcal{K} \right]  = Z \mathcal{K}\,,\qquad Z = 1+ \frac{z_{1;1}}{\epsilon}g^2 + O(g^4)\,.
\end{equation}
The $Z$-factor subtracts the poles in $\epsilon$, so that one is left with
a finite result.
From (\ref{Konishi-div}) we can see that one should take $z_{1;1} = (3 N)/(8\pi^2)$.
Expanding in $\epsilon$, we find for the renormalised operator $\left[ \mathcal{K} \right]$, to one loop
\be \label{Konishi-final}
 \VEV{ [\mathcal{K}](x_{1}) \,[\mathcal{K}](x_{2})} = \frac{N^2-1}{(2\pi)^4}\frac{1}{x^4_{12}}\left[ {c(g^2) -g^2 \frac{N}{4\pi^2} \ln\left(x_{12}^2 \mu^2 \right)}\right] +O(g^4)+O(\epsilon)  \,,
\ee
with $c(g^2)={1}/{3}+g^2 {N}/({8 \pi^2})$.
One might think that the appearance of logarithms is surprising in a conformal field theory.
The reason is that to a given order in the coupling constant, the expansion of (\ref{two-point}) leads to these logarithms.
Indeed, one can rewrite (\ref{Konishi-final}) in the manifestly conformal form
(\ref{two-point}),
\be
\VEV{ [\mathcal{K}](x_{1}) [\mathcal{K}](x_{2})} =  \frac{N^2-1}{(2\pi)^4} \left[ c(g^2) \left(x^2_{12}\right)^{-\Delta_{\mathcal{K}}}  \left(\mu^2 \right)^{-\gamma_{\mathcal{K}}} \right] +O(g^4)+O(\epsilon) \,,
\ee
with the conformal dimension of $\mathcal{K}$ and anomalous dimension $\gamma_{\mathcal{K}}$ at one loop being
\be \label{Kdim}
\Delta_{\mathcal{K}}(g^2) = 2 + \gamma_{\mathcal{K}}(g^2)\,,\qquad \gamma_{\mathcal{K}} = \frac{g^2 N}{8\pi^2}{6} + O(g^4)\,.
\ee

\subsubsection{Operator mixing}
Let us briefly remark on the topic of operator mixing (cf. e.g. \cite{Collins:1984xc}) in the context of
a conformal field theory \cite{Makeenko:1980bh,Ohrndorf:1981qv,Craigie:1983fb,Brodsky:1984xk,Mueller:1993hg}.
Operator mixing occurs, generally speaking, when
there are several
 operators that have the same quantum numbers.
A particular example that often appears in the literature are operators in
the $sl(2)$-sector of $\cN=4$ SYM, see e.g. \cite{Beisert:2003jj,Eden:2005bt}.
Of these, the twist-two operators\footnote{The twist of an operator is defined as the difference between its
classical dimension and its spin.} are built from two elementary fields, e.g.
$\phi^{1}$ and an arbitrary number $j$ of (covariant) derivatives $\mathcal{D}^{\mu}$.
For $j=0$, we already encountered one of these operators in the form of $C^{11} = {\rm} \left( \phi^{1} \phi^{1} \right)$.
For general (even) $j$, they read
\be \label{twist2}
\mathcal{O}_{j} = \sum_{k=0}^{j} c_{jk} O_{jk}\,,\qquad O_{jk} = \,{\rm tr} \left( \mathcal{D}^{\{ \mu_{1}} \ldots \mathcal{D}^{\mu_{k}} \, \phi^{1} \, \mathcal{D}^{\mu_{k+1}} \ldots \mathcal{D}^{\mu_{j} \}}\, \phi^{1}  \right)\,.
\ee
Here, the curly brackets stand for traceless symmetrisation of the vector indices, so that $\mathcal{O}_{j}$
is in a representation of spin $j$.
In a generic situation, there is  more than one possible distribution
of the derivatives on the two fields $\phi^{1}$, which is reflected by the sum in (\ref{twist2}) with \textit{a priori}
arbitrary coefficients $c_{jk} $. As a consequence, the operators ${O}_{jk}$
for a given $j$ mix under renormalisation.
\\

In a generic quantum field theory, the mixing matrix $c_{jk}$ at order $g^n$ can be determined
by a calculation at order $g^{(n+2)}$. In a conformal field theory, this mixing problem can be resolved
using conformal methods that significantly reduce the complexity of the calculations \cite{Makeenko:1980bh,Ohrndorf:1981qv}.
We developed these methods further in \cite{Belitsky:2007jp}. As a particular example, we were able
to determine the mixing matrix $c_{jk}$ at order $g^2$ by evaluating only order
$g$ Feynman graphs, compared to a conventional calculation at order $g^4$.
Similarly, the computation of the anomalous dimensions of the twist two operators in (\ref{twist2}) can also be reduced in loop order (by one unit of $g^2$).
\\

One might object that the resolution of the mixing problem is not relevant,
since it is known that mixing matrices are scheme dependent.
This
scheme dependence, however, is under control and is governed by a renormalisation group equation.
Moreover, one can define a conformal scheme, in which the properties
of conformal correlation functions are realised (see e.g. the review \cite{Braun:2003rp}).
In contrast, in a generic scheme, such as for example minimal subtraction, the conformal properties discussed in section \ref{section-conformalcorr} are in general not present.
\\

The anomalous dimensions of the operators in (\ref{twist2}) have received considerable
attention recently. They are known up to two loops and conjectured at three loops \cite{Kotikov:2004er}.
They are also predicted by a conjectured integrable model \cite{Beisert:2006ez}. It would be desirable to
confirm these conjectures to three loops and beyond. A promising technique to achieve this
is the determination of the anomalous dimensions through the operator product expansion (OPE)
of conformal four-point functions. We conclude this introductory section with a one-loop example
of this technique.

\subsubsection{Conformal four-point functions}
\label{N=1four-point1loop}
Let us consider the four-point function of $\mathcal{Q}^{ij}_{\bf 20'}$.
We already saw in section \ref{section-andim} that
$\mathcal{Q}^{ij}_{\bf 20'}$ does not need to be renormalised.
Its conformal dimension is equal
to its classical dimension, $\Delta=2$.
In this case, from formula (\ref{conf-4point})
we find for the four-point correlator
\be \label{chiral4point}
\VEV{C^{11}(x_{1})\,C^{22}(x_{2})\,{C^{\dagger}}_{11}(x_{3})\,{C^{\dagger}}_{22}(x_{4})} = \frac{1}{x_{12}^2 x_{34}^2 x_{13}^2 x_{24}^2}
f(u,v;g^2)
\ee
The correlator (\ref{chiral4point}) is known up to two loops in perturbation theory
\cite{Eden:2000mv,Bianchi:2000hn}. One reason for being interested in it is that the function $f(u,v;g^2)$ allows to determine the anomalous dimensions of the twist-two operators (\ref{twist2}) via the operator
product expansion (OPE). This relation has been worked out in \cite{Arutyunov:2000ku,Dolan:2004iy}.
\begin{figure}[htbp]
\psfrag{x1}[cc][cc]{$x_{1}$}
\psfrag{x2}[cc][cc]{$x_{2}$}
\psfrag{x4}[cc][cc]{$x_{3}$}
\psfrag{x3}[cc][cc]{$x_{4}$}
\psfrag{x5}[cc][cc]{$x_{5}$}
\psfrag{x6}[cc][cc]{$x_{6}$}
\centerline{\epsfxsize 2.0 truein \epsfbox{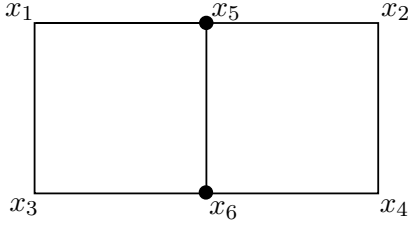}}
\caption{\small The only connected Feynman graph contributing to $\VEV{C^{11}(x_{1})\,C^{22}(x_{2})\,{C^{\dagger}}_{11}(x_{3})\,{C^{\dagger}}_{22}(x_{4})}$ at
one loop. Solid lines denote the chiral superfield propagator (\ref{N=1Phiprop}). The dots represent
the (chiral) superspace integrations at points $x_{5}$ and $x_{6}$.}
\label{Fig:chirals-four}
\end{figure}

As was already discussed, the two-point
functions of chiral primary operators that could in principle
contribute to the disconnected piece of (\ref{chiral4point})
vanish (up to possible contact terms, see e.g. \cite{Penati:2000zv} for a
discussion).
The only nonvanishing Feynman graph contributing to (\ref{chiral4point}) at one loop is shown in Fig. \ref{Fig:chirals-four}. For (\ref{chiral4point}) we need
its lowest component only, i.e. we can set all external $\theta$'s and $\bar{\theta}$'s to
zero. In that case, using the superfield propagator (\ref{N=1Phiprop}), we obtain
\be \label{B1step1}
{\rm Fig.}\, \ref{Fig:chirals-four} = \frac{g^2 N (N^2-1)}{2 (2\pi)^{12}}\frac{1}{ x_{12}^2 x_{34}^2} \, \int d^{4}x_{5} d^{4}x_{6} \int d^2\theta_{5} d^2\bar{\theta}_{6} \, \exp{\left(i \theta_{5} \partial_{56} \bar{\theta}_{6}\right)}\, \frac{1}{x_{15}^2 x_{25}^2 x_{56}^2 x_{63}^2 x_{64}^2}\,.
\ee
The notation $\partial_{56}$ means that the derivative acts on the term $1/x_{56}^2$ only.
The integrals over $\theta_{5}$ and $\bar{\theta}_{6}$ `see' only the quadratic term in
the exponential, i.e.
\be
\int d^2\theta_{5} d^2\bar{\theta}_{6} \, \exp{\left(i \theta_{5} \partial_{56} \bar{\theta}_{6} \right)} = \frac{1}{2}\, \square_{56}\,,
\ee
where $\square = \partial^{\mu} \partial_{\mu}$.
Then, we use that
\be
\square \frac{1}{x^2} = -4 \pi^2 \delta^{(4)}(x)
\ee
to undo one of the space-time integrals in (\ref{B1step1}). We arrive at
\be
{\rm Fig.}\, \ref{Fig:chirals-four} = -\frac{g^2 N (N^2-1)}{2 (2\pi)^{11}}\frac{1}{ x_{12}^2 x_{34}^2} \, h^{(1)}(x_{1},x_{2},x_{3},x_{4}) \,,
\ee
with
\be
h^{(1)}(x_1,x_2,x_3,x_4) = \int \frac{d^4 x_5}{x_{15}^2 x_{25}^2 x_{35}^2
x_{45}^2} = \frac{1}{x_{13}^2 x_{24}^2} \Phi^{(1)}(u,v)\,.
\label{phi1}
\ee
Here $x_{ij} = x_i - x_j$ and the conformal cross-ratios $u$ and $v$
were defined in (\ref{cro}).
\begin{figure}[htbp]
\begin{center}
\ \psfig{file=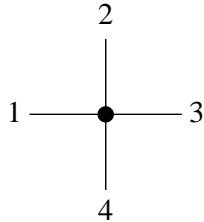}
\end{center}
\caption{\small The one-loop ladder integral. Each line represents a propagator
with the integration point given by a solid vertex. The reason for the
names ladder and box is clearer in the momentum representation of the
same integral.}
\label{figure:1ladder}
\end{figure}
The fact that the integral is characterised by a single
function of two variables follows from its conformal covariance
\cite{Broadhurst:1993ib}. Indeed, performing a conformal inversion on
all points,
\be
x^\m \longrightarrow  \frac{x^\m}{x^2} \implies x_{ij}^2
\longrightarrow \frac{x_{ij}^2}{x_i^2 x_j^2},\hspace{10pt}
d^4 x_5 \longrightarrow \frac{d^4 x_5}{x_5^8} ,
\ee
we find that the integral transforms
covariantly with weight one at each point,
\be
h^{(1)}(x_1,x_2,x_3,x_4) \longrightarrow x_1^2 x_2^2 x_3^2 x_4^2
h^{(1)}(x_1,x_2,x_3,x_4) .
\ee
Since rotation and translation invariance are manifest, we
conclude
that the integral is given by a conformally covariant combination of
propagators multiplied by a function of the conformally
invariant cross-ratios (\ref{cro}), in agreement with (\ref{chiral4point}).
\\

The function $\Phi^{(1)}(u,v)$ has been calculated in
\cite{tHooft:1978xw,Usyukina:1992jd}, where it was also shown that
the same function appears in a three-point integral. The latter
can be obtained  from the four-point one by sending one of the
points to infinity \cite{Broadhurst:1993ib}. We can multiply
equation (\ref{phi1}) by $x_{13}^2$, say, and then take the limit
$x_3 \longrightarrow \infty$. This gives,
\be\label{phi^3pt}
h_{\rm 3pt}^{(1)}(x_1,x_2,x_4) = \lim_{x_3 \rightarrow \infty}
x_{13}^2 h^{(1)}(x_1,x_2,x_3,x_4) = \int \frac{d^4x_5}{x_{15}^2
x_{25}^2
  x_{45}^2} = \frac{1}{x_{24}^2} \Phi^{(1)}(\hat u,\hat v),
\ee where the cross-ratios $u$ and $v$ have become $\hat u$ and
$\hat v$ in the limit, \be u\longrightarrow \hat u =
\frac{x_{12}^2}{x_{24}^2}, \hspace{20pt} v\longrightarrow \hat v =
\frac{x_{14}^2}{x_{24}^2}. \ee Thus the three-point integral
contains the same information as the four-point integral, i.e. the
same function of two variables. The reason is that one can use
translations and conformal inversion to take the point $x_3$ to
infinity and the function of the cross-ratios is invariant under
these transformations. The explicit formula for $\Phi^{(1)}$ is
\cite{Usyukina:1993ch} \footnote{Valid for $u,v>0$, see
\cite{Duplancic:2002dh} for a discussion of the analytic
continuation.}
\begin{eqnarray} \label{phi1formula}
\Phi^{(1)}(u,v) &=& \frac{1}{\lambda} \bigg[ 2 \big( \PLN{2}{-\rho \, u} + \PLN{2}{-\rho \, v}\big) +\ln \frac{v}{u} \ln \frac{1+\rho v}{1+\rho u} + \ln(\rho u) \ln(\rho v) + \frac{\pi^2}{3} \bigg] \,,
\end{eqnarray}
where
\begin{equation}
\lambda(u,v) = \sqrt{(1-u-v)^2-4 uv}\,,\qquad \rho(u,v) = 2 (1-u-v+\lambda)^{-1} \,.
\end{equation}
In order to extract the anomalous dimensions
of twist two operators, we need $\Phi^{(1)}(u,v)$ in the OPE limit $x_{12} \rightarrow 0$, i.e. $u\rightarrow 0$, only.
In this limit, one finds (taking $0<v<1$ for simplicity)
\be \label{phi1OPE}
\Phi^{(1)}(u,v) = \frac{1}{1-v} \left[  \ln\left({u}\right) {\ln(v)} + 2 {\rm Li}_{2}\left(1-v\right)  \right] + O(u)\,,
\ee
Using the results of \cite{Dolan:2004iy}, one can deduce from (\ref{phi1OPE}) the well-known
expression for the one-loop anomalous dimensions of the twist-two operators (\ref{twist2}),
\be \label{twist-two-andim}
\gamma(j) = 4\, a  \sum_{k=1}^{j}\frac{1}{k} + O(a^2)\,,\quad a = \frac{g^2 N}{8 \pi^2}\,.
\ee
One can in principle determine the anomalous dimensions at higher orders
in perturbation theory from the four-point function (\ref{chiral4point}).
It has been computed to two loops in perturbation theory \cite{Eden:2000mv,Bianchi:2000hn}.
It would be very interesting to determine its three-loop, i.e. order $g^6$,
correction in order to test the conjectured formula \cite{Kotikov:2004er} for the
twist-two anomalous dimensions at three loops. Moreover, one may wonder whether
the conjectured integrability of $\mathcal{N}=4$ SYM contrains the four-point function
itself, beyond the contraints coming from conformal symmetry?

\subsection{Conformal four-point integrals}
\label{section-conformalfourpoint}
In section \ref{N=1four-point1loop} we saw that from the four-point function of chiral primary operators  (\ref{chiral4point}) one can deduce the anomalous dimensions of the twist-two operators (\ref{twist2}).
Due to the absence of divergences of the chiral primaries, such four-point functions are finite and do
not require any regulator. Because of its conformality, a calculation at a given loop order
(e.g. with Feynman diagrams) must eventually yield an expression in terms of (finite) conformal
integrals. In \cite{Drummond:2006rz}, we took this as a motivation to study an infinite class of conformal
four-point integrals. We found that there are simple identities between integrals corresponding
to different diagrams. We describe these `magic identities' in this section.
\\

At the same time, let us stress that in this report, the interest in the conformal integrals mainly comes from
an entirely different motivation. We found that, very surprisingly, conformal integrals
also appear in the completely unrelated context of on-shell gluon scattering amplitudes
in momentum space.
As will be explained in section \ref{ch-pseudoconformal}, their appearance suggests that the latter have a (broken) conformal symmetry in a dual
space defined through the gluon momenta.
We hope that the example of (\ref{chiral4point}), which illustrates where
conformal integrals normally appear (off-shell, finite, in $x$-space)
helps clarifying the difference to the integrals that appear when
analysing on-shell gluon scattering amplitudes (which are on-shell, IR-divergent, and defined in momentum space) in section \ref{ch-pseudoconformal}.

\subsubsection{Proof of conformality}
We will discuss an infinite class of conformal four-point integrals in
four dimensions\footnote{In this section we consider and prove
  identities for Euclidean integrals. The corresponding Minkowskian
  version of the identities can be obtained through Wick rotation of the
  integrals. In the Euclidean context we consider integrals with
  separated external points, $x_{ij}\neq 0$. This is the Euclidean
  analogue of the off-shell regime, $x_{ij}^2 \neq 0$, for a
  Minkowskian integral.},
each of which is essentially described by a function of two
variables.
We already discussed the simplest example, the one-loop ladder
integral (\ref{phi1}), in section \ref{N=1four-point1loop}. It is the first in an infinite series of conformal
integrals, the $n$-loop ladder (or scalar box) integrals, which have all
been evaluated \cite{Usyukina:1993ch}.
In particular the 2-loop ladder integral is given by
\be
h^{(2)}(x_1,x_2,x_3,x_4) = x_{24}^2 \int \frac{d^4 x_5 d^4 x_6}{x_{15}^2
  x_{25}^2 x_{45}^2 x_{56}^2 x_{26}^2 x_{46}^2 x_{36}^2} =
\frac{1}{x_{13}^2 x_{24}^2} \Phi^{(2)}(u,v)\,.
\label{2ladder}
\ee
The prefactor $x_{24}^2$ is present to give conformal weight one at
each external point.
\begin{figure}[htbp]
\begin{center}
\ \psfig{file=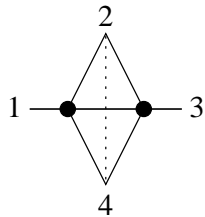}
\end{center}
\caption{\small The two-loop ladder integral. The dashed line represents the
  numerator $x_{24}^2$.}
\label{figure:2ladder}
\end{figure}

\begin{figure}[htbp]
\begin{center}
\ \psfig{file=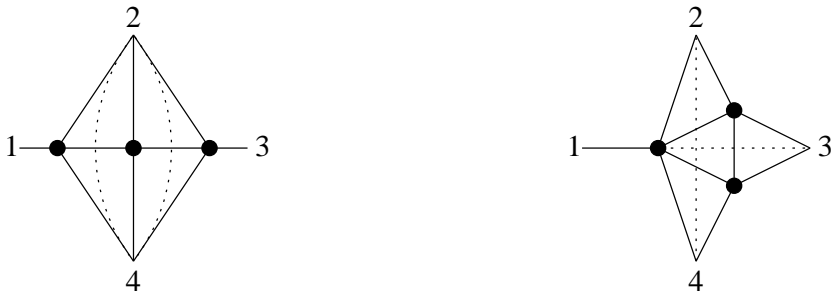}
\end{center}
\caption{\small Two examples of three-loop conformal four-point integrals, the
three-loop ladder and the `tennis-court'.}
\label{figure:3loop-conformal}
\end{figure}

Again conformal transformations can be used to justify the appearance
of the 2-variable function $\Phi^{(2)}$.
The r.h.s. of (\ref{2ladder}) is invariant under the
pairwise swap $x_1 \longleftrightarrow x_2$,
$x_3 \longleftrightarrow x_4$, hence
\be
h^{(2)}(x_2,x_1,x_4,x_3)=h^{(2)}(x_1,x_2,x_3,x_4)\,. \label{turning}
\ee
This symmetry is not immediately evident from the integral. It is its
conformal nature which allows this identification.
\begin{figure}[htbp]
\begin{center}
\ \psfig{file=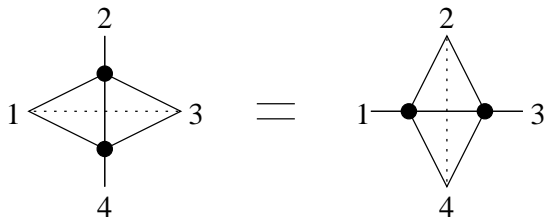}
\end{center}
\caption{\small The two-loop turning identity obtained from the
  pairwise point swap,
  $x_1 \longleftrightarrow x_2$, $x_3 \longleftrightarrow x_4$.}
\label{figure:turning}
\end{figure}
At three loops we consider two conformal integrals, the three-loop ladder,
\be
h^{(3)}(x_1,x_2,x_3,x_4) = x_{24}^4 \int \frac{d^4 x_5 d^4 x_6
  d^4 x_7}{x_{15}^2
  x_{25}^2 x_{45}^2 x_{56}^2 x_{26}^2 x_{46}^2 x_{67}^2 x_{27}^2 x_{47}^2
  x_{37}^2} = \frac{1}{x_{13}^2 x_{24}^2} \Phi^{(3)}(u,v)\,,
\ee
and the so-called `tennis court' \cite{Bern:2005iz},
\be
g^{(3)}(x_1,x_2,x_3,x_4) = x_{24}^2 \int \frac{x_{35}^2\ d^4 x_5 d^4 x_6
  d^4 x_7
  }{x_{15}^2 x_{25}^2 x_{45}^2 x_{56}^2 x_{57}^2 x_{67}^2 x_{26}^2
  x_{47}^2 x_{36}^2 x_{37}^2} = \frac{1}{x_{13}^2 x_{24}^2}
\Psi^{(3)}(u,v)\,, \label{tc}
\ee
which are shown in Fig. \ref{figure:3loop-conformal}.
Notice the presence of the numerator $x_{35}^2$ in the integrand
of the tennis court. It is needed to balance the conformal weight
of the five propagators coming out of point 5.

\subsubsection{`Magic identities' between conformal integrals}
\label{ch-magic}
\begin{figure}[htbp]
\begin{center}
\ \psfig{file=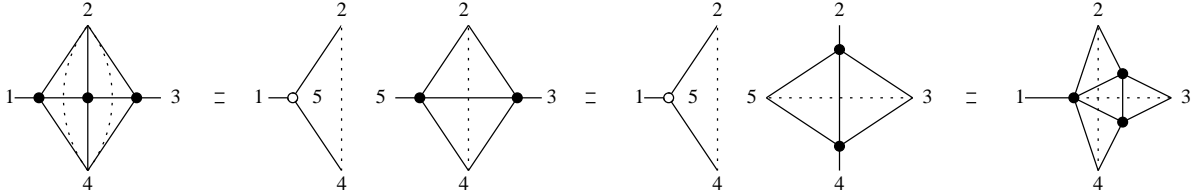}
\end{center}
\caption{\small Diagrammatic representation of the proof of equality of the
  tennis court and the three-loop ladder. The identity follows from the
  turning identity (\ref{turning}) for the two-loop subintegral.}
\label{figure:proof}
\end{figure}

We will show that the three-loop ladder and the
tennis court are in fact the same, i.e. we will prove $\Phi^{(3)} =
\Psi^{(3)}$.  First we shall present a diagrammatic argument. We
consider the $n$-loop ladder as being iteratively constructed
from the $(n-1)$-loop ladder by integrating against a `slingshot' (the
`0-loop' ladder is a product of free propagators). For
example we write the three-loop ladder as
\be
h^{(3)}(x_1,x_2,x_3,x_4) = x_{24}^2 \int \frac{d^4x_5}{x_{15}^2 x_{25}^2
  x_{45}^2} \Bigl( x_{24}^2 \int \frac{d^4x_6 d^4x_7}{x_{56}^2 x_{26}^2
  x_{46}^2 x_{67}^2 x_{27}^2 x_{47}^2 x_{37}^2}\Bigr)\,,
\ee
where inside the parentheses we recognise the two-loop ladder integral
(\ref{2ladder}).
\begin{figure}[htbp]
\begin{center}
\ \psfig{file=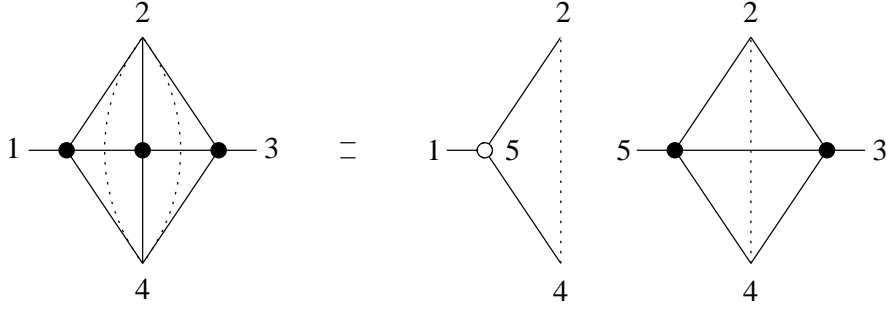}
\end{center}
\caption{\small The three-loop ladder expressed as the integral of the
  two-loop ladder against the `slingshot'. The empty vertex is the point $x_5$
  which must be identified with the point $x_5$ from the two-loop
  ladder sub-integral before being integrated over.}
\label{figure:slingshot}
\end{figure}
We can then show the equality of the three-loop ladder and the tennis court by
using the turning symmetry (\ref{turning}) on the two-loop ladder
sub-integral. Then the tennis court integral (\ref{tc}) can be recognised as
the turned two-loop ladder integrated against the slingshot,
\begin{align}
h^{(3)}(x_1,x_2,x_3,x_4) &= x_{24}^2 \int \frac{d^4x_5}{x_{15}^2 x_{25}^2
  x_{45}^2} h^{(2)}(x_5,x_2,x_3,x_4)\,, \notag \\
&= x_{24}^2 \int \frac{d^4x_5}{x_{15}^2 x_{25}^2 x_{45}^2}
  h^{(2)}(x_2,x_5,x_4,x_3)\,, \notag \\
&= g^{(3)}(x_1,x_2,x_3,x_4)\,. \label{proof}
\end{align}
This proof can be more easily seen in the diagram (Fig. \ref{figure:proof}).
In using the turning identity (\ref{turning}) we have ignored the
possibility of contact terms. These could, in principle, spoil the
derivation of identities like $\Phi^{(3)}=\Psi^{(3)}$ as the proof
(\ref{proof}) involves turning a subintegral. Contact terms could then
generate regular terms upon doing one further integration. We now give
an argument why this cannot happen for any conformal four-point
integral. We again use the example of the 3-loop ladder and tennis
court identity.
\\

Consider inserting the $n$-loop subintegral (the 2-loop ladder in this
case) into an H-shaped frame with a dashed line across the top, as
illustrated below. This generates an $(n+2)$-loop integral which is
conformal with weight 1 at each external point (provided the
subintegral is conformal with weight 1 at each external point).
\begin{figure}[htbp]
\begin{center}
\ \psfig{file=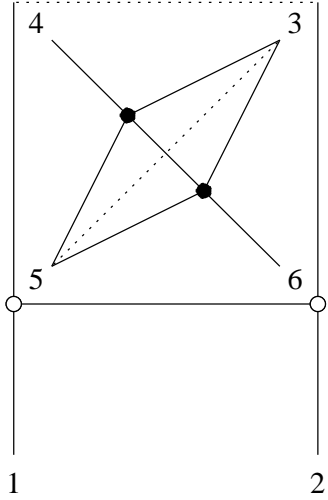}
\end{center}
\caption{\small The 2-loop ladder inserted into an H-shaped frame,
  generating a 4-loop integral.}
\label{figure:Hframe}
\end{figure}
When inserting the 2-loop ladder in this way the 4-loop integral one
obtains is
\begin{align}
f^{(4)}(x_1,x_2,x_3,x_4)&=
x_{34}^2 \int \frac{d^4x_5 d^4x_6} {x_{15}^2 x_{45}^2 x_{56}^2
  x_{26}^2 x_{36}^2} x_{35}^2 \int \frac{d^4x_7 d^4x_8}{x_{67}^2
  x_{57}^2 x_{37}^3 x_{78}^2 x_{58}^2 x_{38}^2 x_{48}^2} \notag \\
&= \frac{1}{x_{13}^2 x_{24}^2}f(u,v)\,.
\label{f4int}
\end{align}
As usual, the second equality follows from conformality.
Now we consider the action of $\Box_1$ on the above integral using
\be
\Box \frac{1}{x^2} = - 4 \pi^2 \delta^{(4)}(x)\,.
\ee
On the integral one obtains
\be
-4\pi^2 \frac{x_{34}^2 x_{13}^2}{x_{14}^2} \int \frac{d^4x_6 d^4x_7
  d^4x_8}{x_{26}^2
  x_{16}^2 x_{36}^2 x_{67}^2 x_{17}^2 x_{37}^2 x_{78}^2 x_{18}^2
  x_{38}^2 x_{48}^2} = -\frac{4 \pi^2 x_{34}^2}{x_{13}^4 x_{14}^2
  x_{24}^2} \Phi^{(3)}(u,v)\,.
\ee On the functional form of (\ref{f4int}) one uses the chain
rule to derive the action of a differential operator on the
function $f$. In this way we find the differential equation, \be
\frac{x_{23}^2 x_{34}^2}{x_{13}^6 x_{24}^4} \D^{(2)}_{uv} f(u,v) =
-\frac{\pi^2 x_{34}^2}{x_{13}^4 x_{14}^2 x_{24}^2}
\Phi^{(3)}(u,v)\,. \label{pdephi} \ee The operator $\D^{(2)}_{uv}$
is given explicitly by \be \D^{(2)}_{uv} = u \del_u^2 + v\del_v^2
+ (u+v-1)\del_u \del_v + 2\del_u + 2\del_v\,. \ee Similarly we can
act with $\Box_2$ on the 4-loop integral to obtain the following
integral, \be -4\pi^2 \frac{x_{34}^2}{x_{23}^2}\int \frac{d^4x_5
d^4x_7 d^4x_8
  x_{35}^2}{x_{15}^2
  x_{25}^2 x_{45}^2 x_{57}^2 x_{58}^2 x_{78}^2 x_{27}^2 x_{48}^2
  x_{37}^2 x_{38}^2} = -\frac{4\pi^2 x_{34}^2}{x_{23}^2 x_{13}^2
  x_{24}^4} \Psi^{(3)}(u,v)\,,
\ee and the corresponding differential equation, \be
\frac{x_{14}^2 x_{34}^2}{x_{24}^6 x_{13}^4} \D^{(2)}_{uv} f(u,v) =
-\frac{\pi^2 x_{34}^2}{x_{23}^2 x_{13}^2 x_{24}^4}
\Psi^{(3)}(u,v)\,. \label{pdepsi} \ee From
(\ref{pdephi},\ref{pdepsi}) it follows that
$\Phi^{(3)}=\Psi^{(3)}$, the point being that one obtains the {\it
  same} differential operator $\D^{(2)}_{st}$ under the two $\Box$
operations. The argument has the obvious generalisation of placing any
conformal integral (in any orientation) inside the frame.
This argument indirectly shows that the previous argument
(\ref{proof}) based on turning the subintegral cannot suffer from
contact term contributions.
\\

\begin{figure}[htbp]
\centerline{\epsfxsize 5.0 truein \epsfbox{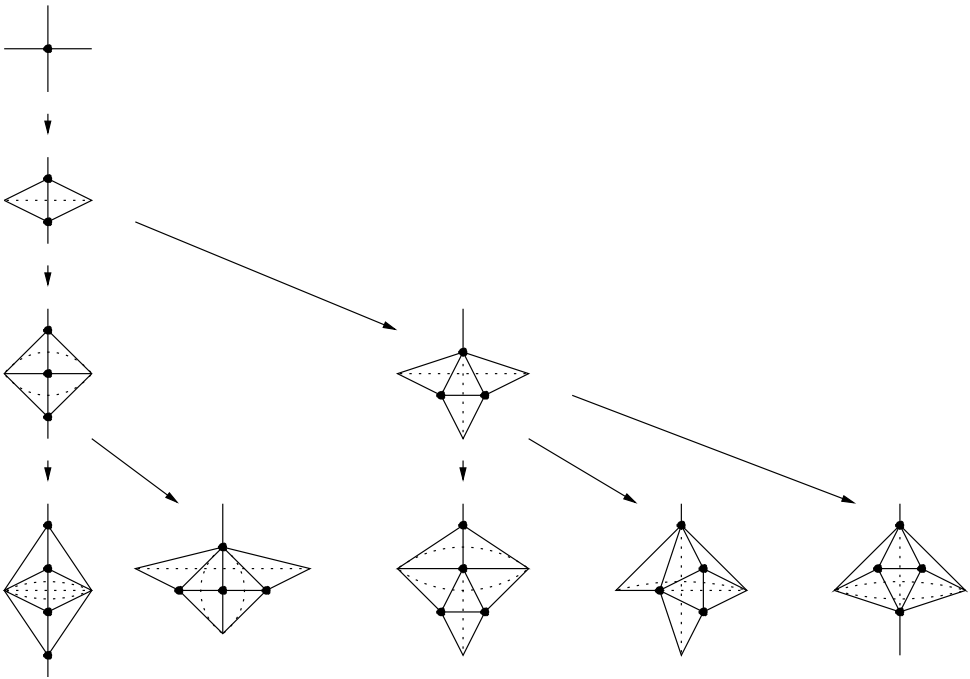}}
\caption{\small The integrals in a given row are all equivalent.
They
  generate the integrals in
the next row by being integrated in all possible orientations
against
  the slingshot attached from above. The ladder series is in the
  left-most column.}
\label{figure:cascadex}
\end{figure}

The identity we have obtained at three loops is just the first example
of an infinite set of identities which all come from the turning
symmetry of subintegrals. We generate $(n+1)$-loop
integrals by integrating $n$-loop integrals against the slingshot in
all possible orientations. The resulting integrals are equal by
turning identities of the form (\ref{turning}). At two loops we get
just one integral (the two-loop ladder). At three loops we have
already seen two equivalent integrals (ladder and tennis
court). At four loops we generate two equivalent integrals from the
three-loop ladder and three equivalent integrals from the tennis
court. Finally, all five four-loop integrals obtained in this way are
equivalent by the three-loop identity for the ladder and tennis court
(see Fig. \ref{figure:cascadex}).
\\

\begin{figure}[htbp]
\hspace{-0.5cm}\centerline{\epsfxsize 4.0 truein
\epsfbox{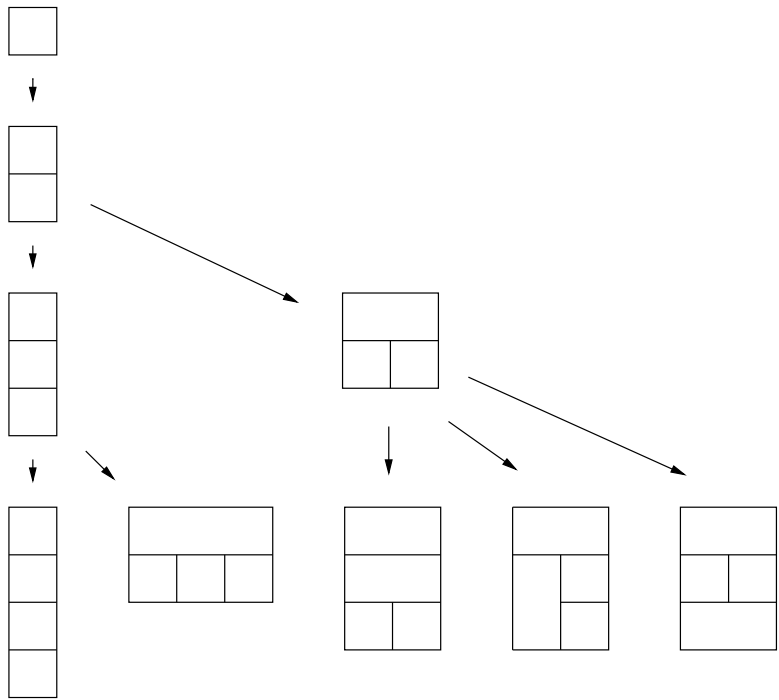}} \caption{\small The momentum notation
for our integrals up to four
  loops. The slingshot translates into the top box in each
  diagram, beneath which are the integrals at one loop lower, arranged
  in all possible orientations. The ladder series is again in the
  left-most column.}
\label{figure:cascadep}
\end{figure}

In general it is more common to give the diagrams in the
`momentum' representation (which has nothing to do with the
Fourier transform) where we regard the integrations as integrals
over loop momenta rather than coordinate space vertices. This
representation is neater but the numerators need to be described
separately as they do not appear in the diagrams.
The momentum-space version of the four generations of integrals
from Fig. \ref{figure:cascadex} is 
given in Fig. \ref{figure:cascadep}. The transition between the
two notations will be discussed in more detail in section
\ref{ch-amplitudes}. As was already mentioned, integrals related to the ones
discussed here will reappear in the analysis of loop corrections
to gluon scattering amplitudes in section \ref{ch-pseudoconformal}.

%
%

\section{Gluon amplitudes in $\cN=4$ SYM}
\label{ch-amplitudes}
%
\begin{figure}[t]
\psfrag{p1}[cc][cc]{$p_{1}$}
\psfrag{p2}[cc][cc]{$p_{2}$}
\psfrag{pN}[cc][cc]{$p_{n}$}
\psfrag{dots}[cc][cc]{$\ldots$}
\psfrag{AN}[cc][cc]{$\mathcal{A}_{n}=$}
\centerline{\epsfxsize 3.8 truein \epsfbox{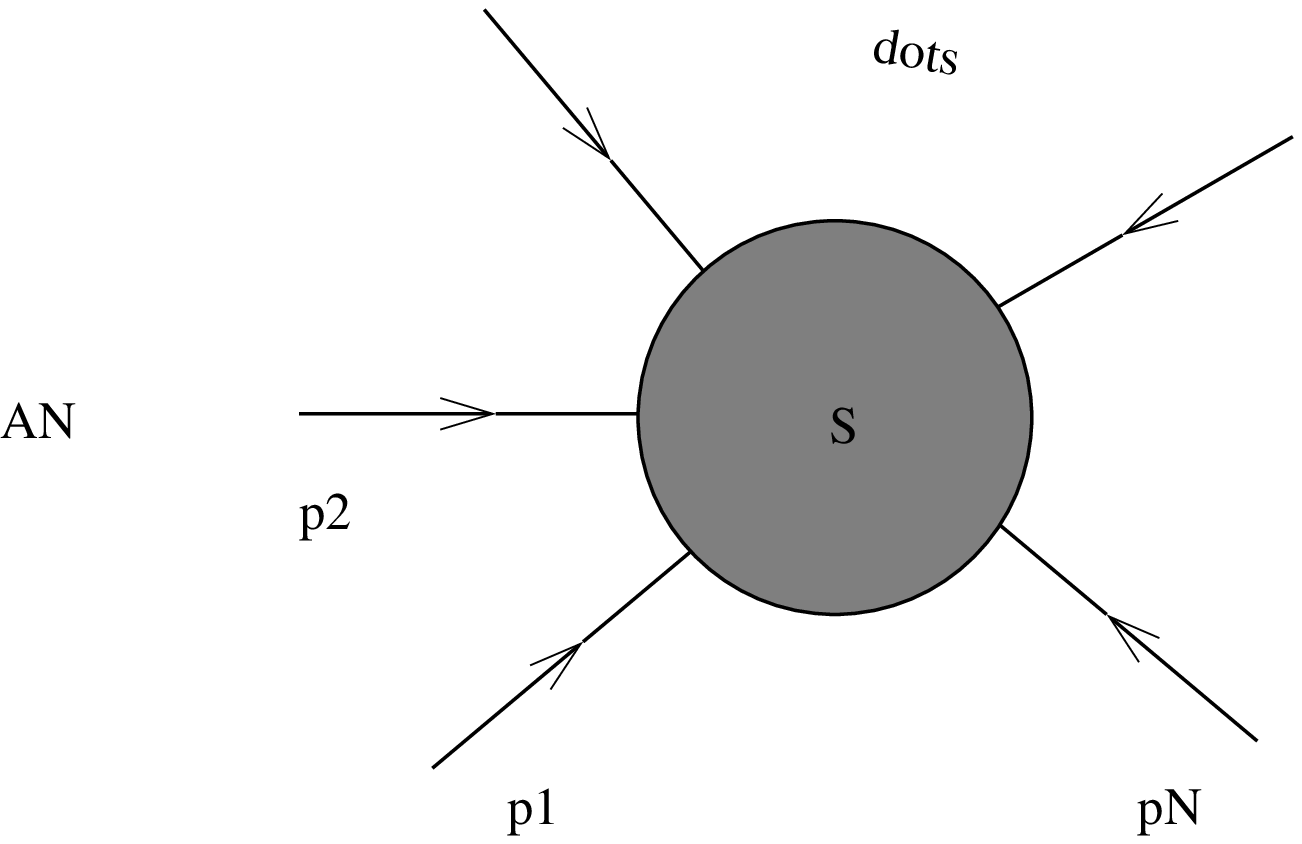}}
\caption{$n$-gluon scattering}
\label{S-mat-Fig}
\end{figure}
\subsection{Introduction}

What are the motivations for computing scattering amplitudes in
$\cN=4$ SYM? There are several answers to this question. Firstly,
being four-dimensional gauge theories, $\cN=4$ SYM and QCD share
some properties, but calculations are generally much harder in the
latter. Having efficient calculational methods for computing QCD
amplitudes is important for collider physics. The reason is that
in searches for new physics, the theoretical uncertainties for the
standard model background processes are often high, in particular
in proton-proton collisions \cite{Bern:2008ef}. $\cN=4$ SYM can be
used as a testing ground for new calculational methods. Moreover,
a part of QCD amplitudes can be predicted from their $\cN=4$
counterparts. For example, tree-level gluon amplitudes in pure
Yang-Mills are identical to those in $\cN=$ SYM. This follows 
simply from the Feynman rules. Further, at one loop, gluon amplitudes in pure
Yang-Mills can be decomposed in the following way:
\be\label{susy-decomp} A_{g} = (\underbrace{A_{g} + 4 A_{f} + 3
A_{s}}_{\cN =4}) - 4 (\underbrace{A_{f} + A_{s}}_{\cN=1}) +
{A_{s}}\,. \ee Here the abbreviations $g,f,s$ stand for gluon,
fermion, and scalar, respectively. They denote the particles
circulating in the one-loop diagrams. In this way, a generic
one-loop Yang-Mills amplitude can be written in terms of two
supersymmetric amplitudes and an amplitude with a scalar
circulating in the loop, which is much simpler than the original
amplitude.
\\

Another motivation comes from the fact that, as we will discuss in
this section, scattering amplitudes contain the cusp anomalous
dimension $\Gamma_{\rm cusp}$, which has received considerable
attention over the last years in the study of the AdS/CFT
correspondence. Its value is predicted (in principle at any given
order) from conjectured integrable models that describe the
spectrum of anomalous dimensions in $\cN=4$ SYM. Therefore,
knowing $\Gamma_{\rm cusp}$ to high orders in perturbation theory
is important to test and fine-tune these models. The three- and
four-loop values of $\Gamma_{\rm cusp}$ were indeed computed via
four-gluon scattering amplitudes.\\

Gluon scattering amplitudes in $\cN=4$ are also studied within
the relation between $\cN=4$ SYM and $\cN=8$ supergravity \cite{Kawai:1985xq}, see
\cite{Bern:2002kj} for a review. This may give new insights into the
ultraviolet properties of $\cN=8$ supergravity, see \cite{Bern:2007hh} and
references therein for a recent discussion.\\

Finally, the scattering amplitudes themselves reveal interesting
properties, on which we will focus in this report. As will be
described in this section, an iterative structure for the
scattering amplitudes in $\cN=4$ SYM was uncovered by Anastasiou,
Bern, Dixon and Kosower (ABDK) \cite{Anastasiou:2003kj} and
generalised to higher loops by Bern, Dixon and Smirnov (BDS)
\cite{Bern:2005iz}. In particular, it turns out that their finite
part seems to be much simpler than could be expected on general
grounds. These discoveries give hope that scattering amplitudes in
the large $N$ limit of $\cN=4$ SYM may be solvable.
\\

Let us consider an $n$-gluon scattering amplitude in $\cN=4$ SYM,
as shown in Fig. \ref{S-mat-Fig}. All gluons are treated as
ingoing, furthermore they are massless and on-shell. They are
characterised by their momenta $p_{i}$ (with $p_{i}^2=0$), their
helicity $h_{i}=\pm 1$ and their colour index (in the adjoint
representation of the gauge group $SU(N)$). There is momentum
conservation, i.e. $\sum_{i=1}^{n} p_{i}^{\mu} = 0$. Taking the
gluon momenta on-shell introduces infrared (IR) divergences that
have to be regularised. For this one usually uses a supersymmetry
preserving regulator, such as for example dimensional reduction
({or a variant of it, the four-dimensional helicity scheme
\cite{Bern:1991aq,Bern:2002zk})}. We will discuss the structure of
IR divergences in section \ref{scat-ir}.
\\

A generic contribution to an $n$-gluon scattering amplitude can consist of
a vast number of terms. For this reason it is very convenient to classify the
amplitudes according to their quantum numbers.
Let us start by considering the colour structure of the amplitudes. It is clear that the
gauge theory factor of a given contribution to the amplitude can be written
as a trace or a product of several traces over colour matrices. We are interested in the
large $N$ limit, so we neglect multiple traces. It is convenient to define
partial amplitudes which correspond to one particular colour structure,
\be\label{defpartial}
\mathcal{A}_n =  Tr \big[ T^{a_1} T^{a_2} \ldots T^{a_n} \big]
A_n^{h_1,h_2,\ldots,h_n}(p_1,p_2,\ldots,p_n) + \ldots\,.
\ee
The (planar) partial amplitudes $A_n^{h_1,h_2,\ldots,h_n}(p_1,p_2,\ldots,p_n)$ depend on the
helicity configuration and on the momenta of the gluons only.

\subsubsection{Helicity structure}\label{scat-helicity}
The next simplification comes from classifying the different
helicity configurations. In order to do this, let us first
introduce the necessary spinor helicity notation. The massless
gluon momenta $p^{\mu} = \sigma^{\mu}_{\alpha \dot{\alpha}}
p^{\alpha \dot{\alpha}}$ can be written as a product of commuting
spinors, \be \label{massless-p} p^{\alpha \dot{\alpha}} =
\lambda_{p}^{\alpha} \bar{\lambda}_{p}^{\dot{\alpha}}\,. \ee We
also introduce a shorthand notation for spinor products, \be
\label{spinor-prod} \langle p \, q \rangle =
\lambda_{p}^{\alpha}\lambda_{q}^{\beta} \epsilon_{\alpha \beta}
\equiv \lambda_{p}^{\alpha} \lambda_{q\,\alpha} \,,\qquad [{p}\,
{q} ] = \bar{\lambda}_{p}^{\dot\alpha}\bar\lambda_{q}^{\dot\beta}
\epsilon_{\dot\alpha \dot\beta} \equiv
\bar{\lambda}_{p\,\dot{\alpha}}
\bar{\lambda}_{q}^{\dot{\alpha}}\,. \ee This allows to write for
example a scalar product between two light-like vectors $p^{\mu}$
and $q^{\mu}$ as \be (p^{\mu}+q^{\mu})^2 = 2 p\cdot q = 2 \langle
p \, q \rangle \, [{p}\, {q} ]\,. \ee The gluons in the partial
amplitudes $A_n^{h_1,h_2,\ldots,h_n}(p_1,p_2,\ldots,p_n)$ are
projected according to their helicities $h_{i}=\pm 1$ with the
following polarisation vectors: \be \label{pol-vectors}
\epsilon_{\mu}^{+}(p) = (\sigma_{\mu})_{\alpha \dot{\alpha}}
\frac{ \lambda_{q}^{\alpha}
\bar{\lambda}_{p}^{\dot{\alpha}}}{\sqrt{2} \langle q \, p\rangle}
\,,\qquad \epsilon_{\mu}^{-}(p) = (\sigma_{\mu})_{\alpha
\dot{\alpha}} \frac{  \lambda_{p}^{\alpha}
\bar{\lambda}_{q}^{\dot{\alpha}}}{\sqrt{2} [ q \, p ]}\,. \ee
Here $q^{\mu}$ is an arbitrary (up to $q^{\mu} \neq p^{\mu}$) reference momentum. A change in
$q^{\mu}$ corresponds to a gauge transformation.
More information on the spinor helicity notation can be found for example in \cite{Risager:2008yz,Dixon:1996wi}.
\\

In $\cN=4$ SYM, one can define analogous scattering amplitudes
where some of the gluons are replaced by fermions or scalars. By
requiring invariance of the $S$-matrix under supersymmetry
transformations and the invariance of the vacuum state $|0
\rangle$, \cite{Grisaru:1976vm,Grisaru:1977px}, one can derive
relations between scattering amplitudes involving different
particles and/or helicities. The action of the $\cN=4$
supersymmetry generators on gluons $g^{\pm}$, fermions
$f^{\pm}_{i}$, and scalars $s_{ij}$ is \cite{Risager:2008yz}
\begin{eqnarray}\label{WI-susy1}
\lbrack Q_{i}(q,\theta) ,g^{+}(k)\rbrack &=& \theta \, \lbrack k q \rbrack \, f_{i}^{+}\,, \nonumber \\
\lbrack Q_{i}(q,\theta) ,f^{+}_{j}(k)\rbrack &=& \theta  \delta_{ij} \, \VEV{ k q } \, g^{+} + \theta \lbrack k q \rbrack s_{ij}\,,  \nonumber \\
\lbrack Q_{i}(q,\theta) ,s_{jk}(k)\rbrack &=& \theta \delta_{ij} \VEV{k q} f_{k}^{+} - \theta \delta_{ik} \VEV{k q} f_{j}^{+} + \theta \lbrack q k \rbrack \epsilon_{ijkl} f_{l}^{-}\,, \nonumber \\
\lbrack Q_{i}(q,\theta) ,f^{-}_{j}(k)\rbrack &=& \theta \delta_{ij} \lbrack q k \rbrack g^{-} + {{1}/{2}}\, \theta \VEV{q k} \epsilon_{ijkl} s_{kl}\,,\nonumber \\
\lbrack Q_{i}(q,\theta) ,g^{-}(k)\rbrack &=& \theta \VEV{q k} f_{i}^{-}\nonumber \,.
\end{eqnarray}
Supersymmetry can be used to derive relations between some amplitudes. For example,
acting with $Q_{i}(q,\theta)$ on $\VEV{0| f^{-}_{j}(1)\, g^{+}(2) \ldots g^{+}(n) |0}$, one finds
\be \label{WI-susy2}
0 = \delta_{ij} \VEV{q 1} \VEV{0|g^{+}(1)\, g^{+}(2) \ldots g^{+}(n)  |0} + \lbrack q 1 \rbrack \VEV{0|s_{ij}(1)\, g^{+}(2) \ldots g^{+}(n) |0}\,.
\ee
Here we have used that amplitudes with two $f^{+}$ vanish because of fermion helicity conservation.
From (\ref{WI-susy2}) it follows \footnote{For $n>3$, because for $n=3$ the prefactor $\VEV{q 1}$ vanishes
for any choice of $q$.} that both terms in it vanish. We are mostly
interested in gluon scattering, so we note
\be\label{WI-susy3}
 \VEV{0|g^{+}(1)\, g^{+}(2) \ldots g^{+}(n)  |0} = A^{++\ldots+} = 0 \,.
\ee In a similar manner, it can be shown by acting with
$Q_{i}(q,\theta)$ on $\VEV{0| g^{-}(1)\, f^{+}_{j}(2)\, g^{+}(3)
\ldots g^{+}(n) |0}$ that \be\label{WI-susy4} A^{-+\ldots+} = 0\,.
\ee Furthermore, relations between amplitudes with different
particles can be obtained from supersymmetry. Given
(\ref{WI-susy3}) and (\ref{WI-susy4}), the first non-zero pure
gluon amplitude has two flipped helicities, and is called
maximally helicity-violating (MHV), \be  A^{\rm (MHV)} =
A^{--+\ldots+}\,, A^{-+-+\ldots+}\,, \quad \ldots \,.\ee 
We remark that for four and
five gluons, all gluon amplitudes are MHV (or
$\overline{\textrm{MHV}}$, which
have all helicities plus except for two, e.g. $A_5^{--+++}$), e.g. 
\be
\{A_4^{++--}\,, \quad A_4^{+-+-}\,, \quad \ldots\} \,,\quad
\{A_5^{+++--}\,, \quad A_5^{+-+--}\,, \quad \ldots \}\,. \ee
Amplitudes with different helicity ordering are related by
cyclicity. It can be shown that in $\cN=4$ SYM, at four and five
gluons there is only one independent MHV amplitude.
\\

Amplitudes with more flipped helicities are called NMHV (next to
maximally helicity-violating), NNMHV, and so on. NMHV amplitudes
(where three helicities are flipped) start appearing at six
points, e.g.
\be
A_6^{+++---}\,,\quad A_6^{-+--++}\,,\ldots\,.
\ee
MHV amplitudes have the special property that for them, there is only one possible helicity structure,
and hence it can be factorised.
 Using the shorthand notation $\langle p_{i} \, p_{j} \rangle = \langle i \, j \rangle $,
 we have for $n \ge 4$ \footnote{One can also define on-shell three-particle scattering for
 complex momenta.}
  \cite{Parke:1986gb,Berends:1987me}, up
 to an inessential normalisation
\be \label{MHVtree}
A^{\rm(tree)}_{n}(1^{+},\ldots,{l}^{-},{(l+1)}^{+},\ldots,{m}^{-},{(m+1)}^{+},\ldots,n^{+})
= i \frac{\langle l \, m \rangle^4}{\langle1 2\rangle \langle2
3\rangle \cdots \langle n 1\rangle}\,, \ee and all loop
corrections are proportional to $A^{\rm(tree)}$. It is then
natural to define the loop corrections to MHV amplitudes by
factorising out the tree amplitude, i.e. \be {A}_{n} =
A_{n}^{\textrm{(tree)}} \, M_{n} = A_{n}^{\textrm{(tree)}} \left[
1 + M_{n}^{(1)}+ M_{n}^{(2)}+\ldots \right] . \ee Here $M_{n}$ is a
function of the momentum invariants $p_{i}\cdot p_{j}$ and of the
dimensional regulator only. We already said that four- and
five-gluon amplitudes are special because all of them are MHV. In
contrast, amplitudes with more than two flipped helicities have a
more complicated form. Their loop corrections involve in
general new helicity structures, and hence they are not
proportional to the tree level term. It remains an interesting
open question how to extend the duality between Wilson loops and
MHV gluon amplitudes discussed in this report to those amplitudes.

\subsubsection{Infrared divergences}\label{scat-ir}
As we already mentioned, the on-shell amplitudes suffer from
infrared divergences. In dimensional regularisation, the
divergences appear as poles in the regulator $\epsilon$. The
structure of these divergences is well understood in any gauge
theory \cite{Akhoury:1978vq,Mueller:1979ih,Collins:1980ih,
Collins:1989gx,Collins:1989bt,Sen:1981sd,Sen:1982bt,Sterman:1986aj,Catani:1989ne,
Catani:1990rp,
Magnea:1990zb,
Catani:1998bh,Sterman:2002qn,
Botts:1989kf,Korchemsky:1988hd,Korchemsky:1988pn}. In order to explain the general structure,
let us proceed by giving an example. Consider the one-loop
corrections to the four-gluon amplitude. They can be written as
\cite{Grisaru:1977px} (with $a = {g^2 N}/(8 \pi^2)$) \be
\label{gluon4point1loop} {A}_4 = {A}_4^{\rm tree} \left[ 1
-\frac{a}{2} s t I^{(1)}(s,t;\epsilon) + O(a^2) \right]\,, \ee
where the one-loop scalar box integral $I^{(1)}$ is defined by \be
\label{scalbox1} I^{(1)} =
\parbox[c]{35mm}{\includegraphics[height = 25mm]{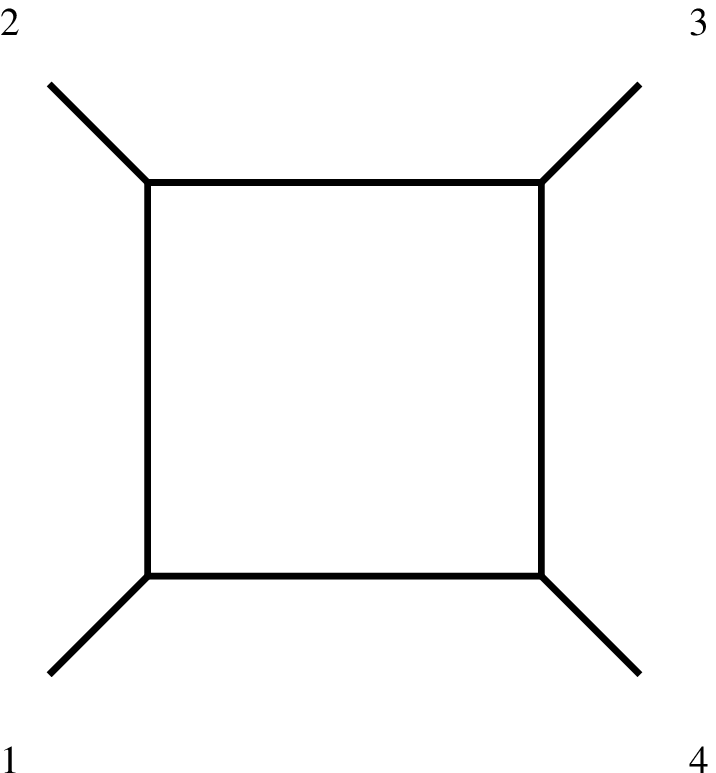}} = C
\int\frac{d^{4-2\epsilon} k
}{k^2(k-p_1)^2(k-p_1-p_2)^2(k+p_4)^2}\,, \ee where $C = \mu^{2 \epsilon}\,
e^{-\epsilon \gamma_{E}} (4\pi)^{-2+\epsilon}$ and
$s=(p_{1}+p_{2})^2$ and $t=(p_{2}+p_{3})^2$ are the usual
Mandelstam variables.
$I^{(1)}$ has infrared
divergences coming from two regimes: the soft regime where $k^2
\sim 0$, and the collinear regime where $k^{\mu} \sim p^{\mu}$,
with $p^{\mu}$ being one of the on-shell momenta entering the
scattering process.
\\

Evaluating (\ref{scalbox1}) one finds, up to terms vanishing as
$O(\epsilon)$, \be \label{MHVexample1loop} {A}_4 = {A}_4^{\rm
tree} \left[ 1 - \frac{a}{ \epsilon^2}
\left(\frac{\mu^2}{-s}\right)^{\epsilon} \right] \left[ 1 -
\frac{a}{\epsilon^2}   \left(\frac{\mu^2}{-t}\right)^{\epsilon}
\right] \left[1+ a\left( {\frac{1}{2} \ln^2 \frac{s}{t} + 4
\zeta_2}\right) \right] + O(a^2,{\epsilon})\,. \ee We can see from
(\ref{MHVexample1loop}) that the loop corrections to $A_{4}$
factorise into divergent pieces, each depending on one kinematical
variable only, and a finite part. This factorised structure is
known to hold to higher loop orders as well (in the planar case). For non-MHV
amplitudes, the only difference compared to
(\ref{MHVexample1loop}) is that the finite part in general depends
on the helicity structure of the amplitude as well. As we already
pointed out, at four points there are only MHV amplitudes, and
this problem does not appear. Their general form is known to
factorise
\cite{Sen:1982bt,Botts:1989kf,Kidonakis:1998bk,Kidonakis:1998nf}
according to \be \label{MHV4all-loop} M_{4} = {A}_4 / {A}_4^{\rm
tree} = \,{\rm div}\left( \frac{\mu^2}{-s}\right)\, {\rm
div}\left( \frac{\mu^2}{-t} \right)\, {\rm
fin}\left(\frac{s}{t}\right)   +  O({\epsilon}) \,, \ee where
${\rm div}(x)$ contains all poles in $\epsilon$ and ${\rm fin}(x)$
is finite as $\epsilon \rightarrow 0$. Notice that the finite part
was chosen such that it is independent of the dimensional
regularisation parameter $\mu$. The dependence on it is contained
in the divergent part and in $O(\epsilon)$ terms only. The
divergent factors in (\ref{MHV4all-loop}) are governed by an
evolution equation
\cite{Korchemsky:1988si,Magnea:1990zb,Korchemsky:1992xv,Catani:1999ss,Sterman:2002qn}.
For theories with vanishing $\beta$ function, such as $\cN=4$ SYM
{}\footnote{This necessarily implies that a regulator is used
where the $\beta$ function is indeed zero} the factors containing
the divergences take the particularly simple form \be
\label{gluon-div}
 { \textrm{div}(x)}  = \exp\left\{-\frac{1}{2} \sum_{l=1}^{\infty} a^l
x^{l\epsilon}\left[\frac{ \Gamma^{(l)}_{\rm cusp}}{ (l\epsilon)^2} + \frac{
G^{(l)}}{ l\epsilon}\right]\right\}\,,
\ee
where
\be\Gamma_{\textrm{cusp}}(a)=\sum_l a^l \Gamma^{(l)}_{\textrm{cusp}}\,,\qquad G(a)=\sum_l a^l G^{(l)}_{\rm cusp}\,.
\ee
Here $\Gamma_{\textrm{cusp}}$ is the cusp anomalous dimension (of Wilson loops), and $G$ is the scheme-dependent collinear anomalous dimension. Their perturbative expansion in the coupling constant
up to two loops is (in the DRED scheme)
\be\label{cusp2loop-MHV}
\Gamma_{\rm cusp}(a) = 2a - 2 \zeta_{2} a^2 + O(a^{3})\,,\qquad G(a) = -\zeta_{3} a^{2} + O(a^{3}) \,.
\ee
We can see that (\ref{MHVexample1loop}) is indeed of the factorised form expected on general grounds,
see formula (\ref{MHV4all-loop}), and the coefficient in front of the double pole agrees with the
known expression (\ref{cusp2loop-MHV}) at one-loop for the cusp anomalous dimension.
\\

By definition \cite{Polyakov:1980ca,Korchemsky:1985xj,Korchemsky:1987wg},
the cusp anomalous dimension $\Gamma_{\rm cusp}(a)$ describes
specific \textit{ultraviolet} divergences of a Wilson loop
evaluated over a contour with a cusp. Its appearance in the
\textit{infrared} divergent part of the scattering amplitude
\re{gluon-div} is not accidental. It has its roots in the deep
relation between scattering amplitudes in gauge theory and Wilson
loops evaluated over specific contours in Minkowski space-time,
defined by the particle momenta
\cite{Korchemsky:1985xj,Korchemsky:1992xv,Korchemskaya:1996je}. A
distinguishing feature of this contour is that it has cusps. As
was found in
Refs. \cite{Korchemsky:1985xj,Korchemsky:1992xv,Korchemskaya:1996je},
the infrared divergences of planar scattering amplitudes are in
one-to-one correspondence with the cusp ultraviolet divergences of
light-like Wilson loops. It should be mentioned that this relation
is not specific to $\mathcal{N}=4$ SYM and it holds in any gauge
theory, including QCD. We explain this relationship in a one-loop
example in section \ref{FF-argument}.
\\

The two-loop expression for $\Gamma_{\rm cusp}(a)$ in a generic (supersymmetric) Yang-Mills theory was found in
Refs.~\cite{Korchemsky:1987wg,Belitsky:2003ys}. We remark that in $\mathcal{N}=4$ SYM theory, $\Gamma_{\rm
cusp}(a)$ is known at weak coupling to four loops \cite{Bern:2006ew}, and it is conjectured to be governed
to all loops by the BES equation \cite{Beisert:2006ez}. At strong coupling, the solution of the latter produces a strong coupling expansion of $\Gamma_{\rm
cusp}(a)$~\cite{Benna:2006nd,Basso:2007wd,Kostov:2008ax}. The first few terms of this expansion are
in agreement with the existing quantum superstring calculation of
Refs.~\cite{Gubser:2002tv,Frolov:2002av,Roiban:2007dq}. The non-universal collinear anomalous
dimension $G(a)$ is known to four loops at weak coupling \cite{Bern:2005iz,Cachazo:2007ad} and there
exists a prediction at strong coupling \cite{Alday:2007hr}.
\\

For planar MHV amplitudes (where the common helicity structure can be factored out),
the generalisation of (\ref{MHV4all-loop})
to an arbitrary number $n$ of gluons is
\be \label{MHVall-loop}
\mathcal{M}_{n} = \ln M_{n}^{\rm(MHV)} = Z_{n} + F_{n} + O(\epsilon)\,,
\ee
with
\begin{equation}\label{Zn-amplitude}
Z_n = -\frac14 \sum_{l=1}^n a^l \lr{\frac{\Gamma^{(l)}_{\rm cusp}}{(l\epsilon)^2}+\frac{G^{(l)}}{l\epsilon}}\sum_{i=1}^n
\lr{-\frac{t_i^{[2]}}{\mu^2}}^{-l\epsilon}\,,
\end{equation}
where $t_i^{[2]} \equiv s_{i,i+1} = (p_i+p_{i+1})^2$ is the
invariant mass of two adjacent gluons with indices $i$ and $i+1$,
and the periodicity condition $i+n\equiv i$ is tacitly implied.
The finite part $F_{n} = F_{n}(t^{[j]}_{i})$ is a dimensionless
function of the momentum invariants $t^{[j]}_{i} = (p_{i} + \ldots
p_{i+j-1})^2$ of the scattering process \footnote{Note that $F_{n}$
does not depend on all $t^{[j]}_{i}$ independently, because
(more than four) four-dimensional vectors have to satisfy Gram determinant
constraints.}. From (\ref{MHVall-loop}) we can see that
MHV amplitudes contain two interesting pieces of information: the
cusp anomalous dimension $\Gamma_{\rm cusp}$ in the double pole
term, and an \textit{a priori} arbitrary function of the momentum
invariants in the finite part. What makes the latter particularly
interesting is that it turns out not to be arbitrary, but seems to
have a very simple form, as we will discuss in section
\ref{ch-amplitudes-results}. In some sense, the finite part
$F_{n}$ is more interesting than $\Gamma_{\rm cusp}$ because at a
given loop order, it is a function (in general of many variables),
whereas $\Gamma^{(l)}_{\rm cusp}$ is merely one number. Finally,
note that $G$ is scheme dependent because one can always
arbitrarily redefine the dimensional regularisation scale $\mu$.
For the same reason $F_{n}$ contains in general a scheme-dependent
constant, but its functional form is scheme independent.

\subsubsection{Calculational technique}

The loop corrections to $M_{4}$ can be represented in terms of
scalar loop integrals, as for example the scalar box integral $I^{(1)}$ encountered in
the previous section. How do these integrals arise? In a standard approach,
one would write down all Feynman graphs contributing to a given amplitude,
involving gluons, fermions, scalars, and
also ghosts.
Then, one has to work out the
numerator algebra in the integrals corresponding to these graphs.
Usually one also has to use integral reduction identities in order to
express the amplitude in terms of master integrals \cite{Passarino:1978jh}.
One can think of the scalar integrals as the result of going through this (lengthy) procedure.
With increasing loop order (or number of gluons) this approach involves writing down a large number of
Feynman graphs and would be very cumbersome if not impossible.
For example, even the tree level formula (\ref{MHVtree}) for large $n$ would require
writing down an astronomical
 number of Feynman graphs, since
their number grows factorially with $n$.
\\

\begin{figure}[t]
\psfrag{im}[cc][cc]{$2\,{\rm Im}$}
\psfrag{equals}[cc][cc]{$ = \sum \int$}
\centerline{\epsfxsize 3.0 truein \epsfbox{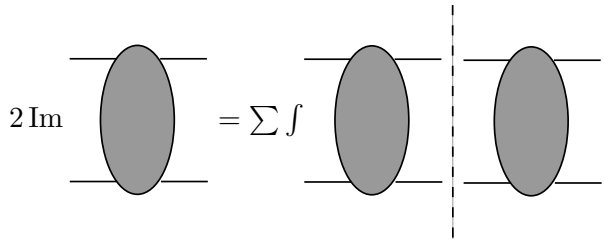}}
\caption{Graphical representation of the unitarity relation $2\,{\rm Im}\left( T \right) =  T T^{\dagger}$ for the four-gluon amplitude.
The sum is over a complete set of intermediate states.}
\label{Fig-unitarity}
\end{figure}

This is why one often employs a different technique based on
the unitarity of the $S$-matrix, see
\cite{Eden-book} and references therein. The latter, which
describes the scattering of particles in a quantum theory, can be
written as \be S = 1 + i T \,.\ee Requiring unitarity of the
$S$-matrix, $S^{\dagger} = S^{-1}$, one obtains \be
\label{unitarity-rel} 2 \, {\rm Im}\left( T \right) = i
(T^{\dagger} - T)  =   T T^{\dagger} \,. \ee If one inserts a
complete set of states between $T$ and $T^{\dagger}$ on the
right-hand side of (\ref{unitarity-rel}), one gets the
diagrammatical relation shown in Fig.~\ref{Fig-unitarity}. There,
the sum is over all allowed intermediate states and the integral
is over their on-shell momenta.
\\

%

Equation (\ref{unitarity-rel}) relates objects at different orders in perturbation theory.
For example, the imaginary part of a one-loop amplitude on its left-hand side is expressed through
tree-level amplitudes on its right-hand side. One can think of the r.h.s. as being a one-loop
amplitude that was cut into two tree-level pieces.
\\

In the unitarity based method of Bern \textit{et al}
\cite{Bern:1994zx,Bern:1994cg}, one does not compute the integral
on the right-hand side of (\ref{unitarity-rel}) directly. Rather,
the idea is to reconstruct the amplitudes from their cuts. More
precisely, one uses (\ref{unitarity-rel}) in order to construct
the loop integrand of integrals that have the correct behaviour
under unitarity cuts. A generic one-loop amplitude can be
expressed in terms of master integrals by integral reduction
techniques \cite{Passarino:1978jh}. Therefore it is sufficient to
know the coefficients with which the master integrals contribute
to the amplitude. Moreover, in supersymmetric field theories, it
is known from integral reduction methods, combined with power
counting arguments, that only a certain subclass of scalar
integrals can appear in the amplitude \cite{Bern:1994cg}. All of
them have either $s$- or $t$-channel cuts, so they can be detected
by the corresponding cuts. Therefore, one-loop amplitudes in
$\cN=4$ SYM are `cut-constructible'.
\\

Let us give a one-loop example of the cutting technique, following
\cite{Dixon:1996wi}. Consider the four-gluon amplitude
$A_{4}(1^{-},2^{-},3^{+},4^{+})$ at one loop. From the $s$-channel
cut, we have
\begin{eqnarray}\label{cut-s1}
A_{4}(1^{-},2^{-},3^{+},4^{+})|_{\rm s-cut} &=& \sum \int \frac{d^{D}k}{(2\pi)^D} 2 \pi \delta^{(+)}({l_{1}^{2}}) 2 \pi \delta^{(+)}({l_{2}^{2}})\\
&&\hspace{1.5cm}\times A_{4}^{\rm tree}(-l_{1},1^{-},2^{-},l_{2})  A_{4}^{\rm tree}(-l_{2},3^{+},4^{+},l_{1})  \nonumber
\end{eqnarray}
Here the cut momenta are $l_{1} = k$ and $l_{2} =
k-p_{1}-p_{2}$, with $k$ being the loop momentum. 
 The sum in (\ref{cut-s1}) is over
all allowed particles and helicity states for the cut lines. In
this example, due to supersymmetric Ward identities, the
contribution of scalars and fermions to (\ref{cut-s1}) vanishes.
Hence, the only non-vanishing contribution comes from gluon tree
amplitudes with the following helicities \footnote{Note the change
of helicity in $l_{1,2}$ depending on whether the lines are
ingoing or outgoing}
\begin{eqnarray}\label{cut-s2}
A_{4}^{\rm tree}(-l_{1}^{+},1^{-},2^{-},l_{2}^{+}) &=& i \frac{\VEV{12}^4}{\VEV{-l_{1} 1}\VEV{12}\VEV{2 l_{2}}\VEV{l_{2} \, -l_{1}}}\,,\nonumber \\
A_{4}^{\rm tree}(-l_{2}^{-},3^{+},4^{+},l_{1}^{-}) &=& i \frac{\VEV{-l_{2} l_{1}}^4}{\VEV{-l_{2} 3}\VEV{34}\VEV{4 l_{1}}\VEV{l_{1} \, -l_{2}}}\,.
\end{eqnarray}
We now replace the delta functions $2\pi \delta^{(+)}(l_{i})$ by
full propagators $i/l_{i}^2$, while still using the conditions
$l_{i}^2 = 0$ in the numerator. In this way, we construct the
integrand of an integral that has the correct behaviour in the
$s$-cut, as was explained above.
\begin{eqnarray}\label{cut-s1a}
A_{4}(1^{-},2^{-},3^{+},4^{+})|_{\rm s-cut} &=& \int \frac{d^{D}k}{(2\pi)^D} \frac{1}{l_{1}^2}\frac{1}{l_{2}^2}
  \frac{\VEV{12}^4}{\VEV{-l_{1} 1}\VEV{12}\VEV{2 l_{2}}\VEV{l_{2} \, -l_{1}}} \nonumber \\&& \hspace{2cm}\times   \frac{\VEV{-l_{1} l_{2}}^4}{\VEV{-l_{2} 3}\VEV{34}\VEV{4 l_{1}}\VEV{l_{1} \, -l_{2}}}  \bigg|_{\rm s-cut}
\end{eqnarray}
By construction, (\ref{cut-s1a}) has the correct properties in the $s$-channel.
Since we want to arrive at usual loop integrals, we make
propagators appear in the denominator by writing e.g. \be
\label{cut-s1b} \frac{1}{\VEV{2 l_{2}}} = -\frac{\lbrack 2 l_{2}
\rbrack}{(k-p_{1})^2}\,. \ee Using identities like
(\ref{cut-s1b}), we see that we can factor out the tree amplitude,
\be A_{4;1}(1^{-},2^{-},3^{+},4^{+})|_{\rm s-cut} = -i  A^{\rm
tree}_{4} \int \frac{d^{D}k}{(2\pi)^D} \frac{\lbrack l_{1}
1\rbrack \VEV{14}\lbrack 4 l_{1}\rbrack \VEV{l_{1}l_{2}}\lbrack
l_{2} 3\rbrack \VEV{32} \lbrack 2 l_{2}\rbrack \VEV{l_{2} l_{1}}}
{k^2 (k-p_{1})^4 (k-p_{1}-p_{2})^2 (k+p_{4})^4} \bigg|_{\rm s-cut}
\,. \ee After some algebra, the numerator in the integrand can be
simplified to $-s t (k-p_{1})^2 (k+p_{4})^2$, up to a term
involving the totally antisymmetric Levi-Civita tensor, which
drops out of a four-point amplitude. Thus, the final answer is
\be\label{cut-s3} A_{4;1}(1^{-},2^{-},3^{+},4^{+})|_{\rm s-cut} =
-s t A^{\rm tree}_{4} \int \frac{d^{D}k}{(2\pi)^D} \frac{1}{k^2
(k-p_{1})^2 (k-p_{1}-p_{2})^2 (k+p_{4})^2} \bigg|_{\rm s-cut} \,.
\ee
The $t$-channel cut can be evaluated in a similar manner. There,
supersymmetry allows scalars and fermions to contribute as well.
Taking all contributions into account, one again finds the scalar
box integral, just as in the $s$-channel. By consistency, the
coefficients obtained in both channels have to match. Thus, we
conclude that in $\cN = 4$ SYM \footnote{Strictly speaking, the
preceding calculation proves (\ref{cut-result})  up to terms
vanishing as $O(\epsilon)$ only, since we used $4$-dimensional
cuts. To extend the derivation to $O(\epsilon)$ terms one could
use $D$-dimensional cuts. However, in \cite{Bern:1994cg} it was
argued that (\ref{cut-result}) is valid to all orders in
$\epsilon$ (see also \cite{Bern:1997nh}).}, \be\label{cut-result}
A_{4;1}(1^{-},2^{-},3^{+},4^{+}) = -s t A^{\rm tree}_{4} \int
\frac{d^{D}k}{(2\pi)^D} \frac{1}{k^2 (k-p_{1})^2 (k-p_{1}-p_{2})^2
(k+p_{4})^2}\,, \ee in agreement with (\ref{gluon4point1loop}).
\\

One-loop amplitudes in supersymmetric theories were shown to be
`cut-constructible' in the above sense. We remark that in
non-supersymmetric theories, there are rational terms in the
amplitude that cannot be detected by cuts, and have to be
determined by other means. This problem can be solved, however, by
using $D$-dimensional cuts. The reason is that in $D$ dimensions,
the rational terms produce cuts that can be detected. This approach
further uses a supersymmetric decomposition of one-loop amplitudes
(\ref{susy-decomp}). This means that only cuts of scalar loops
have to be evaluated in $D$ dimensions. These techniques have
allowed the computation of several QCD amplitudes at one loop.
\\

In $\cN=4$ SYM the cutting techniques were successfully applied to
higher loop amplitudes as well. They allowed the computation of
amplitudes that would have been practically impossible to
determine on the basis of Feynman diagrams. We will summarise the
results for MHV amplitudes in $\cN=4$ SYM in the next section.
Finally, we mention another important development, that of
`generalised cuts', where one requires more than two propagators
to be on shell, see \cite{Britto:2004nc}.

\subsubsection{Results of perturbative calculations}
\label{ch-amplitudes-results} Over the last thirty years,
considerable efforts have been undertaken to compute loop
corrections to scattering amplitudes in QCD and its supersymmetric
extensions, see e.g. the review \cite{Bern:2007dw}. Most progress
has been made for MHV amplitudes in $\cN=4$ SYM. Up to now, the
following MHV amplitudes have been computed in $\cN=4$ SYM:
\begin{itemize}
  \item four gluons at two loops \cite{Bern:1997nh,Anastasiou:2003kj} and three loops \cite{Bern:2005iz};
  the four-gluon amplitude is known in terms of loop integrals at four loops \cite{Bern:2006ew}, and the integrals
  were evaluated (numerically) up to the second pole term, which allowed to determine $\Gamma_{\rm cusp}$ at four loops; a conjecture for the integrals appearing at five loops was put forward in \cite{Bern:2007ct}
  \item five gluons to two loops \cite{Cachazo:2006tj,Bern:2006vw}
  \item six gluons to two loops \cite{Bern:2008ap}
  \item $n$ gluons to one loop \cite{Bern:1994zx}
\end{itemize}
Similar computations in QCD are much harder and up to now are
restricted to lower loop orders.
\\

In \cite{Anastasiou:2003kj}, Anastasiou, Bern, Dixon and Kosower (ABDK) noticed an interesting iterative
structure in the two-loop result for the four-gluon amplitude in $\cN=4$ SYM.
Based on  \cite{Anastasiou:2003kj} and the result of their three-loop calculation, Bern, Dixon and Smirnov (BDS) formulated a conjecture for the $n$-gluon amplitudes in $\cN=4$ SYM to all orders in the coupling constant \cite{Bern:2005iz}.
As we already discussed, the divergent part of the scattering amplitudes is well understood,
so their conjecture is essentially a statement about the finite part.
Their conjecture implies that the finite part $F_{n}$, c.f. (\ref{MHVall-loop}) has the same
functional form as at one loop, $F_{n}^{(1)}$,
multiplied by the (coupling-dependent) cusp anomalous dimension,
\be
F_{n} = \frac{1}{2}{\Gamma_{\rm{cusp}}(a)} F_{n}^{(1)} + \rm{const}\,.
\ee
For example, in application to the four- and five-gluon amplitudes this means that
\begin{eqnarray}
 F_{4} &=& \frac{1}{4} {\Gamma_{\rm{cusp}}(a)} \left[ \ln^2 \frac{s}{t} + \rm{const} \right] \label{BDS4} \\
 F_{5} &=& \frac{1}{4} {\Gamma_{\rm{cusp}}(a)} \left[ \sum_{i=1}^{5} \ln \frac{ s_{i,i+1}}{s_{i+1,i+2}} \ln \frac{s_{i+2,i+3}}{s_{i+3,i+4}} + \rm{const} \right] \label{BDS5}
\end{eqnarray}
where $s,t$ and $s_{i,i+1}=(p_{i}+p_{j})^2$ are the momentum invariants of the scattering processes.
\\

Why should the finite part of the amplitudes be so simple?
One might speculate that some symmetry could be at the origin of this simplicity.
In the next section, we will study the four-gluon amplitude
in more detail. We will see that indeed one can observe `dual' conformal properties
of the integrals contributing to it, thus providing hints for a broken conformal symmetry
governing the gluon amplitudes.
We will see in section \ref{section-CWI} that the conjectured duality with Wilson loops, if true, can explain these observations, and even make them
more quantitative, by providing broken conformal Ward identities.
\\

We should stress that very recently, the BDS conjecture was found to fail at six gluons
and two loops \cite{Bern:2008ap}.
We will see in section \ref{ch-six-point} that it is modified in a way predicted
by the duality (\ref{intro-duality}) between Wilson loops and gluon amplitudes. The duality, if true, also
implies that the BDS conjecture is correct for four and five gluons to all loops
(which is consistent with all available data).

\subsection{Dual conformal properties of the four-gluon amplitude}
\label{ch-pseudoconformal}
Let us focus for now on four-gluon amplitudes, where results from high-order calculations
in perturbation theory are available. We will see in this section that the integrals contributing to the four-gluon
amplitude are all from a set of `pseudo-conformal' integrals, in the sense that will
be defined presently.
\\

Let us come back to the scalar box integral (\ref{scalbox1}).
An important observation made in \cite{Drummond:2006rz}
is that it has particular properties in a dual `coordinate' space,
defined by the change of variables
%
\begin{figure}[t]
\psfrag{a}[cc][cc]{$I^{(1)}$}
\psfrag{b}[cc][cc]{$I^{(2)}$}
\centerline{\epsfxsize 4.0 truein\epsfbox{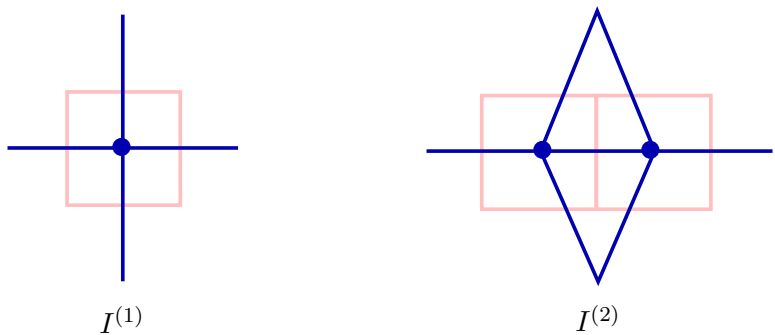}}
\caption{One-loop box integral $I^{(1)}$ and two loop double box integral $I^{(2)}$ and their dual pictures.}
\label{Fig:12-dual}
\end{figure}
\be \label{dualvars}
p_1=x_1-x_2\equiv x_{12}\,,\quad p_2=x_{23}\,,\quad p_3=x_{34}\,,\quad
p_4=x_{41}\,.
\ee
We stress that (\ref{dualvars}) is simply a change of variables, and not a Fourier transform.
Under the identification (\ref{dualvars}) and that of $k=x_{15}$, one obtains the expression
\be
I^{(1)}(s,t;\epsilon)= C \int\frac{d^{4-2\epsilon} k }{k^2(k-p_1)^2(k-p_1-p_2)^2(k+p_4)^2} = C {\int\frac{d^{4-2\epsilon} x_5}{x_{15}^2 x_{25}^2 x_{35}^2 x_{45}^2}}\,.
\ee
The integral together with its dual picture is shown in Fig. \ref{Fig:12-dual}.
The point is that in the dual `coordinate' space defined by (\ref{dualvars}), the integral
$I^{(1)}$ has (broken) conformal properties. Indeed, if it was not for the on-shell conditions
$p_i^2=0$ which lead us to use dimensional regularisation, $I^{(1)}$ would be given by the finite off-shell
conformal integral $\Phi^{(1)}$ encountered in section \ref{N=1four-point1loop}. Recall that $\Phi^{(1)}$ is conformal
because under a conformal inversion, the variation of the integral measure is precisely canceled
by the variation of the four propagators going to the vertex.
In $D=4-2 \epsilon$, we see that the cancelation is no longer exact, and hence the (dual) conformal
symmetry is broken.
So we conclude that $I^{(1)}$ has broken conformal properties in the dual coordinate space.
\\

In order to avoid confusion with (broken) conformal Ward identities for Wilson loops discussed in later parts of this thesis, we will call integrals of this type
`pseudo-conformal' \cite{Bern:2006ew}.
\\

At this stage one might think that the appearance of the pseudo-conformal integral $I^{(1)}$ is just a
one-loop coincidence. However, we will see presently that this pattern continues to hold to higher loops.
At two loops,
\be
M_{4}^{(2)} = \frac{1}{4} \left[ s^2 t I^{(2)}(s,t; \epsilon) + s t^2 I^{(2)}(t,s; \epsilon) \right]\,,
\ee
one encounters one new integral, the double box integral
\be
{I}^{(2)}(s,t;\epsilon) = C^2 \int  \frac{{d^{4-2 \epsilon}p\,d^{4-2 \epsilon}q}}{q^2 (p-k_{1})^2 (p-k_{1}-k_{2})^2
(p+q)^2 q^2 (q-k_{4})^2 (q-k_{3}-k_{4})^2} \,.
\ee
Going to dual coordinates (c.f. Fig. \ref{Fig:12-dual}),
we see that it is the dimensionally continued version of $\Phi^{(2)}$ considered in section \ref{section-conformalfourpoint}.
Notice the dotted line between external points one and three, which is there to make sure that all
external points have the same conformal weight.
There is an easy way to implement the change of variables (\ref{dualvars}) by using the following
pictorial rule:
\begin{itemize}
\item for each loop integration
in $p$-space, draw a dot inside the loop, which corresponds to an integration vertex in the dual coordinate space.
\item the external points in the dual coordinate space are situated between two consecutive external momenta, denote them
by a point.
\item draw a straight line between points that are separated by a momentum line.
\end{itemize}
The numerator lines have to be added by going through the change of variables (\ref{dualvars}). For
pseudo-conformal integrals they are usually easy to find since they have to be chosen such that each integration vertex has conformal weight four, and from the convention that external points have conformal weight one.
\\

We remark that the notion of dual graphs exists for planar graphs only \cite{Nakanishi}. For this reason it is not clear how to apply this notion to non-planar
corrections to gluon scattering amplitudes.
\\

\begin{figure}[htbp]
\psfrag{a}[cc][cc]{(a)}
\psfrag{b}[cc][cc]{(b)}
\centerline{\epsfxsize 4.0 truein \epsfbox{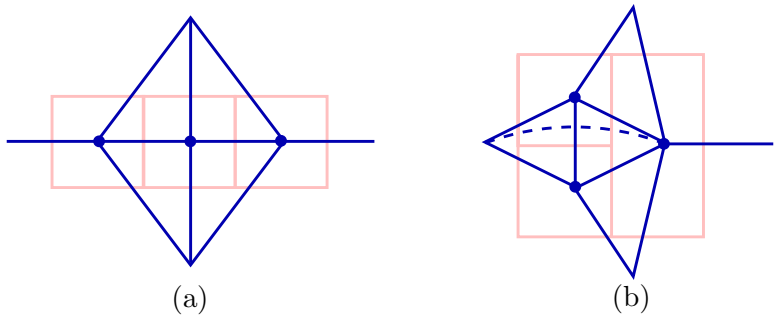}}
\caption{\small Three loop ladder or triple box integral $I_4^{(3) a}$ and the tennis court integral $I_4^{(3) b}$, together with their dual pictures.}
\label{figure:3loop}
\end{figure}

At three loops, the four-gluon amplitude is given by two integrals (see Fig. \ref{figure:3loop}),
\be \label{ThreeloopPlanarResult}
M^{(3)}_{4} =  \frac{1}{8} \left[ s^3 t I^{(3)a}_{4}(s,t;\epsilon) + s t^3 I^{(3)a}_{4}(t,s;\epsilon) + 2 s t^2 I^{(3)b}_{4}(s,t;\epsilon)  + 2 s^2 t I^{(3)b}_{4}(t,s;\epsilon) \right]\,,
\ee
which are again both `pseudo-conformal'.
The two three-loop integrals appearing in the four-point amplitude
(\ref{ThreeloopPlanarResult}), and depicted in Fig. \ref{figure:3loop} are
\begin{eqnarray}
I_4^{(3) a}(s,t;\epsilon) & = &
C^3
 \int
 {\dd^{4-2\epsilon} p\,\dd^{4-2\epsilon} r\,\dd^{4-2\epsilon} q\, \over p^2 \, (p - k_1)^2 \,(p - k_1 - k_2)^2 }
\nonumber \\
&& \null \times
{1\over (p + r)^2 \,
    r^2 \, (q-r)^2 \, (r- k_3 - k_4)^2 \,
    q^2 \,  (q-k_4)^2 \, (q - k_3 - k_4)^2 }\,, \hskip 1 cm
\label{ThreeLoopIntegralA}
\end{eqnarray}
and
\begin{eqnarray}
I_4^{(3) b}(s,t;\epsilon) & = &
C^3
\int
{\dd^{4-2\epsilon} p\,\dd^{4-2\epsilon} r\,\dd^{4-2\epsilon} q\,(p+r)^2 \over p^2 \, q^2 \, r^2 \,
(p - k_1)^2 \,(p +r - k_1)^2 }
\nonumber \\
&& \null \times
{1\over (p+r-k_1-k_2)^2 \, (p+r +k_4)^2 \, (q-k_4)^2
    (r+p+q)^2 \, (p+q)^2 } \,.
\label{ThreeLoopIntegralB}
\end{eqnarray}
\\

The first integral is the triple box integral, which in the dual space corresponds to the
dimensionally continued version of $\Phi^{(3)}$. The other three-loop integral, the so-called `tennis court',
or rather `squash court' integral, exhibits a new interesting feature. Its expression contains a numerator
$(p+r)^2$ (which is not depicted in the momentum space picture), given by $x_{15}^2$ in the dual space
and depicted by a dotted line. We see that five propagators are connected to integration
point five, which is one to many to cancel the conformal weight coming from the transformation of the
measure under a conformal inversion. This extra weight is precisely canceled by the presence of the numerator.
\\

%
\begin{figure}[t]
\centerline{\epsfxsize 6.2 truein \epsfbox{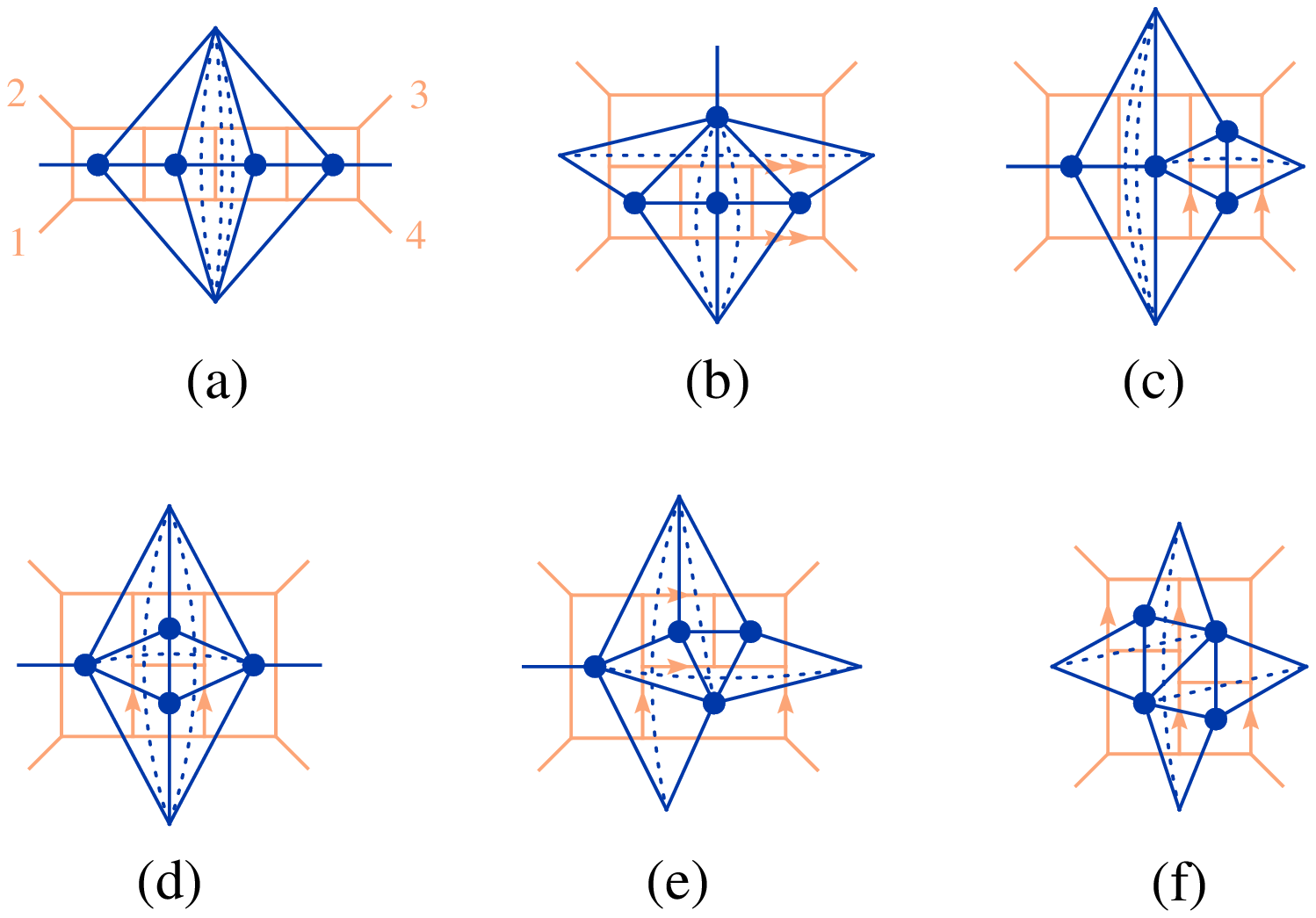}}
\caption{The rung-rule dual diagrams. A factor of $st$ has been
removed.Figure taken from \cite{Bern:2006ew}.}
\label{rrdualFigure}
\end{figure}
%
\begin{figure}[t]
\centerline{\epsfxsize 3.8 truein \epsfbox{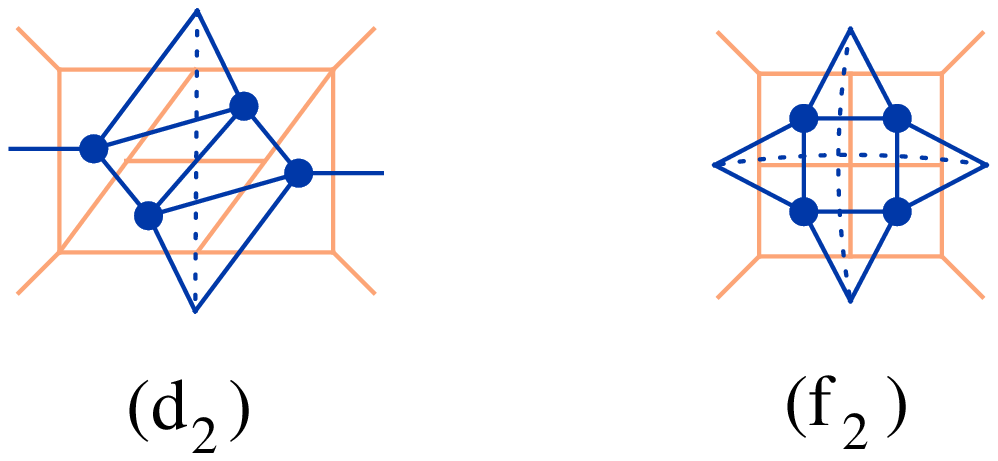}}
\caption{The non-rung rule dual diagrams. A factor of $st$ has been
removed. Figure taken from \cite{Bern:2006ew}.}
\label{nonrrdualFigure}
\end{figure}

At four loops, the number of pseudo-conformal integrals contributing to the amplitude increases
compared to three loops, and one finds eight of them \cite{Bern:2006ew}. They are
shown in Figs. \ref{rrdualFigure},\ref{nonrrdualFigure}, along with their dual conformal pictures.
Let us comment that the integrals shown in Fig. \ref{rrdualFigure} were predicted by
the rung rule \cite{Bern:2005iz}, whereas the integrals shown in Fig. \ref{nonrrdualFigure} cannot be
obtained from it.
The rung rule came from the observation that when constructing $(n+1)$-loop integrals from $n$-loop
integrals by the unitarity method, the two-particle cuts have an iterative structure, and
$(n+1)$-loop integrals can be obtained by adding a `rung' to $n$-loop integrals in all possible ways.
(see also \cite{Cachazo:2008vp}).
It is also interesting to note that in the calculation of \cite{Bern:2006ew}, two integrals that one might
(naively) call pseudo-conformal came with coefficient zero, i.e. they did not appear. A possible
explanation for this was found in \cite{Drummond:2007aua}, where it was shown that those latter integrals are
not finite in four dimensions for off-shell gluons.
It is striking that of all 10 pseudo-conformal integrals, the two
integrals that did not appear are both ill-defined in this sense.
\\

Based on the `conformal hypothesis', i.e. that the expectation that the amplitude should be given
by a linear combination of pseudo-conformal integrals, an amplitude in terms of integrals was conjectured
at five loops \cite{Bern:2007ct}. Strikingly, it was found from unitarity cuts that integrals that
are ill-defined in the sense discussed in the previous paragraph are absent in the five-gluon amplitude.
\\

We remark that, compared to a generic gauge theory, such as for example QCD, the number of integrals
appearing in the final expression for the amplitudes is very small. It seems to be a special property of $\cN=4$ SYM
that the amplitudes can be reduced to such a small set of integrals. In particular,
in calculations done so far, it was observed that no diagrams (representing integrals) with triangle or bubble subgraphs appear \footnote{In the four-loop computation \cite{Bern:2006ew} this was used as an assumption.}
If true in general, this seems to be a special feature of $\cN=4$ SYM.
We remark that if one demands or conjectures that the
amplitude should be built from pseudo-conformal integrals only, then triangle and bubble subgraphs
are automatically excluded.
\\

Based on \cite{Drummond:2006rz,Bern:2006ew}, in  \cite{Drummond:2007aua} it was conjectured
that the four-gluon amplitude is a sum over pseudo-conformal integrals
\begin{equation}
{M}_4 = 1+\sum_\mathcal{I}\, a^{l(\mathcal{I})} \, \sigma_\mathcal{I}
\,\mathcal{I}\,,
\label{all-loop}
\end{equation}
where the sum runs over all pseudo-conformal integrals (which would be finite off-shell in four dimensions), $\mathcal{I}$, $a$ is the coupling constant and $l(\mathcal{I})$ is the loop
order of a given integral.
According to (\ref{all-loop}),  the four-gluon amplitude would be fixed by the relative coefficients
$\sigma_\mathcal{I}$ coming with the pseudo-conformal integrals.
Can these coefficients be predicted as well?
One constraint on them comes from the known exponential form of the divergences of the amplitude, c.f. (\ref{MHV4all-loop}),
\be\label{IRconsistency}
\ln {M}_4 = - \sum_{l=1}^{\infty}\, a^{l} \frac{\Gamma_{\textrm{cusp}}}{(l \epsilon)^2} + 
O(1/\epsilon)
 \,.
\ee
The $\epsilon$ expansion of individual integrals at a given loop order $l$ typically contains
divergences up to $\epsilon^{-2l}$. This means that the coefficients
in (\ref{all-loop}) have to be such that when one takes the logarithm, the leading
divergences cancel, and at most double poles remain.
This relates integrals appearing at different loop orders. A practical problem
is that in order to use this relation, one has to be able
to extract divergences of integrals.
It would be highly desirable to have an efficient way of
determining at least the leading singularities of a given
integral in order to fix some of the relative coefficients (c.f. \cite{Nguyen:2007ya,Cachazo:2008dx,Cachazo:2008vp}
for related work).
\\

Recall that $\Gamma_{\textrm{cusp}}$, the coefficient of
the double pole, is predicted from
a (conjectured) integrable model \cite{Beisert:2006ez}. This implies that so should be the
coefficients $\sigma_\mathcal{I}$ in (\ref{all-loop}), since they implicitly determine $\Gamma_{\textrm{cusp}}$.
At the moment, integrability is mainly applied to
determine the spectrum of anomalous dimensions.
Solving the problem of determining the coefficients $\sigma_\mathcal{I}$ would be an important
step towards extending the use of integrability to the computation of scattering amplitudes.
\\

We studied four-gluon scattering amplitudes.
For more than four gluons, one can
also define the notion of pseudo-conformal integrals, and the integrals
appearing in five- and six-gluon MHV scattering amplitudes up to two loops seem to have this
property \cite{Bern:2006vw,Bern:2008ap}. This is certainly worth further study.
\\

To summarise, the discovery of this hidden structure in
(four-)gluon amplitudes in $\cN=4$ SYM is one of the results of
this thesis. One can say that the `conformal rule' (meaning that
the amplitude is constructed from a linear combination of
`pseudo-conformal' integrals only) is an improvement over the
`rung-rule' \cite{Bern:2005iz}, which fails starting from four
loops \footnote{It still predicts correctly the appearance of the
rung-rule diagrams, together with theirs prefactors, but it misses
other diagrams, as was explained above.}. The conformal rule works
at least to four loops and possibly at five loops
\cite{Bern:2007ct}. Will this new rule also fail at some loop
order? We like to think that the `conformal rule' is correct to
all loops. Support for this expectation comes from the duality
between Wilson loops and gluon amplitudes which will be discussed
in section \ref{ch-duality}. If this duality holds, then the gluon
amplitudes inherit the dual conformal properties of the Wilson
loops. This might explain why only `pseudo-conformal' integrals
appear in the amplitude. Moreover, it would fix (at least some of)
the relative coefficients with which the integrals appear in the
amplitude, since then, in addition to (\ref{IRconsistency}), the
(logarithm of the) sum  of the integrals would have to satisfy the
same conformal Ward identity as
the Wilson loop (see section \ref{section-CWI}).


\section{Wilson loops and gluon amplitudes at strong coupling}
\label{ch-AM}
In this section we summarise work by Alday and Maldacena \cite{Alday:2007hr,Alday:2007he} that
motivated the idea of a duality between Wilson loops and gluon amplitudes in $\cN=4$ SYM.

\subsection{Prescription for computing scattering amplitudes at strong coupling}
In \cite{Alday:2007hr}, Alday and Maldacena proposed a way to calculate
scattering amplitudes in $\cN=4$ SYM at strong coupling via the
AdS/CFT correspondence \cite{Maldacena:1998im}.
According to \cite{Alday:2007hr}, gluon scattering at strong coupling
is described by the scattering of open strings ending on a $D$-brane
in $AdS_{5}$ space with the metric
\begin{equation}
ds^2 = R^2 \left[ \frac{ dx_{3+1}^2 + dz^2}{z^2}\right]\,.
\end{equation}
The $D$-brane is located at $z=z_{\rm IR}$, where $z_{\rm IR}$ can be
thought of as an infrared regulator.
The scattering takes place at fixed angle and at large momentum, and
one finds that the solution is dominated by a saddle point
of the classical action (see also \cite{Gross:1987kza} for a similar
computation  in flat space).
\\

\begin{figure}[t]
\psfrag{p1}[cc][cc]{$p_{2}$}
\psfrag{p2}[cc][cc]{$p_{1}$}
\psfrag{x1}[cc][cc]{$y_{3}$}
\psfrag{x2}[cc][cc]{$y_{2}$}
\psfrag{x3}[cc][cc]{$y_{1}$}
\psfrag{xN}[cc][cc]{}
\centerline{\epsfxsize 2.0 truein \epsfbox{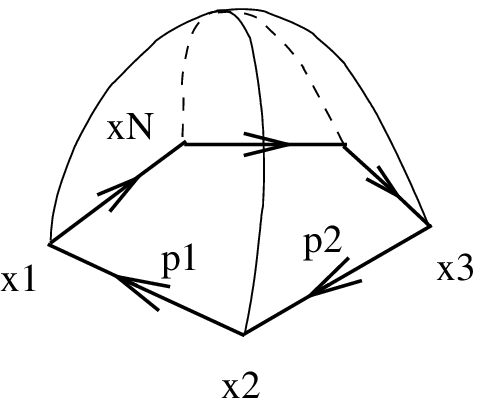}}
\caption{Figure illustrating the minimal surface, whose area $A$ enters Eq. (\ref{AM-prescription}).}
\label{Fig-Min-Area}
\end{figure}

It turns out that the boundary conditions are easier to describe after changing variables to
T-dual coordinates $y^{\mu}$,
for which the boundary conditions are simply
\be \label{AM-dualcoords}
\Delta y^{\mu} = 2 \pi k^{\mu}\,.
\ee
Notice that this is precisely the relation between momenta of
gluons and the dual coordinate space variables \ref{dualvars} from section \ref{ch-pseudoconformal}.
Importantly, after defining $r = R^2 /z$, one finds again an $AdS_{5}$ metric in the T-dual space,
\be \label{AM-T-dual-metric}
d\tilde{s}^2 = R^2 \left[ \frac{ dy_{\mu} dy^{\mu} + dr^2}{r^2}\right]\,.
\ee
So, in summary, the computation of the gluon amplitudes in the original space
was related to a computation in a dual coordinate space, which also has an $AdS_{5}$ metric.
The new boundary conditions imply that one should compute a
minimal \footnote{Strictly speaking, one should say extremal, see
a discussion in \cite{McGreevy:2007kt}.} surface that ends on a
polygon defined by the coordinates $y^{\mu}$ (which are in turn
defined by the momenta of the gluons through
(\ref{AM-dualcoords})), as shown in Fig.~\ref{Fig-Min-Area}.
%
\be \label{AM-prescription}
\mathcal{A} \propto \exp{\left(-\frac{R^2}{2 \pi} A_{\rm min} \right)} = \exp{\left(-\frac{\sqrt{\lambda}}{2\pi} A_{\rm min} \right)}\,.
\ee
The proportionality symbol in (\ref{AM-prescription}) stands for the dependence of
the amplitude $\mathcal{A}$ on the helicity information of the scattering process.
\\

It was further pointed out \cite{Alday:2007hr} that the computation of the minimal surface
in (\ref{AM-prescription}) is formally
similar to the computation of a light-like Wilson loop (defined in the
dual $AdS_{5}$ space with coordinates $y_{\mu}$) at strong coupling.
However, from the AdS (and strong coupling) point of view, one
might expect the relation between gluon scattering amplitudes and Wilson loops
to break down at subleading orders in $1/\sqrt{\lambda}$.

\subsection{Computation of $4$-cusp Wilson loop at strong coupling}
The computation of the minimal surface in (\ref{AM-prescription}) is a non-trivial problem
and so far has been achieved explicitly only in the case of $n=4$ gluons.
Following \cite{Alday:2007hr}, we first review the one-cusp solution obtained in \cite{Kruczenski:2002fb},
and then present the solution for four gluons. The quantities entering in (\ref{AM-prescription}) are
IR divergent, and hence have to be regularised. It turns out that for our purposes it is sufficient
to compute the minimal surface for the unregularised string action, and then compute the regularised minimal
area by using this unregularised solution. Therefore we present here only the solution of the unregularised
minimal surface.

\subsubsection{Single cusp}
\begin{figure}[t]
\centerline{\epsfxsize 2.0 truein \epsfbox{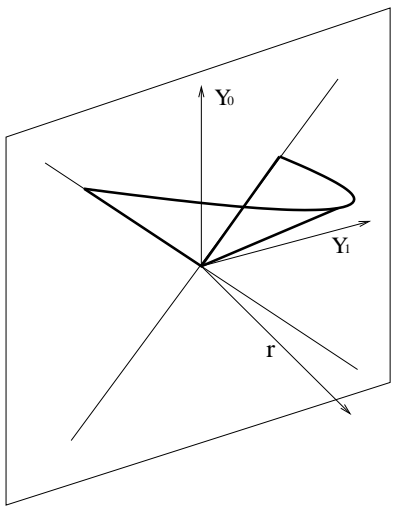}}
\caption{Graphical representation of the one-cusp solution (\ref{AM-cusp-solution}). Figure from \cite{Alday:2007hr}.}
\label{Fig-strong-cusp}
\end{figure}

Let us consider Wilson lines forming a single cusp on the boundary of $AdS_{5}$.
It is sufficient to consider a subspace $AdS_{3}$ of $AdS_{5}$, with the Wilson lines extending
along the light-cone,
\be \label{AM-cusp-boundary}
r=0\,,\qquad y_{1} = \pm y_{0}\,,\qquad y_{0}>0\,.
\ee
The metric is
\be
d^{2}s = \frac{-d^{2}y_{0} + d^{2}y_{1} + d^{2}r}{r^2}\,.
\ee
The Nambu-Goto string action is
\be \label{AM-Nambu-Goto}
S = \frac{R}{2\pi} \int ds_{1} ds_{2} \sqrt{{\rm det}\left(\partial_{a} X^{\mu} \partial_{b} X^{\nu} G_{\mu\nu} \right)}\,.
\ee
In a nonparametric form, where $r$ is expressed through $y_{0}$ and $y_{1}$, $r=r(y_{0},y_{1})$, it reads
\be \label{AM-string-action2}
S = \frac{R}{2\pi} \int dy_{0} dy_{1} \frac{1}{r^{2}} \sqrt{1 + \left(\partial_{0} r\right)^2 + \left( \partial_{1} r \right)^2 }\,.
\ee
It is straightforward to see that
\be\label{AM-cusp-solution}
r(y_{0},y_{1}) = \sqrt{2} \sqrt{y_{0}^2 - y_{1}^2}
\ee
solves the Euler-Lagrange equations following from (\ref{AM-string-action2}) with
the boundary conditions (\ref{AM-cusp-boundary}).
The minimal surface defined by (\ref{AM-cusp-solution}) is shown in Fig. \ref{Fig-strong-cusp}.
%
\\

\subsubsection{Rectangular Wilson loop}
Let us now turn to the rectangular Wilson loop.
In the case $s=t$, the minimal surface has to end on a square on the boundary.
By scale invariance, the size of the square can be chosen to be $1$, and hence the
boundary conditions are
\be \label{AM-square-boundary}
r(\pm 1, y_2 ) = r(y_1 ,\pm 1 ) = 0\,,\qquad y_{0}(\pm 1, y_{2}) = \pm y_{2}\,,\qquad y_{0}(y_{1},\pm 1) = \pm y_{1} \,.
\ee
Inspired by the one-cusp solution, Alday and Maldacena found that
\be \label{AM-sequalst-solution}
y_{0}(y_{1},y_{2}) = y_{1} y_{2}\,,\qquad r(y_{1},y_{2}) = \sqrt{(1-y_{1}^2)(1-y_{2}^2)}\,,
\ee
satisfies the equations of motion following from (\ref{AM-Nambu-Goto}) and solves the boundary condition (\ref{AM-square-boundary}).
By applying conformal transformations on the solution (\ref{AM-sequalst-solution}),
one can find the solution for arbitrary $s$ and $t$ \cite{Alday:2007hr,Ryang:2008hr}.
For its explicit expression and more details, we refer the reader to \cite{Alday:2007hr}.

\subsubsection{Analog of dimensional regularisation at strong coupling}
A subtlety arises when one tries to compute the value of the action (\ref{AM-Nambu-Goto}) by plugging e.g. the
solution (\ref{AM-sequalst-solution}) into it. One finds that the action diverges.
Therefore the action has to be regularised. In order to be able to compare to
gauge theory results, which are usually derived in dimensional regularisation,
Alday and Maldacena introduced an analog of dimensional regularisation at strong
coupling. Their regularisation consists in changing the metric (\ref{AM-T-dual-metric}) to
\be
ds^2 = \sqrt{\lambda_{D} c_{D}}
\left[
\frac{dy^2
 + dr^2}{
r^{2+\epsilon}}
\right]\,,
\ee
with $\lambda_{D} = \lambda \frac{\mu^{2\epsilon}}{(4\pi e^{-\gamma_{E}})^{\epsilon}}$ and $c_{D} = 2^{4\epsilon} \pi^{3\epsilon} \Gamma(2+\epsilon)$ \cite{Alday:2007hr}.
This leads to the modified action
\be \label{AM-reg-action}
S = \frac{\sqrt{\lambda_{D} c_{D}}}{2\pi} \int \frac{ \mathcal{L}_{\epsilon =0}}{r^{\epsilon}}\,,
\ee
where $\mathcal{L}_{\epsilon =0}$ is the Lagrangian density for $AdS_{5}$, and $\epsilon<0$ regularised the infrared divergences \footnote{Note that in \cite{Kruczenski:2007cy} it was found that the dimensional regularisation
procedure used in \cite{Alday:2007hr} has to be modified at subleading orders in $1/\sqrt{\lambda}$.
}.
The Euler-Lagrange equations following from (\ref{AM-reg-action}) now depend explicitly on $\epsilon$.
So, strictly speaking, the solutions of the Euler-Lagrange equations following from (\ref{AM-Nambu-Goto}) obtained in the previous section are no longer valid. In particular, since
conformal invariance is broken by the regulator, conformal arguments
cannot be used to relate the solution at $s=t$ to that of arbitrary $s$ and $t$.
\\

On the other hand, one can imagine the solutions obtained in this way are still useful if one is only
interested in the first few terms of the perturbative expansion in $\epsilon$
of the minimal area. Indeed, Alday and Maldacena argue that one get the accurate value of
the minimal surface by plugging the unregularised solution following from (\ref{AM-Nambu-Goto})
into the regularised action, neglecting terms of order $O(\epsilon)$.
More precisely, since the divergences come from the cusps, Alday and Maldacena modify the solution
near the cusps by an $O(\epsilon)$ term, and in this way they obtain the correct contribution
to the finite part of (\ref{AM-prescription}).
Since this constant is scheme-dependent, we will not be interested
in it here. For a more detailed discussion, see \cite{Alday:2007hr} \footnote{See also \cite{Drukker:1999zq}
for a related discussion of regularisation prescriptions for Wilson loops at strong
coupling, when using a cut-off.}.

One finds \cite{Alday:2007hr} \be \label{AM-final-answer} -
\frac{\sqrt{\lambda}}{2\pi} A_{\rm min} = -
\frac{\sqrt{\lambda}}{2\pi} \left[ A_{\rm div} + \frac{1}{2} \ln^2
\left(\frac{s}{t}\right) + C \right]\,, \ee where $C$ is a
scheme-dependent constant. The divergent part is given by $A_{\rm
div} = 2 A_{\rm div,s} + 2 A_{\rm div, t}$, where \be
\label{AM-div-part} A_{\rm div,s} = \left[ -{\epsilon^{-2}}
-{\epsilon^{-1}}\,\frac{1-\ln 2}{2} \right] \left(
\frac{-s}{\mu^2}\right)^{-\epsilon/2}\,. \ee Together with the
prescription (\ref{AM-prescription}), we see that
(\ref{AM-final-answer}) is in agreement with the general form of
the divergences of gluon scattering amplitudes, c.f. equation
(\ref{Zn-amplitude}). It also reproduces correctly the known
strong-coupling value of $\Gamma_{\rm cusp}$,
 \be
 \Gamma_{\rm cusp}(a) = \sqrt{2} \sqrt{a} + O(1,1/\sqrt{a})\,,\qquad {\rm with}\quad a = \frac{ g^2 N}{8 \pi^2} = \frac{\lambda}{8\pi^2}\,.
 \ee
 As was already explained in section \ref{ch-amplitudes}, the subleading $\epsilon^{-1}$ term and the constant in
 the finite part of (\ref{AM-final-answer}) are scheme-dependent, and therefore we are only interested
 in the functional form of finite part. Most importantly, it is in agreement with the
BDS ansatz for the four-gluon amplitude at strong coupling! This constitutes strong evidence that
the BDS ansatz holds for the four-gluon amplitude.\\

The following comments are in order:
Extending the result (\ref{AM-final-answer}) to $n\ge5$ points turns out to be difficult, and so far
no explicit solution for the minimal surface is known.
In a later paper, Alday and Maldacena found that the problem simplifies when the number of gluons
$n$ is very large. By considering a special contour they were able to relate
this problem to a known spacelike contour, and found a disagreement with the BDS ansatz for $n \rightarrow \infty$ \cite{Alday:2007he}. This suggests that the BDS ansatz should break down at a certain loop level and for a certain
number of gluons.
Finally, note that the strong-coupling calculation of \cite{Alday:2007hr} is insensitive to the helicity configuration of the scattering amplitude under consideration.


\section{Wilson loops}
\label{ch-WL}

Wilson loops were introduced in \cite{Wilson:1974sk} in the context of quark confinement in
strongly coupled QCD.
For a Yang-Mills theory with gauge group $SU(N)$, they are defined as
\be \label{defWL}
 W(C)  = \frac{1}{N}\langle 0 |\,  {\rm Tr}\, {\rm P} \exp \left(i g \oint_{C} dx^{\mu} A_{\mu}\right)   |0\rangle\,.
\ee
where $A_\mu(x)=A_\mu^a(x) t^a$ is the gauge field, $t^a$ are the $SU(N)$
generators in the fundamental representation normalised as $\tr (t^a t^b)=\ft12
\delta^{ab}$, and $\textrm{P}$ indicates the ordering of the $SU(N)$ indices along
the integration contour $C_n$.
$W(C)$ is a gauge invariant functional of the contour $C$, along which the gauge field
$A_{a}^{\mu}$ is integrated.

\subsection{Renormalisation properties}

\subsubsection{Wilson loops defined on smooth contours}
Let us summarise the renormalisation properties of the Wilson
loops (\ref{defWL}). In a generic gauge theory with ultraviolet divergences,
the gauge field $A^{\mu}_{a}$ is renormalised by multiplicative counterterms,
and so is the coupling constant $g$. These counterterms do not depend on the
contour $C$. Let us assume that this standard renormalisation has already been
carried out and focus on the divergences inherent to the Wilson loop (\ref{defWL}).
\\

For smooth contours, the only divergence of (\ref{defWL}) is linear (in the cut-off)
\cite{Polyakov:1980ca,Dotsenko:1979wb}, and proportional to the length
of the contour. It can be absorbed in an overall factor,
\be \label{smoothdiv}
W(C) = e^{- K\, L(C)} \times {\rm finite}\,,
\ee
which can be interpreted as the mass renormalisation of a
test particle moving along the contour $C$ \cite{Polyakov:1980ca,Dotsenko:1979wb}.
In (\ref{smoothdiv}), $L(C)$ is the length of the contour $C$.
\\

For example, the one-loop correction to (\ref{defWL}) is given by \footnote{In (\ref{WLsmoothex1}) and the following we neglect a trivial
normalisation factor
.}
\be\label{WLsmoothex1}
W^{(1)}(C) = \;\;\parbox[c]{15mm}{\includegraphics[height = 10mm]{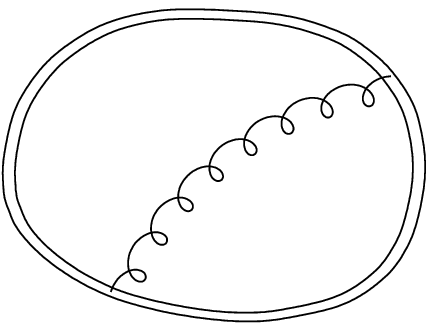}} \propto \oint_{C} \oint_{C} \frac{ dx_{\mu} dx'^{\mu}}{(x-x')^2}    \,,
\ee
where the double line in the picture denotes the integration contour $C$, and the wiggly line is a free gluon propagator.
The integral in (\ref{WLsmoothex1}) has a divergence at $x = x'$, which
can be regularised by introducing a cut-off $a^{-1}$,
\be\label{WLsmoothex2}
W^{(1)}_{\rm reg}(C) \propto \oint_{C} \oint_{C} \frac{ dx_{\mu} dx'^{\mu}}{(x-x')^2+a^2} = \int \frac{\dot{x}(s)\cdot \dot{x}(s+t)\, ds dt}{\left[ x(s+t)-x(s)\right]^2 +a^2}
\ee
Following \cite{Polyakov:1980ca}, we choose a parametrisation of the contour
with $\dot{x}(s) \cdot \dot{x}(s) = {\rm const}$, such that $\dot{x}(s) \cdot \ddot{x}(s)=0$.
Since the divergence in (\ref{WLsmoothex2}) comes from the integration near $t =0$, we can write
\begin{eqnarray}\label{WLsmoothex3}
W_{\rm reg}^{(1)}(C) &\propto& \int ds \, \dot{x}^2(s) \int_{-\Lambda}^{\Lambda} dt \frac{1}{\dot{x}^2(s) t^2 + a^2} + {\rm finite}\nonumber \\
&=& \frac{\pi}{a} \int ds \sqrt{\dot{x}^2(s)} + {\rm finite}= \frac{\pi}{a} L(C) + {\rm finite}\,.
\end{eqnarray}
For smooth contours, there are no divergences apart from the linear divergence \cite{Dotsenko:1979wb,Gervais:1979fv}.

\begin{figure}[htbp]
\psfrag{x}[cc][cc]{$x$}
\psfrag{gamma}[cc][cc]{$\gamma$}
\psfrag{C}[cc][cc]{$C$}
\centerline{{\epsfysize2.5cm \epsfbox{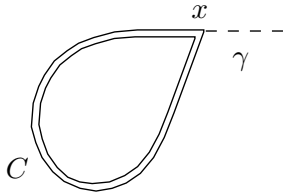}}} \caption{\small An example of a
Wilson loop defined on a contour $C$ having a cusp at the point $x$, the cusp angle being $\gamma$.}
\label{figure:cusp}
\end{figure}
\subsubsection{Contours with cusps}
\label{subsection-ordinary-cusps}
The situation gets more interesting when the contour $C$ is not smooth, but contains
one or several cusps, as shown in Fig. \ref{figure:cusp}. In that case,
UV (short-distance) divergences associated to each cusp appear \cite{Polyakov:1980ca,Brandt:1981kf}.
For example, a one-loop calculation (for gauge group $SU(N)$) gives the result
\be
W(C) = 1 - 2 g^2 C_{F} \left[ \gamma \cot(\gamma) -1 \right] \ln\left( \frac{L}{a}\right)\,,
\ee
where $L$ is the length of the contour, $a$ is a short-distance cut-off, and
$\gamma$ is the cusp angle. $C_F=t^a t^a= (N^2-1)/(2N)$ is the quadratic Casimir of
$SU(N)$ in the fundamental representation.
The result of \cite{Brandt:1981kf} is that to any order in the coupling
constant, the new divergences can be removed by a multiplicative renormalisation,
\be
W_{R}(C) = Z(\gamma) W(C)\,.
\ee
The divergences of $W_{R}(C)$ depend only locally on the contour $C$,
through the cusp angle $\gamma$.
The locality of the counterterms persists when there are several cusps
with cusp angles $\gamma_{i}$. In that case, the multiplicative renormalisation factor
\be
Z(\gamma_{1}, \ldots ,\gamma_{n}) = Z(\gamma_{1})\, \cdots \, Z(\gamma_{n})\,,
\ee
factorises into a product of renormalisation factors, one for each cusp.
Thus, there are no `anomalous' divergences that depend non-locally on the contour \footnote{This can
be also understood in the formalism
of \cite{Arefeva:1980zd,Gervais:1979fv}, where the nonlocal Wilson loop
is reformulated in terms local fermionic degrees of freedom.}.
\\

In \cite{Brandt:1982gz,Korchemsky:1987wg} a renormalisation group
equation for Wilson loops with cusps was derived. It reads \be
\left( \mu \frac{\partial}{\partial \mu} + \beta(g_{\rm R})
\frac{\partial}{\partial g_{\rm R}} + \Gamma_{\rm
cusp}(\gamma,g_{\rm R})\right) W_{\rm R}(L\mu,\gamma,\left\{ \eta
\right\},g_{\rm R}) = 0\,. \ee Here $g_{\rm R}$ is the
renormalised coupling and \be \label{gammadef}
\Gamma(\gamma,g_{\rm R}) = {\rm lim}_{\epsilon \rightarrow 0} Z
\mu \frac{\partial}{\partial \mu} Z^{-1}\,. \ee From
(\ref{gammadef}) one can derive a renormalisation group equation
for $Z^{-1}$, \be \label{RG-WLcusp} \left[ \beta(g_{\rm
R},\epsilon)\frac{\partial}{\partial g_{\rm R}} - \Gamma(\gamma,
g_{\rm R},\epsilon)\right] Z^{-1}(\gamma, g_{\rm R},\epsilon) =
0\,. \ee Its solution is \be \label{Zcusp} Z^{-1}(\gamma, g_{\rm
R},\epsilon) = \exp\left[ \int_{0}^{g_{\rm R}} dg' \Gamma(\gamma,
g',\epsilon)/\beta(g',\epsilon)  \right]\,. \ee In $\cN=4$ SYM in
dimension $D=4-2\epsilon$, and when using a scheme where
$\beta(g_{\rm R},\epsilon) = -g_{\rm R} \epsilon$, the integral in
(\ref{Zcusp}) can be done explicitly, \be \label{Z-WLcusp-N4SYM}
Z(\gamma, g_{\rm R},\epsilon) = \exp\left[ \sum_{n=1}^{\infty}
\frac{g^{2n}}{2n}
\frac{\Gamma^{(n)}(\gamma,\epsilon)}{\epsilon}\right]\,, \ee where
we expanded $\Gamma(\gamma, g_{\rm R},\epsilon) =
\sum_{n=0}^{\infty} g^{2n} \Gamma^{(n)}(\gamma,\epsilon)$. One can
define a $Z$-factor in analogue with the MS-scheme by requiring
that $\Gamma(\gamma, g_{\rm R},\epsilon)=\Gamma(\gamma, g_{\rm
R})$. So we have seen that the renormalisation group equation
(\ref{RG-WLcusp}) implies the exponentiation of the divergent part
of the Wilson loop, and in $\cN=4$ SYM it takes the particular
simple form (\ref{Z-WLcusp-N4SYM}).
\\

The Wilson loops studied in this report are defined in Minkowski space,
with cusp having an infinite cusp angle. These will be studied in more detail in section \ref{subsection-generalities}.
\\

We mention that additional divergences can appear if one considers instead of a closed contour $C$ an
open line. In that case one looses gauge independence, and there are logarithmic divergences
associated to the endpoints \cite{Dotsenko:1979wb}.
\\

We also mention the interesting case of crossed contours
(i.e., with self-intersections) \cite{Brandt:1981kf,Korchemskaya:1994qp}.
For these, there are additional divergences
associated to the self-intersection. It is found that the
loop with the self-intersection can mix with the product
of two similar loops \cite{Brandt:1981kf}, which differ
from the initial one by a different ordering of the colour
indices.

\subsection{Wilson loops in the AdS/CFT correspondence}
The study of Wilson loops in the AdS/CFT correspondence started
with the paper \cite{Maldacena:1998im} (see also
\cite{Rey:1998bq}).
There, Maldacena showed how to compute a particular kind of Wilson loop
in $\cN=4$ SYM at strong coupling, using the AdS/CFT correspondence.
The Maldacena-Wilson loop is similar to (\ref{intro-WL}), but comprises an additional term involving scalars,
\be \label{Malda-WL}
W(C) = \frac{1}{N} \langle 0 | {\rm Tr \, P} \,\exp{ \left(i g \oint_{C} ds \left[ \dot{x}^{\mu} A_{\mu} + |\dot{x}| \theta^{I} \phi_{I} \right] \right) | 0 \rangle }\,.
\ee
The $\theta^{I}$ in (\ref{Malda-WL}) parametrise $S_{5}$ of $AdS_{5} \times S_{5}$, i.e. $\theta^{I} \theta_{I} =1$.
\\

Depending on the shape of the contour, the Maldacena-Wilson loop
can be invariant under a certain number of supersymmetries. For
example, half of the supersymmetries are preserved if
$\dot{x}^{\mu}=\rm{const}$, i.e. if the contour $C$ is a straight
line \footnote{One can imagine that this Wilson `loop' is closed
at infinity.}. Such Wilson loops were studied in
\cite{Erickson:2000af}. Perturbative calculations up to order
$g^4$ and a strong coupling calculation using AdS/CFT suggest that
\be \label{WL-line} W(C)_{\rm \,straight\; line} = 1\,, \ee i.e.
that the loop corrections vanish identically. This is in close
analogy with two- and three- point functions of $1/2$-BPS
operators, see section \ref{ch-conf-pert}, which are also
invariant under half of the supersymmetry.
For Wilson loops that are invariant under less supersymmetry, see
e.g. \cite{Semenoff:2006am}.
\\

In this report we will study Wilson loops with light-like contours, i.e. $|\dot{x}| = 0$. In
this case (\ref{Malda-WL}) reduces to the usual definition of the Wilson loop (\ref{intro-WL}).
Therefore we will not expand on the discussion of the Maldacena-Wilson loop.
For reviews, see \cite{Ooguri:1999ta,Semenoff:2002kk} and also section
$10$ of \cite{Nastase:2007kj}.

\subsection{Loop equations}
Another development concerning Wilson loops is the study of loop
equations. These describe the behaviour of the loops under small
deformations of the contour $C$. We refer the interested reader to
the literature \cite{Makeenko:1979pb,Makeenko:1980vm}.

\subsection{Light-like Wilson loops}\label{subsection-generalities}

\subsubsection{Definitions}
In this report, we consider $\mathcal{N}=4$ SYM theory with $SU(N)$ gauge
group in Minkowski space.
The central object of our consideration is the light-like Wilson loop defined as
\begin{equation}\label{W}
    W\lr{C_n} = \frac1{N}\vev{0|\,{\rm Tr}\, \textrm{P} \exp\lr{i \oint_{C_n} dx^\mu A_\mu(x)}
    |0}\,,
\end{equation}
where $C_{n}$ is a polygon with $n$ cusps.
The $n$ segments of this polygon $C_n=\bigcup_{i=1}^n \ell_i$ joining
the cusp points $x_i^\mu$ (with
$i=1,2,\ldots,n$)
\begin{equation}\label{4'}
 \ell_i=\{x^\mu(\tau_i)= \tau_i x^\mu_i +
(1-\tau_i)x_{i+1}^\mu|\, \tau_i \in[0,1] \}\,,
\end{equation}
are all light-like, i.e. $x_{i,i+1}^2=0$.
Such Wilson loops were considered for the first time in \cite{Korchemskaya:1992je}
for $n=4$ points in the special kinematics $x_{13}^2 = -x_{24}^2$.

\subsubsection{Cusp singularities}\label{ch-cusp-sing}

We already saw in section \ref{subsection-ordinary-cusps} that cusps cause specific
ultraviolet divergences (UV) to appear in Wilson loops. The fact that the cusp edges are light-like
makes these divergences even more severe~\cite{Korchemskaya:1992je}. In order to get some insight into the
structure of these divergences, we start by giving a one-loop example. Later on in
this section we give arguments which lead to the all-order form of the divergences.\\

{\bf One-loop example}\\

\begin{figure}
\psfrag{xi1}[cc][cc]{$x_{i}$} \psfrag{xi3}[cc][cc]{$x_{i+1}$}
\psfrag{xi2}[cc][cc]{$x_{i-1}$} \psfrag{x}[cc][cc]{$x$}
\psfrag{a}[cc][cc]{(a)} \psfrag{b}[cc][cc]{(b)}
\psfrag{c}[cc][cc]{(c)}
\centerline{\includegraphics[height=50mm,keepaspectratio]{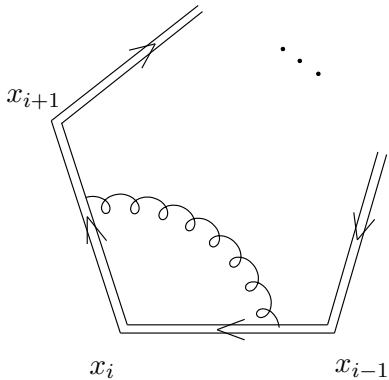}}
\caption[]{\small The Feynman diagram contributing to the one-loop divergence at the cusp point $x_i$. The double
line depicts the integration contour $C_n$, the wiggly line the
gluon propagator.}
\label{Fig:5gon}
\end{figure}

Let us first discuss the origin of the cusp singularities in the $n$-gonal Wilson
loop $W(C_n)$ to the lowest order in the coupling constant. According to
definition \re{W}, it is given by a double contour integral,
\begin{equation}\label{double-sum}
 W(C_n) = 1 + \frac12(ig)^2 C_F
 \oint_{C_n}  dx^\mu \oint_{C_n} dy^\nu\, D_{\mu\nu}(x-y)  +
 O(g^4)  \,.
\end{equation}
Here $C_F= (N^2-1)/(2N)$ is the quadratic Casimir of $SU(N)$ in the
fundamental representation and $D_{\mu\nu}(x-y)$ is the gluon propagator in the
coordinate representation
\begin{equation}\label{a1}
   \vev{A_\m^a(x)\ A_\nu^b(0)}= g^2 \delta^{ab} D_{\mu\nu}(x) \,.
\end{equation}
To regularise the ultraviolet divergences of the integrals entering
\re{double-sum}, we use dimensional regularisation with $D=4-2\epsilon$
and $\epsilon>0$. Also, making use of the gauge
invariance of the Wilson loop \re{W} and for the sake of simplicity, we
perform the calculation in the Feynman gauge, where the gluon
propagator is given by \footnote{\label{foot1} As in
\cite{Drummond:2007cf}, we redefine the conventional dimensional
regularization scale as $ \mu^2\pi{\rm e}^{\gamma_{E}} \mapsto\mu^2$ to
avoid dealing with factors involving $\pi$ and the Euler constant
$\gamma_{E}$.}
\begin{equation}\label{propagator}
D_{\mu\nu}(x)  = \eta_{\mu\nu} D(x)\,,\qquad D(x)= - \frac{\Gamma(1-\vep)}{4\pi^{2}}
(- x^2+i0)^{-1+\vep}\lr{\mu^2 {\rm e}^{-\gamma_{E}}}^{\vep}\,.
\end{equation}
The divergences in \re{double-sum} originate from the integration of the position
of the gluon in the vicinity of a light-like cusp in the Feynman diagram shown in
Fig.~\ref{Fig:5gon} (the other Feynman diagrams at one loop vanish or are finite,
see section \ref{ch-one-loop-duality}).
The calculation of this diagram is straightforward \cite{Korchemskaya:1992je}.
Let us compute it for a gluon attached to the edges $p_{i} := x_{i+1,i}$ and $p_{i-1} := x_{i,i-1}$, as shown in
Fig. \ref{Fig:5gon}.
\ba \label{one-loop-cusp}
V(p_{i-1},p_{i}) &=& -\frac{g^2 C_F}{4 \pi^2}
\lr{-e^{-\gamma_{E}} \mu^2}^\epsilon\Gamma(1-\epsilon) \int_0^1  \frac{dx\, dy\,(p_{i} \cdot
p_{i-1})}{[(p_{i} x + p_{i-1} y)^2]^{1-\epsilon}}\nonumber \\
&=& -\frac{g^2 C_F}{4 \pi^2} \lr{-e^{-\gamma_{E}} \mu^2
x_{i+1,i-1}^2}^\epsilon\frac{\Gamma(1-\epsilon)}{2\epsilon^2}\,.
\ea
In (\ref{one-loop-cusp}) we have dropped the $-i0$ prescription in the gluon propagator
because we are only interested in the kinematical region where all $x_{i+1,i-2}^2<0$.
Adding together the contributions of all cusps we obtain the
following one-loop expression for the divergent part of $W(C_n)$:
\begin{equation}\label{1-loop-div}
W(C_n) = 1 + \frac{g^2}{4\pi^2}C_F \left\{ -\frac{1}{2\vep^2}  \sum_{i=1}^n
\lr{{-x_{i-1,i+1}^2}\,{\mu^2}}^\vep + O(\vep^0)\right\}+O(g^4)\,,
\end{equation}
where the periodicity condition
\begin{equation}
x_{ij}^2=x_{i+n,j}^2=x_{i,j+n}^2 \equiv (x_i-x_j)^2
\end{equation}
is tacitly implied.
We see from (\ref{1-loop-div}) that the leading divergence in $W(C_{n})$ is a double pole
in $\epsilon$, compared to the single pole in (\ref{Z-WLcusp-N4SYM}). This is due to
the light-likeness of the edges defining $C_{n}$, which creates additional divergences.
We will discuss the general structure of divergences of light-like Wilson loops
in the next section.\\

{\bf All-loop structure}\\

Generalising \re{1-loop-div} to higher loops \footnote{For convenience of the presentation,
the following formulae are given in the planar limit, with $a = \frac{g^2 N}{8 \pi^2}$. For general
$N$, the only difference is that $\Gamma_{\rm cusp}$ and $\Gamma$ in (\ref{cusp-2loop-WL}) depend on $N$
as well as on $a$.} we find that the cusp singularities
appear in $W(C_n)$ at the $l$-th loop-order as poles $(a
\mu^{2\epsilon})^l/\epsilon^m$ with $m\le 2l$. Furthermore, $W(C_n)$ can be split
into a divergent (`renormalisation') factor $Z_n$ and a finite (`renormalised')
factor $F_n^{\rm (WL)}$ as
\begin{equation}\label{W=ZF}
\ln W(C_n) = Z_n + F_n^{\rm (WL)}\ .
\end{equation}
From the studies of renormalisation properties of light-like Wilson loops it is
known~\cite{Korchemskaya:1992je,Bassetto:1993xd} that cusp singularities exponentiate to all loops and, as a
consequence,
$Z_n$ has the special form \footnote{ Formula (\ref{Zn-WL})
follows from the evolution equation (8) of \cite{Korchemskaya:1992je}.}
\begin{equation}\label{Zn-WL}
    Z_n = -\frac{1}{4}  \sum_{l\ge 1} a^l\sum_{i=1}^n\lr{-x_{i-1,i+1}^2\mu^2}^{l\ep} \lr{\frac{\Gamma_{\rm
cusp}^{(l)}}{(l\ep)^2}+ \frac{\Gamma^{(l)}}{l\ep}} \,,
\end{equation}
with $a={g^2
N}/{(8\pi^2)}$. Here $\Gamma_{\rm cusp}^{(l)}$ and $\Gamma^{(l)}$ are the expansion coefficients
of the cusp anomalous dimension and the so-called collinear anomalous dimension,
respectively,
\begin{align}\label{cusp-2loop-WL}
& \Gamma_{\rm cusp}(a)=\sum_{l\ge 1} a^l\, \Gamma_{\rm cusp}^{(l)} = 2a -
\frac{\pi^2}3 a^2 + O(a^3)\ ,
\\ \notag
& \Gamma(a) =\sum_{l\ge 1} a^l\, \Gamma^{(l)} = - 7\zeta_3 a^2 + O(a^3)\,.
\end{align}
We have already seen at one-loop order that the divergences in $W(C_n)$ are due
to the presence of the cusps on the integration contour $C_n$. They occur when
the gluon propagator in Fig.~\ref{Fig:5gon} slides along the contour towards a
given cusp point $x_i$. The divergences arise when the propagator becomes
singular. This happens either when a gluon propagates in the
vicinity of the cusp point (short distance divergences), or when a gluon
propagates along the light-like segment adjacent to the cusp point (collinear
divergences). In these two regimes we encounter, respectively, a double and
single pole.
\\

Going to higher orders in the coupling expansion of $W(C_n)$, we immediately
realise that the structure of divergences coming from each individual diagram
becomes much more complicated. Nevertheless, the divergences originate from the same
part of the `phase space' as at one-loop: from short distances in the vicinity of
cusps and from propagation along light-like edges of the contour. The main
difficulty in analysing these divergences is due to the fact that in diagrams
with several gluons attached to different segments of $C_n$ various regimes could
be realized simultaneously, thus enhancing the strength of poles in $\epsilon$.
This implies, in particular, that individual diagrams could generate higher order
poles to $W(C_n)$.
\\

The simplest way to understand the form of $\ln Z_n$ in (\ref{Zn-WL}) is
to analyse the divergences of the Feynman diagrams not in the Feynman gauge,
but in the axial gauge \cite{Leibbrandt:1987qv} defined as
\begin{equation}
n \cdot A(x) = 0\,,
\end{equation}
with $n^\mu$ being an arbitrary vector, $n^2 \neq 0$. The reason for this is that
the contribution from individual Feynman diagrams become less singular in the axial gauge and, most
importantly, the potentially divergent graphs have a much simpler
topology~\cite{Ellis:1978ty,Dokshitzer:1978hw}.\footnote{Note that the contribution of individual Feynman diagrams is
gauge dependent and it is only their total sum that is gauge invariant.} The
axial gauge gluon propagator in the momentum representation is given by the
following expression,
\begin{equation}
\widetilde D_{\mu\nu}(k) = -i\frac{d_{\mu\nu}(k)}{k^2+i0}\,,\qquad
d_{\mu\nu}^{(A)}(k)=
 {\eta_{\mu\nu} - \frac{k_\mu n_\nu + k_\nu n_\mu}{(k n)}+ k_\mu k_\nu
\frac{n^2}{(k  n)^2}}\,.
\end{equation}
The polarisation tensor satisfies
the relation
\begin{equation}\eta^{\mu\nu} d^{(A)}_{\mu\nu}(k) = (D-2) +
\frac{k^2n^2}{(k n)^2}\,,
\end{equation}
(to be compared with the
corresponding relation in the Feynman gauge, which is $g^{\mu\nu}
d_{\mu\nu}^{(F)}(k)=D$) from which we deduce that for $k^2=0$, it
describes only the
physical polarisations of the on-shell gluon.
\footnote{That is the reason why the axial gauge is called physical.}
\\

Let us
consider a graph in which a gluon is attached to the $i$-th segment. In
configuration space, the corresponding effective vertex is described by the
contour integral $\int d\tau_i\, p_i^\mu\, A_\mu(x_i-p_i\tau_i)$. In the momentum
representation, the same vertex reads $\int d\tau_i\, p_i^\mu\, \tilde A_\mu(k)\,
{\rm e}^{ik(x_i-p_i\tau_i) }$ where the field $\tilde A_\mu(k)$ describes all
possible polarisations ($2$ longitudinal and $D-2$ transverse) of the gluon with
momentum $k^\mu$. Let us examine the collinear regime, when the gluon propagates
along the light-like direction $p^\mu_i$. The fact that the gluon momentum is
collinear, $k^\mu \sim p^\mu_i$, implies that it propagates close to the
light-cone and, therefore, has a small virtuality $k^2$. Furthermore, since
$p_i^\mu \tilde A_\mu(k) \sim k^\mu \tilde A_\mu(k)$ we conclude that the
contribution of the transverse polarization of the gluon is suppressed as
compared to the longitudinal ones. In other words, the most singular contribution
in the collinear limit comes from the longitudinal components of the gauge field
$\tilde A_\mu(k)$. The properties of the latter depend on the gauge, however. To
see this, let us examine the form of the emission vertex in the Feynman and in
the axial gauge. In the underlying Feynman integral, the gauge field $\tilde
A_\mu(k)$ will be replaced by the propagator $\tilde D_{\mu\nu}(k)$, with $\nu$
being the polarisation index at the vertex to which the gluon is attached. In
this way, we find that
\begin{equation}
p_i^\mu\, \tilde D_{\mu\nu}(k) \sim k^\mu \tilde D_{\mu\nu}(k) = -i\frac{k^\mu
d_{\mu\nu}(k)}{k^2+i0} = -\frac{i}{k^2}\times \bigg\{\begin{array}{ll}
  k^\nu\,, & \text{Feynman gauge} \\
  k^2\left[\frac{k^\nu n^2}{(kn)^2}-\frac{n^\nu}{(kn)}\right], & \text{axial gauge} \\
\end{array}
\end{equation}
Since $k^2\to 0$ in the collinear limit, we conclude that the vertex is
suppressed in the axial gauge, as compared to the Feynman gauge. This is in
perfect agreement with our physical intuition -- the propagation of longitudinal
polarisations of a gluon with momentum $k^\mu$ is suppressed for $k^2\to 0$. This
property is rather general and it holds not only for gluons attached to the
integration contour, but also for the `genuine' interaction vertices of the
$\mathcal{N}=4$ SYM Lagrangian~\cite{Ellis:1978ty,Dokshitzer:1978hw}. It should not be surprising now that
the collinear divergences in the light-like Wilson loop come from graphs of very
special topology that we shall explain in a moment.
\\

Another piece of information that will be extensively used in our analysis comes
from the non-Abelian exponentiation property of Wilson
loops~\cite{Dotsenko:1979wb,Gatheral:1983cz,Frenkel:1984pz}.
It follows from the combinatorial properties of the
path-ordered exponential and it is not sensitive to the particular form of the
Lagrangian of the underlying gauge theory.
For an arbitrary integration contour
$C$ it can be formulated as follows:
\begin{equation}\label{exponentiation}
\vev{W(C)} = 1+ \sum_{k=1}^\infty
\lr{\frac{g^2}{4\pi^2}}^k W^{(k)} = \exp\left[{\sum_{k=1}^\infty
\lr{\frac{g^2}{4\pi^2}}^k c^{(k)} w^{(k)}}\right]\,.
\end{equation}
Here $W^{(k)}$ denote the perturbative corrections to the Wilson loop, while
$c^{(k)}w^{(k)}$ are given by the contribution to $W^{(k)}$ from `webs' $w^{(k)}$
with the `maximally non-Abelian' color factor $c^{(k)}$. To the first few orders,
$k=1,2,3$, the maximally non-Abelian color factor takes the form $c^{(k)}=C_F
N^{k-1}$, but starting from $k=4$ loops it is not expressible in terms of simple
Casimir operators. Naively, one can think of the `webs' $w^{(k)}$ as of Feynman
diagrams with maximally interconnected gluon lines. For the precise definition of
`webs' we refer the interested reader to \cite{Gatheral:1983cz}.
\\

\begin{figure}
\psfrag{xi1}[cc][cc]{$x_{i}$} \psfrag{xi3}[cc][cc]{$x_{i+1}$}
\psfrag{xi2}[cc][cc]{$x_{i-1}$} \psfrag{x}[cc][cc]{$x$}
\psfrag{a}[cc][cc]{(a)} \psfrag{b}[cc][cc]{(b)}
\psfrag{c}[cc][cc]{(c)}
\centerline{\includegraphics[height=50mm,keepaspectratio]{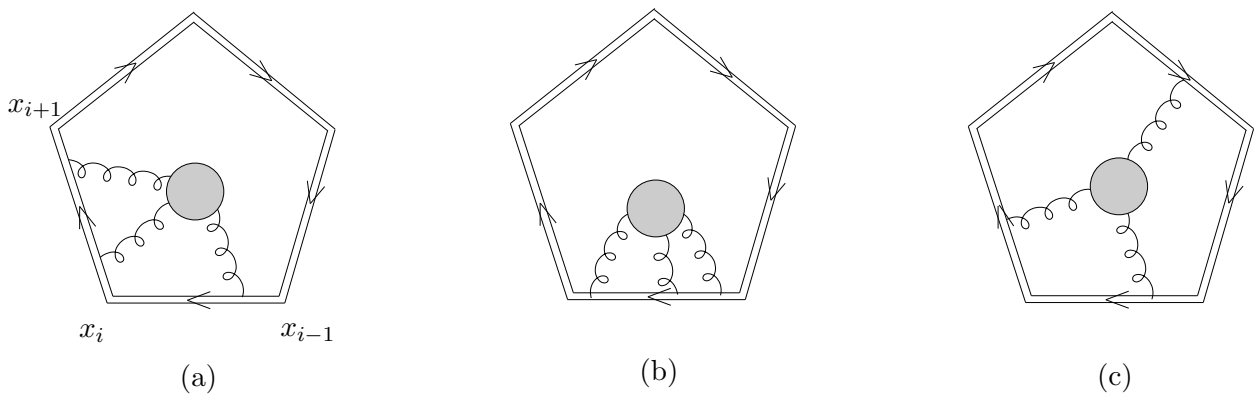}}
\caption[]{\small {Maximally non-Abelian Feynman diagrams of different topology
(`webs') contributing to $\ln W(C_n)$. In the axial gauge, the vertex-type
diagram (a) generates simple pole; the self-energy type diagram (b) generates
double pole in $\epsilon$; the diagram (c) with gluons attached to three and more
segments is finite.} }
\label{Fig:webs}
\end{figure}

Let us now return to the analysis of the cusp divergences of the light-like
Wilson loops and take advantage of both the axial gauge and the exponentiation
\re{exponentiation}. A characteristic feature of the `webs' following from their
maximal non-Abelian nature is that the corresponding Feynman integrals have
`maximally complicated' momentum loop flow, e.g. they cannot be factorised into a
product of integrals. When applied to the light-like Wilson loop, this has the
following remarkable consequences in the axial gauge:
\begin{itemize}
  \item the `webs' do not contain nested divergent subgraphs;
  \item each web produces a double pole in $\epsilon$ at most;
  \item the divergent contribution only comes from
`webs' localised at the cusp points as shown in Fig.~\ref{Fig:webs}(a)
    and (b).
\end{itemize}

These properties imply that the divergent part of $\ln W(C_n)$ is given by a sum
over the cusp points $x_i^\mu$ with each cusp producing a double and single pole
contribution. Moreover, the corresponding residue depend on $x_{i-1,i+1}^2$ --
the only kinematical invariant that one can built out of three vectors
$x_{i-1}^\mu$, $x_i^\mu$ and $x_{i+1}^\mu$ satisfying
$x_{i-1,i}^2=x_{i,i+1}^2=0$. In this way, we arrive at the known relation \re{Zn-WL}
for the divergent part of the light-like Wilson loop.
\\

Our consideration relied on the analysis of Feynman diagrams at weak coupling. It
was recently shown in Refs.~\cite{Kruczenski:2002fb,Buchbinder:2007hm,Alday:2007mf}
that the structure of divergences of $\ln W(C_n)$ remains the same even at strong coupling.


\section{Duality between Wilson loops and gluon amplitudes}\label{ch-duality}

In this section we formulate and discuss the main point of this
thesis -- the proposed duality relation between planar MHV
amplitudes, $\mathcal{M}_n^{\rm (MHV)}$, and light-like Wilson
loops, $W(C_n)$. As was already mentioned in section
\ref{ch-amplitudes}, it is known that the leading infrared (IR)
divergences of gluon amplitudes are in direct equivalence with
the leading ultraviolet divergences of Wilson loops. We will
illustrate this relationship by a one-loop example in section
\ref{FF-argument}. In section \ref{sect-duality} we state the
duality relation between gluon amplitudes and Wilson loops.

\subsection{IR divergences and their relation to Wilson loops}
\label{FF-argument} Infrared divergences of scattering amplitudes
can be understood in terms of Wilson loops. For non-Abelian gauge
theories, this property was found in
\cite{Ivanov:1985bk,Ivanov:1985np,Korchemsky:1985xj}.
It is known
\cite{Magnea:1990zb,Sterman:2002qn} that the IR divergent part of
planar scattering amplitudes factorises into a product of form
factors, and therefore it is sufficient to understand the
relationship between IR divergences of form factors and a
particular kind of Wilson
lines.\\

\begin{figure}[htbp]
\psfrag{p1}[cc][cc]{$p_{1}$}
\psfrag{p2}[cc][cc]{$p_{2}$}
\psfrag{k}[cc][cc]{$k$}
\centerline{\epsfxsize 1.5 truein \epsfbox{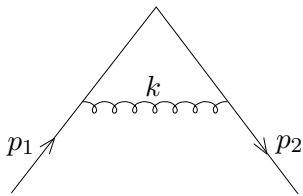}}
\caption{\small Contribution to the quark form factor at one loop.}
\label{Fig:ff-1loop}
\end{figure}

Here, we will give a one-loop
example, following to some extent
\cite{Korchemsky:1988pn} \footnote{I am indebted to Gregory Korchemsky
for enlightening explanations, on which this section is based.}. Let us
take for simplicity a form factor of massive quarks, with on-shell
momenta $p_{1}^{\mu}$ and $-p_{2}^{\mu}$, $p_{1}^2=p_{2}^2 =m^2$.
At tree level, this form factor is simply given by
 \be \label{ff-tree} \Gamma^{\alpha}_{0} = f_{abc}\,
\bar{v}_{2}(p_{2})\gamma^{\alpha} u(p_{1})\,. \ee At one loop, IR
divergences come from the vertex diagram shown in Fig. \ref{Fig:ff-1loop}
only.
Its contribution is given by \be \label{ff-1loop}
\Gamma^{\alpha}_{1} = f_{abc} \frac{g^2}{2} C_{F} \int \frac{d^{D}k}{(2\pi)^D} \frac{
\bar{v}(p_{2}) \gamma^{\mu} ({\not{p}_{2}-\not{k}+m})
\gamma^{\alpha}
({\not{p}_{1}-\not{k}+m})\gamma^{\nu}u(p_{1})}{[(p_{1}-k)^2-m^2+i0][(p_{2}-k)^2-m^2+i0]}
\tilde{D}_{\mu\nu}(k)\,, \ee where $\tilde{D}^{\mu\nu}(k)$ is the
momentum space gluon propagator, e.g.
$\tilde{D}^{\mu\nu}(k)=\eta^{\mu\nu}/k^2$ in the Feynman gauge.
The divergence in (\ref{ff-1loop}) originates from the soft region
of loop momenta, $k \ll m$. In dimensional regularisation, it
manifests itself as a pole in $\epsilon$, $\int_{\Sigma}
{d^{D}k}/{k^4} \sim {1}/{(4-D)}$, where $\Sigma$ denotes the
integration region for small $k$.\footnote{Note that for massless
quarks, $m^2=0$, there are additional collinear divergences, which
is why the leading pole is double in that case.} In that region,
we can neglect $\not k$ in the numerator, and make the
approximation \be (p_{1}-k)^2-m^2 \approx -2 (p_{1} \cdot k)\,,
\ee and similarly for the other quark propagator. Using in
addition the Dirac equations of motion for the on-shell quarks,
e.g.  $({\not{p}_{1}-m})u(p_{1}) = 0$, we obtain \be
\label{ff-1loop1} \Gamma^{\alpha\, {\rm IR}}_{1} = f_{abc} \frac{g^2}{2} C_{F}\,
\bar{v}_{2}(p_{2})\gamma^{\alpha} u(p_{1}) \, \int_{\Sigma}
\frac{d^{D}k}{(2\pi)^D} \frac{ p^{\nu}_{1}}{(-p_{1}\cdot
k+i0)}\frac{  p_{2}^{\mu}}{(-p_{2}\cdot k+i0)}
\tilde{D}_{\mu\nu}(k)\,. \ee We can interpret the integral in (\ref{ff-1loop1}) as
coming from the vertex diagram \ref{Fig:ff-1loop}, with the
quark-antiquark-gluon vertex replaced by the eikonal vertex
\be\label{eikonal vertex}
\parbox[c]{35mm}{\includegraphics[height = 25mm]{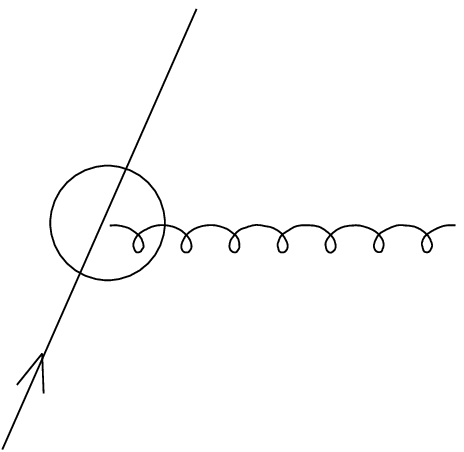}
} = \frac{p_{1}^{\nu}}{-p_{1}\cdot k+i0}\,.
\ee
Notice that the quark helicity is not modified by this eikonal vertex, and $\Gamma^{\alpha\, {\rm IR}}_{1}$
is proportional to the tree level form factor in (\ref{ff-tree}). From now on, we will drop the tree-level factor,
i.e. we consider
\be \label{ff-1loop1a}
I_{\Sigma} =  \frac{g^2}{2} C_{F} \int_{\Sigma} \frac{d^{D}k}{(2\pi)^D} \frac{ p^{\nu}_{1}}{(-p_{1}\cdot k+i0)}\frac{  p_{2}^{\mu}}{(-p_{2}\cdot k+i0)} \tilde{D}_{\mu\nu}(k)\,.
\ee
Note that from the invariance of (\ref{ff-1loop1a}) under rescalings $p_{1} \rightarrow \lambda_{1} p_{1}$
and $p_{2} \rightarrow \lambda_{2} p_{2}$, one can deduce that it depends on $p_{1}^{\mu}$ and
$p_{2}^{\mu}$ through the dimensionless variable $\cosh \gamma = (p_{1} \cdot p_{2})/\sqrt{p_{1}^2 p_{2}^2}$ only.
Geometrically, $\gamma$ corresponds to the cusp angle formed by the momenta $p_{1}$ and $p_{2}$
in Minkowski space.
\\

Now we would like to extend the integration region from $\Sigma$ to the full
phase space. In doing so, the integral in (\ref{ff-1loop1a}) acquires additional,
artificial ultraviolet (UV) divergences. In dimensional regularisation,
the UV and IR divergences cancel, and the integral over the
full phase space is zero,
\be\label{ff-1loop2}
 \frac{g^2}{2} C_{F} \int \frac{d^{D}k}{(2\pi)^D} \frac{ p^{\nu}_{1}}{(-p_{1}\cdot k+i0)}\frac{  p_{2}^{\mu}}{(-p_{2}\cdot k+i0)} \tilde{D}_{\mu\nu}(k) = \Lambda^{\rm IR} + \Lambda^{\rm UV} = 0\,.
\ee
So instead of studying the IR divergences of (\ref{ff-1loop2}) at $k\ll m$, we can equivalently study its
UV divergences at $k\gg m$. These UV divergences are well-understood, because
they correspond to cusp divergences of Wilson loops, as we will see presently.
By using the identity
\be\label{expon-formula}
\frac{1}{-p_{1}\cdot k +i0} = -i \int_{0}^{\infty} ds_{1} e^{i s_{1} (-p_{1}\cdot k +i0)}\,,
\ee
and similarly for ${-p_{2}\cdot k +i0}$, we can rewrite (\ref{ff-1loop2}) as
\begin{eqnarray}\label{ff-1loop3}
I  &=&  \frac{1}{2} (ig)^2 C_{F} \int_{0}^{\infty} ds_{1} \int_{0}^{\infty} ds_{2} \int \frac{d^{D}k}{(2\pi)^D}
p_{1}^{\mu} p_{2}^{\nu} \tilde{D}_{\mu\nu}(k) e^{i\,k\cdot (-s_{1}p_{1}-s_{2}p_{2}+i0)} \nonumber \\
 &=&  \frac{1}{2} (ig)^2 C_{F} \int_{0}^{\infty} ds_{1} \int_{0}^{\infty} ds_{2}
\, p_{1}^{\mu} \, p_{2}^{\nu} \,{D}_{\mu\nu}(p_{1}s_{1}+p_{2}s_{2}) \,,
\end{eqnarray}
where in the second line we introduced the Fourier transform of the gluon propagator,
\be
D_{\mu\nu}(x) = \int \frac{d^{D}k}{(2\pi)^D}
 \tilde{D}_{\mu\nu}(k) e^{i\,k\cdot (x+i0)}\,.
\ee
Equation (\ref{ff-1loop3}) is just the one-loop contribution to a Wilson loop \footnote{Strictly speaking,
it is a Wilson line, but one can imagine it is closed at infinity.}
\be \label{ff-WL}
W(C) = \langle 0 | {\rm Tr}\,{\rm P}\exp{ \left( i g \int_{C} dx_{\mu} A^{\mu}(x) \right)} |0 \rangle\,,
\ee
defined by the
contour $C = C_{1} \cup C_{2}$, where $C_{1} = \{s p_{1}\,, s \in [ -\infty , 0] \}$ and
$C_{2} = \{t p_{2}\,, t \in [ 0,\infty] \}$.
This means that the quarks were effectively replaced by Wilson
lines along their classical trajectory.
\begin{figure}[htbp]
\psfrag{a}[cc][cc]{$(a)$}
\psfrag{b}[cc][cc]{$(b)$}
\psfrag{gamma}[cc][cc]{$\gamma$}
\centerline{\epsfxsize 5.0 truein \epsfbox{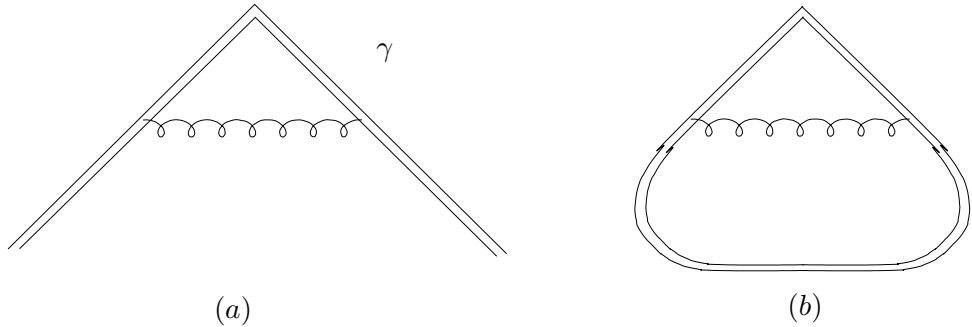}}
\caption{\small One-loop contribution to cusped Wilson loop. The double lines denote the integration contour. The UV divergences of the Wilson loop
correspond to the IR divergences of the form factor.}
\label{Fig:ff-wl-1loop}
\end{figure}

After the Fourier transform, the UV divergences of
(\ref{ff-WL}) come from the region where 
$p_{1}s_{1}+p_{2}s_{2} \ll m$, i.e. from the integration region of
small $s_{1},s_{2}$. This corresponds to the situation where the
gluon propagator is integrated near the cusp of the Wilson loop.
Taking into account the relation between UV and IR divergences
(\ref{ff-1loop2}), we see that the IR divergences of the form
factor (\ref{ff-1loop}) are equivalently described by the UV
divergences of a cusped Wilson loop, as shown in Fig.
\ref{Fig:ff-wl-1loop} (a). The latter satisfies a renormalisation
group equation \cite{Brandt:1982gz} (see also section
\ref{subsection-ordinary-cusps}). As explained in section
\ref{subsection-ordinary-cusps}, for $\cN=4$ SYM this RG equation
implies that the divergent part of the Wilson loop (and hence that
of the form factor as well) takes the form \be Z(\gamma, g_{\rm
R},\epsilon) = \exp\left[ \sum_{n=1}^{\infty} \frac{g^{2n}}{2n}
\frac{\Gamma^{(n)}(\gamma,\epsilon)}{\epsilon}\right]\,. \ee For
an explicit two-loop calculation, see \cite{Korchemsky:1991zp}.
\\

We remark that since we need only the UV divergences of the Wilson
loop shown in Fig. \ref{Fig:ff-wl-1loop} (a), which arise from the
integration of gluons near the cusp, we can choose a modified
integration contour, as shown in Fig. \ref{Fig:ff-wl-1loop} (b).
In this way, the new Wilson loop has only the desired UV
divergences, but no IR divergences.\\

Another important remark is that the definition of the Wilson loop
(\ref{ff-WL}) contains an unusual mixture of momenta and
coordinates. Indeed, the `coordinates' of the Wilson loop (\ref{ff-WL}) are
defined by the gluon momenta (and formally have `wrong' canonical
dimension). The same is true for the Wilson loops that
enter the duality with gluon amplitudes, as we will see in the next section.
These Wilson loops `live' in a dual coordinate space defined by the gluon momenta.

\subsection{Duality relation}\label{sect-duality}
The conjectured duality states that in the planar $\mathcal{N}=4$
SYM theory the finite part of the logarithms of the gluon amplitude and of the Wilson
loop are equal (up to an inessential additive constant),
\begin{equation} \label{duality-4}
{F}_n^{(\mathrm{MHV})} =  F^{(\textrm{WL})}_n + \mathrm{const} \,,
\end{equation}
upon the formal identification of the external on-shell gluon momenta in the amplitude with the
light-like segments forming the closed polygon $C_n$ (the contour of the Wilson loop),
\begin{equation} \label{duality-1}
p_i^{\mu} := x_{i+1}^{\mu} - x_{i}^{\mu}\,.
\end{equation}
Thus, the Mandelstam variables for the scattering amplitudes $t_{i}^{[j]}=(p_{i} + \ldots p_{i+j-1})^2$ are related
to the distances $x_{ij}^2$ between two cusp points on the integration contour of $W(C_{n})$ are follows,
\begin{equation}
t_{i}^{[j]}/t_{k}^{[l]} := x_{i,i+j}^2 / x_{k,k+l}^2\,.
\end{equation}
The divergent parts of the scattering amplitudes and the light-like Wilson loops are also related to each other
but the relationship is more subtle since the two objects are defined in two different schemes (infrared regularisation
for the amplitudes and ultraviolet regularisation for the Wilson loops), both based on dimensional regularisation.
From the previous section (see also the more detailed discussion in \cite{Drummond:2008aq}) we know
that the leading IR divergence of the amplitude (the coefficient of the double pole $\epsilon_{\rm IR}^{-2}$ in 
equation (\ref{Zn-amplitude})) coincides with the leading UV divergence of the Wilson loop (the coefficient of
the double pole $\epsilon_{\rm UV}^{-2}$ in equation (\ref{Zn-WL})), since both are controlled by the universal cusp 
anomalous dimension  $\Gamma_{\rm cusp}(a)$. One can also achieve a matching of the coefficients of the subleading simple
poles corresponding to the (non-universal) collinear anomalous dimensions $G(a)$ and $\Gamma(a)$.
\footnote{We thank Paul Heslop for turning our attention to the incomplete discussion of this point in the first version of \cite{Drummond:2008aq}. 
We are also grateful to Lance Dixon for a discussion of the different physical interpretations of the IR and UV simple poles \cite{Dixon:2008gr}.  } 
To this end one relates the parameters
of the two different renormalization schemes as follows,
\begin{equation} \label{duality-2}
 x_{i,i+2}^2\,\mu^2_{\rm{UV}} :=
 t_i^{[2]}/\mu_{\rm{IR}}^2\, e^{\gamma(a)}\,,\qquad
 \epsilon_{\rm  UV} := -\epsilon_{\mathrm{IR}} \, e^{\epsilon_{\rm{IR}}\delta(a)}\,.
\end{equation}
Here the functions $\gamma(a)$ and $\delta(a)$ are chosen in a way to compensate the mismatch between  $G(a)$ and $\Gamma(a)$, 
without creating extra $\mu$-dependent finite terms \cite{Drummond:2008aq}.
It should be stressed that this procedure is not analogous to comparing two different renormalisation schemes for the 
computation of the same divergent object. It is rather a change of variables (regularisation parameters) which allows 
us to compare two different objects computed in two different schemes. Another such change of variables is the 
identification (\ref{duality-1}) of the particle momenta with the light-like segments on the Wilson loop contour.\\

The duality relation (\ref{duality-4}) was inspired by the
prescription of Alday and Maldacena for computing gluon
scattering amplitudes at strong coupling \cite{Alday:2007hr}. This
prescription essentially recasts the amplitudes into light-like
Wilson loops in the dual variables (\ref{duality-1}), as discussed
in section \ref{ch-AM}. {\sl A priori}, one would expect the
strong coupling relation between gluon amplitudes and Wilson loops
to receive $1/\sqrt{\lambda}$ corrections, which might spoil the
relation at weak coupling \cite{Alday:2007hr}. Nevertheless, in
\cite{Drummond:2007aua} it was found that at one loop and for four
points the Wilson loop and the gluon amplitude agree, which lead
to the idea that the duality might also be true perturbatively. In
\cite{Brandhuber:2007yx} the duality at one loop was shown to
apply to $n$-point amplitudes as well. These one-loop results will be
reviewed in section \ref{ch-one-loop-duality}.

\subsection{Duality at one loop}\label{ch-one-loop-duality}
We already discussed results of perturbative calculations for gluon amplitudes,
in section \ref{ch-amplitudes-results}. Here, we review the Wilson loop calculation at one loop.
Let us start by computing for $n=4$ points, following \cite{Drummond:2007aua}.
The perturbative expansion of the Wilson loop
\be
W(C_{4}) = \frac{1}{N}\vev{0| {\rm Tr}\,{\rm P}\,\exp\lr{i\oint_{C_{4}} dx^\mu A_\mu(x)} |0}
\ee
to the lowest order in the coupling is given by
\be \label{WL4-1-loop}
W(C_{4}) = 1 + \frac12(ig)^2C_F \oint_{C_{4}} dx^\mu \oint_{C_{4}} dy^\nu \, D^{\mu\nu}(x-y) +
O(g^4)\,,
\ee
where $D^{\mu\nu}$ is the free gluon propagator and $C_F=(N_c^2-1)/(2N_c)$.
Its expression in the Feynman gauge was given in (\ref{propagator}).
Recall that the contour $C_{4}$ consists of four light-like segments $x_{i,i+1}^\mu$, $x_{i,i+1}^2=0$ (with
the cyclicity condition $i + 4 \equiv i$ for the indices).
Hence, the Wilson loop can only depend on the two available Lorentz invariant
quantities, $x_{13}^2$ and $x_{24}^2$. Through (\ref{duality-1}) they correspond
to the Mandelstam variables $s$ and $t$ of the four-gluon amplitude, i.e.
$x_{13}^2 = (p_{1}+p_{2})^2 =s$ and $x_{24}^2 = (p_{2}+p_{3})^2 = t$.

\begin{figure}[h]
\psfrag{x1}[cc][cc]{$x_3$}
\psfrag{x2}[cc][cc]{$x_2$}
\psfrag{x3}[cc][cc]{$x_1$}
\psfrag{x4}[cc][cc]{$x_4$}
\psfrag{a}[cc][cc]{(a)}
\psfrag{b}[cc][cc]{(b)}
\psfrag{c}[cc][cc]{(c)}
\centerline{{\epsfysize4cm \epsfbox{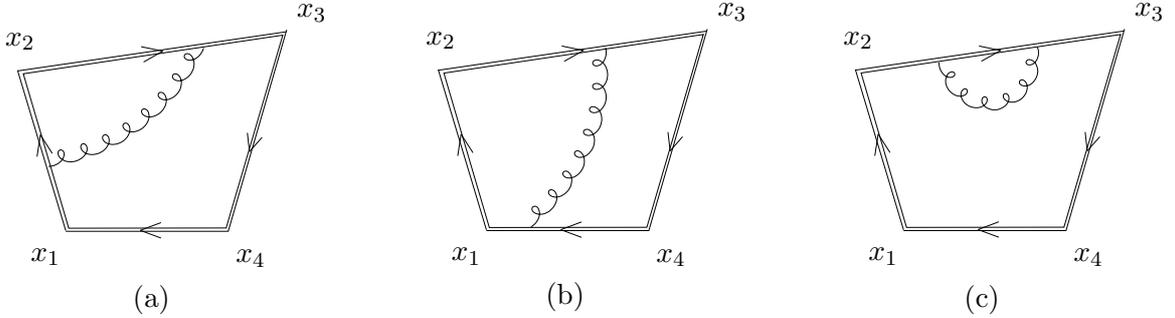}}} \caption[]{\small The Feynman diagrams
representation of the integrals contributing to (\ref{WL4-1-loop}).
The double line depicts the integration contour $C$ and the
wiggly line the gluon propagator. Figure adapted from \cite{Drummond:2007aua}.}
\label{Fig-WL-4-1loop}
\end{figure}

The Feynman diagrams contribution to (\ref{WL4-1-loop}) are shown in Fig. \ref{Fig-WL-4-1loop}.
In the Feynman gauge, diagram (c) vanishes due to the light-likeness of the edge the gluon
is attached to.
The vertex type diagram (a) was already calculated in section \ref{ch-cusp-sing}.
In application to the four-point case we find from (\ref{1-loop-div})
\be
V(p_1\,,p_2) + V(p_2\,,p_3)+ V(p_3\,,p_4)+ V(p_4\,,p_1) = - \frac{g^2 C_F}{4 \pi^2}
\frac{\Gamma(1-\epsilon)}{\epsilon^2}\left[\lr{- e^{-\gamma_{E}}  \mu^2
s}^\epsilon+\lr{- e^{-\gamma_{E}} \mu^2 t}^\epsilon \right]
\ee
Let us now turn to diagram (b), in which the gluon is attached to segments $p_2$
and $p_4$. This diagram is finite for $\varepsilon \to 0$, so we evaluate
it in $D=4$ dimensions
\be
I(p_2,p_4) = -\frac{g^2 C_F}{4 \pi^2} \int_0^1  \frac{dx \,dy\,(p_2 \cdot p_4)}{ (p_2 x
+ p_4 y+p_3)^2 } =  \frac{g^2 C_F}{8\pi^2} \int_0^1  \frac{dx\, dy\,(s+t)}{
tx+sy -(s+t)xy}\,.
\ee
Integration yields
\ba
I(p_2,p_4) &=& -\frac{g^2 C_F}{8\pi^2}\int_0^1
\frac{dx}{x-\frac{s}{s+t}}\left[ \ln\frac{s}{t}+\ln\frac{1-x}{x} \right]=\frac{g^2 C_F}{16\pi^2}  \left[\ln^2 ({s}/{t})+\pi^2\right]\,,
\ea
where we assumed that $s,t <0$.
The other diagram of the same topology, with the gluon going from segment $p_{1}$ to $p_{3}$ gives
the same contribution, since it is obtained from interchanging $s$ and $t$.
Finally, summing up all contributions and expanding in powers of $\epsilon$, we find
\be \label{WL-4-1loop-fin}
W(C_{4}) = 1+\frac{g^2 C_F}{4\pi^2}\left\{-
\frac{1}{\epsilon^2}\left[ \lr{- \mu^2
x_{13}^2}^\epsilon+\lr{- \mu^2 x_{24}^2}^\epsilon \right]  + \frac{1}{2} \ln^2 \left(\frac{x_{13}^2}{x_{24}^2}\right)+ \frac{\pi^2}{3} \right\}
\ee
In order to compare with the corresponding gluon amplitude, we write the finite part of $\ln W(C_{4})$ at one
loop in the planar limit, which amounts to replacing $C_{F} \rightarrow N/2$, i.e. ${g^2 C_F}/({4\pi^2}) \rightarrow a$,
\begin{equation}\label{WL-4-1loop-fin2}
\ln W(C_{4}) = F_{4}^{\rm (WL)} = a \left\lbrack \frac{1}{2} \ln^2 \left(\frac{x_{13}^2}{x_{24}^2}\right)+ \frac{\pi^2}{3} \right\rbrack  + O(a^2)\,.
\end{equation}
Comparing to the corresponding expression for the four-gluon amplitude (\ref{BDS4})
and the one-loop value of the cusp anomalous dimension (\ref{cusp-2loop-WL}), we see
that (\ref{WL-4-1loop-fin2}) is indeed in agreement with the duality relation (\ref{duality-4}).
Let us stress again that the additive constant in the finite part of (\ref{WL-4-1loop-fin}) is
scheme dependent.
\\

To get the $n$-point result, one has to evaluate the finite diagram (b) in more general
kinematics \cite{Brandhuber:2007yx}. The result is again in agreement with the duality relation (\ref{duality-4}).

\subsection{Checks of the duality at two loops and beyond}

We remark that the one-loop Wilson loop calculations presented in the
previous section did not use special properties of $\cN=4$ SYM.
Indeed, only the free gluon propagator $D^{\mu\nu}(x)$ entered the calculation,
and hence the result in other gauge theories would have been the same.
\\

This motivates to extend the checks of the duality (\ref{duality-4})
beyond one loop, where one can see a difference between e.g. QCD and
 $\cN=4$ SYM. The remaining sections of this thesis are devoted to this.
We report on results that confirm the duality (\ref{duality-4}) in several non-trivial
cases at two loops and beyond.
\\

In \cite{Drummond:2007cf}, Drummond, Korchemsky, Sokatchev and the present author carried
out the two-loop computation of the Wilson loop for $n=4$ points,
which we will present in section \ref{ch-four-point}.
Technical details of this calculation can be found in appendix \ref{ch-WL-appendix},
where the evaluation of each Feynman diagram is shown explicitly.
In section \ref{section-CWI} we present the results of the publication \cite{Drummond:2007au},
where we derive all-order Ward identities for the Wilson loops (\ref{intro-WL}).
In section \ref{ch-six-point}, we perform a calculation at six points and two loops
and compare to the BDS ansatz and a recently available two loop calculation of
the six-gluon MHV amplitude.


\section{Two loop tests of the duality}\label{ch-four-point}

\subsection{Rectangular Wilson loop}
In this section we present the result of the calculation of the rectangular Wilson loop
\be
W(C_{4}) = \frac{1}{N} \vev{0|{\rm Tr}\,{\rm P}\,\exp\lr{i\oint_{C_{4}} dx^\mu A_\mu(x)} |0}\,,
\ee
to two loops. The details of the calculation can be found in the appendix.
\\

\begin{figure}[ht]
\psfrag{x1}[cc][cc]{$x_3$} \psfrag{x2}[cc][cc]{$x_2$} \psfrag{x3}[cc][cc]{$x_1$}
\psfrag{x4}[cc][cc]{$x_4$} \psfrag{a}[cc][cc]{(a)} \psfrag{b}[cc][cc]{(b)}
\psfrag{c}[cc][cc]{(c)} \psfrag{d}[cc][cc]{(d)} \psfrag{e}[cc][cc]{(e)}
\psfrag{f}[cc][cc]{(f)} \psfrag{g}[cc][cc]{(g)} \psfrag{h}[cc][cc]{(h)}
\psfrag{i}[cc][cc]{(i)} \psfrag{j}[cc][cc]{(j)} \psfrag{k}[cc][cc]{(k)}
\psfrag{l}[cc][cc]{(l)} \centerline{{\epsfysize14cm \epsfbox{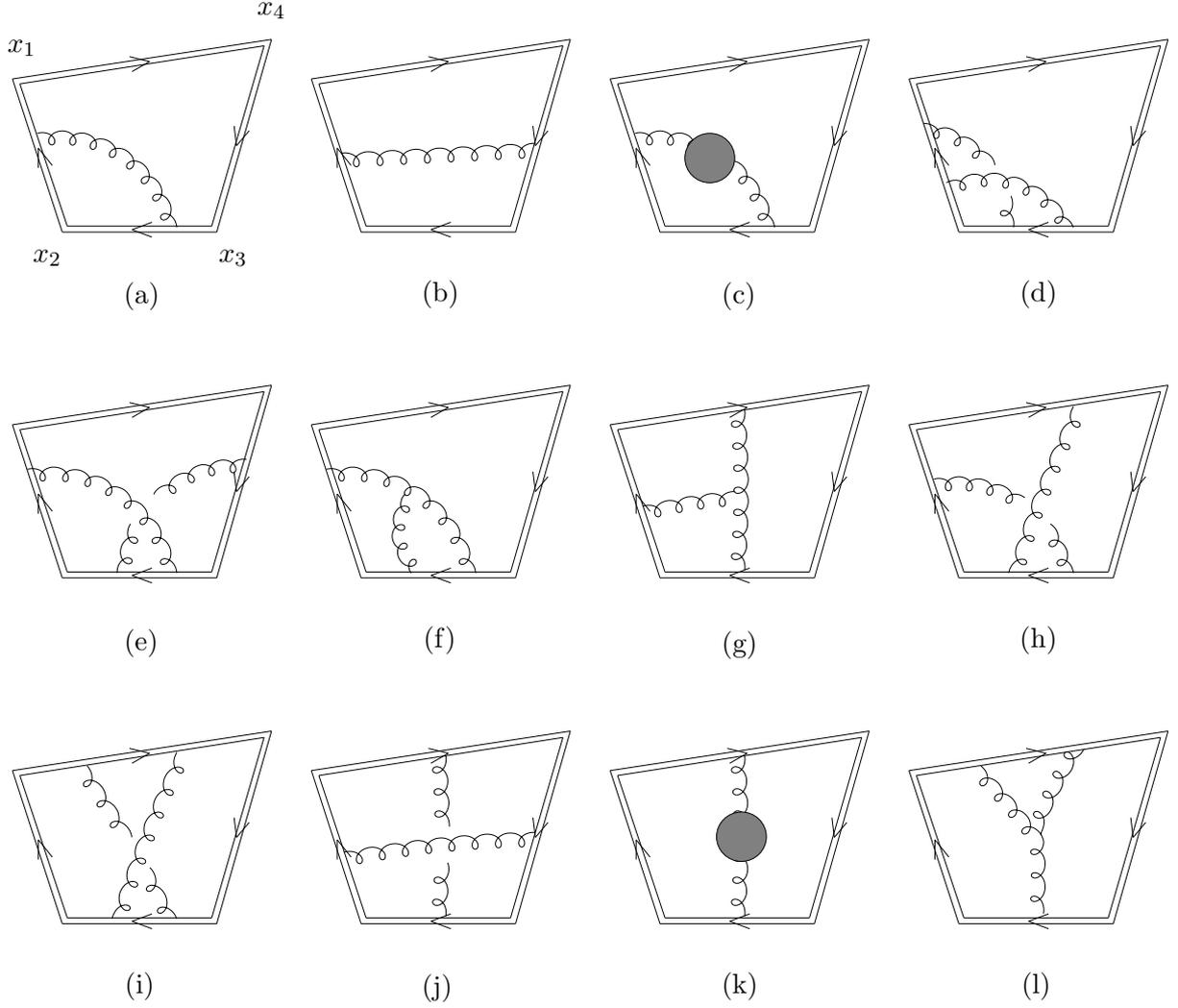}}}
\caption[]{\small The Feynman diagrams contributing to $\ln {W(C_4)}$ to two
loops. The double line depicts the integration contour $C_4$, the wiggly line the
gluon propagator and the blob the one-loop polarization operator.}
\label{all-diags}
\end{figure}

In order to compute the two-loop Feynman diagrams shown in Fig.~\ref{all-diags}
we employ the technique described in detail in Refs.~\cite{Korchemsky:1988si,Korchemsky:1992xv,Korchemskaya:1992je,Bassetto:1993xd}.
Furthermore, the rectangular light-like Wilson loop under consideration has been
already calculated to two loops in \cite{Korchemskaya:1992je,Bassetto:1993xd} in the so-called forward limit
$x_{12}^\mu=-x_{34}^\mu$, or equivalently $x_{13}^2=-x_{24}^2$, in which the
contour $C_4$ takes the form of a rhombus.
\\

The vertex-like diagrams shown in
Fig.~\ref{all-diags}(c), \ref{all-diags}(d) and \ref{all-diags}(f) only depend on
the distance $x_{13}^2$ and hence we can take the results from \cite{Korchemskaya:1992je,Bassetto:1993xd}.
The Feynman diagrams shown in Fig.~\ref{all-diags}(h) -- \ref{all-diags}(l) are
proportional to scalar products $(x_{12}\cdot x_{34})$ and/or $(x_{23}\cdot
x_{14})$ and, therefore, they vanish in the limit $x_{12}^\mu=-x_{34}^\mu$ due to
$x_{i,i+1}^2=0$. For generic forms of $C_4$, these diagrams have to be calculated
anew. Similarly, we also have to reexamine the contribution of diagrams
shown in Figs.~\ref{all-diags}(e) and \ref{all-diags}(g). Finally, the
calculation in Ref.~\cite{Korchemskaya:1992je,Bassetto:1993xd} was performed
within the conventional dimensional
regularisation (DREG) scheme. To preserve supersymmetry, we have to use instead
the dimensional reduction (DRED) scheme. The change of scheme only affects the
diagrams shown in Figs.~\ref{all-diags}(c) and \ref{all-diags}(k) which involve
an internal gluon loop~\cite{Belitsky:2003ys}.
\\

The results of our calculation can be summarised as follows. Thanks to the non-Abelian
exponentiation (\ref{exponentiation}), the two-loop expression for the (unrenormalised) light-like Wilson
loop can be represented as
\be\label{W-decomposition}
\ln W(C_4) = \frac{g^2}{4\pi^2}C_F w^{(1)}  +  \lr{\frac{g^2}{4\pi^2}}^2 C_F N
w^{(2)}  + O(g^6)\,.
\ee
According to (\ref{WL-4-1loop-fin}) the one-loop correction $w^{(1)}$ is given by
\be\label{w1}
w^{(1)}=-\frac{1}{\vep^2} \left[\lr{{-x_{13}^2}\,{\mu^2}}^\vep+\lr{
{-x_{24}^2}\,{\mu^2}}^\vep\right] +\frac12
\ln^2\lr{\frac{x_{13}^2}{x_{24}^2}}+\frac{\pi^2}{3}+ O(\epsilon)\,.
\ee
The two-loop correction $w^{(2)}$ is given by a sum over the individual diagrams
shown in Fig.~\ref{all-diags} plus crossing symmetric diagrams. It is convenient
to expand their contributions in powers of $1/\epsilon$ and separate the UV divergent
and finite parts as follows
\be\label{para}
w^{(2)} = \sum_{\alpha}  \left[(-x_{13}^2\, \mu^2)^{2\epsilon}+ (-x_{24}^2\,
\mu^2)^{2\epsilon}\right] \bigg\{\frac1{\epsilon^4}A_{-4}^{(\alpha)} +
\frac1{\epsilon^3}A_{-3}^{(\alpha)} + \frac1{\epsilon^2}A_{-2}^{(\alpha)} +
\frac1{\epsilon}A_{-1}^{(\alpha)}\bigg\} + A_0^{(\alpha)} + O(\epsilon)\,,
\ee
where the sum goes over the two-loop Feynman diagrams shown in
Fig.~\ref{all-diags}(c)--(l). Here $A^{(\alpha)}_{-n}$ (with $0\le n \le 4$) are
dimensionless functions of the ratio of distances $x_{13}^2/x_{24}^2$. Making use
of \re{para}, we can parametrise the contribution of each individual diagram to
the Wilson loop by the set of coefficient functions $A^{(\alpha)}_{-n}$.

\begin{itemize}

\item UV divergent $O(1/\epsilon^4)$ terms only come from the two Feynman diagrams
shown in Figs.~\ref{all-diags}(d) and \ref{all-diags}(f)
\be\label{A4}
A_{-4}^{\rm (d)}=-\frac1{16}\,,\qquad A_{-4}^{\rm (f)}= \frac1{16}
\ee

\item  UV divergent $O(1/\epsilon^3)$ terms only come from the two Feynman diagrams
shown in Figs.~\ref{all-diags}(c) and \ref{all-diags}(f)
\be\label{A3}
A_{-3}^{\rm (c)}=\frac1{8}\,,\qquad A_{-3}^{\rm (f)}=-\frac1{8}
\ee

\item   UV divergent $O(1/\epsilon^2)$ terms only come from the Feynman diagrams
shown in Figs.~\ref{all-diags}(c)-- \ref{all-diags}(g)
\be\label{A2}
A_{{-2}}^{\rm (c)}=\frac14\,,\qquad A_{{-2}}^{\rm (d)}=-{\frac
{\pi^2}{96}}\,,\qquad A_{{-2}}^{\rm (e)}=-\frac{\pi^2}{24} \,,\qquad
A_{{-2}}^{\rm (f)}=-\frac14+{\frac {5}{96}}\,{\pi }^{2}\,,\qquad A_{{-2}}^{\rm
(g)}= \frac{{\pi }^{2}}{48}
\ee

\item  UV divergent $O(1/\epsilon^1)$ terms come from the Feynman diagrams shown in
Figs.~\ref{all-diags}(c)--\ref{all-diags}(h),\ref{all-diags}(k) and
\ref{all-diags}(l)
\begin{align} \nonumber
&  A_{{-1}}^{\rm (c)}=\frac12+\frac{{\pi }^{2}}{48}\,,  & & A_{{-1}}^{\rm
(d)}=-\frac1{24}\zeta_3\,, &
\\ \nonumber
 & A_{{-1}}^{\rm (e)}=\frac12 \zeta_3\,, & &
A_{{-1}}^{\rm (f)}=-\frac12-\frac{{\pi }^{2}}{48} +{\frac {7}{24}} \zeta_3\,, &
\\ \nonumber
& A_{{-1}}^{\rm (g)}=-\frac18 M_{{2}}+\frac18 \zeta_3\,, & & A_{{-1}}^{\rm
(h)}=\frac14 M_{{2}}\,,  &
\\ \label{A1}
 & A_{{ -1}}^{\rm (k)}=\frac14 M_{{1}}\,,
& & A_{{-1}}^{\rm (l)}=-\frac14 M_{{1}}-\frac18 M_{{2}} &
\end{align}

\item  Finite $O(\epsilon^0)$ terms come from all Feynman diagrams shown in
Figs.~\ref{all-diags}(c)--\ref{all-diags}(l)
\begin{align} \nonumber
&  A_{{0}}^{\rm (c)}= 2+\frac{{\pi }^{2}}{12}+\frac16\zeta_3\,, & & A_{{0}}^{\rm
(d)}=-{\frac {7}{2880}}\,{\pi }^{4} \,, &
\\ \nonumber
& A_{{0}}^{\rm (e)}=-\frac{{\pi }^{2}}{12} M_{{1}}-{ \frac {49}{720}}\,{\pi }^{4}
\,, & & A_{{0}}^{\rm (f)}=-2-\frac{{\pi }^{2}}{12}+{\frac {119 }{2880}}\,{\pi
}^{4}-\frac16 \zeta_3 \,, &
\\ \nonumber
& A_{{0}}^{\rm (g)}=\frac1{24} M_ {{1}}^{2}-\frac14 M_{{3}}+{\frac
{7}{360}}\,{\pi }^{4} \,, & & A_{{0}}^{\rm (h)}= \frac18 M _{{1}}^{2}+\frac38
M_{{3}}+\frac{\pi^2}8 M_{{1}} \,,  &
\\ \nonumber
& A_{{0}}^{\rm (i)}=-\frac1{24} M_{{1 }}^{2} \,, &&  A_{{0}}^{\rm
(j)}=-\frac1{8}M_{{1}}^{2} &
\\ \label{A0}
& A_{{0}}^{\rm (k)}= M_{{1}}+\frac12\,M_{{2}}\,, & & A_{{0}}^{\rm (l)}=
-M_{{1}}+\frac{\pi^2}{24} M_{{1}}-\frac12 M_{{2}}-\frac18\,M_{{3}}\,.  &
\end{align}

\end{itemize}
Here the notation was introduced for the integrals $M_i=M_i(x_{13}^2/x_{24}^2)$
\ba \nonumber
M_{1} &=& \int_{0}^{1}\frac{d\beta}{\beta-\bar{\alpha}}
\ln\left(\frac{\bar{\alpha}
\bar{\beta}}{\alpha \beta}\right) = -\frac{1}{2} \left[ \pi^2 +
\ln^2\left(\frac{x_{13}^2}{x_{24}^2}\right) \right],
\\ \nonumber
M_{2} &=& \int_{0}^{1}\frac{d\beta}{\beta-\bar{\alpha}}
\ln\left(\frac{\bar{\alpha}
\bar{\beta}}{\alpha \beta}\right) \ln(\beta
\bar{\beta})\,,
\\ \label{M-integrals}
M_{3} &=& \int_{0}^{1}\frac{d\beta}{\beta-\bar{\alpha}}
\ln\left(\frac{\bar{\alpha}
\bar{\beta}}{\alpha \beta}\right) \ln^2(\beta
\bar{\beta})\,,
\ea
with $\bar \beta=1-\beta$, $\bar\alpha=1-\alpha$ and
$\bar\alpha/\alpha=x_{13}^2/x_{24}^2$. We do not need the explicit expressions
for the integrals $M_{2}$ and $M_{3}$ since, as we will see shortly, the
contributions proportional to $M_{2}$ and $M_{3}$ cancel in the sum of all
diagrams (for completeness, their explicit form can be found in Appendix \ref{Appendix-integrals}).
Note that the integrals \re{M-integrals} vanish in the
forward limit $x_{13}^2=-x_{24}^2$.
\\

We would like to stress that the above results were obtained in the Feynman
gauge. Despite the fact that the contribution of each individual
Feynman diagram
to the light-like Wilson loop (or equivalently, the
$A_{-n}^{(\alpha)}-$functions) is gauge-dependent, their sum is gauge-invariant.
\\

Next, we substitute the obtained expressions for the coefficient functions,
Eqs.~\re{A4} -- \re{A0}, into \re{para} and finally arrive at the following remarkably
simple expression for the two-loop correction,
\be\label{para1}
w^{(2)} =  \left[(-x_{13}^2\,\mu^2)^{2\epsilon}+
(-x_{24}^2\,\mu^2)^{2\epsilon}\right] \bigg\{ \epsilon^{-2} \frac{\pi^2}{48} +
\epsilon^{-1}\frac{7}{8}\zeta_3\bigg\} -
\frac{\pi^2}{24}\ln^2\lr{\frac{x_{13}^2}{x_{24}^2}} - \frac{37}{720}\pi^4 +
O(\epsilon)\,.
\ee
We verify that in the forward limit, i.e. for ${x_{13}^2}=-{x_{24}^2}$, this relation is in agreement with
the previous calculations of Refs.~\cite{Korchemskaya:1992je,Bassetto:1993xd}.
The following comments are in order.
\\

Arriving at \re{para1} we notice that the
leading UV divergent $O(1/\epsilon^4)$ and $O(1/\epsilon^3)$  terms  cancel in
the sum of all diagrams, in agreement with (\ref{Zn-WL}).
According to \re{A2}, the coefficients in front of
$1/\epsilon^2$ are given by a sum of a rational number and $\pi^2-$term. The
rational terms cancel in the sum of all diagrams. As a consequence, the residue
of the double pole in $\epsilon$ of $w^{(2)}$ in Eq.~\re{para1} is proportional
to $\zeta_2$. In a similar manner, the residue of $w^{(2)}$ at the single pole in
$\epsilon$ is proportional to $\zeta_3$ and this comes about as the result of a
cancelation between various terms in \re{A1} containing rational numbers,
$\pi^2-$terms as well as the integrals $M_1$ and $M_2$. The most striking
simplifications occur in the sum of finite $O(\epsilon^0)$ terms \re{A0}. We find
that the integrals $M_2$, $M_3$, $M_1^2$ as well as the rational corrections and
the terms proportional to $\pi^2$ and $\zeta_3$ cancel in the sum of all diagrams
leading to $-\ft7{720}\pi^4+\ft1{12}\pi^2 M_1$.
\\

We would like to stress that the two-loop expression \re{para1} satisfies the
``maximal transcendentality principle'' in $\mathcal{N}=4$ SYM~\cite{Kotikov:2004er}.
Let us assign transcendentality $n$ to functions ${\rm Li}_{n}$ \footnote{E.g. $\ln(x) = {\rm Li}_{1}(1-x)$ and $\zeta_{2} = \pi^2/6 = {\rm Li}_{2}(1)$ are assigned transcendentality $1$ and $2$, respectively.} and transcendentality $1$ to a
single pole $1/\epsilon$. Then it is easy
to see from \re{para1} that the coefficient in front of $1/\epsilon^n$ (including
the finite $O(\epsilon^0)$ term) has transcendentality equal to $4-n$. In this
way, each term in the two-loop expression \re{para1} has transcendentality $4$.
For the same reason, the one-loop correction to the Wilson loop, Eq.~\re{w1} has
transcendentality $2$. Generalising this remarkable property to higher loops
in planar $\mathcal{N}=4$ SYM, we expect that the perturbative
correction to the Wilson loop \re{W} to order $O(g^{2n})$ should have transcendentality
$2n$.
\\

Notice that in our calculation of the two-loop Wilson loop we did not
rely on the multi-colour limit. In fact, due to the special form of the maximally
non-Abelian colour
factors, $c^{(n)} = C_F N^{n-1}$, relation \re{W-decomposition} is exact for
arbitrary $N$. As was already mentioned, these colour factors become more
complicated starting from $n=4$ loops, where we should expect terms subleading in
$N$.

\subsection{Pentagonal Wilson loop}

In \cite{Drummond:2007au}, the calculation of the pentagonal Wilson loop
at two loops was carried out in order to further test the duality (\ref{duality-4}).
The Feynman diagrams are similar to those shown in Fig. \ref{all-diags}.
Individual diagrams can depend on ratios of the five Lorentz invariants available
from five light-like distances, $x_{13}^2, x_{24}^2, x_{35}^2, x^2_{41}$ and $x_{52}^2$.
We derived (multiple) parameter integral representations for all diagrams.
The pole part of all diagrams could be evaluated analytically, and for the finite
part we found within good numerical precision the following result:
\begin{eqnarray} \label{WL-pentagon-2loop}
w^{(2)}_{5} = \left\lbrack \frac{1}{\epsilon^{2}}\frac{\pi^2}{96}+\frac{1}{\epsilon} \frac{7}{16}\zeta_{3} \right\rbrack \sum_{i=1}^5 (-x_{i,i+2}^2\,\mu^{2})^{2\epsilon} + \frac{\pi^2}{48} \sum_{i=1}^5
\ln
 \Bigl(\frac{x_{i,i+2}^2}{x_{i,i+3}^2}\Bigr) \ln
 \Bigl(\frac{x_{i+1,i+3}^2}{x_{i+2,i+4}^2}\Bigr) - \frac{\pi^4}{144}\,.
\end{eqnarray}
We would like to mention that the calculation of the pole part in \cite{Drummond:2007au} was carried out analytically for
arbitrary $n$ at two loops, in agreement with the general formula (\ref{Zn-WL}). As explained in that paper, one can
define `auxiliary' diagrams to subtract the pole terms, thereby providing a definition of the finite part of the two-loop $n$-cusp
Wilson loop in terms of a set of {\it finite} parametric integrals.

\subsection{Check of the duality at two loops}
Let us introduce the following notation for the perturbative expansion of $\ln W(C_{n}) = Z_{n} + F_{n}^{\rm (WL)}$ in the
planar limit:
\begin{equation}
F_{n}^{\rm (WL)} = a \, F^{\rm (WL)}_{n;1} + a^2 \, F^{\rm (WL)}_{n;2} + \ldots\,.
\end{equation}
Then we see from (\ref{para1}) and (\ref{WL-pentagon-2loop}) that we have
\begin{eqnarray}
F^{\rm (WL)}_{4;2} &=& -
\frac{\pi^2}{24}\ln^2\lr{\frac{x_{13}^2}{x_{24}^2}} + {\rm const}\,, \label{F-WL-4-2l}\\
F^{\rm (WL)}_{5;2} &=& \frac{\pi^2}{48} \sum_{i=1}^5
\ln
 \Bigl(\frac{x_{i,i+2}^2}{x_{i,i+3}^2}\Bigr) \ln
 \Bigl(\frac{x_{i+1,i+3}^2}{x_{i+2,i+4}^2}\Bigr) + {\rm const}\,.  \label{F-WL-5-2l}
\end{eqnarray}
Comparing (\ref{F-WL-4-2l}),(\ref{F-WL-5-2l}) to
(\ref{BDS4}),(\ref{BDS5}) it is straightforward to see that the duality relation (\ref{duality-4}) is verified
for $n=4,5$ at two loops.
\\

One may wonder, independently of the duality with gluon amplitudes, why the two-loop
results for (the finite part of the logarithm of) the rectangular and pentagonal Wilson loops (\ref{F-WL-4-2l}) and (\ref{F-WL-5-2l})
are so simple. For example, one could have expected nontrivial multi-variable functions
to appear in the $5$-point case, such as harmonic polylogarithms.
In section \ref{section-CWI} we will see that the simplicity observed here is explained by conformal symmetry.
There, we will derive all-order broken conformal Ward identities that have stringent implications for
$F_{n}^{\rm (WL)}$.


\section{Conformal symmetry of light-like Wilson loops }\label{section-CWI}

In this section we review the derivation of all-order
broken conformal Ward identities for the Wilson loops $W\lr{C_n}$ \cite{Drummond:2007cf,Drummond:2007au}.
\\

Recall that $C_{n}$ is a polygonal contour with $n$ light-like edges.
The decisive observation is that such contours are stable under conformal
transformations.
That is, the contour $C'_n$, which is the image of the contour $C_n$ upon a conformal
transformation, is also made of $n$ light-like segments with new cusp points ${x_i'}^{\mu}$ obtained as the images of
the old ones $x_i^\mu$.
This property is rather obvious for translations,
rotations and dilatations, but we have to check it for special conformal
transformations. Performing a conformal inversion~\footnote{A special conformal
transformation (boost) is equivalent to an inversion followed by a translation
and then another inversion (cf. section \ref{ch-conformal}).}, ${x^\mu}' = {x^\mu}/{x^2}$, of all points belonging
to the segment
\begin{equation}
x^\mu(\tau_i)= \tau_i x^\mu_i +
(1-\tau_i)x_{i+1}^\mu\,,
\end{equation}
with $\tau \in \lbrack 0, 1\rbrack$, we obtain another segment of the same type,
\begin{equation}
{x'}^\mu(\tau_i')= \tau_i' {x'}_i^\mu + (1-\tau_i') {x'}_{i+1}^\mu\,,
\end{equation}
with
$(x'_{i+1}-x'_i)^2=0$ and $\tau_i' = \tau_i/[\tau_i+
(1-\tau_i)(x'_{i})^2/(x'_{i+1})^2]$.
\\

Were the Wilson loop $W\lr{C_n}$ well defined in $D=4$ dimensional Minkowski
space-time, then it would enjoy the (super)conformal invariance of the underlying
$\mathcal{N}=4$ SYM theory. More precisely, in the absence of conformal anomalies
we would conclude that
\begin{equation}\label{cin}
    W\lr{C'_n} = W\lr{C_n}\,.
\end{equation}
However, as we already pointed out in section \ref{subsection-generalities}, the light-like
cusps of $C_{n}$ cause specific ultraviolet divergences to appear in the Wilson loop (\ref{W}). For this
reason we use dimensional regularisation with $D=4-2\epsilon$ and $\epsilon>0$, which breaks
the conformal invariance of the action, as we will see presently.
In the dimensionally regularised $\mathcal{N}=4$ SYM theory, the Wilson loop
$W\lr{C_n}\equiv \vev{W_n}$ is given by a functional integral
\begin{align}\label{W-path-integral}
\vev{W_n} =  \int {\cal D}A\, {\cal D}\lambda \, {\cal D}\phi\ {\rm e}^{i
S_{\epsilon}[A,\,\lambda,\,\phi] }\ {\rm Tr}\left[ \textrm{P} \exp\left(i \oint_{C_n}
dx\cdot A(x) \right)\right]\,,
\end{align}
where the integration goes over gauge fields, $A$, gaugino, $\lambda$, and scalars,
$\phi$, with the action
\begin{equation}\label{N=4action-reg}
    S_{\epsilon}= \frac{1}{g^2\mu^{2\ep}}\int d^{D}x\ {\cal L}(x)\,, \qquad
    {\cal L} = \mbox{Tr} { \left[-\ft{1}{2}F^2_{\mu\nu}\right]} + \mbox{gaugino}+\text{scalars}+\text{gauge
    fixing}+\text{ghosts}.
\end{equation}
Here $\m$ is the regularisation scale and we redefined all fields in such a way
that $g$ does not appear inside the Lagrangian ${\cal L}(x)$. This allows us to
keep the {\it canonical} dimension of all fields, in particular of the gauge
field $A^\mu(x)$, and hence to preserve the conformal invariance of the
path-ordered exponential entering the functional integral in
\re{W-path-integral}. However, due to the change of dimension of the measure
$\int d^D x$  in (\ref{N=4action-reg}) the action $S_\epsilon$ is not invariant under
dilatations and conformal boosts, which yields an {\it anomalous} contribution to
the Ward identities.

\subsection{Anomalous conformal Ward identities}

The conformal Ward identities for the light-like Wilson loop $W(C_n)$ can be
derived following the standard method
\cite{Sarkar:1974xh,Mueller:1993hg,Braun:2003rp}, by acting on both sides
of (\ref{N=4action-reg}) with generators of conformal transformations..
The expressions of the generators of conformal transformations
acting on fields are given in equation (\ref{conf-trans-fields}).
\\

Let us start with the dilatations and perform a change of variables in the
functional integral \re{W-path-integral}, $\phi_I'(x) = \phi_I(x)
+\varepsilon\,\mathbb{D} \phi_I(x)$. This change of variables could be
compensated by a coordinate transformation ${x^{\mu}}' = (1-\varepsilon)x^\m$. We
recall that the path-ordered exponential is invariant under dilatations, whereas
the Lagrangian is covariant with canonical weight $\Delta_{\cal L}=4$. However, the
measure $\int d^Dx$ with $D=4-2\ep$ does not match the weight of the Lagrangian,
which results in a non-vanishing variation of the action $S_\epsilon$,
\begin{equation} \delta_\mathbb{D} S_\epsilon = \frac{2\ep}{g^2\mu^{2\ep}}\int
d^{D}x\ {\cal L}(x)\,.
\end{equation}This variation generates an operator insertion into the
expectation value, $\vev{\delta_\mathbb{D} S_\epsilon\ W_n}$, and yields an
anomalous term in the action of the dilatation generator on $\vev{W_n}$, i.e.,
\begin{equation}\label{D-var}
    \mathbb{D} \vev{W_n} = \sum_{i=1}^n \lr{x_i\cdot \pa_i} \ \vev{W_n} =
     {-\frac{2i\ep}{g^2\mu^{2\ep}}}\int d^{D}x\ \langle{\cal L}(x) W_n \rangle\ .
\end{equation}
In a similar manner,  the anomalous special conformal (or conformal boost) Ward
identity is derived by performing transformations generated by the operator
$\mathbb{K}^\nu$, Eq.~(\ref{conf-trans-fields}), on both sides of \re{W-path-integral}. In this
case the nonvanishing variation of the action $\delta_{\mathbb{K}^\mu}
S_\epsilon$ again comes from the mismatch of the conformal weights of the
Lagrangian and of
the measure $\int d^D x$.
\footnote{Another source of non-invariance of the action $S_\epsilon$ is the
gauge-fixing term (and the associated ghost term of the non-Abelian theory) which
is not conformally invariant even in four dimensions. However, due to gauge
invariance of the Wilson loop, such anomalous terms do not appear on the
right-hand side of (\ref{K-var}).} This amounts to considering just the $\Delta_\phi$
term in (\ref{conf-trans-fields}) with $\Delta_\phi = \Delta_{\cal L} - D = 2\ep$, and hence
\begin{equation}\label{K-var}
     \mathbb{K}^\nu \vev{W_n} = \sum^n_{i=1} (2x_i^\nu x_i\cdot\pa_i - x_i^2 \pa_i^\nu) \vev{W_n}
     = - \frac{4i\ep}{g^2\mu^{2\ep}}\int d^{D}x\ x^\nu\ \langle {\cal L}(x) W_n \rangle \ .
\end{equation}
The relations \re{D-var} and \re{K-var} can be rewritten for $\ln \VEV{W_{n}}$ as
\begin{align}\label{D-var2}
& \mathbb{D} \ln\vev{W_n} =
     -\frac{2i\ep}{g^2\mu^{2\ep}}\int d^{D}x\ \frac{\langle{\cal L}(x) W_n \rangle}{\vev{W_n}}\
     ,
\\ \notag
&    \mathbb{K}^\nu \ln\vev{W_n}   = - \frac{4i\ep}{g^2\mu^{2\ep}}\int d^{D}x\
x^\nu\
    \frac{\langle{\cal L}(x) W_n \rangle}{\vev{W_n}}\ .
\end{align}
To make use of these relations we have to evaluate the ratio $\langle{\cal L}(x)
W_n \rangle/\vev{W_n}$ obtained by inserting the Lagrangian into the Wilson loop
expectation value. Due to the presence of $\epsilon$ on the right-hand side of
\re{D-var2}, it is sufficient to know its divergent part only.

\subsection{Dilatation Ward identity}

As we will now show, the dilatation Ward identity can be derived by dimensional
arguments and this provides a consistency condition for the right-hand side of
(\ref{D-var2}). By definition \re{W-path-integral}, the dimensionally regularised
light-like Wilson loop $\vev{W_n}$ is a dimensionless scalar function of the cusp
points $x_i^\nu$ and, as a consequence, it satisfies the relation
\begin{equation}\label{Wn8}
     \left(\sum_{i=1}^n (x_i\cdot \pa_i) -\mu \frac{\pa}{\pa \mu}   \right) \ln\vev{W_n}= 0\ .
\end{equation}
In addition, its perturbative expansion is expressed in powers of the coupling
$g^2\mu^{2\ep}$ and, therefore,
\begin{equation}\label{Wn7}
  \mu \frac{\pa}{\pa \mu} \vev{W_n} =   2\ep g^2 \frac{\pa}{\pa g^2} \vev{W_n}
  = -\frac{2i \ep}{g^2\mu^{2\ep}}\int d^{D}x\ \langle{\cal L}(x) W_n \rangle\ ,
\end{equation}
where the last relation follows from \re{W-path-integral}.
Recall that according to \re{W=ZF}, the Wilson loop $\vev{W_n}$ can be split into the product of divergent and
finite parts
\begin{equation}\label{chapter-CWI-W=ZF}
\ln\,\vev{W_n} = Z_n +  F_n\,.
\end{equation}
Notice that the definition of the divergent part
is ambiguous as one can always add to $Z_n$ a term finite for $\epsilon\to
0$. Our definition \re{Zn-WL} is similar to the conventional $\rm MS$ scheme with the
only difference that we choose the expansion parameter to be $a\mu^{2\epsilon}$
instead of $a$. The reason for this is that $Z_n$ satisfies, in our scheme, the
same relation $\mu {\pa}_\mu Z_n = 2\ep g^2 {\pa}_{g^2} Z_n $ as $\vev{W_n}$, see
\p{Wn7}. Together with (\ref{chapter-CWI-W=ZF}) and \re{Wn8}, this implies that the finite part of
the Wilson loop does not depend on the renormalisation scale, i.e.
\begin{equation}\label{Fn-indep-eps}
    \mu{\pa_\mu} F_n = O(\epsilon)\,.
\end{equation}
Using
(\ref{chapter-CWI-W=ZF}) and the explicit
form of $Z_n$ in (\ref{Zn-WL}), the relation (\ref{Wn7}) leads to the following dilatation Ward identity%
\footnote{In what follows, we shall systematically neglect corrections to $F_n$
vanishing as $\epsilon\to 0$.}
\begin{equation}\label{F-dilatations}
 \sum_{i=1}^n (x_i\cdot \pa_{x_i}) \, F_n= 0\,.
\end{equation}
Adding to this the obvious requirement of Poincar\'{e}
invariance, we conclude that the finite part ${F}_n$ of the light-like Wilson
loop can depend on the dimensionless ratios $x^2_{ij}/x^2_{kl}$ only. In
particular, for $n=4$ there is only one independent ratio, i.e. ${F}_4 =
{F}_4\left( {x^2_{13}}/{x^2_{24}}\right)$.
\\

Furthermore, making use of (\ref{chapter-CWI-W=ZF}), \re{Zn-WL} and \re{F-dilatations} we find that the all-loop
dilatation Ward identity for $\vev{W_n}$ takes the form
\begin{equation}\label{D-all-loop}
\mathbb{D} \ln \vev{W_n}= \sum_{i=1}^n
(x_i\cdot \pa_i) \, \ln\vev{W_n} = -\frac{1}{2} \sum_{l\ge 1}
a^l\sum_{i=1}^n\lr{-x_{i-1,i+1}^2\mu^2}^{l\ep} \lr{\frac{\Gamma_{\rm
cusp}^{(l)}}{l\ep}+ {\Gamma^{(l)}}} \ .
\end{equation}
This relation provides a constraint on the form of the Lagrangian insertion on
the right-hand side of (\ref{D-var2}).

\subsection{One-loop calculation of the anomaly}

The derivation of the dilatation Ward identity \re{F-dilatations} relied on the
known structure of cusp singularities (\ref{Zn-WL}) of the light-like Wilson loop and
did not require a detailed knowledge of the properties of Lagrangian insertion
$\langle{\cal L}(x) W_n \rangle/\vev{W_n}$. This is not the case anymore for the
special conformal Ward identity.
\\

\begin{figure}
\psfrag{xi1}[cc][cc]{$x_{i}$} \psfrag{xi3}[cc][cc]{$x_{i+1}$}
\psfrag{xi2}[cc][cc]{$x_{i-1}$} \psfrag{x}[cc][cc]{$x$}
\psfrag{a}[cc][cc]{(a)} \psfrag{b}[cc][cc]{(b)}
\psfrag{c}[cc][cc]{(c)}
\centerline{\includegraphics[height=50mm,keepaspectratio]{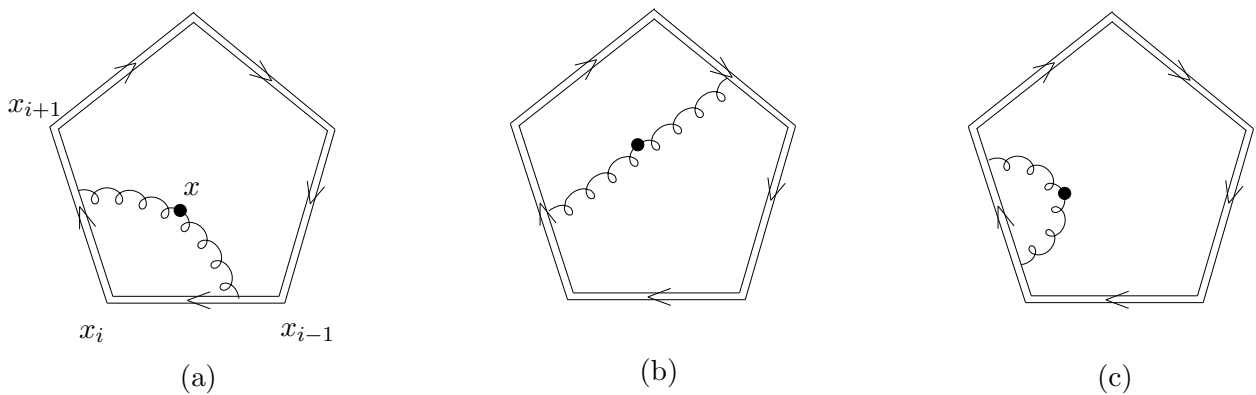}}
\caption[]{\small The Feynman diagrams contributing to $\langle{\cal L}(x) W_n
\rangle$ to the lowest order in the coupling. The double line depicts the
integration contour $C_n$, the wiggly line the gluon propagator and the blob the
insertion point.}
\label{Fig:insertion}
\end{figure}

To start with, let us perform an explicit one-loop computation of $\langle{\cal
L}(x) W_n \rangle/\vev{W_n}$. To the lowest order in the coupling, we substitute
$\vev{W_n}=1 +O(g^2)$ in the denominator and retain inside ${\cal L}(x)$ and $W_n$ terms
quadratic in gauge field only. The result is
\begin{equation}\label{insertion}
\frac{\langle{\cal L}(x) W_n \rangle}{\vev{W_n}} = {-\frac{1}{4N}\VEV{\mbox{Tr}
\left[(\partial_\mu A_\nu(x) - \partial_\nu A_\mu(x))^2\right]\Tr\left[ \big(i{\mbox{$\oint_{C_n}$} dy \cdot
A(y)}\big)^2\right] }+ O(g^6)}\,.
\end{equation}
The Wick
contractions between gauge fields coming from the Lagrangian and the path-ordered
exponential yield a product of two gluon propagators \re{a1}, each connecting the
point $x$ with an arbitrary point $y$ on the integration contour $C_n$. Gauge
invariance allows us to choose e.g. the Feynman gauge, in which the gluon
propagator is given by \re{propagator}. To the lowest order in the coupling
constant, the right-hand side of \re{insertion} receives non-vanishing
contributions from Feynman diagrams of three different topologies shown in
Figs.~\ref{Fig:insertion}(a)  -- (c).
\\

In the Feynman gauge, only the vertex-like diagram shown in
Fig.~\ref{Fig:insertion}(a)
develops poles in $\epsilon$. Performing the calculation, we find after some algebra%
\footnote{It is advantageous to perform the calculation by taking a Fourier
transform with respect to $x$ and later take the inverse transform of the final
result.}
\begin{align}\label{fin222}
 {\frac{2i}{g^2\mu^{2\ep}}\frac{\langle{\cal L}(x) W_n \rangle}{\vev{W_n}}}  =  a \sum_{i=1}^n
\lr{-x_{i-1,i+1}^2\mu^2}^\ep\bigg\{\ep^{-2}\delta^{(D)}(x-x_i)+\ep^{-1} \Upsilon^{(1)}(x|
x_{i-1},x_i,x_{i+1})+ O(\epsilon^0)\bigg\},
\end{align}
where $a=g^2N/(8\pi^2)$ and the notation was introduced for
\begin{equation}
\Upsilon^{(1)}(x| x_{i-1},x_i,x_{i+1}) =\int_0^1 \frac{ds}{s}
\bigg[{\delta^{(D)}(x-x_i- s x_{i-1,i})}+{\delta^{(D)}(x-x_i+ s
x_{i,i+1})-2\delta^{(D)}(x-x_i)} \bigg].
\end{equation}
We see that the leading
double-pole singularities are localised at the cusp points $x=x_i$. The
subleading single poles are still localised on the contour, but they are
`smeared' along the light-like edges adjacent to the cusp.
\\

Substitution of \re{fin222} into \re{D-var2} yields
\begin{align}\label{CWI-1loop}
& \mathbb{D} \ln\vev{W_n} = -a \sum_{i=1}^n \lr{-x_{i-1,i+1}^2\mu^2}^\ep \ep^{-1}
+O(a^2)
     ,
\\ \notag
&    \mathbb{K}^\nu \ln\vev{W_n}   = -2a \sum_{i=1}^n
x_i^\nu\lr{-x_{i-1,i+1}^2\mu^2}^\ep \ep^{-1}  +O(a^2) \ .
\end{align}
Notice that $\sum_{i=1}^n \Upsilon^{(1)}(x| x_{i-1},x_i,x_{i+1})$ does not
contribute to the right-hand sides of these relations by virtue of
\begin{align}\label{Y-identity}
& \int d^D x\, \Upsilon^{(1)}(x| x_{i-1},x_i,x_{i+1}) = 0\,,
\\ \notag
& \int d^D x\,x^\nu \Upsilon^{(1)}(x| x_{i-1},x_i,x_{i+1}) =
(x_{i-1}+x_{i+1}-2x_i)^\nu\,.
\end{align}
As was already mentioned, the right-hand sides of the Ward identities
\re{CWI-1loop} are different from zero due to the fact that the light-like Wilson
loop has cusp singularities.
\\

We verify with the help of \re{cusp-2loop-WL} that to the lowest order in the
coupling, the first relation in \re{CWI-1loop} is in agreement with
\re{D-all-loop}.

\subsection{Structure of the anomaly to all loops}

To extend the analysis of the special conformal Ward identity to all loops we
examine the all-loop structure of divergences of $\langle{\cal L}(x) W_n
\rangle/\vev{W_n}$. They arise in a very similar way to those of the Wilson loop
itself, which were discussed in section \ref{ch-cusp-sing}.
\\

It is convenient to couple ${\cal L}(x)$ to an auxiliary `source' $J(x)$ and
rewrite the insertion of the Lagrangian into $\vev{W_n}$ as a functional
derivative
\begin{equation}
\langle{\cal L}(x) W_n \rangle/\vev{W_n} =
-i \frac{\delta }{\delta J(x)}\ln \vev{W_n}_J\bigg|_{J=0},
\end{equation}
where the subscript $J$ indicates that the expectation value is taken in the
$\mathcal{N}=4$ SYM theory with the additional term $\int d^D x J(x){\cal L}(x)$
added to the action. This generates new interaction vertices inside Feynman
diagrams for $\vev{W_n}_J$, but does not affect the non-Abelian
exponentiation property \re{exponentiation}.\footnote{We recall that the non-Abelian exponentiation is not sensitive to the
form of the action.} The only difference compared to \re{exponentiation} is
that the webs $w^{(k)}$ now depend on $J(x)$ through a new interaction vertex.
Making use of the non-Abelian exponentiation we obtain
\begin{equation}\langle{\cal L}(x) W_n \rangle/\vev{W_n} = {\sum_{k=1}^\infty
\lr{\frac{g^2}{4\pi^2}}^k c^{(k)} \left[-i\frac{\delta w^{(k)}}{\delta
J(x)}\right] \bigg|_{J=0} }\,.
\label{jderiv}
\end{equation}
As for $\ln \vev{W_n}$, the webs produce at most double poles in $\epsilon$, and are
localised at a given cusp. This allows us to write down a general expression for
the divergent part of the Lagrangian insertion,
\begin{eqnarray}\label{L-insertion}
\frac{2i\ep}{g^2\mu^{2\ep}} \ \frac{\langle{\cal L}(x) W_n \rangle}{\vev{W_n}}
&=& \sum_{l\ge 1} a^l \sum_{i=1}^n \lr{-x_{i-1,i+1}^2\mu^2}^{l\ep} \times
 \label{25} \\
   &&
   \left\{ \frac12\lr{\frac{\Gamma_{\rm cusp}^{(l)}}{ l\ep
}+  \Gamma^{(l)}}\delta^{(D)}(x-x_i)  + \Upsilon^{(l)}(x; x_{i-1},x_i,x_{i+1})
\right\} + O(\ep)\ ,  \nn
\end{eqnarray}
which generalises \re{fin222} to all loops. The following comments are in order.
\\

In Eq.~\re{L-insertion}, the term proportional to $\delta^{(D)}(x-x_i)$ comes
from the double pole contribution to the web $w^{(k)}$, which is indeed located at
short distances in the vicinity of the cusp $x_i^\mu$. The residue of the
simple pole in the right-hand side of \re{L-insertion} is given by the cusp
anomalous dimension. This can be shown by substituting \re{L-insertion} into the
dilatation Ward identity \p{D-var} and comparing with (\ref{D-all-loop}).
\\

The contact nature of the leading singularity in \p{L-insertion} can also be
understood in the following way. The correlator on the left-hand side can be
viewed as a conformal $(n+1)$-point function with the Lagrangian at one point and
the rest corresponding to the cusps. In the $\cN=4$ SYM theory the Lagrangian
belongs to the protected stress-tensor multiplet, therefore it has a fixed
conformal dimension four. The Wilson loop itself, if it were not divergent, would
be conformally invariant. This means that the $n$ cusp points can be regarded as
having vanishing conformal weights. Of course, the presence of divergences might
make the conformal properties anomalous. However, the  conformal behaviour of the
leading singularity in \p{L-insertion} cannot be corrected by an
anomaly.\footnote{Such an anomalous contribution should come from a $1/\ep^3$
pole in the correlator with two insertions of the Lagrangian, but repeated use of
the argument above shows that the order of the poles does not increase with the
number of insertions.} Then we can argue that the only function of space-time
points, which has conformal weight four at one point and zero at all other
points, is the linear combination of delta functions appearing in
\p{L-insertion}.\footnote{The mismatch of the conformal weight four of the
Lagrangian and $D=4-2\ep$ of the delta functions does not affect the leading
singularity in \p{L-insertion}.}
\\

As was already emphasised, the residue of the simple pole of $\delta
w^{(k)}/\delta J(x)$  at the cusp point $x_i$ depends on its position and at most
on its two nearest neighbours $x_{i-1}$ and $x_{i+1}$, as well as on the insertion
point. It gives rise to a function $\Upsilon^{(l)}(x; x_{i-1},x_i,x_{i+1})$,
which is the same for all cusp points due to the cyclic symmetry of the Wilson
loop.
\\

Notice that in \p{25} we have chosen to separate the terms with the collinear
anomalous dimension $\Gamma^{(l)}$ from the rest of the finite terms. This choice
has implications for the function $\Upsilon^{(l)}(x; x_{i-1},x_i,x_{i+1})$.
Substituting the known factor $Z_n$ \p{Zn-WL} and the particular form of \p{25} into
the dilatation Ward identity \p{D-var}, we derive
\begin{equation}\label{26}
    \sum_{i=1}^n \int d^Dx\ \Upsilon^{(l)}(x; x_{i-1},x_i,x_{i+1}) = 0\ .
\end{equation}
We can argue that in fact each term in this sum vanishes. Indeed, each term in
the sum is a Poincar\'{e} invariant dimensionless function of three points
$x_{i-1}$, $x_i$ and $x_{i+1}$. Given the light-like separation of the
neighbouring points, the only available invariant is $x_{i-1,i+1}^2$. Further,
$\Upsilon^{(l)}(x; x_{i-1},x_i,x_{i+1})$ cannot depend on the regularisation
scale because $\m$ always comes in the combination $a\m^{2\ep}$ and thus
contributes to the $O(\ep)$ terms in \p{25}. The dimensionless Poincar\'{e}
invariant $\int d^Dx\ \Upsilon^{(l)}(x; x_{i-1},x_i,x_{i+1})$ depends on a single
scale and, therefore, it must be a constant, after which \p{26} implies
\begin{equation}\label{26'}
    \int d^Dx\ \Upsilon^{(l)}(x; x_{i-1},x_i,x_{i+1}) = 0\ .
\end{equation}
For $l=1$ this relation is in agreement with the one-loop result \re{Y-identity}.

\subsection{Special conformal Ward identity}

We are now ready to investigate the special conformal Ward identity \p{K-var}.
Inserting \p{25} into its right-hand side, and integrating over $x$ we obtain
\begin{eqnarray}
  \mathbb{K}^\nu \ln W_n &=& \sum^n_{i=1} (2x_i^\nu x_i\cdot\pa_i - x_i^2 \pa_i^\nu) \ln W_n
  \label{28}\\
  &=& -\sum_{l\ge 1} a^l \lr{\frac{\Gamma_{\rm cusp}^{(l)}}{ l\ep
}+  {\Gamma^{(l)}} }\sum_{i=1}^n\ \lr{-x_{i-1,i+1}^2\mu^2}^{l\ep} \ x^\nu_i  -2
\sum_{i=1}^n \Upsilon^\nu(x_{i-1},x_i,x_{i+1}) + O(\ep) \ ,   \nn
\end{eqnarray}
where
\begin{equation}\label{29}
    \Upsilon^\nu(x_{i-1},x_i,x_{i+1})  = \sum_{l\ge 1}a^l \int d^D x\, x^\nu \Upsilon^{(l)}(x; x_{i-1},x_i,x_{i+1})\ .
\end{equation}
Next, we substitute $\ln W_n =Z_n + F_n$ into \p{28}, replace $\ln Z_n$
by its explicit form \p{Zn-WL} and expand the right-hand side in powers of $\ep$ to
rewrite \p{28} as follows:
\begin{equation}\label{32}
    \mathbb{K}^\nu {F}_n  =  \frac{1}{2} \Gamma_{\rm cusp}(a) \sum_{i=1}^n
    \ln \frac{x_{i,i+2}^2}{x_{i-1,i+1}^2} x^\nu_{i,i+1} - 2 \sum_{i=1}^n
    \Upsilon^\nu(x_{i-1},x_i,x_{i+1}) + O(\ep)\ .
\end{equation}
Note that the quantities $\Upsilon^\nu(x_{i-1},x_i,x_{i+1})$ are translation
invariant. Indeed, a translation under the integral in \p{29} only affects the
factor $x^\nu$ (the functions $\Upsilon^{(l)}(x; x_{i-1},x_i,x_{i+1})$ are
translation invariant), but the result vanishes as a consequence of \p{26'}.
Furthermore, $\Upsilon^\nu(x_{i-1},x_i,x_{i+1})$ only depends on two neighbouring
light-like vectors $x^\m_{i-1,i}$ and $x^\m_{i,i+1}$, from which we can form only
one non-vanishing Poincar\'{e} invariant, $x^2_{i-1,i+1}$. We have already argued
that $\Upsilon^{(l)}$ are independent of $\mu$, and so must be $\Upsilon^\nu$.
Taking into account the scaling dimension one of $\Upsilon^\nu$, we conclude that
\begin{equation}\label{34}
    \Upsilon^\nu(x_{i-1},x_i,x_{i+1})  = \alpha x^\nu_{i-1,i} + \beta x^\nu_{i,i+1}\ ,
\end{equation}
where $\alpha,\ \beta $ only depend on the coupling. The symmetry of the Wilson
loop $W_n$ under mirror exchange of the cusp points
translates
into symmetry of \re{34} under exchange of the neighbours $x_{i-1}$ and $x_{i+1}$
of the cusp point $x_i$, which reduces (\ref{34}) to
\begin{equation}\label{35}
    \Upsilon^\nu(x_{i-1},x_i,x_{i+1})   = \alpha \, (x^\nu_{i-1} + x^\nu_{i+1} - 2 x^\nu_i)\
    ,
\end{equation}
with $\alpha=a + O(a^2)$ according to \re{Y-identity}.
 Substituting this relation into \re{32} we find
\begin{equation}\label{36}
    \sum_{i=1}^n \Upsilon^\nu(x_{i-1},x_i,x_{i+1}) = 0\ .
\end{equation}
This concludes the derivation of the special conformal Ward identity. In the
limit $\ep\to0$ it takes the form (cf. \cite{Drummond:2007cf})
\begin{equation}\label{cwi37}
    \sum^n_{i=1} (2x_i^\nu x_i\cdot\pa_i - x_i^2 \pa_i^\nu)\, {F}_n  =  \frac{1}{2} \Gamma_{\rm cusp}(a) \sum_{i=1}^n  \ln \frac{x_{i,i+2}^2}{x_{i-1,i+1}^2} x^\nu_{i,i+1}\ .
\end{equation}

\subsection{Solution and implications for $F_n$}\label{crrrr}

Let us now examine the consequences of the conformal Ward identity \p{cwi37} for the
finite part of the Wilson loop $W_n$. We find that the cases of $n=4$ and $n=5$
are special because here the Ward identity \p{cwi37} has a unique solution up to an
additive constant. The solutions are, respectively,
\begin{align}
 {F}_4 &= \frac{1}{4}\Gamma_{\rm cusp}(a)
\ln^2\left(\frac{x_{13}^2}{x_{24}^2}\right) + \text{
  const }\,, \label{remarkable1}\\
 {F}_5 &= - \frac{1}{8}\Gamma_{\rm cusp}(a) \sum_{i=1}^5 \ln
  \left(\frac{x_{i,i+2}^2}{x_{i,i+3}^2}\right) \ln
  \left(\frac{x_{i+1,i+3}^2}{x_{i+2,i+4}^2}\right) + \text{ const }\, , \label{remarkable2}
\end{align}
as can be easily verified by making use of the identity
\begin{equation}\label{Kactsonxij}
\mathbb{K}^{\mu}
x_{ij}^2 = 2 (x_i^\mu + x_j^\mu) x_{ij}^2\,.
\end{equation}
We find that, upon identification
of the kinematical invariants
\begin{equation}\label{xx}
x_{k,k+r}^2 := (p_k+\ldots + p_{k+r-1})^2\,,
\end{equation}
the relations (\ref{remarkable1}) and (\ref{remarkable2}) are exactly the functional forms of the ansatz of
\cite{Bern:2005iz} for the finite parts of the four- and five-point MHV amplitudes (or
rather the ratio of the amplitude to the corresponding tree amplitude).
\\

The reason why the functional form of ${F}_4$ and ${F}_5$ is fixed up to an
additive constant is that there are no conformal invariants one can build from
four or five points $x_i$ with light-like separations $x_{i,i+1}^2=0$. Such
invariants take the form of cross-ratios
\begin{equation}\label{crr}
    \frac{x^2_{ij}x^2_{kl}}{x^2_{ik}x^2_{jl}}\ .
\end{equation}
It is obvious that with four or five points they cannot be constructed. This
becomes possible starting from six points, where there are three such
cross-ratios, e.g.,
\begin{equation}\label{uvars-CWI}
u_1 = \frac{x_{13}^2 x_{46}^2}{x_{14}^2
x_{36}^2}, \qquad u_2 = \frac{x_{24}^2 x_{15}^2}{x_{25}^2 x_{14}^2}, \qquad u_3 =
\frac{x_{35}^2 x_{26}^2}{x_{36}^2 x_{25}^2}\ .
\end{equation}
Hence the general
solution of the Ward identity at six cusp points and higher will contain an
arbitrary function of the conformal cross-ratios.
\\

Nevertheless, one can prove that the functional form of the ansatz of \cite{Bern:2005iz} still
provides a particular solution to the Ward identity (\ref{cwi37}) \cite{Drummond:2007cf}.
The ansatz of \cite{Bern:2005iz} for the
logarithm of the ratio of the amplitude to the tree amplitude reads
\be\label{ansatz1}
\ln \mathcal{M}_n^{\rm (MHV)} = Z_n + F^{\rm (BDS)}_n + C_n
+O(\epsilon)\,,
\ee
and
\be \label{BDS-ngluons}
F^{\rm (BDS)}_{n} = \frac{1}{2} \C_{\rm cusp}(a) \mathcal{F}_n
\ee
where $Z_n$ is the IR divergent part, $F_n$ is the finite part depending on the
Mandelstam variables and $C_n$ is the constant term. At four points the proposed
form of the finite part, written in the dual notation (\ref{duality-1}), is
\be
\mathcal{F}_4 = \frac{1}{2} \ln^2\Bigl(\frac{x_{13}^2}{x_{24}^2}\Bigr) + 4 \z_2\ ,
\label{F4}
\ee
while for $n\geq 5$ it is
\be
\mathcal{F}_n = \frac{1}{2} \sum_{i=1}^n g_{n,i}\,,\quad
g_{n,i} = - \sum_{r=2}^{\lfloor \tfrac{n}{2} \rfloor -1} \ln
\Bigl(\frac{x_{i,i+r}^2}{x_{i,i+r+1}^2}\Bigr) \ln
\Bigl(\frac{x_{i+1,i+r+1}^2}{x_{i,i+r+1}^2}\Bigr) + D_{n,i} + L_{n,i} +
\frac{3}{2} \z_2\ .
\ee
The functions $D_{n,i}$ and $L_{n,i}$ depend on whether $n$ is odd or even. For
$n$ odd, $n=2m+1$, they are
\begin{align} \label{di1}
D_{n,i} &= - \sum_{r=2}^{m-1} {\rm Li}_2 \Bigl(1 - \frac{x_{i,i+r}^2
 x_{i-1,i+r+1}^2}{x_{i,i+r+1}^2 x_{i-1,i+r}^2} \Bigr)\ , \\ \nn
L_{n,i} &= - \frac{1}{2} \ln
 \Bigl(\frac{x_{i,i+m}^2}{x_{i,i+m+1}^2}\Bigr) \ln
\Bigl(\frac{x_{i+1,i+m+1}^2}{x_{i+m,i+2m}^2}\Bigr)\ .
\end{align}
For $n$ even, $n=2m$, they are
\begin{align}  \label{di2}
D_{n,i} &= - \sum_{r=2}^{m-2} {\rm Li}_2 \Bigl(1 - \frac{x_{i,i+r}^2
 x_{i-1,i+r+1}^2}{x_{i,i+r+1}^2 x_{i-1,i+r}^2} \Bigr) -
 \frac{1}{2}{\rm Li}_2\Bigl(1 - \frac{x_{i,i+m-1}^2
 x_{i-1,i+m}^2}{x_{i,i+m}^2 x_{i-1,i+m-1}^2} \Bigr)\ ,\\  \nn
L_{n,i} &= \frac{1}{4} \ln^2 \Big(\frac{x_{i,i+m}^2}{x_{i+1,i+m+1}^2}\Bigr)\ .
\end{align}
We have already seen that at four points and five points the general solution to
the Ward identity coincides with \re{ansatz1}. We now show that the ansatz
\re{ansatz1} is a solution of the Ward identity for arbitrary $n$.
\\

First we observe that the dilogarithmic contributions in \re{di1} and \re{di2}
are functions of conformal cross-ratios of the form (\ref{crr}). They are
therefore invariant under conformal transformations and we have immediately
\be
K^\m D_{n,i} = 0\,.
\ee
For the logarithmic contributions we use the identity (\ref{Kactsonxij}). When $n$ is
odd we then find
\ba
&& \hspace*{-6mm} K^\m g_{n,i} = - 2\sum_{r=2}^{m-1}\bigl[ x_{i+r,i+r+1}^\m(\ln
 x_{i+1,i+r+1}^2 - \ln x_{i,i+r+1}^2) - x_{i,i+1}^\m(\ln x_{i,i+r}^2
 - \ln x_{i,i+r,2}^2)\bigr]
\\ \nn
&& -  x_{i+m,i+m+1}^\m(\ln x_{i+1,i+m+1}^2 - \ln
 x_{i+m,i+2m}^2)
- (x_{i+1,i+2m}^\m -
 x_{i+m,i+m+1}^\m)(\ln x_{i,i+m}^2 - \ln x_{i+m+1,i}^2)\, .
\ea
Changing variables term by term in the sum over $i$ one finds that only the $\ln
x_{i,i+2}^2$ terms remain and indeed (\ref{cwi37}) is satisfied. The proof for $n$
even goes exactly the same way except that one obtains
\be
K^\m \mathcal{F}_n = \sum_{i=1}^n \bigl[\ln x_{i,i+2}^2 (x_i^\m +
 x_{i+2}^\m - 2 x_{i+1}^\m) + \frac{1}{2} \ln x_{i,i+m}^2
 (x_{i+m-1,i+m+1}^\m - x_{i-1,i+1}^\m) \bigr]
\ee
and one has to use the fact that $n=2m$ to see that the $\ln x_{i,i+m}^2$ term
vanishes under the sum.
\\

Thus we have seen that the BDS ansatz for the $n$-gluon MHV amplitudes satisfies the
conformal Ward identity for the Wilson loop.


\section{Hexagon Wilson loop and six-gluon MHV amplitude}
\label{ch-six-point}

In section \ref{ch-four-point} we saw that the duality
(\ref{intro-duality}) is valid at two loops for $n=4,5$ points (or gluons). 
For $n=4,5$, the Ward identities derived in section
\ref{section-CWI} for the Wilson loops provide a possible
explanation of the duality to all orders in the coupling constant,
assuming the BDS ansatz is correct in these cases. 
This speculation receives support from strong
coupling, where the result of \cite{Alday:2007hr} is in agreement 
with the BDS ansatz for $n=4$.
\\

Recall that the authors of \cite{Alday:2007he} found a disagreement 
with the BDS ansatz at strong coupling for the number of gluons $n$ very large.
Given this result it seems natural to ask whether the ansatz already breaks down at some
perturbative level and for a finite number of gluons. If we assume
the duality with Wilson loops, then $n=4$ and $n=5$ are completely
fixed by conformal symmetry. 
In fact, one might even suspect that the observed validity of the duality 
relation for $n=4,5$ is true only because both objects have the same
conformal symmetry. To put it differently, the proposed duality
might be reduced to the weaker (but still highly non-trivial)
statement that the MHV gluon amplitudes are governed by dual conformal
symmetry. The first real test of the stronger form of the duality
(\ref{intro-duality}) is provided by the case $n=6$.
In order to distinguish between these possibilities we performed a two-loop
calculation of the hexagonal Wilson loop.
\footnote{This section is based on the publications
\cite{Drummond:2007bm} and \cite{Drummond:2008aq}.}
\\

\subsection{Finite part of the hexagon Wilson loop}

The finite part of the hexagon Wilson loop, $F_6^{\rm (WL)}$, does
not depend on the renormalisation scale and it is a dimensionless
function of the distances $x_{ij}^2$. Since the edges of $C_6$ are
light-like, $x_{i,i+1}^2=0$, the only nonzero distances are
$x_{i,i+2}^2$ and $x_{i,i+3}^2$ (with $i=1,\ldots, 6$ and the
periodicity condition $x_{i+6}=x_i$). We saw in section
\ref{section-CWI} that it has to satisfy the conformal Ward
identity (\ref{cwi37}).
 Specified
to $n=6$, its general solution is given by \cite{Drummond:2007cf}
\be\label{solutionWI} F_6^{\rm (WL)} = F_6^{\rm (BDS)}
+R_6(u_1,u_2,u_3;a)\ . \ee Here, upon the identification
$p_i=x_{i+1}-x_{i}$,
\begin{align}\label{BDS6pointxnotation}
  F_{6} ^{\rm (BDS)}   =  \frac{1}{4} \Gamma_{\rm
 cusp}(a)\sum_{i=1}^{6} &\bigg[
    - \ln\Bigl(
\frac{x_{i,i+2}^2}{x_{i,i+3}^2} \Bigr)\ln\Bigl(
\frac{x_{i+1,i+3}^2}{x_{i,i+3}^2} \Bigr)
\nonumber\\
 &  +\frac{1}{4} \ln^2 \Bigl( \frac{x_{i,i+3}^2}{x_{i+1,i+4}^2}
\Bigr)   -\frac{1}{2} {\rm{Li}}_{2}\Bigl(1-
 \frac{x_{i,i+2}^2  x_{i+3,i+5}^2}{x_{i,i+3}^2 x_{i+2,i+5}^2}
\Bigr) \bigg]\,,
\end{align}
while $R_{6}(u_1,u_2,u_3;a)$ is an arbitrary function of the three
cross-ratios \footnote{The last term in \re{BDS6pointxnotation} is
a function of cross-ratios only, but we keep it in $F_{6} ^{\rm
(BDS)}$, because it is part of the BDS conjecture.}
\begin{equation}u_1 = \frac{x_{13}^2 x_{46}^2}{x_{14}^2
x_{36}^2}, \qquad u_2 = \frac{x_{24}^2 x_{15}^2}{x_{25}^2 x_{14}^2},
\qquad u_3 = \frac{x_{35}^2 x_{26}^2}{x_{36}^2 x_{25}^2}\ . \label{u1u2u3}
\end{equation}
These variables are invariant under conformal transformations of
the coordinates $x_i^\mu$, and therefore they are annihilated by
the conformal boost operator entering the left-hand side of
\re{cwi37}. In addition, the Wilson loop $W(C_6)$ is invariant
under cyclic ($x_i \to x_{i+1}$) and mirror ($x_i\to x_{6-i}$)
permutations of the cusp points~\cite{Drummond:2007au}. This
implies that $R(u_1,u_2,u_3)$ is a totally symmetric function of
three variables.
\\

The one-loop Wilson loop calculation of \cite{Brandhuber:2007yx}
has shown that the `remainder' function $R_6$ is just a constant
at one loop, which is a confirmation of the Wilson loop/scattering
amplitude duality going beyond the scope of conformal symmetry.
However, one might suspect just a low loop-order `accident'. The
point is that the function $R_6$ must satisfy a further, rather
powerful constraint, coming from the collinear limit of gluon amplitudes
\cite{Bern:1994zx,Anastasiou:2003kj}. It could be that due to the
limited choice of loop integrals at this low perturbative level,
the function \re{BDS6pointxnotation} made of them is the only one
satisfying both the conformal Ward identity \re{cwi37} and the
collinear limit. If so, at some higher perturbative level new
functions with these properties might appear which could spoil the
BDS ansatz and/or the Wilson loop/scattering amplitude duality.
\\

Combining together \re{duality-4} and \re{solutionWI}, we conclude that were
the BDS conjecture {\it and}  the duality relation \re{duality-4} correct for
$n=6$, we would expect that $R(u_1,u_2,u_3)=\text{const}$. The explicit two-loop
calculation we report on here shows that this is not true.
\\

For the sake of simplicity we performed the calculation of $W(C_6)$ in the Feynman
gauge. In addition, we made  use of the non-Abelian exponentiation
property of Wilson loops \cite{Dotsenko:1979wb,Gatheral:1983cz,Frenkel:1984pz} to reduce the number of
relevant Feynman diagrams. In application to  $R(u_1,u_2,u_3)$ this property can
be formulated as follows (the same property also holds for $F_6^{\rm (WL)}$)
\begin{equation}\label{W-decomposition6}
R = \frac{g^2}{4\pi^2}C_F\,
R^{(1)}  +  \lr{\frac{g^2}{4\pi^2}}^2 C_F N\, R^{(2)}  +
O(g^6)\,,
\end{equation}
where $C_F=(N^2-1)/(2N)$ is the Casimir in the fundamental
representation of $SU(N)$. The functions $R^{(1)}$ and $R^{(2)}$
do not involve the colour factors and only depend on the distances
 between the cusp points on $C_6$.
At one loop, $R^{(1)}(u_1,u_2,u_3)$ is in fact a constant \cite{Brandhuber:2007yx}.
\\

\begin{figure}[h]
\psfrag{1}{}\psfrag{2}{}\psfrag{3}{}\psfrag{4}{}\psfrag{5}{}\psfrag{6}{}\psfrag{7}{}\psfrag{8}{}
\psfrag{9}{}\psfrag{10}{}\psfrag{11}{}
\psfrag{12}{}\psfrag{13}{}\psfrag{14}{}\psfrag{15}{}\psfrag{16}{}\psfrag{17}{}\psfrag{18}{}
\psfrag{19}{}\psfrag{20}{}\psfrag{21}{}
\psfrag{x1}[rc][cc]{$\scriptscriptstyle x_1$}
\psfrag{x2}[rc][cc]{$\scriptscriptstyle x_2$}
\psfrag{x3}[rc][cc]{$\scriptscriptstyle x_3$}
\psfrag{x4}[lc][cc]{$\scriptscriptstyle x_4$}
\psfrag{x5}[lc][cc]{$\scriptscriptstyle x_5$}
\psfrag{x6}[lc][cc]{$\scriptscriptstyle x_6$}
\centerline{{\epsfxsize17cm \epsfbox{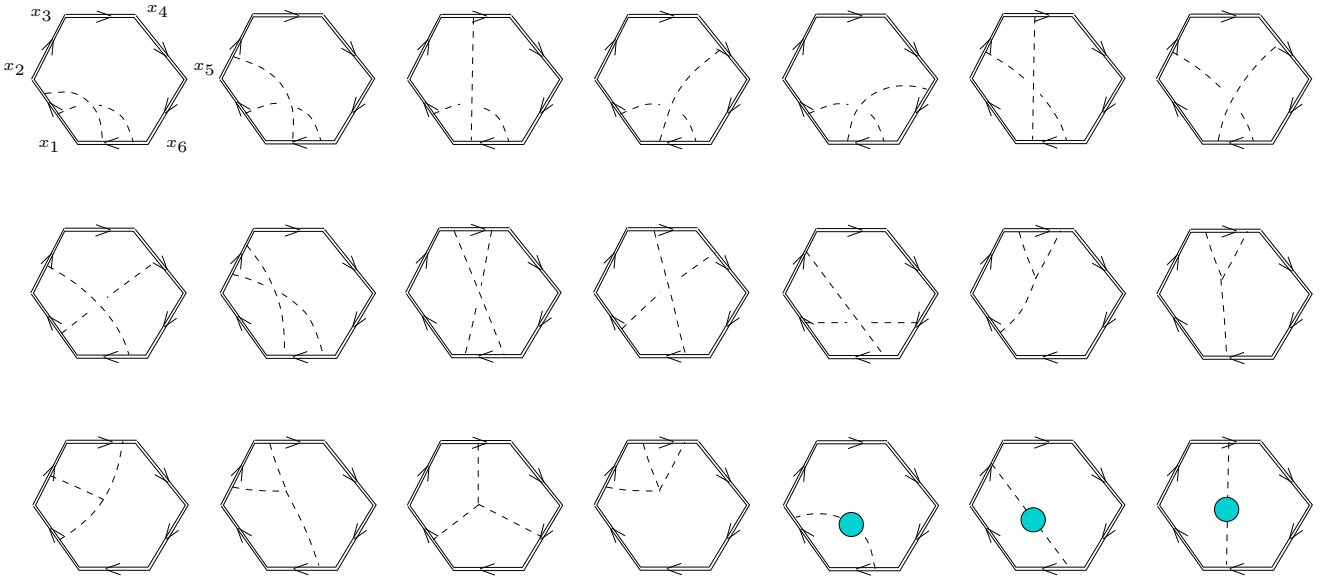}}} \caption[]{\small The
maximally non-Abelian Feynman diagrams of different topology contributing to
$F_6^{\rm (WL)}$. The double lines depict the integration contour $C_6$, the
dashed lines depict the gluon propagator and the blob stands for the one-loop polarisation operator.
}
\label{Fig-WL6}
\end{figure}

As explained in \cite{Drummond:2007cf}, the relation \re{W-decomposition6} implies
that in order to determine the function $F_6^{\rm (WL)}$ at two loops (and hence $f^{(2)}$) it is sufficient to calculate
the contribution to $W(C_6)$ from two-loop diagrams containing the `maximally
non-Abelian' colour factor $C_F N$ only. All relevant two-loop graphs are shown in
Fig.~\ref{Fig-WL6}.
The finite part of $W(C_6)$ receives contributions from all Feynman diagrams shown in
Fig.~\ref{Fig-WL6}.  We derived parameter integral representations for them, which are denoted
by $A_{0}^{(\alpha)}$, so that the two-loop contribution to $F_{6}^{\rm (WL)} =\sum_{p\ge 1} a^p F_{6;p}^{\rm
(WL)} $ is given by
\be\label{F62}
F_{6;2}^{\rm (WL)} = 2 \sum_{\alpha={\rm a,\ldots, u}} A_0^{(\alpha)} \,.
\ee
The integrals $A_0^{(\alpha)}$ are difficult to compute analytically and so we evaluated them
numerically, as described in the next section.

\subsection{Numerical evaluation}\label{hwlvbds}

For a given set of kinematical invariants,
\begin{equation}\label{distances}
K=\{x_{13}^2, x_{14}^2, x_{15}^2,x_{24}^2, x_{25}^2,
x_{35}^2,x_{26}^2,x_{36}^2,x_{46}^2\}\,,
\end{equation}
defined as the distances between the vertices of the hexagon $C_6$,
it is straightforward to evaluate the sum on the right-hand side of \re{F62} numerically. 
We should take into account,
however, that in $D=4$ dimensions there exists a relationship between the distances \re{distances},
such that only eight of them are independent. This relation reflects the fact that the five
four-dimensional vectors  $p_i^\mu = x_{i+1}^\mu - x_{i}^\mu$ (with $i=1,\ldots,5$) are linearly
dependent, and hence their Gram determinant vanishes
\be\label{Gram}
G[K] = \det \| (p_i \cdot p_j) \| = 0\,,\qquad (i,j=1,\ldots,5)\,.
\ee
Using $p_i=x_{i+1}-x_{i}$, the entries of the matrix can be written as linear combinations of the
distances \re{distances}.
\\

We would like to note that the sum on the right-hand side of \re{F62} defines a function of
$x_{ij}^2$ even for configurations $K$ which do not satisfy (\ref{Gram}). This function can be
viewed as a particular continuation of the finite part of the hexagon Wilson loop off the
hypersurface defined by (\ref{Gram}).
\\

Let us now consider the relation between the hexagon Wilson loop
and the BDS ansatz, Eq.~\re{solutionWI}. To two-loop accuracy it
takes the form
\begin{align}\label{WL-BDS1}
 R_{\rm W}(u_1,u_2,u_3)= F_{6;2}^{\rm (WL)} - F_{6;2}^{\rm (BDS)}  \,,
\end{align}
where $R_{\rm W}$  and  $F_{6;2}^{\rm (BDS)}$ denote the two-loop {contributions} to the remainder
function \re{W-decomposition6} and to the BDS ansatz \re{BDS6pointxnotation}, respectively. We recall that the functions
$F_{6;2}^{\rm (WL)}$ and  $F_{6;2}^{\rm (BDS)}$ satisfy the Ward identity \re{K-var}, and hence $R_{\rm
W}$ is a function of the three conformal cross-ratios \re{u1u2u3} only.
\\

The simplest way to check relation \re{WL-BDS1} is to evaluate the difference $F_{6;2}^{\rm (WL)} -
F_{6;2}^{\rm (BDS)} $ for different kinematical configurations $K$ and $K'$, related to each
other by a conformal transformation of the coordinates $x_i^\mu$ (with $i=1,\ldots,6$).
Since $F_{6;2}^{\rm (WL)}$ and $F_{6;2}^{\rm (BDS)}$ are dimensionless functions of the distances
$x_{ij}^2$, they are automatically invariant under translations, Lorentz rotations and dilatations
of the coordinates $x^\mu_i$. The only non-trivial transformations are the special conformal
transformations (boosts), which are combinations of an inversion, a translation and another
inversion.
\\

Let us start with a kinematical configuration $K=K(x_{ij}^2)$, Eq.~\re{distances}, and perform an
inversion of the coordinates, $x_i^\mu \to x_i^\mu/x_i^2$, to define the new configuration
\be\label{inverse}
K' = K  \left(  {x_{ij}^2}/{(x_i^2 x_j^2)}  \right).
\ee
Since the variables $u$ \re{u1u2u3} are invariant under such transformations, $u_a[K] = u_a[K']$,
the difference  $F_{6;2}^{\rm (WL)} - F_{6;2}^{\rm (BDS)} $ should also be invariant.
\\

As an example, let us consider six light-like four-dimensional vectors $p_i^\mu$,
\begin{align}
& p_1=(1,1,0,0)\,,&&  p_2 = (-1,p,p,0)\,,&& p_3=(1,-p,p,0)\,,
\\ \notag
& p_4=(-1,-1,0,0)\,,&& p_5=(1,-p,-p,0)\,,&& p_6=(-1,p,-p,0)\,,
\end{align}
with $p=1/\sqrt{2}$  and $\sum_i p_i^\mu=0$. These vectors define the
external momenta of the gluons in the six-gluon amplitude. Applying the duality relation
$p_i=x_{i+1}-x_{i}$, we evaluate the corresponding distances \re{distances},
\begin{align} \notag
K^{\rm (a)}: \quad & x_{14}^2= x_{15}^2=x_{24}^2= x_{25}^2=-2\,,\quad  x_{36}^2=-2-2\sqrt{2}\,,
\\ \label{Ka}
& x_{13}^2=x_{35}^2=x_{26}^2=x_{46}^2=-2-\sqrt{2}\,,
\end{align}
and the conformal cross-ratios \re{u1u2u3}, \be\label{u-ex}
u_1=u_3=\ft12 +\ft12\sqrt{2} \,,\quad u_2=1\,. \ee By
construction, this kinematical configuration satisfies the Gram
determinant constraint \re{Gram}. To define the conformal
transformations \re{inverse}, we choose an arbitrary reference
four-vector $x_1^\mu=(x_1^0,x_1^1,x_1^2,x_1^3)$ and reconstruct
the remaining $x-$vectors {according to} $x_{i+1}^\mu = x_{i}^\mu
+ p_i^\mu$ {(due to translation invariance, the Wilson loop does
not depend on the choice of $x_1^\mu$)}. Then, relation
\re{WL-BDS1} implies that the function $R_{\rm W}$ evaluated for
the kinematical configuration \re{inverse} should be the same as
for the
original configuration $K$.\\

As an example, we choose $x_1^\mu=(1,1,1,1)$ and apply the conformal transformation \re{inverse} to
$K^{(a)}$ defined in \re{Ka} to obtain the new kinematical configuration
\begin{align} \notag
K^{\rm (b)}: \quad & x_{14}^2= x_{15}^2=x_{24}^2= x_{25}^2=-\ft12-\ft14\,\sqrt {2}\,,\quad
x_{36}^2=-1-\ft34\,\sqrt {2}\,,
\\ \notag
&  x_{13}^2=-\ft32-\sqrt {2}\,,\quad x_{35}^2=-\ft52-\ft74\,\sqrt {2}\,,\quad
x_{26}^2=-\ft14-\ft18\,\sqrt {2}\,,
\\ \label{Kb}
&
x_{46}^2=-\ft38-\ft14\,\sqrt {2}\,.
\end{align}
The results of our numerical tests are summarised in Table~\ref{tab:Test}. They clearly show that
$F_{6;2}^{\rm (WL)}$ and $F_{6;2}^{\rm (BDS)} $ vary under conformal transformations, whereas their
difference $R_{\rm W} = F_{6;2}^{\rm (WL)} - F_{6;2}^{\rm (BDS)} $ stays invariant.
\\

We recall that in four dimensions the kinematical invariants \re{distances} have to verify the Gram
determinant constraint \re{Gram}. This relation $G[K]=0$ is invariant under the conformal
transformations \re{inverse}, simply because the conformal boosts map six light-like vectors
$p_i^\mu$ into another set of light-like vectors. There exist, however, certain kinematical
configurations $K'$ for which  $u_i[K] = u_i[K']$ but $G[K']\neq 0$. Since the difference function
$R_{\rm W}$ only depends on the $u-$variables, its value should be insensitive to the Gram
determinant condition~\footnote{We recall that the functions entering \re{WL-BDS1} can be defined
for configurations $K'$ satisfying $G[K']\neq 0$.}. For example, consider the following kinematical
configuration
\begin{align} \notag
K^{\rm (c)}: \quad & x_{14}^2= x_{15}^2=-1\,,\quad x_{24}^2= x_{25}^2=-2\,,\quad
x_{36}^2=-2-2\sqrt{2}\,,
\\ \label{Kc}
& x_{13}^2=-1-1/\sqrt{2}\,,\quad x_{35}^2=x_{26}^2=x_{46}^2=-2-\sqrt{2}\,.
\end{align}
The corresponding conformal cross-ratios \re{u1u2u3} are given by \re{u-ex}, but $G[K^{\rm
(c)}]\neq 0$. We verified numerically that $R_{\rm W}[K^{\rm (a)}]=R_{\rm W}[K^{\rm (b)}]=R_{\rm
W}[K^{\rm (c)}]$ with accuracy $< 10^{-5}$ (see Table \ref{tab:Test}). {This observation allows us
to study the function $R_{\rm W}(u_1,u_2,u_3)$ without any reference to the Gram determinant
condition.}

\renewcommand{\baselinestretch}{1.5}
\begin{table}[th]
\begin{center}
\begin{tabular}{||c||c|c|c|| }
\hline \hline Kinematical point &
 $F_{6;2}^
{\rm (WL)}$ & $F_{6;2}^ {\rm (BDS)}$ & $R_{\rm W}$
\\[1mm]
\hline \hline   $K^{\rm (a)}$ & $-5.014825  $ & $-14.294864 $ & $9.280039$
\\
\hline   $K^{\rm (b)}$& $-6.414907 $ & $ -15.694947$ & $9.280040$
\\
\hline   $K^{\rm (c)}$ & $ -5.714868$ & $ -14.994906 $ & $ 9.280038$
\\
\hline
\end{tabular}
\end{center}
\renewcommand{\baselinestretch}{1}
\caption{ Two-loop contributions to the hexagon Wilson loop, $F_{6;2}^{\rm (WL)}$, to the BDS
ansatz, $F_{6;2}^{\rm (BDS)}$,  and to their difference, $R_{\rm W}$,  evaluated for three
kinematical configurations \re{Ka}, \re{Kb} and \re{Kc} corresponding to the same values of $u_1$,
$u_2$ and $u_3$, Eq.~\re{u-ex}. } \label{tab:Test}
\end{table}
\renewcommand{\baselinestretch}{1}

\subsection{Collinear behaviour}
We recall that for the six-gluon amplitude $\mathcal{M}_6$ depending on
light-like momenta, $\sum_{i=1}^6 p_i^\mu=0$ and $p_i^2=0$, the collinear limit
amounts to letting, e.g. $p^{\mu}_{5}$ and $p^{\mu}_{6}$ be nearly
collinear (see e.g. \cite{Bern:1994zx} for more details), so that
$(p_5+p_6)^2 \to 0$ and
\be
p_5^\mu \to z P^\mu\,,\qquad p_6^\mu \to (1-z)  P^\mu\,,
\ee
with $P^2=0$ and $0< z < 1$ being the momentum fraction. Using the identification
$p_i^\mu = x_{i+1}^\mu - x_i^\mu$, we translate these relations into
properties of
the corresponding Wilson loop $W(C_6)$. We find that the cusp at point
$6$ is `flattened' in the collinear limit and the contour $C_6$
reduces to one with five cusps. In terms of the distances
$x_{ij}^2$, the collinear limit amounts to
\begin{align}\label{collinear1}
x_{15}^2 & \rightarrow   0 \,, & & x_{36}^2 \rightarrow z x_{13}^2 + (1-z)
x_{35}^2 \nonumber  \,, \\[2mm]
 x_{46}^2 &\rightarrow  z
x_{14}^2\,, & & x_{26}^2 \rightarrow (1-z) x_{25}^2 \,,
\end{align}
while the other distances $x_{13}^2,x_{24}^2,x_{25}^2,x_{35}^2$ remain
unchanged.
For the conformal cross-ratios the relation \re{collinear1} implies
\begin{equation}\label{collinear2}
u_{1} \rightarrow u \,,\quad u_{2} \rightarrow 0 \,,\quad u_{3} \rightarrow
1-u\,,
\end{equation}
with $u= {z x_{13}^2}/{(z x_{13}^2 +(1-z) x_{35}^2)}$ fixed. As was already
mentioned, the relation \re{solutionWI} is consistent with the
collinear limit of
the six-gluon amplitude provided that, in the limit \re{collinear2},
the function
$R(u_1,u_2,u_3)$ approaches a finite value independent of the kinematical
invariants. The same property can be expressed as follows (we recall that the function $R(u_1,u_2,u_3)$ is totally symmetric)
\be\label{const}
R(0,u,1-u) = c \,,
\ee
with $c$ being a constant.
Using our two-loop results for the finite part $F_6$, we performed thorough
numerical tests of the relation \re{const} for different kinematical
configurations of the contour $C_6$.
\\

\begin{figure}[h]%
\psfrag{gamma}[cc][cc]{$\gamma$}
\psfrag{f}[lc][cc]{}
\centerline{{\epsfxsize9cm \epsfbox{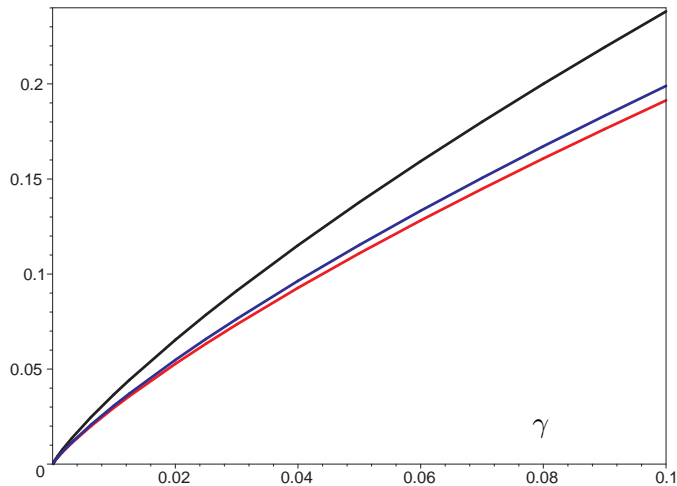}}} \caption[]{\small The
$\gamma-$dependence of the function $\widehat R^{(2)}(\gamma,u,1-u)$,
Eq.~\re{sub}, for
  different values of
the parameter $u=0.5$ (lower curve), $u=0.3$ (middle curve) and $u=0.1$ (upper
curve).}
\label{Fig-f}
\end{figure}

\begin{figure}[h]%
\psfrag{u}[cc][cc]{\vspace*{-10mm}$u$}
\psfrag{f}[lc][cc]{ }
\centerline{{\epsfxsize9cm \epsfbox{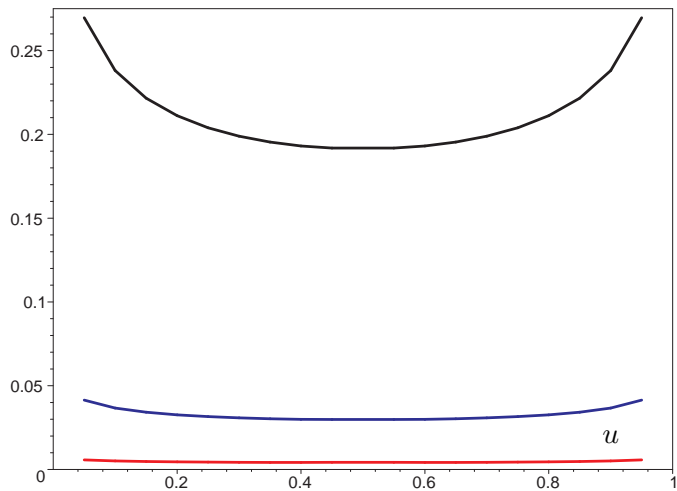}}} \caption[]{\small The
$u-$dependence of the function $\widehat R^{(2)}(\gamma,u,1-u)$, Eq.~\re{sub},
for different values of the parameter $\gamma=0.001$ (lower curve), $\gamma=0.01$
(middle curve) and $\gamma=0.1$ (upper curve).}
\label{Fig-f2}
\end{figure}

We found that, in agreement with \re{const}, the limiting value of the
function  $R^{(2)}(\gamma,u,1-u)$ as $\gamma\to 0$ does not depend on $u$. Since
the duality relation \re{duality-2} is not sensitive to the value of this constant, it is
convenient to subtract it from $R^{(2)}(\gamma,u,1-u)$ and introduce the function
\be\label{sub}
\widehat R^{(2)}(\gamma,u,1-u) = c - R^{(2)}(\gamma,u,1-u)
\ee
which satisfies  $\widehat R^{(2)}(0,u,1-u)=0$. To summarise our findings,  in Fig.~\ref{Fig-f} we
plot the function $\widehat R^{(2)}(\gamma,u,1-u)$ against
$\gamma$ for different choices of the parameter $0<u<1$ and in Fig.~\ref{Fig-f2}
the same function against $u$ for different choices of the parameter $\gamma$.
The important region for the collinear limit is where $\gamma$ is close to zero.
We also give numerical values for a  range of values of $\gamma$ such that
one can see how the function $R^{(2)}(u_1 ,u_2 , u_3 )$ varies in the particular
parametrisation $u_{1} = \gamma, u_{2} =u , u_{3}=1-u$.\\

So, in conclusion, our explicit $n=6$ Wilson loop calculation \cite{Drummond:2007bm} has shown that at
two loops there exists a non-trivial `remainder' function $R_6$ satisfying the conformal
Ward identities and the collinear limit condition. The
crucial test then was to compare the results of our Wilson loop calculation with a  parallel
two-loop six-gluon amplitude calculation, in order to check whether the proposed duality between
Wilson loops and gluon amplitudes continues to hold at this level (which, if true, automatically
implies the breakdown of the BDS ansatz). The results of the six-gluon calculation have become
available very recently \cite{Bern:2008ap}, and we present the comparison in section~\ref{WLvsgluons}.

\subsection{The hexagon Wilson loop versus the six-gluon MHV amplitude}\label{WLvsgluons}
The recently completed two-loop six-gluon amplitude calculation \cite{Bern:2008ap} found that
indeed the BDS ansatz fails at that level, i.e.
\be \label{defRA}
R_{\rm A} = F_{6;2}^{\rm (MHV)}  - F_{6;2}^{\rm (BDS)} \,.
\ee
with a nontrivial `remainder function'.
We now compare numerically the `remainder' function $R_{\rm A}$ with the analogous
quantity defined for the hexagonal Wilson loop in (\ref{WL-BDS1}).
In order to test the duality relation (\ref{duality-4}), we have to show that
for general kinematical configurations $K$ in  (\ref{distances}),
we have
\be\label{diff-diffJJ}
R_{\rm W}[K]-R_{\rm A}[K]=c_{\rm W}  \,.
\ee
Since $R_{\rm W}$ is a function of $u_{1},u_{2},u_{3}$ only, this relation would imply that so is
$R_{\rm A}$.
\\

To get rid of the constant $c_{\rm W}$, which
has lower numerical precision than
the evaluations for generic kinematics, we subtract from \re{diff-diffJJ} the same relation
evaluated for some reference kinematical configuration $K^{(0)}$,
\be
R_{\rm A}[K]  - R_{\rm A}[K^{(0)}] = R_{\rm W}[K]  - R_{\rm W}[K^{(0)}] \,.
\ee
The numerical tests of this relation for different kinematical configurations are summarised in
Table~\ref{tab:A0}.
\renewcommand{\baselinestretch}{1.5}
\begin{table}[th]
\begin{center}
\begin{tabular}{||c|c||c|c|| }
\hline \hline Kinematical point & $(u_1,u_2,u_3)$ & $R_{\rm W}  - R_{\rm W} ^{(0)} $ & $R_{\rm A} -
R_{\rm A} ^{(0)}$
\\[1mm]
\hline
\hline   $K^{(1)}$ & $(1/4,1/4,1/4)$ & $< 10^{-5}$ &
$-0.018 \pm 0.023$
\\
\hline   $K^{(2)}$& $(0.547253,0.203822,0.88127)$ & $-2.75533 $ & $-2.753\pm 0.015 $
\\
\hline   $K^{(3)}$ & $(28/17,16/5,112/85)$ & $-4.74460 $ &
$ -4.7445\pm 0.0075$
\\
\hline $K^{(4)}$  & $(1/9,1/9,1/9)$ & $4.09138$
& $4.12\pm 0.10$
\\
\hline  $K^{(5)}$  & $(4/81,4/81,4/81)$ & $ 9.72553 $
& $10.00 \pm 0.50$
\\
\hline
\end{tabular}
\end{center}
\renewcommand{\baselinestretch}{1}%
\caption{Comparison of the deviation from the BDS ansatz of the Wilson loop, $R_{\rm W}$, and of
the six-gluon amplitude, $R_{\rm A}$,  evaluated for the kinematical configurations \re{Ks}. Here,
$R_{\rm W}^{(0)} =R_{\rm W}\lr{1/4,1/4,1/4} = 13.26530$ and $R_{\rm A}^{(0)}=R_{\rm
A}\lr{1/4,1/4,1/4} =1.0937\pm 0.0057$ denote the same quantities evaluated at the reference
kinematical point $K^{(0)}$. The numerical results for $R_{\rm A}$ and $R_{\rm A}^{(0)}$ are taken
from Ref.~\cite{Bern:2008ap}. } \label{tab:A0}
\end{table}%
\renewcommand{\baselinestretch}{1}%
\begin{align} \notag
& K^{(0)}: && x_{i,i+2}^2 = -1\,,\qquad x_{i,i+3}^2 = -2\,;
\\[3mm] \notag
& K^{(1)}: && x_{13}^2=-0.7236200\,,\quad x_{24}^2 = -0.9213500\,,\quad x_{35}^2=-0.2723200\,,\quad
x_{46}^2=-0.3582300\,,
\\ \notag
& && x_{15}^2=-0.4235500\,,\quad x_{26}^2= -0.3218573\,,\quad x_{14}^2=-2.1486192\,,\quad
x_{25}^2=-0.7264904\,,
\\  \notag
& && x_{36}^2=-0.4825841\,;
\\[3mm] \notag
& K^{(2)}: && x_{13}^2=-0.3223100\,,\quad x_{24}^2 = -0.2323220\,,\quad x_{35}^2=-0.5238300\,,\quad
x_{46}^2=-0.8237640\,,
\\ \notag
& && x_{15}^2=-0.5323200\,,\quad x_{26}^2= -0.9237600\,,\quad x_{14}^2=-0.7322000\,,\quad
x_{25}^2=-0.8286700\,,
\\   \notag
& && x_{36}^2=-0.6626116\,;
\\[3mm]  \notag
& K^{(3)}: && x_{i,i+2}^2 = -1\,,\qquad x_{14}^2 = -1/2\,,\quad x_{25}^2=-5/8\,,\quad x_{36}^2=-17/14\,;
\\[3mm]  \notag
& K^{(4)}: && x_{i,i+2}^2 = -1\,,\qquad x_{i,i+3}^2 = -3\,;
\\[3mm] \label{Ks}
& K^{(5)}: && x_{i,i+2}^2 = -1\,,\qquad x_{i,i+3}^2 = -9/2\,.
\end{align}
Among the six configurations in equation (\ref{Ks}) only the first four verify the Gram determinant
condition \ref{Gram}. Also, the configurations $K^{(0)}$ and $K^{(1)}$ are related to each other by
a conformal transformation. This explains why the first entry in the third column of
Table~\ref{tab:A0} is almost zero, and also reflects the high precision of the numerical evaluation
for the Wilson loop. Comparing the numerical values for the six-gluon amplitude and the hexagon
Wilson loop, we observe that their finite parts coincide within the error bars. Therefore, we
conclude that the duality relation \re{duality-4} is satisfied to two loops at least.
We consider this very strong evidence that the duality should hold to all orders in the coupling,
although the `remainder' function $R_n$ is likely to receive corrections at each loop order.


\section{Conclusions and outlook}\label{ch-conclusions}

In this report, we studied a new duality
between Wilson loops and gluon amplitudes. The latter was
motivated mainly by work of
Alday and Maldacena, which showed that there is a relation between
gluon amplitudes and Wilson loops at strong coupling, using the
AdS/CFT correspondence. Unexpectedly, the duality seems to hold also
at weak coupling. Previous studies of the duality were restricted to one loop.
In this thesis we computed the expectation value of the rectangular
and pentagonal Wilson loops at two loops. A comparison with the
corresponding MHV gluon amplitudes shows that the duality holds.
\\

As was emphasised in this report, broken (dual) conformal symmetry plays
an important role on both sides of the duality.
We showed that the integrals contributing to the loop corrections of
the four-gluon amplitude up to four loops are all `pseudo-conformal' in the sense
explained in section \ref{ch-pseudoconformal}.
The origin of the dual conformal symmetry for gluon amplitudes remains puzzling.
For the Wilson loops, however, it is completely understood and controlled by an all-orders
Ward identity, which we derived in this thesis.
\\

The case $n=6$ provides a test of the duality relation beyond (dual)
conformal symmetry. What makes the hexagonal Wilson loops and the
six-gluon MHV amplitude agree? It seems natural to conjecture that
(dual) conformal symmetry is part of a larger symmetry group, which
is yet to be discovered. Such a speculation is further supported by
the fact that both Wilson loops and gluon amplitudes are intimately
related to the cusp anomalous relation, which in turn is governed
by integrability.
\\

The results obtained give strong support for
the duality between Wilson loops and gluon scattering amplitudes
in $\cN=4$ SYM.
In our opinion, given the evidence presented here, it is likely that
the duality is true to all orders in the coupling constant, and
for an arbitrary number of points/gluons.
This duality gives
rise to several intriguing possibilities.\\

The calculation of the pole part of the $n$-cusp Wilson loop in \cite{Drummond:2007au} was carried 
out analytically for arbitrary $n$ at two loops, in agreement with the general formula (\ref{Zn-WL}). 
As explained in \cite{Drummond:2007au}, one can define `auxiliary' diagrams to subtract the pole terms, 
thereby providing a definition of the finite part of the two-loop $n$-cusp
Wilson loop in terms of a set of {\it finite} parametric integrals.
We studied the latter numerically in the $n=5,6$ case, as discussed in this report,
and a generalisation to $n>6$ should not be difficult.\\

It would be very interesting if the arguments of Alday and Maldacena
that related gluon scattering amplitudes and Wilson loops at strong coupling
could be extended to arbitrary values of the coupling constant.
This would constitute a proof of the duality via the AdS/CFT correspondence,
or conversely, such a derivation would be an impressive test of the AdS/CFT
correspondence.\\

So far, the duality is limited to maximally helicity violating (MHV)
amplitudes. It is natural to wonder whether or how the duality extends
to amplitudes with different helicity configurations, as for example NMHV.
An obvious question is then how to match the helicity structure of the
NMHV amplitudes to the Wilson loops, which are helicity independent.
One may speculate that there could exist suitable modifications of the Wilson
loops such that they match those amplitudes. For example, one could
consider Wilson loops with local operator insertions.
\\

A related question is what the dual conformal properties of NMHV
amplitudes are. This will be the subject of a forthcoming paper
together with J. Drummond, G. Korchemsky and E. Soktachev.
\\

Another interesting possibility, in our opinion, would be to consider
a different point of the moduli space of $\cN=4$ SYM, such that some
particles become massive. One can ask the question whether scattering
amplitudes of massive on-shell gluons are still dual to
the corresponding Wilson loops, whose sides
are no longer light-like (because of the mass). On the other hand, for finite
masses the conformal symmetry
that was important in our analysis would be lost for such a configuration.


\section{Acknowledgements}

First of all, it is a pleasure to thank my advisor, Emery Sokatchev, for initiating me to
this interesting area of research and for sharing his knowledge.
Special thanks are also due to Andrei Belitsky, James Drummond and Gregory Korchemsky
for numerous discussions and encouragement.
I have also profitted from discussions with Costas Bachas, Benjamin Basso, Niklas Beisert, Massimo Bianchi, Fawzi Boudjema, Lance Dixon, Vladimir Dobrev,
Burkhard Eden, Johanna Erdmenger, Giovanni Feverati, Laurent Gallot, Fran\c{c}ois Gieres, Jean-Phillipe Guillet, Paul Heslop, Candide Jarczak, 
Dieter M\"uller, Eric Pilon, Jan Plefka, Eric Ragoucy, Paul Sorba, Yassen Stanev, Kelly Stelle, Raymond Stora, Frank Thuiller, Ivan Todorov and Arkady Tseytlin. 
I am grateful to Nans Baro, Livia Ferro, Charlotte Grosse-Wiesmann and Konstantin Wiegandt for pointing out typos in the draft to me. 
\\

I would like to thank the Laboratoire d'Annecy-le-Vieux de Physique Th\'eorique (LAPTH)
and its director, Patrick Aurenche, for providing a stimulating
working environment during the last three years.
The secretaries, Nicole Berger, V\'eronique Jonnery, Dominique Turc and Virginie Malaval
have helped considerably to simplify the administrative part of this thesis.
It has been great to share the experience of doing a PhD with the fellow students
Nans Baro, Samuel Belliard, Christophe Bernicot, Candide Jarczak, Ninh Le Duc, Gr\'egory Sanguinetti, Wessel Valkenburg.
They and the other members of LAPTH made the three years much more enjoyable and worthwile.
\\

This research was supported in part by the
French Agence Nationale de la Recherche under grant ANR-06-BLAN-0142.
Partial financial support from the Universit\'e de Savoie through an \emph{Explora'Doc}
grant and travel money through an APS grant (2005) is gratefully acknowledged.
It is a pleasure to acknowledge the warm hospitality extended to me by the
Theory Group of the Dipartimento di Fisica, Universit\`a di Roma ``Tor Vergata'',
where part of this work was done (from April 2007 to September 2007).
Special thanks are due to Yassen Stanev for his hospitality and numerous enlightening discussions.
\\

I would like to thank the theoretical physics sections of
the Universit\`a di Roma ``Tor Vergata'', the Universit\`a Torino,
the Univerisit\`a Milano Bicocca, the ETH Z\"{u}rich,
and Imperial College London, where part of this work was
presented, for their seminar invitations. In particular I would
like to thank the participants of these seminars for their comments
and for discussions, which certainly improved the way the material
is presented in this thesis.\\

Finally, this thesis would not have been written had it not been for
the university education I received. At the 
Universit\"at Augsburg I would like to thank in particular Reinhard Schertz and
Gert-Ludwig Ingold for their excellent lectures that aroused my 
interest in mathematics and theoretical physics, and similarly Daniel D\'ecamp
at the Univerit\'e de Savoie for initiating me to the world of elementary 
particle physics. 
Finally, I would like to thank Luc Frappat, Frank Thuiller and Jean-Claude Le Guillou
for their help in organising my \textit{Erasmus} year at the Universit\'e de Savoie, and
Frain\c{c}ois Delduc for the organisation of the theoretical physics master's programme at the Ecole Normale Sup\'erieure de Lyon.


\appendix


\section{Alternative proof of $\Phi^{(3)}=\Psi^{(3)}$ using the Mellin--Barnes representation}
\label{ch-MB}
In this section we show how the above identity between the off shell triple box and tennis
court integrals derived in section \ref{section-conformalfourpoint} can also be obtained by means of
the method of Mellin--Barnes (MB) representation.

\subsection{Introduction to the Mellin--Barnes technique}
\begin{figure}[t]
\psfrag{imz}[cc][cc]{${\rm Im}(z)$}
\psfrag{rez}[cc][cc]{${\rm Re}(z)$}
\psfrag{lambda}[cc][cc]{$-\lambda$}
\psfrag{beta}[cc][cc]{$\beta$}
\psfrag{one}[cc][cc]{$1$}
\psfrag{two}[cc][cc]{$2$}
\centerline{\epsfxsize 3.0 truein \epsfbox{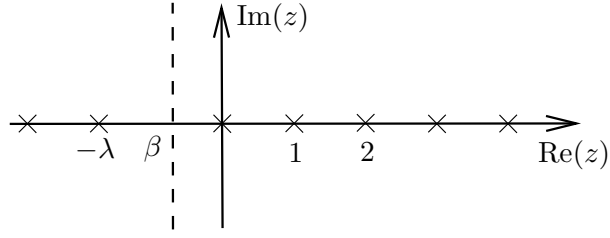}}
\caption{Position of the integration contour (dashed line) and poles of the integrand (crosses) in
formula (\ref{MB}), for real $\lambda$.}
\label{Fig-MB-technique}
\end{figure}

This method is one of the most powerful
methods of evaluating individual Feynman integrals.
It is especially successful for evaluating four-point Feynman
integrals.
For massless off-shell four-point integrals, first results were obtained by
means of MB representation in \cite{Usyukina:1992jd,Usyukina:1993ch}. In the
context of dimensional regularisation, with the space-time dimension
$d=4-2\ep$ as
a regularisation parameter, two alternative strategies
for resolving the structure of singularities in $\ep$ were
suggested in \cite{Smirnov:1999gc,Tausk:1999vh} where first results on evaluating
four-point on-shell massless Feynman integrals were obtained.
Then these strategies
were successfully applied to
evaluate massless on-shell double \cite{Smirnov:1999gc,Tausk:1999vh,
Glover:2000zu,Gehrmann:2000xj,Anastasiou:2000kp,Smirnov:1999wz,Anastasiou:2000mf}
and triple \cite{Smirnov:2002mg,Smirnov:2003vi,Bern:2005iz} boxes, with results written in terms of
harmonic polylogarithms \cite{Remiddi:1999ew}, double boxes with one leg off
shell \cite{Smirnov:2000vy,Smirnov:2000ie} and massive on-shell double boxes
\cite{Smirnov:2001cm,Smirnov:2004ip,Heinrich:2004iq,Czakon:2005gi} (see also chapter~5 of \cite{Smirnov:2004ym}).
It is based on the MB representation
\bea
\frac{1}{(X+Y)^{\lm}} = \frac{1}{\Gm(\lm)} \frac{1}{2\pi  i}
\int_{\beta- i  \infty}^{\beta + i  \infty} \frac{Y^z}{X^{\lm+z}}
\Gm(\lm+z) \Gm(-z)\, \dd z \,,
\label{MB}
\eea
applied to replace a sum of terms raised to some power by
their product to some powers.
The integration in (\ref{MB}) goes along a straight line parallel to the imaginary
axis, with $-{\rm Re}(\lambda) < \beta < 0$, so that the poles of $\Gamma(\lambda+z)$ lie on its left
and those of $\Gamma(-z)$ on its right, as shown in Fig.~\ref{Fig-MB-technique}.
%
Formula (\ref{MB}) is easy to understand. If $Y<X$, we can close the integration
contour on the right. By the residue theorem, we will pick up
contributions from the poles of $\Gamma(-z)$, leading to
\be\label{MB-expl1}
\frac{1}{\Gm(\lm)} \frac{1}{2\pi  i}
\int_{\beta- i  \infty}^{\beta + i  \infty} \frac{Y^z}{X^{\lm+z}}
\Gm(\lm+z) \Gm(-z)\, \dd z = X^{-\lambda} \sum_{n=0}^{\infty} (-1)^{n} \frac{\Gamma(\lambda+n)}{\Gamma(\lambda)} \left(\frac{Y}{X}\right)^{n}\,.
\ee
The right-hand side of (\ref{MB-expl1}) is indeed the series expansion of $(X+Y)^{-\lambda}$ for $Y<X$.
However equation (\ref{MB}) is also valid for $X<Y$, which can be seen by closing the integration
contour on the left.
\\

The first step of the method is the derivation of an appropriate
MB representation. It is very desirable to do this for general
powers of the propagators (indices) and irreducible numerators.
On the one hand, this provides crucial checks of a given
MB representation using simple partial cases. (For example, one
can shrink either horizontal or vertical lines to points, i.e.
set the corresponding indices to zero, and obtain simple diagrams
quite often expressed in terms of gamma functions.)
On the other hand, such a general derivation provides unambiguous
prescriptions for choosing integration contours (see details in
\cite{Smirnov:2004ym}).

\subsection{Example: One-loop integral}
As an example, let us come back to the one-loop integral $\Phi^{(1)}$. Using
the relation to the three-point integral (\ref{phi^3pt}) it is straightforward to
obtain the following parameter integral representation:
\be \label{phi1param}
\Phi^{(1)}(u,v) = \left(2\pi i\right)^{-2} \int_{0}^{1} \frac{ d\beta_{1} d\beta_{2} d\beta_{3} \delta(1-\beta_{1} -\beta_{2} -\beta_{3})}{\beta_{1}\beta_{2} u + \beta_{2} \beta_{3} v + \beta_{1} \beta_{3}} \,,
\ee
or, after a change of variables,
\be \label{phi1param2}
\Phi^{(1)}(u,v) = \left(2\pi i\right)^{-2} \int_{0}^{1} dx \int_{0}^{1} d\lambda \frac{\lambda}{ \lambda (x u + \bar{x} v) + \bar{\lambda} x \bar{x} } \,,
\ee
where $\bar{x}=1-x$ and $\bar{\lambda} = 1-\lambda$.
Then, one uses the Mellin
Barnes formula (\ref{MB}) twice in order to factorise the denominator in (\ref{phi1param}),
i.e.
\be \label{denom-MB}
\frac{1}{a+b+c} =  \int_{-i\infty}^{i\infty} dz_{1}  \int_{-i\infty}^{i\infty} dz_{2} \Gamma(-z_{1}) \Gamma(-z_{2}) \Gamma(1+z_{1}+z_{2}) a^{z_{1}} \,b^{z_{2}} \,c^{(-1-z_{1}-z_{2})}\,.
\ee
As usual, the integration contours are straight vertical lines, and they are chosen between the poles of the $\Gamma$ functions, i.e.
\be
{\rm Re}(z_{1}) <0 \,,\qquad {\rm Re}(z_{2}) <0 \,,\qquad {\rm Re}(z_{1})+{\rm Re}(z_{2}) >-1 \,,
\ee
so one could choose $ {\rm Re}(z_{1})  = -0.2$ and ${\rm Re}(z_{2})  = -0.4$ for example.
The integration over the parametric integrals is now straightforward,
and we obtain the MB representation \cite{Usyukina:1992jd} (see also \cite{Smirnov:2004ym})
\be\label{phi1-MB1}
\Phi^{(1)}(u,v) = \left( {2\pi i}\right)^{-2} \int_{- i \infty}^{i \infty} dz_{1}  \int_{- i \infty}^{i \infty} dz_{2} u^{z_{1}} v^{z_{2}} \Gamma^2(-z_{1}) \Gamma^2(-z_{2}) \Gamma^2(1+z_{1}+z_{2})\,.
\ee
For $0<u,v<1$ one can close the integration contours in (\ref{phi1-MB1}) on the right and use the residue theorem to
compute the integrals. This yields the result as a series expansion in $u$ and $v$.
\\

The representation (\ref{phi1-MB1}) is also useful for studying the limiting behaviour of $\Phi^{(1)}$. As an example, consider the OPE limit $x_{12} \rightarrow 0$, i.e. $u \rightarrow 0$, cf. (\ref{phi1OPE}). Since $ {\rm Re}(z_{1})  <0 $
the limit cannot be taken naively in the integrand of (\ref{phi1-MB1}). Therefore, we
deform the contour such that $0 < {\rm Re}(z_{1})  < 1 $. In doing so, we cross a pole at $u=0$ and hence
pick up a contribution from the residue of $u^{z_{1}} \Gamma^2(-z_{1})\Gamma^2(1+z_{1}+z_{2})$. Neglecting terms that vanish as $u \rightarrow 0$, we obtain
\begin{eqnarray}\label{phi1-ope1}
\Phi^{(1)}(u,v) &=& - \left(2\pi i\right)^{-1} \ln\left({u}\right) \int_{-i\infty}^{i \infty}
v^{z_{2}} \Gamma^{2}(-z_{2}) \Gamma^{2}(1+z_{2}) dz_{2} \nonumber \\
&&  - \left(2\pi i\right)^{-1}  \int_{-i\infty}^{i \infty}
v^{z_{2}} \Gamma^{2}(-z_{2}) \Gamma^{2}(1+z_{2}) \left[ \Psi(1+z_{2})+\gamma_{E}\right] dz_{2}  + O(u)\,.
\end{eqnarray}
In the second line $\gamma_{E}$ is the Euler constant and for integer $v$ the expression $\Psi(1+v)+\gamma_{E} $ is
equal to an harmonic sum $S_{v}=\sum_{n=1}^{v}1/n$.
Computing the integrals in (\ref{phi1-ope1}) by closing the contour on the right we pick up residues from $v=0,1,\ldots$
\begin{eqnarray}
\textrm{Res}_{z_{2}=n}\left[ v^{z_{2}} \Gamma^{2}(-z_{2}) \Gamma^{2}(1+z_{2}) \right] &=& v^n \ln v  \\
\textrm{Res}_{z_{2}=n}\left[ v^{z_{2}} \Gamma^{2}(-z_{2}) \Gamma^{2}(1+z_{2}) (\Psi(1+z_{2})+\gamma_{E}) \right] &=& 2 v^n \left[\Psi{'}(1+n)+ S_{n}\ln v \right] \,.
\end{eqnarray}
The resulting series can be summed. The first is just a geometrical series, and the second one can be done
as well (e.g. using Mathematica). We obtain
\be
\Phi^{(1)}(u,v) = \frac{1}{1-v} \left[  \ln\left({u}\right) {\ln(v)}- 2 \ln(v) \ln(1-u) +\frac{\pi^2}{3} - 2 {\rm Li}_{2}\left(v\right)  \right] + O(u)\,,
\ee
or equivalently (still for $0<v<1$),
\be
\Phi^{(1)}(u,v) = \frac{1}{1-v} \left[  \ln\left({u}\right) {\ln(v)} + 2 {\rm Li}_{2}\left(1-v\right)  \right] + O(u)\,,
\ee
in agreement with (\ref{phi1OPE}).
Although this expression was obtained for $0<v<1$, it can be analytically continued to all $v>0$.
\\

Many other examples of the application of the Mellin--Barnes technique to the evaluation
of Feynman integrals can be found in the book \cite{Smirnov:2004ym}.

\subsection{Alternative proof of the magic identity at three loops}
\begin{figure}[htbp]
\begin{center}
\ \psfig{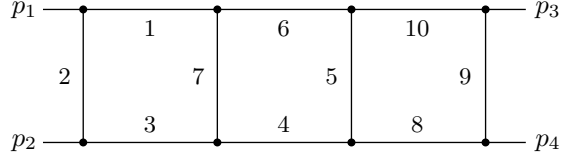}
\end{center}
\caption{Labelled triple box.}
\label{figure:tbox}
\end{figure}

\begin{figure}[htbp]
\begin{center}
\ \psfig{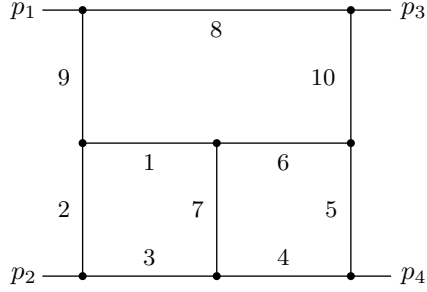}
\end{center}
\caption{Labelled tennis court.}
\label{figure:tcourt}
\end{figure}

Let us now use the MB technique to find an alternative proof of the `turning identity' $\Phi^{(3)}=\Psi^{(3)}$.
We consider the off shell triple box and tennis
court labelled as shown in
Figs.~\ref{figure:tbox} and~\ref{figure:tcourt},
with general powers of the propagators and one irreducible
numerator in the tennis court chosen as
$[(l_1+ l_3)^2]^{-a_{11}}$, where $l_{1,3}$  are the
momenta flowing through lines $1$ and $3$ in the same direction.
\\

Experience shows that a minimal number of MB integrations for
planar diagrams is achieved if one introduces MB integrations loop
by loop, i.e. one derives a MB representation for a one-loop
subintegral, inserts it into a higher two-loop integral, etc. This
straightforward strategy provides the following 15-fold MB
representations for the dimensionally regularised off-shell triple
box and tennis court with general indices:
\bea
T_1(a_1,\ldots,a_{10};s,t,p_1^2,p_2^2,p_3^2,p_4^2;\ep)=
\frac{\left(i\pi^{d/2} \right)^3 (-1)^a(-s)^{6-a-3\ep}}{
\prod_{j=2,4,5,6,7,9}\Gm(a_j)
\Gm(4 - 2 \ep- a_{4,5,6,7} )}
&& \nn \\ && \hspace*{-125mm}
\times \frac{1}{(2\pi i)^{15}}
\int_{-i\infty}^{+i\infty}  \prod_{j=1}^{15} \dd z_j
\frac{(-p_1^2)^{z_{12}} (-p_2^2)^{z_{13}}
(-p_3^2)^{z_{4,9,14}} (-p_4^2)^{z_{5,10,15}}
(-t)^{z_{11}}}{(-s)^{z_{4,5,9,10,11,12,13,14,15}}}
\nn \\ && \hspace*{-125mm} \times
\frac{\Gm(a_9 + z_{11,12,13})
\Gm(a_7 + z_{1,2,3})
\Gm(2 - \ep- a_{5,6,7}  - z_{1,2,4})
\Gm(2 - \ep- a_{4,5,7}  - z_{1,3,5})}{ \Gm(a_1 - z_{2}) \Gm(a_3 - z_{3})
\Gm(4 - 2 \ep- a_{1,2,3}  + z_{1,2,3})}
\nn \\ &&  \hspace*{-125mm} \times
\frac{\Gm(a_5 + z_{1,4,5})
\Gm(a_{4,5,6,7} + \ep-2  + z_{1,2,3,4,5})
\Gm(z_{11,14,15} - z_{6})}{ \Gm(a_8 - z_{7}) \Gm(a_{10} - z_{8})
\Gm(4 - 2 \ep - a_{8,9,10} + z_{6,7,8})}
\nn \\ &&  \hspace*{-125mm} \times
\Gm(2- \ep - a_{8,9} + z_{6,7} - z_{11,12,14})
\Gm(2 - a_{2,3} - \ep + z_{1,3} - z_{6,8,10})
\nn \\ &&  \hspace*{-125mm} \times
\Gm(a_{8,9,10}+\ep-2 + z_{11,12,13,14,15}- z_{6,7,8})
\Gm(2- \ep - a_{9,10}  + z_{6,8} - z_{11,13,15})
\nn \\ &&  \hspace*{-125mm} \times
\Gm(a_2 + z_{6,7,8})
\Gm(2- \ep - a_{1,2}  + z_{1,2} - z_{6,7,9})
\nn \\ &&  \hspace*{-125mm} \times
\Gm(z_{6,9,10}-z_{1} )
\Gm(a_{1,2,3} + \ep-2 - z_{1,2,3}  + z_{6,7,8,9,10})
\prod_{j=2,3,4,5,7,\ldots,15} \Gm(-z_j)\;;
\label{TB}
\eea
\bea
T_2(a_1,\ldots,a_{11};s,t,p_1^2,p_2^2,p_3^2,p_4^2;\ep)=
\frac{\left(i\pi^{d/2} \right)^3 (-1)^a (-s)^{6-a-3\ep}}{
\prod_{j=2,4,5,6,7,9}\Gm(a_j)\Gm(4 - 2 \ep- a_{4,5,6,7} )}
&& \nn \\ && \hspace*{-130mm}
\times \frac{1}{(2\pi i)^{15}}
\int_{-i\infty}^{+i\infty}  \prod_{j=1}^{15} \dd z_j
\frac{ (-p_1^2)^{z_{12}} (-p_2^2)^{z_{13}}
(-p_3^2)^{z_{5,10,14}}
(-p_4^2)^{z_{15} + z_{8}}
(-t)^{z_{11}}}{(-s)^{z_{5,8,10,11,12,13,14,15}}} \prod_{j=2}^{15} \Gm(-z_j)
\nn \\ && \hspace*{-130mm} \times
\frac{\Gm(a_9 + z_{11,12,13})
\Gm(a_7 + z_{1,2,3})
\Gm(2 - a_{5,6,7} - \ep - z_{1,2,4})
\Gm(2 - a_{4,5,7} - \ep - z_{1,3,5})}{\Gm(a_1 - z_{2}) \Gm(a_3 - z_{3})
\Gm(4- 2 \ep - a_{1,2,3}  + z_{1,2,3}) \Gm(a_{10} - z_{7})}
\nn \\ && \hspace*{-130mm} \times
\frac{\Gm(a_5 + z_{1,4,5})
\Gm(a_{4,5,6,7} + \ep -2+ z_{1,2,3,4,5})
\Gm(2 - a_{2,3} - \ep + z_{1,3} - z_{6,8,10})}
{\Gm(8 - 4 \ep- a - z_{5,6,8,10})
\Gm(a_8 - z_{4,9})
\Gm(a_{1,2,3,4,5,6,7,11}+ 2 \ep-4  + z_{4,5,6,7,8,9,10})}
\nn \\ && \hspace*{-130mm} \times
\Gm(6 -a+ a_{10}- 3 \ep - z_{5,6,7,8,10,11,12,14})
\Gm(a+ 3 \ep-6 + z_{5,6,8,10,11,12,13,14,15})
\nn \\ && \hspace*{-130mm} \times
\Gm(a_2 + z_{6,7,8})
\Gm(2 - \ep- a_{1,2}  + z_{1,2} - z_{6,7,9})
\Gm(6- 3 \ep -a+a_{8}- z_{4,5,6,8,9,10,11,13,15})
\nn \\ && \hspace*{-130mm} \times
\Gm(z_{6,9,10} -z_{1})
\Gm(a_{1,2,3} + \ep-2 - z_{1,2,3} + z_{6,7,8,9,10})
\nn \\ && \hspace*{-130mm} \times
\Gm(a_{1,2,3,4,5,6,7,11}+ 2 \ep-4+ z_{4,5,6,7,8,9,10,11,14,15})\;.
\label{TC}
\eea
Here
$a_{4,5,6,7}=a_4+a_5+a_6+a_7, a=\sum a_i, z_{11,12,13}=z_{11}+z_{12}+z_{13}$,
etc. Moreover, 
the letters $s$ and $t$ denote
the usual Mandelstam variables $s=(p_1+p_2)^2$ and $t=(p_1+p_3)^2$.
\\

These representations are written for the Feynman integrals in
Minkowski space. (This is rather convenient, in particular this allows
one to put
some of the legs on-shell.) The corresponding Euclidean versions are
obtained by the replacements $-s \to s, -t \to t, -p_1^2 \to
p_1^2,\ldots$ and by omitting the prefactors $(-1)^a$ and $i^3$.
\\

To calculate the triple box we need, i.e.
$T_1^{(0)}=T_1(1,\ldots,1)$ at $d=4$, we simply set all the
indices $a_i$ to one. We cannot immediately set $\ep=0$ because
there is $\Gm(- 2\ep)$ in the denominator. The value of the
integral is, of course, non-zero, so that some poles in $\ep$
arise due to the integration. To resolve the structure of poles
one can apply Czakon's code \cite{Czakon:2005rk}, which provides the
following value of the integral in the limit $\ep\to 0$ after
relabelling  the variables by
$z_{10} \to  z_{2}, z_{14} \to  z_{3}, z_{15} \to  z_{4}, z_{11} \to
z_{5}, z_{12} \to z_{6}$:
\bea
T_1^{(0)}=
\frac{\left(i\pi^{2} \right)^3}{(2\pi i)^{6}}
\int_{-i\infty}^{+i\infty}  \prod_{j=1}^{6} \dd z_j
\frac{ (-p_1^2)^{z_{6}} (-p_2^2)^{-1 - z_{5,6}}
(-p_3^2)^{-1 - z_{5,6}} (-p_4^2)^{z_{6}} (-t)^{z_{5}}}{(-s)^{2 - z_{5}}}
&& \nn \\ &&  \hspace*{-125mm} \times
\frac{\Gm(1 + z_{3,4}) \Gm(1 + z_{1} - z_{3,4,5})
\Gm(z_{2,3,4,5}-z_{1})\Gm(z_{4} - z_{6})}{\Gm(1 + z_{4} - z_{6}) \Gm(1 + z_{2,4} - z_{6})
\Gm(2 + z_{1,6} - z_{2,4})\Gm(2 + z_{3,5,6})}
\prod_{j} \Gm(-z_j)
\nn \\ &&  \hspace*{-125mm} \times
 \Gm(z_{2,4} - z_{6})^2
\Gm(1 + z_{1,6} - z_{2,4})^2 \Gm(1 + z_{5,6})
\Gm(1 + z_{3,5,6})\;.
\label{TB0}
\eea
To calculate the tennis court we need, i.e.
$T_2^{(0)}=T_2(1,\ldots,1,-1)$ at $d=4$, we proceed like in the
previous case. Czakon's code provides the following integral
(after relabelling $z_{10} \to  z_{2}, z_{14} \to  z_{3}, z_{15} \to  z_{4},
z_{11} \to  z_{5}, z_{12} \to  z_{6}$):
\bea
T_2^{(0)}=
\frac{\left(i\pi^{2} \right)^3}{(2\pi i)^{6}}
\int_{-i\infty}^{+i\infty}  \prod_{j=1}^{6} \dd z_j
\frac{(-p_1^2)^{z_{6}} (-p_2^2)^{-1 - z_{5} - z_{6}}
(-p_3^2)^{-1 - z_{5} - z_{6}} (-p_4^2)^{z_{6}}(-t)^{z_{5}} }{(-s)^{1 - z_{5}} }
\nn \\ &&  \hspace*{-125mm} \times
\frac{\Gm(1 + z_{3,4})
\Gm(1 + z_{1} - z_{3,4,5})  \Gm(z_{2,3,4,5}-z_{1})
\Gm(z_{4} - z_{6})}{\Gm(1 + z_{4} - z_{6}) \Gm(1 + z_{1} - z_{2,3,5,6})
\Gm(2 + z_{3,5,6}) \Gm(2 + z_{2,3,5,6})}
\prod_{j} \Gm(-z_j)
\nn \\ &&  \hspace*{-125mm} \times
\Gm(z_{1} - z_{2,3,5,6})^2
\Gm(1 + z_{5,6}) \Gm(1 + z_{3,5,6})
\Gm(1 + z_{2,3,5,6})^2 \;.
\label{TC0}
\eea
Now the simple change of variables $z_{2} \to  -z_{2} + z_{1} - z_{3}
- z_{4} - z_{5}$
in (\ref{TC0}) leads to an expression identical to (\ref{TB0}) up
to a factor of $u$ and we obtain the identity
$T_2^{(0)}= s T_1^{(0)}$, which corresponds to the identity
$\Phi^{(3)} = \Psi^{(3)}$ of section \ref{ch-magic}. (Observe that
the factor $s$ here appears because the general integrals
(\ref{TB}) and (\ref{TC}) are defined without the appropriate
prefactors present in the definitions of $\Phi^{(3)}$ and
$\Psi^{(3)}$.
\\

Let us stress that one can also apply the technique of MB
representation in a similar way in various situations where a given
four-point off-shell Feynman integral cannot be reduced to ladder integrals.


\section{Two-loop calculation of the rectangular light-like Wilson loop}
\label{ch-WL-appendix}
In this appendix we present the detailed calculation \cite{DHKSunpublished,Drummond:2007cf} of the two-loop four-point Wilson loop.

\subsection{Computation of individual diagrams}
Due to non-Abelian exponentiation, we only need the diagrams containing
then non-Abelian colour factor $\sim C_F N_c$.
The two-loop correction $w^{(2)}$ is given by a sum over the individual diagrams
shown in Fig.~\ref{all-diags} plus crossing symmetric diagrams. It is convenient
to expand their contributions in powers of $1/\epsilon$ and separate the UV divergent
and finite parts as follows
\begin{equation}\label{A-para-app}
w^{(2)} = \sum_{\alpha}  \left[(-x_{13}^2\, \tilde{\mu}^2)^{2\epsilon}+ (-x_{24}^2\,
\tilde{\mu}^2)^{2\epsilon}\right] \bigg\{\frac1{\epsilon^4}A_{-4}^{(\alpha)} +
\frac1{\epsilon^3}A_{-3}^{(\alpha)} + \frac1{\epsilon^2}A_{-2}^{(\alpha)} +
\frac1{\epsilon}A_{-1}^{(\alpha)}\bigg\} + A_0^{(\alpha)} + O(\epsilon)\,,
\end{equation}
where the sum goes over the two-loop Feynman diagrams shown in
Fig.~\ref{all-diags}(c)--(l).
Here $A^{(\alpha)}_{-n}$ (with $0\le n \le 4$) are
dimensionless functions of the ratio of distances $x_{13}^2/x_{24}^2$.
Making use
of \re{A-para-app}, we can parameterise the contribution of each individual diagram to
the Wilson loop by the set of coefficient functions $A^{(\alpha)}_{-n}$.
In (\ref{A-para-app}) $\tilde{\mu}$ is defined by $\tilde{\mu}^2 = \mu^2 e^{\gamma_{\rm E}} {\pi}$, where
$\gamma_{\rm E}$ is the Euler constant.
The gluon propagator in the Feynman gauge is
\be\label{gluonprop}
D^{\mu \nu}(x) = \eta^{\mu\nu} D(x)\,,\qquad D(x) = -\frac{\Gamma(1-\epsilon)}{4\pi^2} (e^{-\gamma_{\rm E}} \tilde{\mu}^2)^\epsilon
(-x^2+i0)^{-1+\epsilon}\,.
\ee
For simplicity we will consider the Wilson loop in the region where $x_{13}^2>0,\; x_{24}^2>0$,
which allows us to drop the $+i0$ prescription in the gluon propagator (\ref{gluonprop}).
Let us also introduce some notations that will be used throughout this section:
\be \alpha = x_{13}^{2}/(x_{13}^{2}+x_{24}^{2})\,,\qquad \bar{\alpha} = 1-\alpha \,,\qquad \gamma = \alpha / \bar{\alpha}\,.
\ee
In addition, we will sometimes use the notation
\be p_{i}^{\mu}  = x_{i+1}^{\mu} - x_{i}^{\mu}\,.
\ee
Finally, we will drop a factor of $(-\mu^2)^{2\epsilon}$ in intermediate steps of the calculation, since
it can be easily reinstated at the end.

\subsubsection{Contributions independent of the kinematics}

In \cite{Korchemskaya:1992je}, the two-loop computation was
carried out in the special kinematics $x_{13}^2 = - x_{24}^2 = 1$.
Since diagrams $I^{(c)},I^{(d)}$ and $I^{(f)}$ only depend in the
kinematics through trivial prefactors like $(-x_{13}^2
\tilde{\mu}^2)^{2\epsilon}$, we can use the results of
\cite{Korchemskaya:1992je}, and find the following
contributions\footnote{For diagram $I^{(c)}$ we take into account
the difference between the DREG scheme used in
\cite{Korchemskaya:1992je} and DRED scheme used here, see
\cite{Belitsky:2003ys}.} to the coefficient functions
$A^{(\alpha)}_{-n}$ defined in (\ref{A-para-app}).
\begin{eqnarray}
&& \hspace{-1cm} A^{(c)}_{-4} = 0\,,\quad A^{(c)}_{-3} = \frac{1}{8} \,,\quad A^{(c)}_{-2} = \frac{1}{4}\,,\quad A^{(c)}_{-1} = \frac{1}{2}+\frac{\pi^2}{48} \,,\quad A^{(c)}_{0} = 2+\frac{\pi^2}{12}+\frac{1}{6} \zeta_{3}  \,.
\\
&& \hspace{-1cm} A^{(d)}_{-4} = -\frac{1}{16}\,,\quad A^{(d)}_{-3} = 0 \,,\quad A^{(d)}_{-2} = -\frac{\pi^2}{96}\,,\quad A^{(d)}_{-1} = -\frac{1}{24}\zeta_{3} \,,\quad A^{(d)}_{0} = -\frac{7}{2880} \pi^4  \,.
\\
&& \hspace{-1cm} A^{(f)}_{-4} = \frac{1}{16}\,,\quad A^{(f)}_{-3} = -\frac{1}{8} \,,\quad A^{(f)}_{-2} = -\frac{1}{4}+\frac{5}{96}\pi^2\,,\quad A^{(f)}_{-1} = -\frac{1}{2} - \frac{\pi^2}{48} +\frac{7}{24}\zeta_{3} \,,
\\
&& \hspace{-1.35cm} \quad A^{(f)}_{0} = -2-\frac{\pi^2}{12} +\frac{119}{2880}\pi^4-\frac{1}{6} \zeta_{3}  \,.
\end{eqnarray}

Let us now consider the diagrams containing two crossed gluon propagators, i.e. $I^{(e)},I^{(h)},I^{(i)},I^{(j)}$.
Their group theory factor is $\textrm{Tr}\left( t_{a} t_{b}t_{a}t_{b} \right)= C_{F} \left( C_{F} -\frac{1}{2} N_{c}\right)$, of which we keep the part $-\frac{1}{2} C_{F} N_{c}$ only because of non-Abelian exponentiation.

\subsubsection{Diagram $I^{(e)}$}

 We consider the crossed diagram $[p_1,p_4][p_4,p_3]$ (the square brackets indicate the two segments to which the gluon is attached) shown in Fig.~\ref{all-diags}(e). The other diagrams of the same topology will be
 taken account of by including a symmetry factor $4$ for the contribution to $w^{(2)}$.
 The contribution of the diagram takes the form
\baa
I^{(e)} &=& -\ft12 C_{F} N_c (ig)^4  \int_0^1 dt_1 \int_{t_1}^1 dt_2 \int_0^1
ds_1\int_0^1 ds_2\,
\\
&\times&(p_1\cdot p_4) (p_3\cdot p_4) D(p_1s_2+p_4t_2) D(p_4(1-t_1)+p_3 s_1)\,.
\eaa
Taking into account the light-likeness of the edges of the Wilson loop are light-like, i.e. $p_{i}^2 = x_{i,i+1}^2=0$,
we find
\be
I^{(e)}= -\frac{C_F N_c  g^4 }{2 (8 \pi^2)^2}  {\Gamma^2(1-\epsilon)} (x_{13}^{2} x_{24}^{2} { e^{- 2\gamma_{\rm E}})^\epsilon}
 \int_0^1 dt_1 \int_{t_1}^1 dt_2 \int_0^1
ds_1\int_0^1 ds_2\, \lr{s_2 t_2 (1-t_1) s_1}^{-1+\epsilon}
\ee
Doing the integrals, one finds that the contribution of the diagram is
\be
I^{(e)}= -\frac{ C_F N_c g^4}{2 (8\pi^2)^2} \Gamma^2(1-\epsilon)(x_{13}^{2} x_{24}^{2} { e^{-2 \gamma_{\rm E}})^\epsilon} \frac{1}{\epsilon^4}\left[1-\frac{\Gamma^2(1+\epsilon)}{\Gamma(1+2\epsilon)}
\right]
\ee
It develops a double pole in $\epsilon$.
Including the symmetry factor, its contribution to $w^{(2)}$ is
\be
A^{(e)}_{-2} = -\frac{\pi^2}{24}\,,\qquad A^{(e)}_{-1} = \frac{1}{2} \zeta_{3} \,,\qquad A^{(e)}_{0} = -\frac{\pi^2}{12} M_{1} - \frac{49}{720} \pi^4\,.
\ee

\subsubsection{ Diagram $I^{(h)}$}

The contribution of this diagram takes the form
\baa
I^{(h)} &=& -\ft12 N_c C_F (ig)^4   \int_0^1 dt_1 \int_{t_1}^1 dt_2 \int_0^1
ds_1\int_0^1 ds_2\,
\\
&\times&(p_1\cdot p_4) (p_2\cdot p_4) D(p_1s_2+p_4t_2) D(p_1+p_2 s_1 + p_4t_1)
\eaa
Integration over $s_1, s_2$ and $t_2$ yields
\ba\nonumber
I^{(h)} &=&\ft12 N_c C_F
\frac{g^4}{(8\pi^2)^2} \Gamma^2(1-\epsilon)(e^{-2\gamma_{\rm E}} x_{24}^{2})^\epsilon (x_{13}^{2}+x_{24}^{2})
\\
&\times&\frac1{\epsilon^3}\int_0^1\frac{dt_1\,(1-t_1^\epsilon)}{x_{13}^{2}-(x_{13}^{2}+x_{24}^{2})t_1}
\left[\lr{x_{13}^2 (1-t_1)}^\epsilon - \lr{x_{24}^{2} t_1}^\epsilon
\right]
\ea
This expression in not symmetric under $\alpha\to 1-\alpha$. Its symmetrisation
gives (upon substitution $x \to 1-x$)
\ba
 I^{(h) {\rm (sym)}} &=& \ft12 C_F N_c \frac{g^4}{ (8 \pi^2)^2}  \bigg\{\frac1{4\epsilon}\left[ (x_{13}^2)^{2\,\epsilon}+ (x_{24}^2)^{2\,\epsilon}
 \right]\int_0^1\frac{dx\, }{x-\alpha}\ln \left( {\frac {\alpha\bar x }{ \bar\alpha  x}}
\right) \ln  \left( x \bar x \right)
\\
\nonumber && +\frac14\int_0^1\frac{dx\, }{x-\alpha}\ln \left( {\frac {\alpha\bar
x }{
\bar\alpha x}} \right)  \left[ \frac32 \ln^2  \left( x \bar x \right)
-\frac12 \ln^2  \left( {\frac {\bar\alpha}{\alpha}} \right) + \frac12  \ln^2
\left( { \frac {\alpha\bar x }{ \bar \alpha x}}
 \right)    \right] + O(\epsilon)
\bigg\}\,,
\ea
and in the notation of the basic integrals of section \ref{Appendix-integrals}
\begin{eqnarray}
I^{(h) {\rm (sym)}} &=& \frac{1}{8} C_F
N_c \left( \frac{g^2}{8 \pi^2} \right)^2 \bigg\{\frac1{\epsilon}\left[ (x_{13}^2)^{2\,\epsilon}+ (x_{24}^2)^{2\,\epsilon}
 \right] M_2 \\
&&  +  \left[ \frac32 M_3 -\frac12 \ln^2  \left( {\frac
{\bar\alpha}{\alpha}} \right) M_1 + \frac12  M_9   \right] + O(\epsilon) \bigg\} \nonumber\,.
\end{eqnarray}
Including the combinatorial factor corresponding to this diagram, $c^{(h)} = 8$, we have
\be
A^{(h)}_{-1} = \frac{1}{4} M_{2} \,,\qquad A^{(h)}_{0} = \frac{1}{8} M^2_{1} +\frac{3}{8}M_{3} +\frac{\pi^2}{8} M_{1}\,.
\ee

\subsubsection{ Diagram $I^{(i)}$}

The contribution of this diagram takes the form
\baa
I^{(i)} &=& -\ft12 N_c C_F (ig)^4   \int_0^1 dt_1 \int_{t_1}^1 dt_2 \int_0^1
ds_2\int_{s_2}^1 ds_1\,
\\
&\times&(p_2\cdot p_4)^2 D(p_1+p_2 s_1 + p_4t_1) D(p_1+p_2 s_2 + p_4t_2)
\eaa
It is finite for $\epsilon\to 0$ and, therefore, the calculation can be
performed in $D=4$ dimensions
\be
I^{(i)} = -\ft12 N_c C_F \frac{g^4}{(8\pi^2)^2}   J
\ee
with
$$
J=\int_0^1 dt_1 \int_{t_1}^1 dt_2 \int_0^1 ds_2\int_{s_2}^1 ds_1\,\left[(\alpha
s_1 + \bar{\alpha} t_1 - s_1 t_1) (\alpha s_2 + \bar{\alpha} t_2 -  s_2 t_2)\right]^{-1}
$$
Then, after integrating over $s_2$ and $t_1$ and rescaling $s_1 \to \bar\alpha s_1$ and $t_2 \to \alpha t_2$ one gets
\be
J=\int_0^{1/\alpha}\int_0^{1/\bar\alpha} \frac{ds_1 \,dt_2}{(1-s_1)(1-t_2)}\ln
\lr{1+s_1\frac{1-t_2}{t_2}}\ln \lr{1+t_2\frac{1-s_1}{ s_1}}
\ee
It seems natural to simplify the integrand by changing integration variables
\be
x =s_1\frac{1-t_2}{t_2}\,,\qquad y=t_2\frac{1-s_1}{ s_1}\,,
\ee
which leads to
\begin{equation}
J = \int_{\Sigma} dx\, dy\, f(x,y)\,,\qquad f(x,y) = \frac{1-xy}{x (1+x) y (1+y)} \ln(1+x) \ln(1+y)
\end{equation}
where the integration domain $\Sigma$ is defined by
\begin{equation}
0 \leq \frac{1-xy}{1+y} \leq 1+\frac{1}{\gamma}\,,\qquad 0 \leq
\frac{1-xy}{1+x} \leq 1+\gamma\,,\qquad \gamma = \frac{\alpha}{\bar{\alpha}}=\frac{s}{t}>0\,.
\end{equation}
This can be written more conveniently as
\begin{eqnarray} y&\leq&
\frac{1}{x}\,,\qquad (x>0)\\
y&\leq&-(1+\gamma+\frac{\gamma}{x})\,,\qquad
(-1<x<0)\\y&\geq&-\frac{1}{1+\gamma+\gamma x}\,.
\end{eqnarray}
Thus the integration over $\Sigma$ can be parametrised as follows:
\be
J = \underbrace{\int_{0}^{\infty}dx \int_{0}^{1/x}dy f(x,y)}_{J_{1}} + \underbrace{\int_{0}^{\infty}dx \int_{-\frac{1}{1+\gamma+\gamma
x}}^{0}dy f(x,y)}_{J_{2}} +\underbrace{\int_{-1}^{0} dx \int_{\frac{-1}{1+\gamma+\gamma
x}}^{-(1+\gamma+\gamma/x)}dy f(x,y)}_{J_{34}}\,.
\ee
The integration regions lie in the first, second and third and fourth quadrant (clockwise).
The first integration can be easily done using
\be
\int^{y} dz f(x,z)  = \frac{\ln(1+x)}{x (1+x)}\left[
-\frac{1}{2}(1+x)\ln^2(1+y)-\textrm{Li}_{2}(-y) \right]\,.
\ee
The final integration is harder, but all integrals can be expressed in terms
of polylogarithms \cite{Lewin}.
After some algebra we find
\ba
J_{1} &=& \frac{\pi^4}{180}\\
J_{2} &=& \frac{19 \pi^4}{360} +\frac{\pi^2}{12}\ln^2(\gamma)+\frac{1}{24}\ln^4(\gamma) +\frac{\pi^2}{12} \ln^2(1+\gamma)+\frac{1}{6} \ln(\gamma)\ln^3(1+\gamma)-\frac{1}{12}\ln^4(1+\gamma) \nonumber \\
&&- \left[ -\frac{\pi^2}{3} -\frac{1}{2} \ln^2(\gamma) +\frac{1}{2} \textrm{Li}_{2}(-\gamma) \right] \textrm{Li}_{2}(-\gamma) \nonumber \\
&& - \ln\left(\frac{\gamma}{1+\gamma}\right) \left[\textrm{Li}_{3}(-\gamma) + \zeta_{3} \right] -\textrm{Li}_{4}\left(\frac{1}{1+\gamma}\right)-\textrm{Li}_{4}\left(\frac{\gamma}{1+\gamma}\right)\\
J_{34} &=& -\frac{\pi^4}{60}-\frac{\pi^2}{12} \ln^2(1+\gamma)-\frac{1}{6} \ln(\gamma)\ln^3(1+\gamma)+\frac{1}{12}\ln^4(1+\gamma) \nonumber \\
&&+ \left[ -\frac{\pi^2}{3} -\frac{1}{2} \ln^2(\gamma) +\frac{1}{2} \textrm{Li}_{2}(-\gamma) \right] \textrm{Li}_{2}(-\gamma) \nonumber \\
&& + \ln\left(\frac{\gamma}{1+\gamma}\right) \left[\textrm{Li}_{3}(-\gamma) + \zeta_{3} \right] +\textrm{Li}_{4}\left(\frac{1}{1+\gamma}\right)+\textrm{Li}_{4}\left(\frac{\gamma}{1+\gamma}\right)\,.
\ea
Here we used frequently polylogarithm identities like those given in appendix \ref{Appendix-polylogs} in order to simplify the individual expressions.
Summing them up, we see that most terms cancel between $J_{2}$ and $J_{34}$, and we obtain the simple result
\begin{equation}
J = \frac{1}{24}\left[ \pi^2 + \ln^2(\gamma) \right]^2 \,.
\end{equation}
Including the symmetry factor $2$, we find
\be
A^{(i)}_{0} = -\frac{1}{24} M^2_{1}\,.
\ee

\subsubsection{Factorised diagram $I^{(j)}$}

The contribution of this diagram takes the form
\baa
I^{(j)} &=& -\ft12 C_F  N_c (ig)^4  \int_0^1 dt_1 \int_{0}^1 dt_2 \int_0^1
ds_2\int_{0}^1 ds_1\,
\\
&\times&(p_2\cdot p_4) (p_1 \cdot p_3) D(p_2+p_1 s_1 + p_3 t_1) D(p_1+p_2 s_2 +
p_4t_2)
\eaa
and the integral factorises into the product of finite one-loop integrals,
\ba
I^{(j)}
&=&-\ft12 N_c C_F \lr{\frac{g^2}{4\pi^2}}^2 \left[\frac12 \int_0^1  \frac{dx
dy\,(x_{13}^2+x_{24}^2)}{ x_{24}^2 x+ x_{13}^2 y -(x_{13}^2+x_{24}^2)xy} \right]^2\,.
\ea
Evaluating the integral we find
\be
A^{(j)}_{0} = -\frac{1}{8} M^2_{1}\,.
\ee

\subsubsection{Self-energy diagram $I^{(k)}$}

The contribution of this diagram reads
\be
I^{(k)} = -g^2 C_F (p_2\cdot p_4) \int_0^1 ds_1 \int_0^1 dt_1\,
D^{(1)}(p_1+p_2 s_1 + p_4 t_1)\,,
\ee
where the gluon propagator in the Feynman gauge, with the one-loop correction included,
is
\be
D^{(1)}(x) = \frac{g^2}{64\pi^D} \left[(3D-2-2\epsilon) N_c - 2(D-2) n_f
-n_s\right]\frac{\Gamma^2(D/2-1)}{(D-4)(D-3)(D-1)} (-x^2+i0)^{3-D}
\ee
with $D=4-2\epsilon$. Here we added the contribution both of
$\epsilon-$scalars in the DRED scheme and of $n_s$ scalars. Then,
\be
I^{(k)} =\frac{g^4}{2\cdot 64\pi^D} C_F\left[(3D-2-2\epsilon) N_c - 2(D-2)
n_f -n_s\right]\frac{\Gamma^2(D/2-1)}{(D-4)(D-3)(D-1)} J^{(k)}\,,\ee
with
\be
J^{(k)} = (x_{13}^2+x_{24}^2)\int_0^1 ds_1 \int_0^1 dt_1\, [-x_{13}^2 s_1 - x_{24}^2 t_1 + (x_{13}^2+x_{24}^2) s_1 t_1
]^{3-D}
\ee
The integration is straightforward.
In $\mathcal{N}=4$ SYM, for $n_f=4 N_c$ and $n_s=6 N_c$,
\ba
I^{(k)}
&=&\frac18\lr{\frac{g^2}{4\pi^2}}^2C_FN_c\left\{\epsilon^{-1}  \left[ { (x_{13}^2)^{2\epsilon}
+ (x_{24}^2)^{2\epsilon}} \right] M_1+4M_1+2M_2 + O(\epsilon) \right\}\,.
\ea
The combinatorial factor corresponding to this diagram is
$
c^{(k)} = 2\,,
$ so we have
\be
A^{(k)}_{-1} = \frac{1}{4} M_{1}\,,\qquad A^{(k)}_{0} = M_{1} + \frac{1}{2} M_{2} \,.
\ee

\subsubsection{`Mercedes-Benz' diagram $I^{(l)}$}

We assume that two gluon legs are attached to the segment $p_2$ and the third leg
is attached to the segment $p_4$
\begin{eqnarray}
I^{(l)}&=& \frac12 g^4 C_F N_c \int_0^1dt_1 \int_0^1d t_2 \int_{t_2}^1dt_3 \\
&& \times p_4^{\mu_1}
p_2^{\mu_2} p_2^{\mu_3}
\Gamma_{\mu_1\mu_2\mu_3}(-i\partial_{z_1},-i\partial_{z_2},-i\partial_{z_3})\int
d^D z_4 \prod_{i=1}^3 D(z_i-z_4) \nonumber\,,
\end{eqnarray}
where
\be \label{zvars}
z_1^{\mu}=-p^{\mu}_1-p^{\mu}_4 t_1\,,\qquad z^{\mu}_2 = p^{\mu}_2 t_2\,,\qquad z^{\mu}_3 = p^{\mu}_2 t_3\,,
\ee
and the three-gluon vertex is given by
\be
p_4^{\mu_1}p_2^{\mu_2} p_2^{\mu_3}
\Gamma_{\mu_1\mu_2\mu_3}(-i\partial_{z_1},-i\partial_{z_2},-i\partial_{z_3}) =
i(p_2 \cdot p_4) \lr{p_2 \partial_{z_2} - p_2 \partial_{z_3}} = i (p_2 \cdot p_4)
\lr{\partial_{t_2}-\partial_{t_3}}\,.
\ee
Then,
\be
I^{(l)} = \frac{i}2 g^4 C_F N (p_2 \cdot p_4)\int_0^1dt_1 \int_0^1d t_2 \int_{t_2}^1dt_3
\,\lr{\partial_{t_2}-\partial_{t_3}} J(z_1,z_2,z_3)\,,
\ee
where the three-point integral $J(z_1,z_2,z_3)$ is defined in appendix \ref{Appendix-integralJ}. By virtue of $z_{23}^2=0$
\be \label{Jlightlike}
J(z_1,z_2,z_3) =  \frac{i^{1-2D}}{32\pi^D}\frac{\Gamma(D-3)}{4-D}\int_0^1 d\tau\,
(\tau\bar \tau)^{D/2-2} \left[(-z_1+\tau z_2 + \bar \tau z_3)^2 \right]^{3-D}
\ee
Using the explicit expression for the $z-$coordinates in (\ref{zvars})
and equation (\ref{Jlightlike}) we find
\baa
I^{(l)}&=&i^{2-2D}\frac{g^4 C_F N}{128 \pi^D}
\frac{\Gamma(D-3)}{(4-D)^2}\int_0^1\frac{dt_1\, (-x_{13}^2-x_{24}^2)}{x_{13}^2 \bar{t}_{1} -x_{24}^2
t_1}\int_0^1 {d\tau} (\tau\bar \tau)^{D/2-2}
\\
&\times& \left\{\lr{2-\frac1{\tau}}\left[ \lr{x_{13}^2 \bar{t}_1}^{4-D}-\lr{x_{24}^2
t_1}^{4-D}\right] +\lr{\frac1{\tau}-\frac1{\bar\tau}}{\lr{x_{13}^2 \bar
\tau \bar{t}_1}^{4-D}}\right\}\,.
\eaa
Notice that in the second term, one can integrate by parts using
\be
(\tau\bar \tau)^{D/2-2}\lr{\frac1{\tau}-\frac1{\bar\tau}}
=(D/2-2)^{-1}\frac{d}{d\tau}(\tau\bar \tau)^{D/2-2}\,.
\ee
After some algebra one arrives at
\baa
I^{(l)}&=&  \frac{g^4 C_F N}{64 \pi^4}
\frac{\Gamma^2(1-\epsilon)}{(2\epsilon)^3(1-2\epsilon)} e^{-2\epsilon \gamma_{\rm E}}
\\
&\times& \int_0^1\frac{dt}{\alpha-t}\bigg[{(-
x_{13}^2)^{2\epsilon}(1-t)^{2\epsilon}-(-x_{24}^2
)^{2\epsilon}t^{2\epsilon}}\bigg]\bigg[1  - \frac{\Gamma(2-2\epsilon)
}{\Gamma^2(1-\epsilon)}(t(1-t))^{-\epsilon} \bigg]\,,
\eaa
where we replaced $D$ by $4-2 \epsilon$. Expanding in powers of $\epsilon$ we find
\be
I^{(l)}=-\frac{g^4 C_F N}{64 \pi^4}  \bigg\{
\frac{1}{8\epsilon}  \left[ { (x_{13}^2)^{2\epsilon}
+ (x_{24}^2)^{2\epsilon}} \right]  \lr{2M_1+M_2}
+\lr{1-\frac1{24}\,{\pi }^{2}}M_1+\frac12 M_2 +\frac18\, M_3 +
O(\epsilon)\bigg\}\,,
\ee
where we used the basic integrals of section \ref{Appendix-integrals}.
The combinatorial factor corresponding to this diagram is $c^{(l)} = 4\,,$
so we have
\be
A^{(l)}_{-1} = -\frac{1}{4} M_{1}-\frac{1}{8} M_{2} \,,\qquad A^{(l)}_{0} = -M_{1} + \frac{\pi^2}{24} M_{1} -\frac{1}{2} M_{2} -\frac{1}{8} M_{3} \,.
\ee

\subsubsection{T-shaped diagram}

We assume that two gluon legs are attached to the segment $p_2$ and the third leg
is attached to the segment $p_4$
\begin{eqnarray}
I^{(g)}&=& \frac12 g^4 c_F N_c \int_0^1dt_1 \int_0^1d t_2 \int_0^1dt_3 \\
&&\times p_4^{\mu_1}
p_1^{\mu_2} p_2^{\mu_3}
\Gamma_{\mu_1\mu_2\mu_3}(-i\partial_{z_1},-i\partial_{z_2},-i\partial_{z_3})\int
d^D z_4 \prod_{i=1}^3 D(z_i-z_4)\,,\nonumber
\end{eqnarray}
where
\be
z_1=-p_4 t_1\,,\qquad z_2 = p_1 t_2\,,\qquad z_3 = p_1+p_2 t_3
\ee
and three-gluon vertex is given by
\baa
\Gamma_3 &\equiv& p_4^{\mu_1}p_1^{\mu_2} p_2^{\mu_3}
\Gamma_{\mu_1\mu_2\mu_3}(-i\partial_{z_1},-i\partial_{z_2},-i\partial_{z_3})
\\
&=&i(p_1 \cdot p_4) \lr{p_2 \cdot (\partial_{z_2} - \partial_{z_1})}
+i(p_1 \cdot p_2) \lr{p_4 \cdot(\partial_{z_3} - \partial_{z_2})}
+ i(p_2 \cdot p_4) \lr{p_1 \cdot(\partial_{z_1} -
\partial_{z_3})}\,.
\eaa
Translation invariance of the $z_4-$integral can be used to simplify the vertex to
\be
\Gamma_3=-i(p_1\cdot  p_4)\partial_{t_3}-i(p_1 \cdot p_2)\partial_{t_1}-2i(p_1 \cdot p_4) \lr{p_2 \cdot
\partial_{z_1}} +2i(p_1\cdot p_2) \lr{p_4 \cdot \partial_{z_3}} + i(p_2 \cdot p_4) \lr{p_1\cdot (\partial_{z_1} -
\partial_{z_3})}
\ee
Let us rewrite
the integral as
$I^{(g)}=I^{(g)}_1+I^{(g)}_2$, where
\baa
I^{(g)}_1 &=& \frac{i}2 g^4 c_F N_c \int_0^1dt_1 \int_0^1d t_2 \int_0^1dt_3 \,[\ft12
x_{24}^2
\partial_{t_3}+\ft12 x_{13}^2 \partial_{t_1} ]V(z_1,z_2,z_3)
\\
I^{(g)}_2 &=& \frac{i}2 g^4 c_F N_c \int_0^1dt_1 \int_0^1d t_2 \int_0^1dt_3 \,
\\
&& \hspace{-2cm}\times \left[ -x_{24}^2 \lr{\lr{p_2
\partial_{z_1}}+\partial_{t_3}} +x_{13}^2 \lr{\lr{p_4 \cdot \partial_{z_3}}-\partial_{t_1}} - \ft12 (x_{13}^2+x_{24}^2) \lr{p_1 \cdot (\partial_{z_1} -
\partial_{z_3})}\right]V(z_1,z_2,z_3)
\eaa
The first contribution $I^{(g)}_1$ can be split into a divergent and a finite part,
$I^{(g)}_1 = I^{(g)}_{1d} + I^{(g)}_{1f}$
\baa
I^{(g)}_{1d} &=&\frac{i}2 g^4 c_F N_c \int_0^1dt_2 \bigg[ -\ft12 x_{13}^2 \int_0^1dt_3 \,
V(0,z_2,z_3)-\ft12 x_{24}^2 \int_0^1d t_1\,V(z_1,z_2,p_1) \bigg]\,,
\\
I^{(g)}_{1f} &=& \frac{i}2 g^4 c_F N_c \int_0^1dt_2 \bigg[+ \ft12 x_{13}^2 \int_0^1dt_3 \, V(-p_4,z_2,z_3)+\ft12 x_{24}^2 \int_0^1d t_1
 V(z_1,z_2,p_1+p_2) \bigg]\,.
\eaa
Divergences come from the two integrals in the first line only since the three arguments of
the $V-$functions in second line cannot be light-like.
The divergent integrals can be easily evaluated as
\be \label{Ig1d}
I^{(g)}_{1d} =  \lr{\frac{g^2}{32\pi^2}}^2C_F N
\frac{\Gamma^2(1-\epsilon) e^{-2\epsilon \gamma_{\rm E}}}{\epsilon^4}\left[\lr{x_{13}^2}^{2\epsilon}
+ \lr{x_{24}^2 }^{2\epsilon}
\right]\left[1-\frac{\Gamma(1+\epsilon)\Gamma(1-2\epsilon)}{\Gamma(1-\epsilon)}
\right]\,.
\ee
Next, we examine the finite part of $I_1$. it can be written as
\be
I^{(g)}_{1f} = \frac{g^4 C_F N_c}{4\cdot 64 \pi^4}\int [d\beta]_3 \int_0^1dt_2
 \int_0^1dt_1 \frac{x_{24}^2}{
\beta_3(\beta_2 \bar t_2 + \beta_1 \bar t_1)x_{13}^2+\beta_1 t_1 \beta_2 t_2
x_{24}^2}+\big[ x_{14}^2 \leftrightarrows x_{13}^2 \big]\,.
\ee
Here $[d\beta]_3 = d\beta_{1} d\beta_{2} d\beta_{3} \delta(1-\beta_{1}-\beta_{2}-\beta_{3})$.
After changing variables, we have
\be
I^{(g)}_{1f} = \frac{g^4 C_F N_c}{4\cdot 64 \pi^4} \, J^{(g)}_{1f} \,,
\ee
where
\be
J^{(g)}_{1f} =\int_0^1 dx \int_0^1 d\lambda \int_0^1dt_2 \int_0^1dt_1
\frac{1}{
\bar\lambda(x \bar t_2 + \bar x \bar t_1)\gamma+\lambda x\bar x t_1
t_2}+\left(\gamma\to 1/\gamma \right)
\ee
After some algebra one finds 
\baa \label{Jg1f}
J^{(g)}_{1f} &=& \bigg\{  \frac{7}{60}\pi^4 + \frac{\pi^2}{3} \LNP{\gamma}{2}+\frac{1}{12} \LNP{\gamma}{4} -\frac{2}{3}\pi^2 \LN{\gamma} \LN{1+\gamma}-\frac{1}{3} \LNP{\gamma}{3} \LN{1+\gamma}\nonumber  \\
&& +\frac{2}{3} \pi^2 \LNP{1+\gamma}{2} +\frac{1}{2}\LNP{\gamma}{2}\LNP{1+\gamma}{2}-\frac{1}{6}\LNP{1+\gamma}{4}+\frac{\pi^2}{3} \PLN{2}{-\gamma}\nonumber  \\
&&+2 \LN{\gamma}
 \PLN{3}{\frac{\gamma}{1+\gamma}}-2 \LN{\gamma} \zeta_{3}
+2 \PLN{4}{-\gamma}-2 \PLN{4}{\frac{1}{1+\gamma}}-2 \PLN{4}{\frac{\gamma}{1+\gamma}}
 \bigg\}\nonumber   \\ &&+ \left( \gamma \to  1/\gamma \right) \,.
\eaa
Let us now study $I_{2}^{(g)}$. For $D=4-2\epsilon$ it is given by
\begin{eqnarray*}
 I^{(g)}_2 &=& -\frac{g^4 C_F N_c}{4\cdot 64\pi^4} J^{(g)}_{2}\,,\qquad
\\
  J^{(g)}_{2} &=& -(1-2 \epsilon)\Gamma(1-2 \epsilon)  \int_0^1dt_1
\int_0^1d t_2 \int_0^1dt_3 \int_0^1
[d\beta]_3(\beta_{1}\beta_{2}\beta_{3})^{-\epsilon}
\\
&\times&\frac {(x_{13}^2+x_{24}^2)\big[\beta_1 t_1\lr{
\beta_2-2\beta_3} x_{24}^2 + \beta_3 t_3\lr{
\beta_2-2\beta_1}x_{13}^2\big]}{[\beta_3 t_3(\beta_2\bar t_2+\beta_1\bar
t_1)x_{13}^2+ \beta_1t_1(\beta_2 t_2+\beta_3\bar t_3)x_{24}^2]^{2-2\epsilon}}\,.
\end{eqnarray*}
Then, we change the integration variables according to
$$
\beta_1=\lambda x\,,\quad \beta_2=\bar\lambda\,,\quad \beta_3=\lambda\bar x\,,
$$
and integrate over $t_{2}$, which leads to
\baa
J^{(g)}_{2}&=&{(x_{13}^2)}^{2\epsilon}(1+\gamma)\int_0^1dt_1dt_3\int_0^1d\lambda
\bar\lambda^{-1-\epsilon} \int_0^1dx (x\bar x)^{-\epsilon}\frac{\bar xt_3 (\bar\lambda-2\lambda x)+x t_1 (\bar\lambda -2 \lambda
\bar x)\gamma}{xt_1\gamma-\bar xt_3}
\\
&\times&\bigg\{x^{-1+2\epsilon} [\bar xt_3\lambda\bar t_1+t_1(\bar\lambda
+\lambda\bar x \bar t_3)\gamma]^{-1+2\epsilon} - \bar x^{-1+2\epsilon}[
t_3(\bar\lambda+\lambda x \bar t_1) +xt_1 \lambda
\bar t_3 \gamma]^{-1+2\epsilon}\bigg\}
\eaa
We notice that integration over small $x$ and $\bar x$ produces a single pole in
$\epsilon$. We split up $J^{(g)}_{2}$ into a simplified divergent integral and
several convergent integrals, in which we neglect $O(\epsilon)$ terms.
\baa
J^{(g)}_{2d}&=&{(x_{13}^2)}^{2\epsilon}(1+\gamma)\int_0^1dt_1dt_3\int_0^1d\lambda
\bar\lambda^{-\epsilon} \int_0^1dx
\\
&\times&\bigg\{-x^{-1+\epsilon} [t_3\lambda\bar t_1+t_1(\bar\lambda +\lambda\bar
t_3)\gamma]^{-1+2\epsilon} - \bar x^{-1+\epsilon}[ t_3(\bar\lambda+\lambda \bar
t_1) + t_1 \lambda
\bar t_3 \gamma]^{-1+2\epsilon}\bigg\}\\
J^{(g)}_{2f}&=& (1+\gamma)\int_0^1dt_1dt_3\int_0^1d\lambda \int_0^1dx  \\
&\times& \bigg\{ -2 \, \frac{\lambda}{\bar\lambda} \, \frac{t_3+t_1\gamma}{xt_1\gamma-\bar xt_3}
 \left[ \frac{\bar x}{\bar xt_3\lambda\bar t_1+t_1(\bar\lambda
+\lambda\bar x
\bar t_3)\gamma} - \frac{x}{ t_3(\bar\lambda+\lambda x
\bar t_1) +xt_1 \lambda
\bar t_3 \gamma} \right] \\
&\phantom{\times}& \phantom{\bigg\{}+2\, \frac{1}{xt_1\gamma-\bar xt_3}
\, \bigg[ \frac{t_1\gamma}{\bar xt_3\lambda\bar t_1+t_1(\bar\lambda
+\lambda\bar x
\bar t_3)\gamma} - \frac{t_3}{ t_3(\bar\lambda+\lambda x
\bar t_1) +xt_1 \lambda
\bar t_3 \gamma}\bigg]\\
&\phantom{\times}& \phantom{\bigg\{}
-\frac1{x} \left[\frac1{\bar xt_3\lambda\bar t_1+t_1(\bar\lambda
+\lambda\bar x
\bar t_3)\gamma}-\frac1{t_3\lambda\bar t_1+t_1(\bar\lambda
 \lambda
\bar t_3)\gamma}\right]
\\
&&\quad - \frac1{\bar x}\left[\frac1{ t_3(\bar\lambda+\lambda x
\bar t_1) +xt_1 \lambda
\bar t_3 \gamma}-\frac1{ t_3(\bar\lambda+\lambda
\bar t_1) + t_1 \lambda
\bar t_3 \gamma}\right]\bigg\}
\eaa
To summarize,
\be
J_{2}^{(g)}=J^{(g)}_{2d}+J^{(g)}_{2f} + O(\epsilon)\,,
\ee
where only the first term produces poles in $\epsilon$. Calculating $J^{(g)}_{2d}$ we get
\ba
J^{(g)}_{2d}
&=&\frac1{\epsilon}\lr{x_{13}^2+x_{24}^2}^{2\epsilon}\int_0^1
\frac{dx}{x-\alpha}\ln\frac{x\bar\alpha}{\bar x\alpha} \ln(x\bar x)
\\ \nonumber
&& - \int_0^1\frac{dx}{x-\alpha}\bigg[\ln(x\bar x) \lr{\ln^2(\bar x\alpha)-\ln^2(
x\bar\alpha)}+ \ln\frac{x\bar\alpha}{\bar x\alpha} \lr{\frac{\pi^2}{6}+\ln
x\ln\bar x} \bigg]\,,
\ea
or equivalently, writing $J^{(g)}_{2d} = J^{(g)}_{2dd}+J^{(g)}_{2df}$,
\ba \label{Ig2d}
J^{(g)}_{2dd}&=&\frac1{2\epsilon}\left[(x_{13}^2)^{2\epsilon}+(x_{24}^2)^{2\epsilon}\right]
M_{2}\\
J^{(g)}_{2df}&=&
-\ln(\alpha\bar\alpha)
M_{2}
 - \int_0^1\frac{dx}{x-\alpha}\bigg[\ln(x\bar x) \lr{\ln^2(\bar x\alpha)-\ln^2(
x\bar\alpha)}+ \ln\frac{x\bar\alpha}{\bar x\alpha} \lr{\frac{\pi^2}{6}+\ln
x\ln\bar x} \bigg]\,,
\ea
where the integral $M_{2}$ is defined in (\ref{integrals}).
From (\ref{Ig1d}) and (\ref{Ig2d}) we can read off the contribution of $I^{(g)}$ to the pole
coefficients (taking into account the combinatorial factor $4$ for this diagram),
\be
A^{(g)}_{-2} = \frac{\pi^2}{48} \,,\qquad A^{(g)}_{-1} = -\frac{1}{8} M_{2}+\frac{1}{8} \zeta_{3}\,.
\ee
Let us now turn to the finite part.
All integrals appearing in it can be evaluated in terms of polylogarithms.
After a lot of algebra, one finds
\begin{eqnarray} \label{Jg2df}
J^{(g)}_{2df} &=& \bigg\{ \frac{11}{60}\pi^4 +\frac{\pi^2}{4}\LNP{\gamma}{2}+\frac{1}{12}\LNP{\gamma}{4}-\frac{3}{2}\pi^2\LN{\gamma}\LN{1+\gamma}
-\frac{1}{2}\LNP{\gamma}{3}\LN{1+\gamma} \nonumber \\
&&+2\pi^2\LNP{1+\gamma}{2}+\frac{3}{2}\LNP{\gamma}{2}\LNP{1+\gamma}{2}-\frac{1}{2}\LNP{1+\gamma}{4}
+3\LN{\gamma}\PLN{3}{-\gamma}\nonumber \\
&& 6 \LN{\gamma}\PLN{3}{\frac{\gamma}{1+\gamma}}-3 \LN{\gamma} \zeta_{3} -6 \PLN{4}{\frac{1}{1+\gamma}}-6 \PLN{4}{\frac{\gamma}{1+\gamma}}
 \bigg\}
\end{eqnarray}
and
\begin{eqnarray} \label{Jg2f}
J^{(g)}_{2f} &=& \bigg\{  \frac{47}{120}\pi^2 +\frac{\pi^2}{2}\LNP{\gamma}{2}+\frac{1}{8}\LNP{\gamma}{4}
-\frac{3}{2}\pi^2\LN{\gamma} \LN{1+\gamma}-\frac{1}{2}\LNP{\gamma}{3}\LN{1+\gamma} \nonumber \\
&&+ 2\pi^2 \LNP{1+\gamma}{2} +\frac{3}{2}\LNP{\gamma}{2} \LNP{1+\gamma}{2}-\frac{1}{2}\LNP{1+\gamma}{4}+3\LN{\gamma}\PLN{3}{-\gamma} \nonumber \\
&&+6\LN{\gamma} \PLN{3}{\frac{\gamma}{1+\gamma}}-3 \LN{\gamma} \zeta_{3}-6\PLN{4}{\frac{1}{1+\gamma}}-6\PLN{4}{\frac{\gamma}{1+\gamma}} \bigg\}
\end{eqnarray}
Summing up (\ref{Jg1f}),(\ref{Jg2df}) and (\ref{Jg2f}) we find
\be
-J^{(g)}_{1f}+J^{(g)}_{2df}+J^{(g)}_{2f}
=\frac1{6} M_{1}^2+\frac{11}{90}\pi^4 -M_3\,.
\ee
Thus, taking into account the symmetry factor $4$ and a constant contribution coming from equation (\ref{Ig2d}), we find the total contribution to the finite part of
this diagram to be
\be
A^{(g)}_{0} = \frac{1}{24} M^2_{1} - \frac{1}{4} M_{3} +\frac{7}{360}\pi^4 \,.
\ee
This completes the calculation. Summing up all contributions, we find formula (\ref{A-para-app}).

\subsection{Useful formulae for diagrams with three-gluon vertex}
\label{Appendix-integralJ}

In configuration space, these diagrams involve three propagators joined at the
same point $z$ which is integrated out
\be
V(z_1,z_2,z_3) = \mu^{-2\epsilon}\int d^D z\, D(z-z_1) D(z-z_2) D(z-z_3)\,,
\ee
with $D=4-2\epsilon$ and $D(z)$ being the gluon propagator in the Feynman gauge
\be
D(z) =  -\frac{i^{-1+\epsilon}}{4\pi^2} (\pi \mu^2)^\epsilon \int_0^\infty
ds\, s^{-\epsilon} \e^{-is z^2}\,.
\ee
Shifting variable $z\to z+z_1$ and performing the integration one gets
\be
V(z_1,z_2,z_3) = \frac{i^{2\epsilon}}{64\pi^4} (\pi
\mu^2)^{2\epsilon}\int_0^\infty ds_1 ds_2
ds_3\frac{(s_1s_2s_3)^{-\epsilon}}{(s_1+s_2+s_3)^{2-\epsilon}}\exp\lr{i
A(s_1,s_2,s_3)}\,,
\ee
with
\be
A(s_1,s_2,s_3) = \frac{(s_2 z_{21}+s_3 z_{31})^2}{s_1+s_2+s_3} - s_2 z_{21}^2
-s_3 z_{31}^2 = -\frac{s_1s_2 z_{12}^2+s_2s_3 z_{23}^2+s_3 s_1
z_{31}^2}{s_1+s_2+s_3}\,.
\ee
Changing integration variables according to $s_i=\lambda\beta_i$ with $\sum_i\beta_i=1$ and
$0\le \beta_i\le 1$, one obtains
\be
V(z_1,z_2,z_3) = - \frac{i^{1+4\epsilon}}{64\pi^4} (\pi \mu^2)^{2\epsilon}
{\Gamma(1-2\epsilon)} \int_0^1 \frac{[d\beta]_3\,
(\beta_1\beta_2\beta_3)^{-\epsilon}}{[\beta_1\beta_2 z_{12}^2+\beta_2\beta_3
z_{23}^2+\beta_3\beta_1 z_{31}^2 ]^{1-2\epsilon}}\,.
\ee
Let us consider the integral in the kinematics $(z_2- z_3)^2=0$, where it can be further simplified.
In that case,
\be
A(s_1,s_2,s_3) = -\frac{s_1(s_2+s_3)}{s_1+s_2+s_3} \left[{ z_1-\frac{s_2 z_2 +s_3
z_3}{s_2+s_3}}\right]^2
\ee
Then, we introduce standard parametrization
\be
s_i = \rho \beta_i \,,\quad \beta_1=1-\lambda\,,\qquad \beta_2=\lambda x\,,\qquad
\beta_2=\lambda (1-x)
\ee
and find
\be
V(z_1,z_2,z_3) = - \frac{i^{1+4\epsilon}}{64\pi^4} (\pi
\mu^2)^{2\epsilon}\frac{\Gamma(1-2\epsilon)}{\epsilon}\int_0^1
\frac{dx\, (x\bar x)^{-\epsilon}}{[(xz_{21}+\bar xz_{31})^2]^{1-2\epsilon}}\,.
\ee

\subsection{Basic integrals}\label{Appendix-integrals}
Here we give explicit expressions for integrals encountered in the two-loop calculation.
\begin{eqnarray}\label{integrals}
M_{1} &=& \int_{0}^{1}\frac{dx}{x-\bar{\alpha}} \ln\left(\frac{\bar{\alpha}
\bar{x}}{\alpha x}\right) = -\frac{1}{2}
\left[ \pi^2 + \ln^2\left(\frac{\alpha}{\bar{\alpha}}\right) \right] \\
M_{2}
&=& \int_{0}^{1}\frac{dx}{x-\bar{\alpha}}
\ln\left(\frac{\bar{\alpha} \bar{x}}{\alpha x}\right) \ln(x \bar{x})\\
&=& -\frac{\pi^2}{2} \ln(\alpha \bar{\alpha}) + 2 \rm{Li}_{3}(1)-
\rm{Li}_{3}\left(-\frac{\alpha}{\bar{\alpha}}\right)
-\rm{Li}_{3}\left(-\frac{\bar{\alpha}}{\alpha}\right) -
\ln\left(\frac{\alpha}{\bar{\alpha}}\right)
\left[ \rm{Li}_{2}(\alpha)-\rm{Li}_{2}(\bar{\alpha}) \right] \nonumber \\
M_{3}
&=& \int_{0}^{1}\frac{dx}{x-\bar{\alpha}} \ln\left(\frac{\bar{\alpha}
\bar{x}}{\alpha x}\right) \ln^2(x \bar{x})\\
&=&-\frac{49}{180}\pi^4 -\frac{1}{3}\pi^2 \left[ \ln^2(\alpha) +6 \ln(\alpha)
\ln(\bar{\alpha}) + \ln^2(\bar{\alpha}) \right] \nonumber
\\
&&-\frac{1}{12} \left[ \ln^4(\alpha) +\ln^4(\bar{\alpha}) +4
\ln(\bar{\alpha})\ln^3(\alpha)-18\ln^2(\alpha)\ln^2(\bar{\alpha})
 +4 \ln({\alpha})\ln^3(\bar{\alpha}) \right]
 \nonumber \\
 &&-4 \ln\left(\frac{\alpha}{\bar{\alpha}}\right) \left[
\rm{Li}_{3}\left(\alpha\right) - \rm{Li}_{3}\left(\bar{\alpha}\right) \right] + 8
\left[ \rm{Li}_{4}\left(\alpha\right)+ \rm{Li}_{4}\left(\bar{ \alpha}\right)
\right] \nonumber\\
M_9 &=& \int_0^1\frac{dx\, }{x-\alpha} \ln^3 \left( { \frac {\alpha\bar x }{ \bar
\alpha x}}
 \right)=-\frac14\lr{\ln^2\frac{\alpha}{\bar\alpha}+\pi^2}^2 = - M_1^2\,,
\end{eqnarray}
where $\bar\alpha=1-\alpha$ and ${\rm Li}_{n}(z)$ (with $n=2,3,4$) are
polylogarithms \cite{Lewin}.

\subsection{Identities for polylogarithms of related arguments}\label{Appendix-polylogs}
Most (or all) of the following polylogarithm identities can be found in Lewin's
book \cite{Lewin}. They can be easily proved by differentiating (the integration constant
can be fixed by evaluating the identity for a particular value of $z$).
\begin{equation}\label{polylog-1-1}
{\rm{Li}}_2\left(-z\right) +{\rm{Li}}_2\left(-\frac{1}{z}\right) =
-\frac{1}{2} {\rm ln}^2(z)  -\frac{\pi^2}{6} \,,\qquad  z>0\,.
\end{equation}
\begin{equation}\label{polylog-1-2}
{\rm{Li}}_2\left(z\right) +{\rm{Li}}_2\left(1-z\right) = -{\rm ln}(z)
{\rm ln}(1-z) +\frac{\pi^2}{6} \,,\qquad  0<z<1\,.
\end{equation}
It follows from (\ref{polylog-1-2}), by putting $z\rightarrow1/(1+z)$,
\begin{equation}\label{polylog-1-3}
{\rm{Li}}_2\left(\frac{1}{1+z}\right)
+{\rm{Li}}_2\left(\frac{z}{1+z}\right) =
-{\rm ln}\left(\frac{1}{1+z}\right) {\rm ln}\left(\frac{z}{1+z}\right)
+\frac{\pi^2}{6} \,,\qquad  z>0\,.
\end{equation}
Similar useful identities are
\begin{equation}\label{polylog-1-4}
{\rm{Li}}_2\left({1-z}\right) +{\rm{Li}}_2\left(1-\frac{1}{z}\right) = -
\frac{1}{2}{\rm ln}^2\left({z}\right) \,,\qquad  z>0\,.
\end{equation}
and
\begin{equation}\label{polylog-1-5}
{\rm{Li}}_2\left({-z}\right) +{\rm{Li}}_2\left(\frac{z}{1+z}\right) = -
\frac{1}{2}{\rm ln}^2\left({1+z}\right) \,,\qquad  z>0\,.
\end{equation}
Using (\ref{polylog-1-1}),(\ref{polylog-1-3}) and (\ref{polylog-1-5}) the functions
${\rm{Li}}_2\left(-\frac{1}{z}\right),{\rm{Li}}_2\left(\frac{1}{1+z}\right),
{\rm{Li}}_2\left(\frac{z}{1+z}\right)$ can all be reexpressed in terms
of ${\rm{Li}}_2\left({-z}\right)$ and logarithms. The inversion identity for ${\rm{Li}}_{3}$ reads
\begin{equation}\label{polylog-1-6}
{\rm{Li}}_3\left(-z\right) -{\rm{Li}}_3\left(-\frac{1}{z}\right) =
-\frac{\pi^2}{6} {\rm ln}(z) - \frac{1}{6}{\rm ln}^3(z) \,,\qquad  z>0\,.
\end{equation}
There is also
\begin{eqnarray}\label{polylog-1-7}
&& {\rm{Li}}_3\left(-z\right)
+{\rm{Li}}_3\left(\frac{1}{1+z}\right)+{\rm{Li}}_3\left(\frac{z}{1+z}\right)-\zeta_{3}\nonumber
\\
&=& \frac{1}{3} {\rm ln}^3(1+z)
-\frac{1}{2}{\rm ln}(z){\rm ln}^2(1+z)-\frac{\pi^2}{6}{\rm ln}(1+z) \,,\qquad
z>0\,.
\end{eqnarray}
For ${\rm{Li}}_{4}$ there is just the inversion identity:
\begin{equation}\label{polylog-1-8}
{\rm{Li}}_4\left(-z\right) +{\rm{Li}}_4\left(-\frac{1}{z}\right) =
-\frac{\pi^2}{12} {\rm ln}^2(z) - \frac{1}{24}{\rm ln}^4(z) -\frac{7
\pi^4}{360}\,,\qquad  z>0\,.
\end{equation}




\end{document}